\documentclass[a4paper, 10pt]{article}

\usepackage{latexsym} 
\usepackage{amsfonts} 
\usepackage{amsmath}
\usepackage{amsthm}
\usepackage{amssymb}
\usepackage{mathrsfs} 
\usepackage{dsfont}
\usepackage{bbold}
\usepackage[english]{babel}
\usepackage{caption}
\usepackage{epsfig} 
\usepackage{float} 
\usepackage{subfigure} 
\usepackage{psfrag} 
\usepackage{graphicx}
\usepackage{amsbsy}
\usepackage{xcolor}
\usepackage{mathtools}
\usepackage{marginnote}
\usepackage{calc}
\usepackage{accents}
\usepackage[numbers]{natbib}

\usepackage{setspace}

\usepackage[normalem]{ulem}

\usepackage[hidelinks]{hyperref}

\numberwithin{equation}{section}

\let\circo\circ
\renewcommand{\circ}{{\circo}}

\newtheorem{theorem}{Theorem}
\newtheorem*{prop*}{Theorem} 
\newtheorem*{hyp*}{Hypothesis} 
\newtheorem{thm}{Theorem}
\newtheorem{coro}[theorem]{Corollary}
 
\newtheorem{lemma}[theorem]{Lemma}
\newtheorem*{lemma*}{Lemma} 
\newtheorem{prop}[theorem]{Proposition} 
\newtheorem{rmk}[theorem]{Remark}

\newtheorem{hyp}{Hypothesis}

\numberwithin{theorem}{section}

\newcommand{\zerarcounters}{}

\renewenvironment{thebibliography}[1]{
  \begin{oldthebibliography}{#1}
    \setlength{\itemsep}{0.1em}
    \setlength{\parskip}{0.3em}
}
{
  \end{oldthebibliography}
}

\newcommand{\ovl}{\overline}

\newcommand{\ZZZ}{\mathds{Z}} 
 
\newcommand{\NNN}{\mathds{N}} 
 
\newcommand{\RRR}{\mathds{R}} 
\newcommand{\TTT}{\mathds{T}} 
\newcommand{\uno}{\mathds{1}} 
 
\newcommand{\AAA}{{\mathcal A}}

\newcommand{\DD}{{\mathcal D}} 
 
\newcommand{\calF}{{\mathcal F}} 
 
\newcommand{\calH}{{\mathcal H}}

\newcommand{\LL}{{\mathcal L}} 
\newcommand{\HH}{{\mathcal H}} 
 
\newcommand{\NN}{{\mathcal N}} 
\newcommand{\calO}{{\mathcal O}} 
\newcommand{\calP}{{\mathcal P}} 
\newcommand{\calQ}{{\mathcal Q}}

\newcommand{\calU}{{\mathcal U}} 
 
\newcommand{\calW}{{\mathcal W}} 
\newcommand{\calX}{{\mathcal X}}

\newcommand\calmB{{\mathscr B}}

\newcommand{\EEE}{{\mathscr E}}

\newcommand{\HHH}{{\mathscr H}}

\newcommand{\SSS}{{\mathscr S}} 
\newcommand{\calmT}{{\mathscr T}}

\newcommand{\gota}{{\mathfrak a}} 
 
\newcommand{\gotc}{{\mathfrak c}} 
\newcommand{\ccc}{{\mathfrak c}}

\newcommand{\gotf}{{\mathfrak f}} 
\newcommand{\gotg}{{\mathfrak g}}

\newcommand{\pp}{{\mathfrak p}}
 
\newcommand{\qq}{{\mathfrak q}}

\newcommand{\uu}{{\mathfrak u}} 
\newcommand{\vv}{{\mathfrak v}}

\newcommand{\ww}{{\mathfrak w}}

\newcommand{\gotB}{{\mathfrak B}} 
\newcommand{\gotC}{{\mathfrak C}} 
\newcommand{\gotD}{{\mathfrak D}}

\newcommand{\PP}{{\mathfrak P}} 
 
\newcommand{\gotR}{{\mathfrak R}} 
 
\newcommand{\SSSS}{{\mathfrak S}}

\newcommand{\ol}{\overline} 
\newcommand{\ul}{\underline}

\newcommand{\norm}{\vert{\hskip-.03cm}\vert{\hskip-.03cm}\vert}

\newcommand{\prova}{\noindent{\it Proof. }} 
\newcommand{\io}{\infty} 
\newcommand{\e}{\varepsilon} 
\newcommand{\al}{\alpha} 
\newcommand{\als}{\alpha_*} 
\newcommand{\de}{\delta} 
\newcommand{\be}{\beta} 
\newcommand{\n}{\nu} 
\newcommand{\m}{\mu}

\newcommand{\ka}{\kappa} 
\newcommand{\ga}{\gamma} 
 
\newcommand{\h}{\eta} 
 
\newcommand{\la}{\lambda} 
\newcommand{\f}{\varphi}

\newcommand{\der}{{\rm d}}

\newcommand{\XXXX}{\calX} 
\newcommand{\HHHH}{\HH'} 
\newcommand{\FFFF}{F'} 
\newcommand{\SSSSS}{\SSS'} 
\newcommand{\HHHHH}{{\mathscr H}'} 
\newcommand{\PPPPhi}{\Phi'} 
\newcommand{\ffff}{f'}

\def\ins#1#2#3{\vbox to0pt{\kern-#2 \hbox{\kern#1 #3}\vss}\nointerlineskip} 

\DeclarePairedDelimiter{\aver}{\langle}{\rangle}
\def\average#1{\big\langle{#1}\big\rangle}
\def\Average#1{\Big\langle{#1}\Big\rangle}

\begin{document}


\title
{\bf Synchronization and averaging in partially hyperbolic systems with fast and slow variables}

\author{{\bf Federico Bonetto$^{1}$ and Guido Gentile$^2$}\\
\small $^1$ School of Mathematics, Georgia Institute of Technology,  
Atlanta, GA 30332, USA\\
\small $^2$ Dipartimento di Matematica e Fisica, Universit\`a Roma Tre, Roma,
00146, Italy\\
\small e-mail: fb49@gatech.edu (corresponding author), 
guido.gentile@uniroma3.it}

\date{} 


\maketitle 

\vspace{-1cm}

\begin{abstract} 
We study a family of dynamical systems obtained by coupling an Anosov map on the two-dimensional
torus -- the \emph{chaotic system} -- with the identity map on the one-dimensional torus -- the
\emph{neutral system} -- through a dissipative interaction. We show that the two systems
synchronize: the trajectories evolve toward an attracting invariant manifold, and the full dynamics
is conjugated to its linearization around the invariant manifold. As a byproduct, we obtain that
there exists a unique exponentially mixing physical measure. When the interaction is
small, the evolution of the variable which describes the neutral system is very close to the
identity; hence, it appears as a \emph{slow} variable with respect to the variable which describes
the chaotic system, and which is wherefore named the \emph{fast} variable. We demonstrate that,
seen on a suitably long time scale, the slow variable effectively follows the solution of a
deterministic differential equation obtained by averaging over the fast variable. More precisely, we
prove that the invariant manifold is in probability close to the fixed point of the averaged
dynamics and that the difference between the exact evolution of the slow variable, seen from the
invariant manifold, and its averaged evolution is in probability exponentially decreasing for
arbitrarily large
times.
\end{abstract} 




\zerarcounters 
\section{Introduction}
\label{sec:1} 

In this paper, we study the long time evolution of a dynamical system obtained by coupling a chaotic
system on the two-dimensional torus with a one-dimensional neutral system\footnote{By `neutral
system' we mean a system which in the absence of the interaction is at rest, so that the Lyapunov
exponent of the central direction is zero -- and remains close to zero when the interaction is
switched on and is small.} through a weakly dissipative interaction. Our goal is to show the
occurrence of two related phenomena of interest.

The first one is the unfolding of an attracting invariant manifold, on which the dynamics is
essentially a copy of the evolution of the chaotic system. This is what is usually referred to as
\emph{synchronization}. In this respect, we extend the results in \cite{GGG} to a more
general class of systems and, in addition, we show that in a large neighborhood of the invariant
manifold the dynamics is conjugated to its linearization around the invariant manifold.

The second phenomenon can be described as follows. When the interaction between the two systems is
very small, the evolution of the neutral variable becomes independent of the chaotic variable and is
governed only by the average of the interaction over the chaotic variable.\footnote{The `chaotic
variable' and the `neutral variable' are the variables that, in the absence of interaction, describe
the chaotic system and the neutral system, respectively.} More precisely, for an asymptotically full
measure set of initial conditions for the chaotic variable, the evolution of the neutral variable,
on a suitable time scale, is described by a differential equation that does not involve the chaotic
variable. This phenomenon is known as \emph{averaging} and it is studied for systems related to ours
in \cite{DL3,DL2}, among others. In the class  of systems we consider, we
show that, excluding a set of initial condition whose measure vanishes with the size of the interaction,
the averaged differential equation describes the evolution of the neutral variables
toward the invariant manifold up to an error which vanishes with the interaction and remains
uniformly bounded in time for all times. Moreover, when the dynamics is seen from the invariant
manifold, the averaged differential equation describes the evolution of the neutral variables
with an error exponentially decreasing in time, still after excluding a set of
initial conditions which vanishes with the interaction.

We prove the existence of both the invariant manifold and the conjugation via an explicit
construction relying on the hyperbolic properties of the chaotic system. To obtain averaging we
prove several, progressively more complex  correlation inequalities for the chaotic evolution that,
combined with the explicit construction, allow us to control effectively the fluctuations of the
full evolution around the averaged one.

We now give a more detailed description of the results briefly outlined above, without insisting on
the technical aspects. Formal statements are provided in Section \ref{sec:model}, after introducing
all the appropriate notations and definitions.

\subsection{Deterministic description:~synchronization} \label{subsec:synchro}

Synchronization in quasi-integrable systems is well known to occur in the presence of dissipation; a
typical example is the orbital resonance in celestial mechanics \cite{MD}. By contrast, in chaotic
systems, where trajectories starting at close initial conditions tend to diverge from each other,
synchronization may appear as an unlikely phenomenon. Nonetheless, the presence of negative Lypaunov
exponents due to a dissipative coupling can still produce synchronization \cite{PC1}.

One of the simplest models one can think of is the partially hyperbolic system obtained by coupling
a chaotic system, for instance an Anosov automorphism on the two-dimensional torus $\TTT^2$, such as
the Arnold's cat map, with a one-dimensional neutral system on $\TTT$ through a dissipative
perturbation which is unidirectional, that is which affects only the motion of the neutral system.
Such a model has been explicitly considered in \cite{GGG}, by fully exploiting the perturbative nature
of the coupling and  under suitable assumptions which
simplify the analysis to a great extent. 
More complicated and realistic models can be easily envisaged \cite{PCJMH,PRK,BKOVZ,G,ADKMZ},
however the main advantage of the simple model studied in \cite{GGG} is that its solution
can be explicitly worked out and studied in great detail, without resorting to numerical simulations
or heuristic arguments. What is found in \cite{GGG} is that a two-dimensional invariant manifold appears on which
the dynamics is conjugated to that of the unperturbed automorphism: as a consequence, the two
systems synchronize asymptotically, in the sense that they tend to realize a drive-response
configuration, with the originally neutral system slaved to follow the dynamics of the chaotic
system (see \cite{BKOVZ} and references therein for an introduction to the topics). The
invariant manifold is no more than H\"older continuous, but, with the hypotheses considered in
\cite{GGG}, its oscillations are small, that is of the same order as the perturbation -- a
property which is not expected to hold in general.

In the present paper we study a class of dynamical systems that include those considered in
\cite{GGG}, and show that an invariant manifold exists in a more general setting, and it is
still the graph of a H\"older continuous function over the two-dimensional torus (see
Theorem~\ref{thm:1}). Moreover, we extend the analysis beyond the perturbative regime, by requiring
the coupling only to be dissipative in a finite region. The oscillations of the  invariant manifold
may be rather large in general, even in the perturbative regime, albeit large oscillations are rare
in the latter case, given that the variance of the function whose
graph describes the invariant manifold is of the order of the perturbation (see
Theorem~\ref{thm:3}).
We also provide a detailed description of the dynamics away from the invariant manifold, by
demonstrating that it is conjugated to its linearization around the invariant manifold (see
Theorem~\ref{thm:2}). In particular, the invariant manifold is proved to be an attractor. The
conjugation too, in general, is no more than H\"older continuous.
The variance of the conjugating function, seen as a function on $\TTT^2$ for fixed value
of the neutral variable, is found to be of the order of the coupling, despite the fact that
deviations of order 1 from its average are possible even in the case of small coupling (see Theorem~\ref{thm:exthm3}).
As a byproduct, we obtain that any system in the class we consider admits a unique physical measure given by the
lift to the invariant manifold of the normalised Lebesgue measure of $\TTT^2$. This measure is
exponentially mixing, with mixing rate limited by the low regularity of the invariant manifold (see
Theorem~\ref{thm:mixing}).

\subsection{Probabilistic description:~averaging} \label{subsec:ave}

When the perturbation is very small, two time scales naturally appear in the evolution: the
\emph{fast} time scale of the chaotic dynamics on $\TTT^2$ as opposed to the \emph{slow} time scale
of the neutral system, whose evolution is driven only by the perturbation. The study of systems with
fast and slow variables is well established for quasi-integrable systems where the motion of the
slow variable is close to periodic, and probably originated with Lagrange's analysis of the secular
variations of the orbital elements of planets \cite{Lag}. In such 
systems, the
slow variable, for a very long time, only feels the average over one period of its interaction with
the fast variables. Thus, to study the drift of the slow variable, one applies
the so-called \emph{averaging method} \cite{Hale,AKN,SVM}: once the oscillations of the fast variables have been
integrated out, one finds an approximate solution which provides, in general, a reliable description
of the dynamics up to a time inversely proportional to the slow time scale. However, this
is only the first step in the study of the dynamics of the system, and one has to control the
corrections in a deeper way in order to make the analysis rigorous and extend it up to longer -- even infinite -- time scales.

In a similar spirit, we investigate the dynamics of the systems we are considering  when the
size $\rho$ of  the perturbation is very small. We find that, in this situation, in order to describe the
evolution of the slow variable, the fast variable, despite being deterministic,
can somehow be treated as a random variable. More precisely, the evolution of
the slow variable becomes essentially independent of the dynamics on the torus and is effectively
described by an {\it averaged dynamics} in which the function describing the influence of the fast
variable on the slow variable is replaced by its average on the torus. We show that,
taking a random initial condition for the fast
variable, the probability of seeing a sizable deviation from the deterministic averaged evolution is
of the order of the perturbation (see Theorem~\ref{thm:4}, complemented by Lemma \ref{lem:Phik},
and the ensuing Corollary \ref{noname}). Thanks to the 
dissipation, which makes the
trajectories to evolve toward the invariant manifold, the probability of the deviations can be
controlled and proved to remain uniformly small along the full trajectory for arbitrarily large
times. Furthermore, for most initial values of the fast variable, the evolution of the slow variable,
when seen from the invariant manifold, is approximated by the averaged evolution 
up to corrections decreasing exponentially in time (see Theorem~\ref{thm:6}).
We can interpret the above results in the form of a \emph{scaling limit},
in the sense that we can fix a finite time $t$ and study, when the size $\rho$ of  the perturbation
is very
small, what happens after $\lfloor t/\rho \rfloor$ iterations of the map. In this regime,
henceforth called the \emph{scaling regime}, the
averaged dynamics 
is described by
a suitable ordinary differential equation.
This implies that the evolution of the slow variable, seen as a continuous time stochastic
process in the rescaled time $t$, converges to the solution of an ordinary differential equation in
probability in the topology of the uniform convergence (see Theorem~\ref{thm:prob}).

Properties 
as those discussed above are 
exhibited by the class of models 
studied in \cite{DL3,DLPV,DL2,L},
where the one-dimensional system described by the slow variable is
coupled with an expanding circle map and the coupling is dissipative only in average. 
When the system is not uniformly dissipative, the dynamics may become much more complicated.
This is especially true when the averaged system admits more than one attractor.
If this happens, starting close to one of them does not prevent the
dynamics to move toward some other attractor, and, as a consequence, a further
time scale naturally appears, on which
the dynamics appears as a random walk between metastable states.
A result of this kind has been
worked out in the aforementioned models based on expanding maps \cite{DL3},
where the rate of decay of correlations, in the case of more coexisting sinks
for the one-dimensional averaged system, is proved to be much slower than in the
case of only one attractor. An analogous behaviour is expected
to occur also in the case of the Anosov automorphisms, 
when the interaction is dissipative only in average. On the other hand, in
the fully dissipative case, where synchronization is achieved at exponential
rate, the class of models we consider is more general, because of the
presence of both 
stable and unstable directions for the fast chaotic dynamics on $\TTT^2$;
furthermore, considering Anosov systems instead of expanding
maps has the advantage that the systems one deals with are reversible and
exhibit symplectic structure \cite{L}. Noteworthily, the stronger
assumption on the dissipation allows to identify the stable state with the
metastable state (in particular, the last time scale listed in the introduction
of \cite{DLPV} does not arise), and to obtain stronger bounds for the
decay of the correlations, with no logarithmic corrections.

\subsection{More general settings:~a brief overview of possible applications} \label{subsec:future}

Systems of the kind described above, usually in more complex and articulated variants,
appear in many applications. 
In fact, the class of systems we study in this paper can be considered as
a prototype -- or at least a simplified version -- of physically more structured models, in the same spirit as in \cite{GGG,DL2}.

%
%

One of the first problems to be studied, where fast and slow variables are
coupled to each other, were the planetary motions in celestial mechanics. A
well-known example are the effects of the revolution of the Moon around the
Earth (the fast motion) on the revolution of the Earth around the Sun (the slow motion).
Krylov-Bogolyubov theory provides a useful tool to deal with such a kind of
problems and, more generally, to study the behaviour of oscillating systems where
at least two very different time scales are involved: an averaged equation for
the slow variables is obtained after integrating out the motions of the fast variables \cite{KB,BM,Hol}.
The theory has been successfully applied to a wide class of dynamical systems,
which range from very simple two-dimensional systems,
such as the Van der Pol equation or the inverted pendulum, to much
more complicated ones, such as the stability of the Solar system, where,
because of the complexity of the equations, numerical analysis plays a dominant role.

Recently the averaging method has been applied to study the stability of the
$H_2^+$ ion within the framework of classical mechanics \cite{CGGG}.
For certain initial conditions of the electron coordinates,
the protons are found numerically to be captured in an oscillatory state.
On the other hand, the numerical simulations also show that for other initial
conditions the motion of the electron becomes chaotic.
It would be interesting to investigate further the chaotic regime, in the light of the
increasing results in the literature showing that synchronization may still occur
when the dynamics of the fast variable moves from regular to chaotic; for
instance, a behaviour of this kind is observed numerically in electromechanical
systems with flexible arms (see \cite{KW} and references quoted therein).
Another recent application of the theory of averaging is provided by the derivation of
Haff's law for the system formed by two hard disks on a two-dimensional torus
that dissipate energy due to inelastic collisions \cite{Grigo}: by rewriting
the dynamics in terms of a dispersing billiard, in the limit of
vanishing dissipation, the speed of the particle is approximated uniformly by the solution
of a suitable ordinary differential equation.


Undoubtedly, also in the light of the relation of the problem of climate change
with society and life on our planet,
one of the most significant examples of chaotic systems where fast and slow variables interact
with each other is the Earth's climate system:
while weather processes, such as the atmospheric and ocean dynamics, behave
as fast motions, what one is really interested in is the slow evolution of
the climate \cite{Ki}; moreover, one has to take into account also
intentional and unintentional human-induced perturbations, such as the global
warming due to human activities.
Because of the wide range of external forces
involved in the climate system, the mathematical models which are used to treat
the problem in full generality are inevitably complicated, and the corresponding
differential equations are mainly studied numerically. 
However, analytically more accessible models have been studied,
not only because there are problems which admit a simpler description, but also on the grounds that
obtaining analytical results allows us to improve our general understanding of the problem.
A class of simple climate models are the \emph{energy balance models}, where only a few
variables appear. An example is provided by a two-dimensional model where
the evolution of the mean surface temperature and of the mean deep ocean
temperature is governed by a system of two stochastic differential equations:
the climate system response is characterised by two time scales, with the deep
ocean temperature reacting much more slowly \cite{SBR}.
The use of mathematical models in order to deal with the climate change has
intensified in the last few years, thanks to the recent developments in
dynamical systems theory as well as in statistical mechanics and probability
(see \cite{Dij,LBHRPW,GL} for reviews on the topics).
The tools we use in the present paper provide a possible path to follow in order
to address the analysis of climate models. Studying coupled Anosov systems,
which in principle could appear as mathematical abstractions, is justified in
consideration of Gallavotti-Cohen chaotic hypothesis \cite{GC}. In this regard,
we stress that the results obtained by relying on Ruelle response theory
\cite{LRF,GL} exploit the very same assumption.



Another well established -- and more ambitious -- line of research aims at deriving rigorously  the macroscopic law
of transport of energy in a crystal, that is the \emph{heat equation}, starting from
the deterministic microscopic dynamics (see \cite{BLRB} for a review).
Since one is interested in the long-time dynamics, in order to avoid the difficulties inherent in perturbations
of integrable Hamiltonian systems one may either add some randomness, which leads to stochastic equations,
or turn to a different approach and move toward strongly chaotic dynamics.
In this perspective, as pointed out in \cite{L}, one would like to investigate the
case of a large number $N$ of chaotic systems, locally coupled with an equal number of neutral systems
and weakly interacting with each other, and look for results
which are uniform in $N$. One is ultimately interested in
taking the \emph{hydrodynamical limit}, that is considering the system
obtained by a scaling limit in which both $N$ and the (discrete) time $k$ go to infinity
in a suitable way (see \cite{Spohn} for a review).
A model of this kind, with a different approach with respect to ours, is studied in \cite{BK},
where the local systems are chaotic maps,
the neutral variables play the role of local energies
and a further conservative small interaction is introduced between the local systems; then,
for initial conditions with the energies confined in a very small region, a
diffusion equation is proved to be satisfied by the local energies at finite time.

Stated in its full generality, the problem is too hard for our present
knowledge. As discussed in \cite{DL3,L}, a possible strategy to
pursue is to split the study into two separate steps:
\vspace{-.2cm}
\begin{enumerate}
\itemsep-.2em
\item First one studies a single local system weakly interacting with a neutral
variable in the limit in which the size $\rho$ of the interaction
vanishes (the \emph{scaling regime}). In order to obtain a non-trivial evolution one
studies the behaviour of the system for times $k$ diverging with a law which
depends on $\rho$. In this way one obtains a differential equation (the
\emph{mesoscopic equation}) describing the dynamics of the neutral variable.
\item Next, one couples a large number $N$ of such systems and takes the limit
in which both $N$ and the rescaled time go to infinity according to the
hydrodynamic limit. The heat equation should emerge as the partial differential
equation describing the evolution of the local energy concentration, that is
the average of the neutral variable in a small region.
\vspace{-.2cm}
\end{enumerate}
As far as the first step is concerned, one wishes the dynamics to be well understood in the absence of interaction.
Since integrable systems have to be excluded because they are too special and
are expected to display a non-typical behaviour, 
the local systems are usually assumed to be chaotic.
A further simplifying hypothesis is to
consider an interaction which makes non-conservative the evolution of the
neutral variable, so that local attractors appear 
and the dynamics is controlled
over long times. In such a situation the scaling limit
requires $k \sim \rho^{-1}$ and the mesoscopic equation is an ordinary
differential equation. In the more difficult conservative case, a different
scaling law $k \sim \rho^{-2}$ is looked for and the mesoscopic equation
is expected to be a stochastic differential equation.


The present paper deals with the first step of the strategy outlined above, in the fully dissipative case.
Of course, the problem one has in mind is the general case of partially hyperbolic systems,
where all manifolds -- unstable, central and stable -- exist,
in the presence of a dissipative interaction along the central direction.
Thus, the fact that the {neutral system} does not influence the chaotic evolution on the torus,
although it still remains highly non-trivial, makes the analysis easier.
Notwithstanding this simplification, we think that the model we study here
contains most of the relevant features we need to control the statistical properties
over long times of models with more general couplings (see Section~\ref{sec:outlook}).
At the same time, it has the advantage of being well suited for explicit computations
and eliminating details which would introduce technical
complications without really adding anything to the underlying physics.\footnote{Similar
considerations are at the basis of both the models studied in \cite{GGG,DGG}
and the skew products given in \cite{DL3,DL2} as examples for the general theory in the case of expanding maps.}
Therefore, in our opinion, results obtained for such a model represent a first step toward a full mathematical understanding of the problem
of synchronization and averaging in chaotic systems, before considering more realistic situations.

\zerarcounters
\section{Model and Results} \label{sec:model}

In this section, first,  we introduce the basic ingredients that will be used in the rest of the paper.
Then, we give the formal definition of the model we study and present our main results,
referring to Sections~\ref{sec:2} to \ref{last} -- and the Appendices -- for the proofs.
A more detailed plan of the paper is given in Subsection~\ref{subsec:content}.

\subsection{Basic Ingredients} \label{subsec:norms}

Set $\TTT:=\RRR/2\pi\ZZZ$ and consider an Anosov automorphism $A_0$ on $\TTT^2$ 
\cite{AW,BS}, such as Arnold's cat map \cite{AV}. Call $\lambda_{\pm}$ the eigenvalues
and $v_{\pm}$ the eigenvectors of $A_0$, with $|v_\pm|=1$, 
$|\lambda_+|>1>|\lambda_-|$ and $|\lambda_-\lambda_+|=1$, and set $\lambda:=|\lambda_+|$.

Consider $\Omega:=\calU \times \TTT^2$, where $\calU$ is a non-empty closed interval
of either $\TTT$ or $\RRR$, and, for any function $f\!:\Omega\to \RRR$ and any
$\f\in\calU$, set
\begin{equation}  \label{eq:1.222}
\langle f \rangle (\f) := \aver{ f(\f,\cdot)} :=
\int_{\TTT^2} f(\f,\psi) \, 
m_0(\der \psi) , \qquad \tilde f(\f,\psi):=f(\f,\psi)-\langle f \rangle (\f) ,
\end{equation}
where $m_0(\der\psi) := \der\psi/(2\pi)^2$ is the normalised Lebesgue measure on $\TTT^2$. 

Let 
%
$\|f\|_\infty:=\sup_{(\f,\psi)\in\Omega} |f(\f,\psi)|$ be the supremum norm on $\Omega$
%
and let $\calmB(\Omega,\RRR)$ denote the Banach space of the bounded continuous
functions $f$ equipped with the norm $\|f\|_{\io}$. For $\al\in (0,1]$,
consider the H\"older seminorm and the two directional H\"older seminorms
\begin{equation}  \label{eq:1.5}
|f|_{\al} := \!\!\!\!\!\!\! \sup_{(\f,\psi),(\f,\psi')\in\Omega} \!\!\!\!\!\!\! \frac{|f(\f,\psi')-f(\f,\psi)|}{ |\psi'-\psi|^\al} , \qquad
%
%
%
|f|_{\al}^{\pm} := \!\!\!\! \sup_{(\f,\psi)\in\Omega}\sup_{x\in\RRR} \frac{|f(\f,\psi+x v_\pm)-f(\f,\psi)|}{ 
|x|^\al} ,
\end{equation}
that satisfy the inequalities
%
$|f|_{\al}^{\pm} \leq|f|_{\al}\leq |f|_{\al}^{+}+|f|_{\al}^{-}$ and 
$\|f\|_\infty \leq c_{\al} |f|_{\alpha} + \|\langle f \rangle \|_\infty$,
%
with $c_{\al}:=(\pi\sqrt{2})^\alpha$.
Introduce also the norms
\begin{equation}   
\| f \|_{\al_-,\al_+} := \| f \|_\io + \al_-|f|_{\al_-}^{-} + \al_+|f|_{\al_+}^{+} ,
\qquad
\| f\|_{\al}^- := \| f\|_{\al,0} ,
\qquad
\| f\|_{\al}^+ := \| f\|_{0,\al} ,
\end{equation}
and let $\calmB^*_{\al_-,\al_+}\!(\Omega,\RRR)$,
$\calmB_{\al}^+(\Omega,\RRR)$ and
$\calmB_{\al}^-(\Omega,\RRR)$ denote the anisotropic Banach spaces of the functions
defined on $\Omega$ equipped with the norms $\|\cdot\|_{\al_-,\al_+}$,
$\|\cdot\|_{\al}^+$ and $\|\cdot\|_{\al}^-$, respectively. If $\al_-=\al_+=\al$ the norm
$\|\cdot\|_{\al,\al}$ is equivalent to the norm $\|\cdot\|_\al := \|\cdot\|_\io + \al|\cdot|_{\al}$
of the $\al$-H\"older continuous functions, so that
$\calmB^*_{\al,\al}(\Omega,\RRR)=\calmB_{\al}(\Omega,\RRR)$,
where $\calmB_{\al}(\Omega,\RRR)$ is the Banach space defined by the norm $\|\cdot\|_\al$;
on the other hand, using anisotropic Banach spaces allows to treat differently
the stable and unstable manifolds \cite{BKLiv,Dem}, and this will be exploited in what follows.
Finally, observe that $\calmB_{0}(\Omega,\RRR)=\calmB(\Omega,\RRR)$.

It is easy to see that
\vspace{-.1cm}
\begin{subequations} \label{eq:1.6}
\begin{align}
& |fg|_{\alpha} \le \|f\|_\io |g|_{\al} + |f|_{\al} \|g\|_\io ,  
& |fg|_{\alpha}^{\pm} \le\|f\|_\io |g|_{\al}^{\pm} + |f|_{\al}^{\pm} \|g\|_\io ,
\label{eq:1.6a} \\
%
%
& \| fg \|_{\al}  \le \| f \|_{\al} \| g \|_ {\al}  , 
& \| fg \|_{\al_-,\al_+}  \le \| f \|_{\al_-,\al_+} \| g \|_ {\al_-,\al_+} .
\label{eq:1.6b}
\end{align}
\end{subequations}

For $k \ge 1$, define also $\calmB_{\al,k}(\Omega,\RRR)$ as the Banach space of
the functions $f\!:\Omega \to \RRR$ which are $k$-times continuously
differentiable in the first variable, that is in the variable $\f\in\calU$, and
such that the first $k-1$ derivatives are $\al$-H\"older continuous in the
second one, that is in the variable $\psi\in\TTT^2$, equipped with the norm
\begin{equation} \label{eq:1.11110}
\norm f\norm_{\al,k} := \sum_{n=0}^{k-1} \|\partial_\f^n f \|_{\al} + \| \partial_\f^k f\|_{\io} 
=\sum_{n=0}^k\|\partial_\f^n f \|_{\io} + \al \sum_{n=0}^{k-1} |\partial_\f^n f 
|_{\al} \, ,
\end{equation}
where $\partial_\f$ denotes the derivative with respect to the first variable.
Similarly, define the norms $\norm \cdot \norm_{\al,k}^+$ and $\norm \cdot\norm_{\al,k}^-$,
as in \eqref{eq:1.11110} with $|\cdot|_\al$ replaced with $|\cdot|_\al^+$ and $|\cdot|_\al^-$,
and call the respective Banach spaces $\calmB_{\al,k}^+(\Omega,\RRR)$ and
$\calmB_{\al,k}^-(\Omega,\RRR)$. As in \eqref{eq:1.6b} we get
\begin{equation}\label{eq:normalk}
\norm f g\norm_{\al,k}\leq \norm f\norm_{\al,k}\norm g\norm_{\al,k}  ,
\qquad
\norm f g\norm_{\al,k}^\pm \leq \norm f\norm_{\al,k}^\pm\norm g\norm_{\al,k}^\pm .
\end{equation}

We call $\gotB^*_{\al_+,\al_-}\!(\TTT^2,\RRR)$ the subspace of
$\calmB^*_{\al_+,\al_-}\!(\Omega,\RRR)$
whose functions do not depend on $\f$,
and
similarly for $\gotB_{\al}^\pm(\TTT^2,\RRR)$, $\gotB_{\al}(\TTT^2,\RRR)$ and
$\gotB(\TTT^2,\RRR)$. For such functions, the  seminorms $|\cdot|_\al^\pm$ and
the norms $\|\cdot\|_\io$, $\|\cdot\|_{\al_,\al_+}$ and $\|\cdot\|_\al^\pm$ are
defined as previously, with the supremum taken over $\psi\in\TTT^2$ only.
The behavior of the seminorms \eqref{eq:1.5} for $f\in\gotB_{\al}(\TTT^2,\RRR)$ under
the action of $A_0^n$, for $n\in\ZZZ$, is given by
\begin{equation}\label{eq:1.6p}
|f \circ A_0^{n}|_{\al} \leq \lambda^{\al |n|} |f|_{\al}  \, ,
\qquad
|f \circ A_0^{n}|_{\al}^{+} \leq \lambda^{\al n} |f|_{\al}^{+} \, ,
\qquad
|f \circ  A_0^{n}|_{\al}^{-} \leq \lambda^{-\al n} |f|_{\al}^{-} \, .
\end{equation}
If the functions $g_{0},\ldots,g_{n-1}$ are in $\gotB_{\al}^\pm(\TTT^2,\RRR)$, for some $\al\in(0,1]$, then
\begin{equation} \label{gAn1}
\Bigl| \prod_{i=0}^{n-1} g_{i} \circ A_0^{\mp i} \Bigr|_{\al}^\pm \le
\sum_{i=0}^{n-1}\lambda^{-\al i} |g_{i}|_{\al}^\pm
\prod_{\substack{ j=0 \\ j\not=i}}^{n-1} \|g_{i}\|_\infty .
\vspace{-.1cm}
\end{equation}

\begin{rmk} \label{moregeneral}
\emph{
In the following we also consider sets of the form
%
$\AAA=\{(\f,\psi) : a_-(\psi)\leq\f\leq a_+(\psi)\}$ in $\RRR\times \TTT^2$,
%
where $a_\pm \! :\TTT^2\to\RRR$ are H\"older continuous functions.
All the definitions of the norms given above extend naturally if the set
$\Omega$ is replaced with any other closed subset of $\AAA$ 
as above.
We only need to take the supremum over $(\f,\psi)\in\AAA$ and
replace \eqref{eq:1.5} with
\vspace{-.1cm}
\begin{equation*}  
|f|_{\al}  \! := \!\!\!\!\!\! \sup_{(\f,\psi),(\f,\psi')\in\AAA} \!\!\!\!\!\!
\frac{|f(\f,\psi')-f(\f,\psi)|}{ |\psi'-\psi|^\al} , 
\qquad
|f|_{\al}^{\pm} \!:= \!\!\!
\sup_{(\f,\psi)\in\AAA} \sup_{\genfrac{}{}{0pt}{1}{x\in\RRR}{(\f,\psi+x v_\pm)\in\AAA}} \!\!\!\!\!\!
\frac{|f(\f,\psi+x v_\pm)-f(\f,\psi)|}{ |x|^\al} .
\vspace{-.1cm}
\end{equation*}
Then, we may define the corresponding Banach spaces in
the same way as before, with $\AAA$ instead of $\Omega$.
}
\end{rmk}

In order not to introduce  further symbols, we use the notation $\norm f \norm_{0,k}$
also to denote the $C^k$-norm of any function $f$ depending only on the first variable $\f$.
We thus identify $C^k(\DD,\RRR)$, for any given subset $\DD\subseteq\RRR$, with
the subspace of $\calmB_{0,k}(\DD\times\TTT^2,\RRR)$ of the functions
independent of $\psi$.

A crucial role in our analysis is played by the correlation functions and their
decay due to the hyperbolicity of $A_0$. 
The following estimate is proved in Appendix~\ref{app:corre1}.

\begin{prop} \label{prop:4.3}
Let the functions $g_+$ and $g_-$ be, respectively, in
$\gotB_{\al}^+(\TTT^2,\RRR)$ and in $\gotB_{\al}^-(\TTT^2,\RRR)$,
for some $\al\in(0,1]$. Then for all $n\in\NNN$ one has
\vspace{-.1cm}
\begin{equation*}
\left| \aver*{g_+ \, g_-\circ A_0^{n}}-\aver{g_+}\aver{g_-}\right| 
\le C_{0} (1+\al n) \lambda^{-\al n} \|\tilde g_+\|_{\al}^+ \|\tilde g_-\|_{\al}^- ,
\vspace{-.1cm}
\end{equation*}
for a suitable positive constant $C_{0}$ independent of $n$, $\al$, $g_-$ and $g_+$. 
\end{prop}

\begin{rmk} \label{unia}
\emph{
Proposition~\ref{prop:4.3} implies that, under the same assumptions, for every
$\al'<\al$ one has
$\left| \aver*{g_+ \, g_-\circ A_0^{n}}-\aver{g_+}\aver{g_-}\right| \le C_{0}' \lambda^{-\al' n} \|\tilde g_+\|_{\al}^+ \|\tilde g_-\|_{\al}^- $,
with the positive constant $C_{0}'$ depending only on $\alpha'$.
}
\end{rmk}

\subsection{The model}\label{subsec:model}

We consider the dynamical system defined by the map $\SSS$ on 
$\TTT\!\times\!\TTT^2$ given by
\vspace{-.1cm}
\begin{equation} \label{eq:1.1}
\SSS(\f,\psi) = (\SSS_\f(\f,\psi),\SSS_\psi(\f,\psi)) =(G(\f,\psi),A_0\psi) , \qquad \qquad
G(\f,\psi) := \f+F(\f,\psi) \, ,
\vspace{-.1cm}
\end{equation}
with $F\in\calmB_{\alpha_0,6}(\TTT\times\TTT^2,\RRR)$, for some $\alpha_0\in(0,1]$.
%
The assumption of H\"older continuity 
in 
$\psi$ is optimal, inasmuch as 
it cannot be weakened and 
requiring stronger
regularity does not provide stronger results.

\begin{rmk} \label{Sn}
\emph{
Throughout the paper, for any map $\SSS$, the notation $\SSS^n$ means $\SSS\circ
\SSS^{n-1}=\SSS \circ \ldots \circ \SSS$, that is the composition of $\SSS$ with itself $n$ times.
}
\end{rmk}

We assume $\SSS$ in \eqref{eq:1.1} to satisfy the following further hypotheses.

\begin{hyp}\label{hyp:1} 
There exists a non-empty closed interval
$\calU\subsetneq\TTT$ such that $\partial_\f \SSS_\f(\f,\psi)>0$ for all $(\f,\psi)\in\Omega:= \calU\!\times\! \TTT^2$.
If, after a suitable parametrization of $\TTT$, one writes $\calU= [\phi_m,\phi_M]$, with $\phi_m<\phi_M$,
then $\SSS_\f(\phi_m,\psi)>\phi_m$ and $\SSS_\f(\phi_M,\psi)<\phi_M$ for all $\psi\in\TTT^2$.

\end{hyp}

\begin{hyp}\label{hyp:2} 
For every $\psi\in\TTT^2$ there is a unique point $S(\psi) \in {\rm int}\, \calU$ such that $F(S(\psi),\psi)=0$.
\end{hyp}

Set also
\begin{equation} \label{rhoetal} 
S_m:= \!\inf_{\psi\in\TTT^2} \! S(\psi) , \quad  S_M:= \! \sup_{\psi\in\TTT^2} \! S(\psi) , \quad
%
%
\Lambda :=[S_m,S_M]\times \TTT^2 , \quad \Gamma := - \!\!\sup_{(\f,\psi)\in\Lambda} \!\!
\partial_\f F(\f,\psi) .
\end{equation}

\begin{hyp}\label{hyp:3} 
One has $\Gamma > 0$.
\end{hyp}

A map $\SSS$ satisfying Hypotheses~\ref{hyp:1}--\ref{hyp:3} defines a partially hyperbolic
system 
\cite{Do}.
%
By Hypothesis~\ref{hyp:1}, $\SSS$ is injective on $\Omega$ and  $\Gamma<1$.
By Hypotheses~\ref{hyp:1} and \ref{hyp:2}, $\SSS(\Omega)\subsetneq {\rm int}\, \Omega$
and $S(\psi)\in(\phi_m,\phi_M)$ $\forall\psi\in\TTT^2$, and hence  $[S_m,S_M] \subsetneq (\phi_m,\phi_M)$.
%
Hypothesis~\ref{hyp:3} ensures uniform dissipation.

The following lemma lists a few immediate consequences of Hypotheses~\ref{hyp:1}--\ref{hyp:3}.

\begin{lemma}\label{lem:easy}
If $\SSS$ satisfies Hypotheses~\ref{hyp:1}--\ref{hyp:3}, then the following
properties hold:
\vspace{-.2cm}
\begin{enumerate}
\itemsep-.1em
\item one has $F(\f,\psi)>0$ for $\f < S_m$, while $F(\f,\psi)<0$ for $\f > S_M$;
\item one has $\partial_\f F(\f,\psi)>-1$ for $(\f,\psi)\in\Omega$ and hence
$-1<\partial_\f  F(\f,\psi)\le -\Gamma$ for $(\f,\psi)\in\Lambda$;
\item\label{prop-Nr} for any $r>0$ there exists $N_r\in\NNN$ such that
$\SSS^{N_r}(\Omega)\subset \Lambda_r:=[S_m-r,S_M+r] \times \TTT^2$, with
%
\begin{equation} \label{Nr}
N_r\leq \frac{\displaystyle{\max\{\phi_M-S_M,S_m-\phi_m\} }}{\displaystyle{\inf_{\Omega\backslash\Lambda_r} |F(\f,\psi)|}}  \, ;
\vspace{-.2cm}
\end{equation}
\item\label{prop-rG} for any $\Gamma'\in(0,\Gamma)$ there exists $r=r(\Gamma')$ such that
$\partial_\f F(\f,\psi) \le - \Gamma'$ for all $(\f,\psi)\in\Lambda_{r}$;
\item the set $\Lambda$ is positively invariant under $\SSS$ and is attracting for $\SSS$ on $\Omega$;
\item there exists a unique $\overline\f \in (S_m,S_M)$
such that $\langle F(\overline\f,\cdot)\rangle=0$.
\end{enumerate}
\end{lemma} 

\begin{rmk} \label{barf0}
\emph{
Thanks to property 6 in Lemma~\ref{lem:easy}, without loss of generality we choose the
parametrization in Hypothesis~\ref{hyp:2} in such a way that $\overline\f=0$.
}
\end{rmk}

\begin{rmk}\label{betadelta}
\emph{
By Remark \ref{barf0},
we can write
%
$F(\f,\psi)=\beta(\psi)-\nu(\psi) \, \f+\delta(\f,\psi) \, \f^2$,
%
with 
$\langle\beta\rangle=0$. 
Conversely,
if we replace $F$ with the function 
%
$F_0(\f,\psi):=\beta(\psi)-\nu(\psi)\,\f+\delta(\psi)\,\f^2$,
%
with $\aver{\beta}=0$ and $4\|\delta\|_{\infty}\|\beta\|_{\infty}<  \nu_0^2$, where we have set
$\n_0:=\inf_{\psi\in\TTT^2}|\n(\psi)|$,
then it is easy to see that \eqref{eq:1.1} defined by $F_0$ instead of $F$ satisfies
Hypotheses~\ref{hyp:1}--\ref{hyp:3} with $\phi_M=\|\beta\|_\infty/2\nu_0$ and  $\phi_m=-\phi_M$.
}
\end{rmk}

\begin{rmk} \label{ext}
\emph{ 
Although $\SSS$ is defined as a map on $\TTT\times\TTT^2$,
given that $\SSS(\Omega)\subsetneq \Omega$ what we are really interested in 
only the action of $\SSS$ on $\Omega$, that we identify as a subset of $\RRR\times\TTT^2$.
For technical reasons, we also need to extend $\SSS|_\Omega$ to a map $\SSS_{\rm ext}$ on
$\Omega_{\rm ext}=\mathcal{U}_{\rm ext}\times\TTT^2$,
with $\calU_{\rm ext}\supset\calU$ a  closed interval,
in such a way that Hypotheses~\ref{hyp:1}--\ref{hyp:3} remain valid (we refer to Subsection~\ref{extension} for further details).
}
\end{rmk}

We call $\f$ the \emph{slow variable} and $\psi$ the \emph{chaotic}
or \emph{fast variable} -- such
a terminology is motivated by the fact that we are mainly interested in the
limit of small $\Gamma$, where
the neutral variable $\f$ moves slowly with respect to the chaotic
variable $\psi$ describing the hyperbolic system.

\subsection{Synchronization} \label{subsec:synch}

\subsubsection{The invariant manifold} \label{invariantmanifold}

If in \eqref{eq:1.1} we replace $A_0$ with the
identity $\uno$, that is if we consider the dynamical system defined by the map
$\SSS_{\uno}(\f,\psi)=(\f+F(\f,\psi),\psi)$, with the function $F$ still satisfying Hypotheses~\ref{hyp:1}--\ref{hyp:3}, then,
setting $\calW_{\uno}=\{(S(\psi),\psi) : \psi\in \TTT^2\}$, we have
$\SSS_\uno(\calW_\uno)=\calW_\uno$. It is natural to ask whether a similar
property remains true for $\SSS$ notwithstanding the chaotic nature of the
evolution generated by $A_0$ on $\TTT^2$.
We say that $\calW:=\{(W(\psi),\psi) : \psi\in \TTT^2\}$ is an \emph{invariant manifold} for $\SSS$ in \eqref{eq:1.1}
if we have, for all $\psi\in\TTT^2$,
\vspace{-.1cm}
\begin{equation}\label{inva}
\SSS(W(\psi),\psi)=(W(A_0\psi),A_0\psi) .
\vspace{-.1cm}
\end{equation}
This means that on $\calW$ the dynamics
generated by $\SSS$ is conjugated to the dynamics generated by $A_0$.

The following result is proved in Subsection~\ref{subsec:proofW}.

\begin{thm}[\textbf{Synchronization}] \label{thm:1}
Consider the dynamical system described by $\SSS$ in \eqref{eq:1.1} satisfying Hypotheses~\ref{hyp:1}--\ref{hyp:3}.
There exists a unique
invariant manifold $\calW=\{(W(\psi),\psi):\psi\in\TTT^2\}\subset \Lambda$ for
$\SSS\!$, with $W\in \gotB^*_{\al_-,\al_+}\!(\TTT^2,\RRR)$ for suitable
$\al_-,\al_+ \in (0,\al_0]$.
\end{thm}

\begin{rmk} \label{size}
\emph{
In general, 
$\calW$ is only H\"older continuous in $\psi$ even if $F$ is smooth in $\psi$
and 
$W$ is 
of order 1 in $\|F\|_{\io}$ even if $F$ is
very close to 0, its 
size depending mainly on $S(\psi)$ (see also Remark \ref{rmk:2.1}).
}
\end{rmk}

\subsubsection{The linearized map and the conjugation}\label{subsec:linear}

To analyze the evolution generated by $\SSS$ outside $\calW$ we can try to 
conjugate it with its linearization around $\calW$, that is with the simpler 
system $\SSS_0$ given by 
\vspace{-.1cm}
\begin{equation}\label{eq:sss0}
\SSS_0(\eta,\psi)=(\ka(\psi)\eta,A_0\psi) , \qquad \qquad \ka(\psi):=1+ \partial_\f F(W(\psi),\psi) ,
\vspace{-.1cm}
\end{equation}
where, by Hypothesis~\ref{hyp:3}, one has $\ka(\psi)\in(0,1-\Gamma)$
for all $\psi\in\TTT^2$.
In other words, we look for  a function $\HHH$ -- 
called the \emph{conjugating function} (or simply \emph{conjugation}) -- such that
\vspace{-.1cm}
\begin{equation} \label{eq:1.11}
\HHH \circ \SSS = \SSS_0 \circ \HHH.
\vspace{-.1cm}
\end{equation}
%

\begin{rmk} \label{ambi}
\emph{
Since $\SSS_0$ is linear in $\eta$, it is easy to see that if $\HHH$ is a
solution to \eqref{eq:1.11} then, for any $a\not=0$, also $\HHH(a\f,A_0\psi)$ is a solution.
Thus, we say that $\HHH$ is \emph{the} conjugation if it solves
\eqref{eq:1.11} and can be written as
%
$\HHH(\f,\psi)=(\HH(\f,\psi),\psi)$,
%
with $\partial_\f \HH(0,\psi)=1$ for every $\psi\in\TTT^2$.
}
\end{rmk}

The following result on the conjugation is proved in Subsection~\ref{subsec:proof2}.

\begin{thm}[\textbf{Conjugation}] \label{thm:2}
Consider the dynamical system described by $\SSS$ in \eqref{eq:1.1}
satisfying Hypotheses~\ref{hyp:1}--\ref{hyp:3}. There exist  a set $\Omega_0
\subset \RRR \times \TTT^2$ and a function $\HHH\!:\Omega\to\Omega_0$ that
conjugates $\SSS$ via \eqref{eq:1.11} to $\SSS_0$ given by \eqref{eq:sss0}.
Moreover $\HHH$ can be written as in  Remark~\ref{ambi} with $\HH\in\calmB_{\als,2}(\Omega,\RRR)$,
for a suitable $\als \in (0,\min\{\al_-,\al_+\}]$ and $\al_-,\al_+$ as in Theorem~\ref{thm:1}.
\end{thm}

%

\begin{rmk} \label{attractor}
\emph{Theorem~\ref{thm:2} implies that, for any initial datum in $\Omega$, the evolution generated by $\SSS$ leads towards $\calW$.
Therefore, $\calW$ is a global attractor for the dynamical system $(\Omega,\SSS)$. }
\end{rmk}

From the proof of Theorem~\ref{thm:2} it is easy to see that the function
$\HHH$ is invertible; indeed, the following result is proved in Subsection
\ref{subsec:proofcor}.

\begin{coro} \label{l}
The map $\HHH$ in Theorem~\ref{thm:2} is invertible, and its inverse 
$\HHH^{-1}\!:\Omega_0 \to \Omega$, which satisfies
%
$\HHH^{-1} \circ \SSS_0 = \SSS \circ \HHH^{-1} $,
%
can be written as $\HHH^{-1}(\h,\psi)=(\LL(\h,\psi),\psi)$,
with $\LL\in\calmB_{\als,1}(\Omega_0,\RRR)$.
\end{coro}

As for $W$, the conjugation and its inverse are no more than
H\"older continuous even if $F$ is very smooth in $\psi$.
%
The proofs of Theorem~\ref{thm:2} and of Corollary~\ref{l}
show that $\HH$ and $\LL$ can be written as
%
\begin{equation} \label{hl}
\HH(\f,\psi) =\f - W(\psi) +(\f - W(\psi))^2h(\f,\psi), \qquad
\LL(\h,\psi) = W(\psi) + \eta + \eta^2 l(\eta,\psi) ,
\end{equation}
with $h \!\in\! \calmB_{\als,1}(\Omega,\RRR)$ and $l \!\in\!\calmB_{\als,1}(\Omega_0,\RRR)$.
These representations are used in Subsection~\ref{subsec:oscillations},
when studying the deviations of the conjugation.

\subsubsection{The physical measure} \label{2.3.3}

Theorems~\ref{thm:1} and \ref{thm:2} imply that, given any observable
$\calO\in\calmB_{\alpha,0}(\Omega,\RRR)$, one has
\vspace{-.1cm}
\[
\lim_{n\to\infty} \frac{1}{n}\sum_{i=0}^{n-1}\calO(\SSS^n(\f_0,\psi_0))=
\int \calO(W(\psi),\psi) \, m_0(\der\psi)=:\nu_W(\calO) 
\vspace{-.1cm}
\]
for almost every $(\f_0,\psi_0)\in\Omega$ with respect to the measure
$\nu_0(\der\f \der\psi) := m_0(\der\psi) \der\f/(\phi_M-\phi_m).$

Hence, $\nu_W$ is the unique {\it physical measure} for $\SSS$ on $\Omega$. It 
is thus interesting to study the mixing property of $\nu_W$ with respect to $\nu_0$.
The following result is proved in Subsection~\ref{proofthmpm}.

\begin{thm}[\textbf{Mixing on the invariant manifold}]\label{thm:mixing}
Consider the dynamical system described by $\SSS$ in \eqref{eq:1.1}
satisfying Hypotheses~\ref{hyp:1}--\ref{hyp:3}. For any observables $\calO_1,\calO_2 \in \calmB_{\als,1}(\Omega,\RRR)$,
with $\als$ as in Theorem~\ref{thm:2}, and any ${\lambda_{*}^{-1}} > \max
\{\lambda^{-\als},{1-\Gamma} \}$, one has
\vspace{-.1cm}
\[
\left|\nu_0\left(\calO_1\,\calO_2\circ\SSS^n\right)-\nu_0\left(
\calO_1\right)\nu_W\left(\calO_2\right)\right| \leq 
C_{1} \, {\norm \calO_1\norm_{\als,0}\norm\calO_2\norm_{\als,1}} { \lambda_{*}^{-n} } ,
\vspace{-.1cm}
\]
where the constant $C_{1}$ depends only on $\la_*$.
\end{thm}

\subsection{Averaging} \label{subsec:scaling}

We are interested in the long time evolution generated by the map $\SSS\!$ in \eqref{eq:1.1},
when the component $\SSS_\f$ is close to the identity.
To this aim, we consider the family of functions $F(\f,\psi)=\rho f(\f,\psi)$,
with $f\in\calmB_{\alpha_0,6}(\TTT\times\TTT^2,\RRR)$ and $\rho>0$ a parameter,
and study the behaviour of the map
\vspace{-.1cm}
\begin{equation} \label{eq:1.1bis}
\SSS(\f,\psi)=(\f+\rho f(\f,\psi),A_0\psi)
\vspace{-.1cm}
\end{equation} 
when $\rho$ is small.
We also define
%
$\gamma:= - \sup_{(\f,\psi)\in\Lambda} \partial_\f f(\f,\psi)$,
%
so that $\Gamma=\rho\,\gamma$ in \eqref{rhoetal}.

\begin{rmk} \label{Psi0ge1}
\emph{
We may and do assume, without loss of generality, that $\norm f
\norm_{\al_0,6}= 1$,
and hence, since $\ga  < \norm f \norm_{\al_0,6}$, that $\ga\in(0,1)$.
%
%
Hypothesis~\ref{hyp:1} requires $\rho$ not to be arbitrarily large.
Therefore, when considering a map $\SSS$ of the form \eqref{eq:1.1bis}, we tacitly assume $\rho$ to be smaller than
a suitable value $\rho_*$, depending on $f$,  such that $\SSS$ satisfies
Hypotheses~\ref{hyp:1}--\ref{hyp:3} for all $\rho\in(0,\rho_*)$.
}
\end{rmk}

We investigate the evolution generated by $\SSS$, as given in \eqref{eq:1.1bis}, when $\rho\to
0^+$. In such a limit the evolution tends to become trivial. For this reason, for given initial
conditions $(\f_0,\psi_0)$, one usually studies the dynamics after a linear \emph{rescaling} of
time, that is one considers
the trajectory $\SSS^n(\f_0,\psi_0)$, with $n=\lfloor t/\rho \rfloor$, where $t$ is fixed and $\rho$ tends to $0^+$.
In the remaining part of this section
we present results for the invariant manifold and the conjugation and
discuss properties
of $\SSS^n$ which are uniform in $n$, when $\rho$ is small
(see Subsections \ref{242} to \ref{subsec:probability}); finally,
we interpret the results in terms of the scaling limit (see Subsection \ref{subsec:continuous}).
Throughout the paper, we refer to the regime where $\rho$ is small as the \emph{scaling regime}
even when we look for properties which do not involve the rescaled variable $t$.

\subsubsection{Heuristic discussion}\label{heuristic}

If $\rho$ is small enough we can take $n$ very large but still much smaller than
${\rho^{-1}}$. Expanding $\SSS^n$ to first order in $\rho$, we write
\vspace{-.2cm}
\[
(\SSS^n)_\f(\f,\psi) = \f + \rho\sum_{i=0}^{n}  f(\f,A_0^i\psi)+ o(n\rho) .
\]
Since $A_0$ is strongly mixing, we further obtain that
$(\SSS^n)_\f(\f,\psi) \simeq  \f + n\rho \, \langle f\rangle(\f)+o(n\rho)$
for most values of $\psi_0$,
where the right hand side has lost any dependence on $\psi$, at least at first
order in $\rho$. Calling $n\rho=t$ and writing $\f(t) :=\SSS^{\lfloor t/\rho\rfloor}(\f, \psi)$, we can read the last expression as
$\f(t)\simeq \f +\langle f\rangle(\f) \, t+ o(t)$.
This propounds that, for $\rho$ small, the evolution $\f(t)$ that starts from a given $\f$ and a randomly chosen $\psi$ is
essentially independent of $\psi$ and agrees at first order in $t$ with the solution $\phi(t)$ of the Cauchy problem
\vspace{-.1cm}
\begin{equation} \label{eq:1.9}
\left\{
\begin{aligned}
& \dot{\phi}=\langle f\rangle( \phi)\\
&  \phi(0)=\f \, .
\end{aligned}\right.
\vspace{-.2cm}
\end{equation}

To see whether we can get a better agreement, we expand $\SSS^n$ to second order in $\rho$ and find
\vspace{-.2cm}
\[
(\SSS^n)_\f(\f,\psi)= \f + \rho\sum_{i=0}^{n}
f(\f,A_0^i\psi)+\rho^2 \!\!\!\! \sum_{0\leq i<j\leq n} \!\!\! \partial_\f f(\f,A_0^j\psi)f(\f,A_0^i\psi) +{o(n^2\rho^2)} \, ,
\vspace{-.2cm}
\]
so that, for $\varphi(t)$ to remain close to $\phi(t)$ up to corrections $o(t^2)$, we need that
\vspace{-.2cm}
\[
\sum_{0\leq i<j\leq n} \partial_\f f(\f,A_0^j\psi)f(\f_0,A_0^i\psi )\simeq \frac {n^2}2 \partial_\f \aver{f} (\f)\aver{f}(\f) ,
\]
that is we need a strong form of decay of correlations for $A_0$.
This provides a simple instance of the idea of averaging induced by the chaotic behavior of $A_0$,
as it suggests that, on the correct time scale, $(\SSS^n)_\f(\f,\psi)$
evolves according to a differential equation involving only the average of $f$.

Clearly, the argument above is only heuristic. In fact, extending the analysis
outlined above to all orders is likely to get too tangled and to require very
high regularity of the map. Moreover such an analysis is not suitable for
dealing with the case of arbitrary $t$, in particular for deriving results
uniform in $t$. Therefore, the heuristic argument only hints what to look for,
but, in order to obtain something rigorous, actually we follow a different approach.

\subsubsection{Synchronization in the scaling regime} \label{242}

The following two lemmas collect the implications of Theorems~\ref{thm:1} and
\ref{thm:2} and their proofs for the dynamical system $\SSS$ in
\eqref{eq:1.1bis} with $\rho$ small -- for the proofs see the end of Subsections~\ref{subsec:proofW}
and \ref{subsec:proof2-2}, respectively.

\begin{coro} \label{lem:1}
Consider the dynamical system described by $\SSS$ in \eqref{eq:1.1bis} satisfying Hypotheses~\ref{hyp:1}--\ref{hyp:3}.
Let $\calW=\{(W(\psi),\psi):\psi\in\TTT^2\}$
be the invariant manifold for $\SSS$ as in Theorem~\ref{thm:1}.
One has $\al_+=O(1)$ and $\alpha_-=O(\rho)$, and, furthermore,
both $\|W\|_\io$ and $|W|_{\al_-}^{-}$ are $O(1)$, while $|W|_{\al_+}^{+}=O(\rho)$.
\end{coro}

\begin{coro} \label{lem:2}
Consider the dynamical system described by $\SSS$ in \eqref{eq:1.1bis} satisfying Hypotheses~\ref{hyp:1}--\ref{hyp:3}.
Let $\HHH$ be the conjugation in Remark~\ref{ambi} 
for $\SSS$ as in Theorem~\ref{thm:2},
and let $h$ and $l$ be defined as in \eqref{hl}.
One has $\als=O(\rho)$, while both $\norm h \norm_{\als,1}$ and $\norm l \norm_{\als,1}$ 
are $O(1)$ in $\rho$.
\end{coro}

Corollary~\ref{lem:1} shows that, when $\rho$ is small, the invariant manifold $\calW$ 
loses most of the smoothness of $f$ in the stable direction while maintaining it and varying slowly in the unstable one.

By using Corollary~\ref{lem:1}, the following  corollary follows immediately from Theorem~\ref{thm:mixing}.

\begin{coro}\label{cor:mixingrho}
Consider the dynamical system described by $\SSS$ in \eqref{eq:1.1bis} satisfying Hypotheses~\ref{hyp:1}--\ref{hyp:3}.
For any observables $\calO_1, \calO_2 \in \calmB_{\als,1}$, with $\als$ as
in Theorem~\ref{thm:2}, and for all $n \in \NNN$, one has
\vspace{-.1cm}
\[
 \left|\nu_0\left(\calO_1\,\calO_2\circ\SSS^n\right)-\nu_0\left(\calO_1\right)\nu_W\left(\calO_2\right)\right|\leq 
C_{2} \, {\norm\calO_1\norm_{\als,0}} \norm \calO_2\norm_{\als,1}e^{-\xi \rho n} ,
\vspace{-.1cm}
\]
where $C_{2}$ and $\xi$ are suitable constants not depending on $\rho$.
\end{coro}

\subsubsection{The averaged map} \label{averaged}

As an intermediate step toward a rigorous justification of the conclusions drawn in Subsection \ref{heuristic},
we consider also the dynamical system described by the \emph{averaged map}
\vspace{-.1cm}
\begin{equation} \label{Sbar}
\overline\SSS(\f,\psi):=(\overline G(\f),A_0\psi)\, , \qquad\qquad
\overline G(\f) := \f+\rho \overline f(\f) , \qquad\qquad \overline f(\f):=\langle f\rangle(\f),
\vspace{-.1cm}
\end{equation}
%
%
and its linearization 
\vspace{-.2cm}
\begin{equation} \label{Sbarlin}
\overline\SSS_0(\f,\psi):=(\ovl\mu \,\f, A_0\psi) , \qquad \qquad
\ovl\mu:= \partial_\f \ovl G(0) = 1+\rho \,\partial_\f \overline f(0) .
\vspace{-.1cm}
\end{equation}
Noting that
%
$\ovl\Gamma = \rho \, \ovl \ga :=  - \rho \sup_{\f\in\calU} \partial_\f \ovl f(\f) \ge \rho \, \ga = \Gamma$,
we see that $\ovl \SSS$ satisfies Hypotheses~\ref{hyp:1}--\ref{hyp:3} if $\SSS$ does.

\begin{rmk} \label{Wbar}
\emph{
Since $\overline f(0)=0$ (see Remark \ref{barf0}), we see that $\overline W(\psi)=0$
solves the equation $\overline\SSS(\overline W(\psi),\psi)$ $= (\overline W(A_0\psi),A_0\psi)$,
and hence $\ol\SSS$ admits the invariant manifold $\overline \calW=\{(0,\psi),\psi\in \TTT^2\}$.
}
\end{rmk}

\begin{rmk}
\emph{
Although the action of $\ovl\SSS$ on each variable is independent of the other one, 
since one has  $(\ol\SSS^n)_\f(\f,\psi)\!=\!\ol G^n(\f)$,
the notation introduced in \eqref{Sbar} helps clarify the forthcoming discussion.
}
\end{rmk}

In 
Subsections~\ref{subsec:oscillations} and \ref{subsec:probability}
we compare the evolution generated by
$\SSS$ with the evolution generated by $\overline \SSS$. In the remainder of 
this subsection, we show that, by adapting the analysis in Subsection~\ref{subsec:linear} to
$\ol\SSS$, the trajectories of the dynamics generated by
$\ol\SSS$ can be compared to the solutions of \eqref{eq:1.9}.

First, we proceed as in Theorem~\ref{thm:2}, and look for an invertible function $\ovl\HHH$ such that
\vspace{-.1cm}
\begin{equation} \label{eq:Hbar2}
 \overline\HHH \circ \overline\SSS= \overline\SSS_0 \circ \overline\HHH\, ,
\vspace{-.1cm}
\end{equation}
that is a function $\overline\HHH$ that conjugates $\ovl\SSS$ to $\ovl\SSS_0$. We write
%
\vspace{-.2cm}
\begin{equation}\label{def:lbar}
\ovl\HHH(\f,\psi)=(\overline\HH(\f), \psi) , \qquad \overline\HH(\f)=\f+\f^2\overline h(\f) , 
%
\qquad
\overline\HHH^{-1}(\eta,\psi)=(\eta+\eta^2\bar l(\eta),\psi) .
\vspace{-.2cm}
\end{equation}

The following result is proved in Subsection~\ref{subsec:barH}.

\begin{lemma}\label{lem:Hbar}
Consider the dynamical system described by $\SSS$ in \eqref{eq:1.1} satisfying Hypotheses~\ref{hyp:1}--\ref{hyp:3},
and define $\ovl\SSS$ as in \eqref{Sbar}.
There exist a closed interval $\calU_0\subset\TTT$ and a function 
$\overline\HHH\!:\calU \times \TTT^2  \to\calU_0\times\TTT^2$ such that \eqref{eq:Hbar2} holds.
Moreover, there exists a constant $C_{3}$ such that one has
$\norm\overline h\norm_{0,3}\leq C_{3}$ and $\norm\bar l\norm_{0,3}\leq C_{3}$.
\end{lemma}

\begin{rmk}
\emph{
Note that $\Omega_0 \neq \calU_0 \times \TTT^2$: in fact, one has $\Omega_0=\HHH(\Omega)$,
while $\calU_0 = \ovl \calH(\calU)$.
}
\end{rmk}

Let $\Phi$ be the flow generated by \eqref{eq:1.9}, that is the set of the solutions of
\vspace{-.1cm}
\begin{equation}\label{flow}
\left\{
\begin{aligned}
 &\frac{d}{dt}\Phi_t(\f)=\overline f(\Phi_t(\f)) , \\
 &\Phi_0(\f)=\f , 
\end{aligned}\right.
\vspace{-.1cm}
\end{equation}
when varying $\f \in\calU$. Because of Hypotheses~\ref{hyp:1}--\ref{hyp:3},
all trajectories $\Phi_t(\f)$ of the system \eqref{flow}, with $\f\in\calU$,
move towards the origin at exponential rate. 

The following result, proved in Subsection~\ref{subsec:proofPhi}, shows
that the trajectories generated by \eqref{Sbar} and \eqref{flow} remain close and, in fact, merge asymptotically.

\begin{lemma} \label{lem:Phik}
Consider the dynamical system described by $\SSS$ in \eqref{eq:1.1bis} satisfying Hypotheses~\ref{hyp:1}--\ref{hyp:3},
and define the map $\ovl\SSS$ as in \eqref{Sbar} and the flow $\Phi$ as in \eqref{flow}.
For any $\ga'\in(0,\ga)$ there exists a positive constant $C_{4}$
such that for all $\f\in\calU$ and all $n >0$ one has
\vspace{-.1cm}
\begin{equation}\label{SbarPhi}
|(\overline\SSS^n)_\f(\f,\psi)-\Phi_{n\rho}(\f)|\le C_{4}\rho\,(1-\rho\ga')^n .
\end{equation}
\end{lemma}

\begin{rmk} \label{rho'}
\emph{
With the notation established after Remark \ref{Psi0ge1}, when introducing the scaling terminology,
we can write \eqref{SbarPhi} as
$|(\overline\SSS{\vphantom{s}}^{\lfloor t/\rho\rfloor})_\f(\f,\psi)-\Phi_{t}(\f)| \le C_{4}'\rho\,e^{-\ga' t}$,
for a suitable constant $C_{4}'$.}
\end{rmk}

\begin{rmk} \label{rho'bis}
\emph{
The decay rate in Lemma~\ref{lem:Phik} is not in general equal to $\ga$ as one could na\"ively
expect. Indeed, the constant $C_4$ in Lemma~\ref{lem:Phik}, as well as the constants $C_{6}$,
$C_{8}$ and $C_{10}$ in the forthcoming Lemma~\ref{lem:flulin}, Theorem~\ref{thm:4} and
Theorem~\ref{thm:6}, respectively, depend on $\ga'$ and may diverge as $\ga'$ tends to $\ga$.
}
\end{rmk}

\subsubsection{Oscillations and deviations in the scaling regime} \label{subsec:oscillations}

In the next subsection we show that $(\SSS^n)_\f$ remains close to $(\ol\SSS^n)_\f$ and hence to $\Phi_{n\rho}$,
in the sense that the first  and second moments, with respect to $\psi$, of their difference are $O(\rho)$
uniformly in $n$. In this subsection we present several preparatory results
that are of interest in their own. The proofs of these
results form the main technical part of the present work and are reported in
Section~\ref{sec:5}. The main tools used in the proofs are the decay of
correlations estimates provided by Propositions~\ref{prop:decay} and \ref{prop:decaynonltot}.

We first show that even though the oscillations of $\calW$ around
$\overline\calW$ can be of order 1 in $\rho$, large oscillations are rare,
in the sense of the following result, proved in Subsection~\ref{subsec:proof3}.

\begin{thm}[\textbf{Oscillations of the invariant manifold}] \label{thm:3}
Consider the dynamical system des\-cri\-bed by $\SSS$ in
\eqref{eq:1.1bis} satisfying Hypotheses~\ref{hyp:1}--\ref{hyp:3}. Let
$\calW=\{(W(\psi),\psi):\psi\in\TTT^2\}$ be the invariant manifold for the map
$\SSS$ as in Theorem~\ref{thm:1}.
There is a constant $C_{5}$ such that
\vspace{-.1cm}
\begin{equation} \label{Wflu}
\left| \langle W\rangle \right| \leq C_{5} \rho , \qquad\qquad  \langle W^2\rangle\leq C_{5} \rho .
\end{equation}
\end{thm}

From Theorem~\ref{thm:3} and Chebyshev inequality we obtain that, for any $\delta>0$,
\vspace{-.1cm}
\begin{equation} \label{0time}
m_0 \Bigl( \Bigl\{\psi \in\TTT^2 :  |W(\psi)|>\delta \Bigr\} \Bigr)\leq \frac{C_{5} \rho}{\delta^2} .
\vspace{-.1cm}
\end{equation}
%
Therefore, the invariant manifold $\calW$ for $\SSS$ converges in probability
to the invariant manifold $\overline \calW$ of $\overline \SSS$.

Next we compare the  linearized maps $\SSS_0$ and $\ovl\SSS_0$, defined in
\eqref{eq:sss0} and \eqref{Sbarlin}, respectively. Observe that
$\overline\SSS_0^n(\f,\psi)=(\ovl\mu^n \f,A_0^n\psi)$,
while $\SSS_0^n(\f,\psi)=(\ka^{(n)}(\psi)\f,A_0^n\psi)$, where
\vspace{-.2cm}
\begin{equation}\label{eq:kan}
 \ka^{(n)}(\psi): =\prod_{i=0}^{n-1}\ka(A_0^{i}\psi)\, ,
\vspace{-.3cm}
\end{equation}
with $\ka(\psi)$ defined in \eqref{eq:sss0}.

\begin{rmk} \label{convention}
\emph{
$\!\!$Throughout the paper we use the convention that a product
over an empty
set of indices is 1,
while a sum over an empty set of  indices is 0. In particular, this means that  $\ka^{(0)} = 1$ in \eqref{eq:kan}.
}
\end{rmk}

The following result, proved in Subsection~\ref{fluk},
shows that the linearized maps $\SSS_0^n$ and $\overline\SSS_0^n$ stay close to each other uniformly in $n$;
what makes the result not trivial is that the function $\ka$ in \eqref{eq:kan}
is only in $\gotB^*_{\al_-,\al_+}(\TTT^2,\RRR)$,
with $\al_-=O(\rho)$ and $|W|^{-}_{\al_-}=O(1)$ in $\rho$.

\begin{lemma}\label{lem:flulin}
Consider the dynamical system described by $\SSS$ in \eqref{eq:1.1bis}
satisfying Hypotheses~\ref{hyp:1}--\ref{hyp:3}.
Let $\ka^{(n)}(\psi)$ and $\ovl \mu$ be defined
in \eqref{eq:kan} and in \eqref{Sbarlin}, respectively.
Then for any $\ga' \in(0,\ga)$ there is a constant $C_{6}$ such that, for all $n\in \NNN$,
\vspace{-.1cm}
\begin{equation} \label{linear}
\bigl|\aver{\kappa^{(n)} - \ovl\mu^{n}}\bigr| \leq C_{6} \rho \, (1-\rho \, \ga')^n , 
\qquad
\bigl|\aver{\bigl(\kappa^{(n)} - \ovl\mu^{n}\bigr)^2}\bigr| \leq C_{6} \rho \, (1-\rho \ga')^{2n} .
\vspace{-.1cm}
\end{equation}
\end{lemma}

Finally, we estimate the deviation of $\calH$ from $\ol\calH$. To this end,
it is enough to study the deviations of $h$ 
from $\ol h$. 
The following result is proved in Subsection~\ref{proof3-2}.

\begin{thm}[\textbf{Deviations of the conjugation}] \label{thm:exthm3}
Consider the dynamical system described by $\SSS$ in \eqref{eq:1.1bis} satisfying Hypotheses~\ref{hyp:1}--\ref{hyp:3}.
Let $h$ and $\ovl h$ be defined as in \eqref{hl} and in \eqref{def:lbar}, respectively.
There is a constant $C_{7}$ such that, for all $\f \in \calU$,
%
\begingroup \addtolength\jot{6pt}
\begin{subequations} \label{bounds-thm3}
\begin{align}
\left| \aver*{h(\f,\cdot) -\overline h (\f) } \right| \leq & C_{7} \rho ,
& \aver*{(h(\f,\cdot)-\overline h(\f))^2} & \leq C_{7}\rho , 
\label{bounds-thm3a} \\
 \left| \aver*{\partial_\f h(\f,\cdot) - \partial_\f\overline h (\f) } \right| & \leq C_{7} \rho ,
&\aver*{( \partial_\f h(\f,\cdot) - \partial_\f \overline
h(\f))^2}  &\leq C_{7} \rho .
\label{bounds-thm3b}
\end{align}
\end{subequations}
\endgroup
\end{thm}
\vspace{-.1cm}

Bounds analogous to \eqref{bounds-thm3} hold for the deviations of
$l$ from $\ol l$, with $l$ and $\ol l$ also defined in \eqref{hl} 
and \eqref{def:lbar}, respectively;
we refer to Subsection~\ref{proof3-2} -- to Proposition~\ref{prop:dascrivere} in particular -- for a precise statement,
which requires introducing a suitable extension of the map $\SSS$ along the lines outlined in Remark \ref{ext}.
%
Note that it is in order to prove the bounds in Theorem~\ref{thm:exthm3} 
and in 
Proposition~\ref{prop:dascrivere} 
that we need $F$ to be in $\calmB_{\al_0,6}(\Omega,\RRR)$ (see also Remark \ref{regularity2}). 

\subsubsection{Summing up:~convergence in square mean and in probability} \label{subsec:probability}

We can now complete the comparison of the evolution generated by $\SSS$ with
that generated by $\overline \SSS$. From \eqref{eq:1.11} and \eqref{eq:Hbar2} we get
\vspace{-.2cm}
\begin{equation} \label{S1n-Sbarn}
\big( \SSS^n \big)_\f (\f,\psi) - ( \overline \SSS^n)_\f
(\f,\psi)=
\HH^{-1}
\bigl(\SSS_0^n(\HH (\f,\psi),\psi)\bigr)-
\overline\HH^{-1} \bigl(\overline
\SSS_0^n( \overline\HH(\f ),\psi) \bigr) .
\vspace{-.1cm}
\end{equation}
Combining the estimates in Lemma~\ref{lem:flulin} with the bounds in Theorem~\ref{thm:exthm3}
and the analogous bounds for the inverse conjugation in Proposition~\ref{prop:dascrivere},
the following result is proved in Subsection~\ref{proof4}.

\begin{thm}[\textbf{Convergence in square mean}] \label{thm:4}
Consider the dynamical system described by $\SSS$ in \eqref{eq:1.1bis}
satisfying Hypotheses~\ref{hyp:1}--\ref{hyp:3}. For any $\ga' \in(0, \ga)$
there exists a constant $C_{8}$ such that, for all $n\in \NNN$ and all $\f\in\calU$,
\vspace{-.3cm}
\begingroup \addtolength\jot{4pt}
\begin{subequations} \label{boundSSS}
\begin{align}
\bigl|  \bigl\langle (\SSS^n)_\f(\f,\cdot)- \bigl((\overline  \SSS^n)_\f(\f,\cdot)+W(A_0^n\cdot)\bigr) \bigr\rangle \bigr| 
& \leq C_{8} \rho \, (1-\rho\,\ga')^{n} ,
\label{mediaSSS} \\
\bigl\langle \bigl((\SSS^n)_\f(\f,\cdot)- \bigl((\overline  \SSS^n)_\f(\f,\cdot)+W(A_0^n\cdot)\bigr)\bigr)^{\!2} \bigr\rangle 
& \leq 
C_{8} \rho \, (1-\rho\,\ga')^{2n} .
\label{varSSS}
\end{align}
\end{subequations}
\endgroup
\end{thm}

\vspace{-.2cm}

\begin{rmk} \label{commentfluc}
\emph{
The bound \eqref{varSSS} for the second moment of the fluctuations and the Cauchy-Schwartz
inequality trivially imply a weaker bound than \eqref{mediaSSS}. By contrast, proving that also the
first moment of the fluctuations is of order $\rho$ requires a substantially greater amount of work.
Apart of its intrinsic interest, the stronger bound \eqref{mediaSSS}
plays a crucial role in view of some of the extensions proposed in Section \ref{sec:outlook}.
A similar comment holds for the results in
Theorems~\ref{thm:3} and \ref{thm:exthm3}, and in Lemma~\ref{lem:flulin} as well.
}
\end{rmk}

The following result is an immediate consequence of Theorems
\ref{thm:3} and \ref{thm:4}, and of Lemma~\ref{lem:Phik}.

\begin{coro} \label{noname}
Under the hypotheses of Theorem~\ref{thm:4},  there exists a constant $C_{9}$
such that, for all $n\in \NNN$ and all $\f\in\calU$, one has
\vspace{-.1cm}
\begin{equation} \label{nonumero}
\bigl| \bigl\langle (\SSS^n)_\f(\f,\cdot)- \Phi_{n\rho} (\f) \bigr\rangle \bigr| \leq
C_{9} \rho  , \qquad \qquad
\bigl\langle \bigl( (\SSS^n)_\f(\f,\cdot)- \Phi_{n\rho} (\f) \bigr)^2 \bigr\rangle \leq C_{9} \rho , 
\end{equation}
with the flow $\Phi$ defined as in \eqref{flow}.
\end{coro}

From Theorem~\ref{thm:4},
for fixed $\f$ and $n$, we obtain that,
for any $\ga'\in(0,\ga)$ and for any $\de>0$,
\vspace{-.2cm}
\begin{equation}\label{1time}
m_0\Bigl(\Bigl\{\psi \in \TTT^2 : (1-\rho\,\ga')^{-n} \bigl|
(\SSS^n)_\f(\f,\psi)-\bigl(\Phi_{n\rho}(\f)+
W(A_0^n\psi)\bigr) \bigr|>\delta
\Bigr\}\Bigr)\leq  \frac{C_{9} \rho}{\delta^2} ,
\vspace{-.1cm}
\end{equation}
with the constant $C_{9} $ as in \eqref{boundSSS}.
Note that the set of angles $\psi$  considered in \eqref{1time} depends on $n$,
even though its measure is bounded independently of the value of $n$.

The following theorem,
proved in Subsection~\ref{proof:prob},
shows that, for most most values of $\psi$, the difference
between $(\SSS^n)_\f(\f,\psi)$ and $\Phi_{n\rho}(\f) + W(A_0^n\psi)$ is small, exponentially in $n$.

\begin{thm}[\textbf{Convergence in probability, I}]\label{thm:6}
Consider the dynamical system described by $\SSS$ in \eqref{eq:1.1bis}
satisfying Hypotheses~\ref{hyp:1}--\ref{hyp:3}. Let the flow $\Phi$ be defined
as in \eqref{flow}. For any $\ga'\in(0,\ga)$ there exists a constant $C_{10}$ such that, for all $\f\in\calU$,
\vspace{-.2cm}
\begin{equation}\label{eq:prob1}
m_0\Bigl(\Bigl\{\psi \in \TTT^2 : \sup_{n\ge 0} \, (1-\rho\,\ga')^{-n} \bigl |
(\SSS^n)_\f(\f,\psi)-\bigl(\Phi_{n\rho}(\f)+W(A_0^n\psi)\bigr) \bigr| >\delta
\Bigr\}\Bigr)\leq  \frac{C_{10} \rho}{\delta^3} .
\vspace{-.2cm}
\end{equation}
\end{thm}

In \eqref{eq:prob1} we can chose $\de$ as a function of $\rho$ in such a way that both $\de$ and $\rho/\de^3$ tend to 0
as $\rho$ tends to 0 (a natural choice is $\de=\rho^{1/4}$ so that also $\rho/\de^{3}=\rho^{1/4}$).
Thus,
Theorem \ref{thm:6} implies that, up to a set of initial conditions for $\psi$ whose measure
vanishes with $\rho$, the difference between the actual evolution of the neutral variable
toward the invariant manifold and its approximate evolution as described by the averaged
differential equation vanishes with $\rho$ as well. Moreover, such a difference remains
uniformly bounded in time for arbitrarily long times.

\subsubsection{Aftermath:~continuous time} \label{subsec:continuous}

We now give a more probabilistic description
of the results in Theorem~\ref{thm:6} on the relations between
$\SSS^{\lfloor t/\rho\rfloor}$ and the solutions of
\eqref{eq:1.9} and express them in terms of the rescaled time $t$. To this aim,
for every $\f \in\TTT$ and every $t\in\RRR^+$, we
consider the random variable $X_t \! :\TTT^2\to\TTT$ defined as
\vspace{-.1cm}
\begin{equation}\label{stpr}
X_t(\psi) := (\SSS^{\lfloor t/\rho \rfloor})_\f(\f,\psi) +(t/\rho-\lfloor t/\rho \rfloor)
\bigl((\SSS^{\lfloor t/\rho \rfloor+1})_\f(\f,\psi)-(\SSS^{\lfloor t/\rho \rfloor})_\f(\f,\psi)\bigr) .
\vspace{-.1cm}
\end{equation}
For any $\psi$, $X_t(\psi)$ is a continuous function of $t$, so that $X_{t}$ can be seen as a
stochastic process with trajectories in $C^0(\RRR^+,\TTT)$.
We consider also the process $\XXXX_t$,
with trajectories in $C^0(\RRR^+,\TTT)$, given by
\vspace{-.1cm}
\begin{equation} \label{stprtilde}
\XXXX_t(\psi)= X_t(\psi)-W(A_0^{\lfloor t/\rho \rfloor}\psi)-(t/\rho-\lfloor t/\rho \rfloor)
\bigl(W(A_0^{\lfloor t/\rho \rfloor+1}\psi)-W(A_0^{\lfloor t/\rho \rfloor}\psi)\bigr) .
\vspace{-.1cm}
\end{equation}

We want to compare the stochastic processes $X_t$ and $\XXXX_t$ with the
flow $\Phi_t$ defined in \eqref{flow}, seen as a stochastic process on
$C^0(\RRR^+,\TTT)$.
To this aim 
we introduce the norm
%
$\|x\|^{\rm exp}_{\xi} := \sup_{t\geq0} \,  e^{\xi t}|x(t)| $,
%
with $\xi \ge 0$. Combining Lemma~\ref{lem:Phik} and Theorem~\ref{thm:6} gives
the following result, proved in Subsection~\ref{proof:probcont}.

\begin{thm}[\textbf{Convergence in probability, II}] \label{thm:prob}
Consider the dynamical system described by $\SSS$ in \eqref{eq:1.1bis} satisfying Hypotheses~\ref{hyp:1}--\ref{hyp:3}.
Let $\XXXX_t$ and $\Phi_t$ the stochastic process \eqref{stprtilde} and the flow 
defined in \eqref{flow}, respectively. One  has
$\lim_{\rho\to0^+}\XXXX_t=\Phi_t $.
where the limit is taken in probability in the topology on
$C^0(\RRR^+,\TTT)$ generated by the norm $\|\cdot\|^{\rm exp}_\xi$
for any $\xi\!\in\![0,\gamma)$.
\end{thm}

We close our results with an immediate consequence of Theorem~\ref{thm:prob}.

\begin{coro}\label{coro:prob}
Assume the hypotheses of Theorem~\ref{thm:prob} and let
$X_t$ be the stochastic process defined in \eqref{stpr}. One has
$\lim_{\rho\to0^+}X_t=\Phi_t$,
where the limit is taken  in probability in the topology of the uniform  convergence in $C^0(\RRR^+,\TTT)$.
\end{coro}

\zerarcounters 
\subsection{Extensions and generalizations}\label{sec:outlook}



Systems of the form \eqref{eq:1.1} are called \emph{skew products} since the
fast variable does not depend on the slow variable.
Skew product systems are widely studied in the literature, with a view to focus on the
major features of the equations and discard the more technical aspects;
for averaging in fast-slow systems where the base is an expanding map we may quote \cite{KKM},
beyond the illustrative examples proposed in \cite{DL3,DL2}.
Here, we sketch a path to generalize our results to the case of a fully coupled system.

As a first step one may consider systems of the form
\vspace{-.1cm}
\begin{equation} \label{eq:7.4a}
\SSS(\f,\psi) = (\f+F_\f(\f,\psi) , \AAA(\psi)) , \qquad \AAA_0(\psi):=A_0 \psi + F_\psi(\psi) ,
\vspace{-.1cm}
\end{equation}
with $F_\psi\!:\TTT^2\to\TTT^2$ and $F_\f(\f,\psi)\!:\TTT\times\TTT^2 \to \TTT$ such that $\AAA_0(\psi)$ 
describes an Anosov diffeomorphism on $\TTT^2$ and $\SSS$ still satisfies
Hypotheses~\ref{hyp:1}--\ref{hyp:3}. However, since any Anosov diffeo\-morphism of the form in \eqref{eq:7.4a}
is conjugated with its linear part $A_0$ (see also \cite{BFG,F,N} for a more general context),
there exists a H\"older continuous map
$\HHHH \! :\TTT^2\to\TTT^2$ such that
$\HHHH\circ \AAA_0=A_0\circ \HHHH$ ,
%
%
so that, setting $\HHHHH(\f,\psi):=(\f,\HHHH(\psi))$,
we may use $\HHHHH$ to conjugate $\SSS$ in \eqref{eq:7.4a} with
\vspace{-.1cm}
\begin{equation}\label{Stilde}
\SSSSS(\f,\psi)=(\f+\FFFF(\f,\psi),A_0 \psi) ,
\vspace{-.1cm}
\end{equation}
with $\FFFF(\f,\HHHH(\psi))=F_\f(\f,\psi)$.
Clearly $\SSSSS$ is of the form \eqref{eq:1.1} and satisfies Hypotheses \ref{hyp:1}--\ref{hyp:3}.
In this situation it is still natural to chose the initial $\psi$ distributed
according to the Lebesgue measure $m_0$ on $\TTT^2$. This is essentially equivalent to
considering the SRB measure $m_{\AAA_0}$ associated with $\AAA_0$, since $(\AAA_0^*)^nm_0$ converges
exponentially fast to $m_{\AAA_0}$ \cite{GBG}.
Thus, to apply the results in Section~\ref{sec:model} to $\SSS$ in \eqref{eq:7.4a},
one needs estimates on the decay of correlations like those in Proposition \ref{prop:4.3}
but with $(\HHHH)^* m_{\AAA_0}$ in place of $m_0$.
Observe that $\HHHH$ is H\"older continuous and $(\HHHH)^* m_{\AAA_0}$ is
invariant under the action of $A_0$; thus $m_{\AAA_0}$ can be represented as a Gibbs
state on the same subshift of finite type used for $A_0$ in Appendix
\ref{app:corre}. One thus need to extend the proof Proposition \ref{prop:4.3} using the properties
of the potential that generates such a Gibbs state as discussed in \cite{GBG}.


The next step is to look at systems with bidirectional perturbations of the form
\vspace{-.1cm}
\begin{equation} \label{eq:7.4}
\SSS(\f,\psi) = (\f+F_\f(\f,\psi) , \mathcal A(\psi;\f) ) , \qquad \mathcal A(\psi;\f) := A_0 \psi +
F_\psi(\f,\psi)) ,
\vspace{-.1cm}
\end{equation}
with $F_\f(\f,\psi)$ satisfying once more Hypotheses~\ref{hyp:1}--\ref{hyp:3} and $F_\psi(\f,\psi)$
such that the
dynamical system on $\TTT^2$  generated by the map 
$\mathcal A(\cdot;\f)$
is an Anosov diffeomorphism for any fixed $\f\in[\phi_m,\phi_M]$.
An invariant manifold is proved to exist for the system \eqref{eq:7.4}  in~\cite{DGG} in
the perturbative regime under stronger conditions.
More generally, one 
looks for a conjugation
$\HHHHH(\f,\psi)=(\f,\HHHH(\f,\psi))$ such that
%
$\HHHHH\circ\SSS=\SSSSS\circ\HHHHH$,
%
with $\SSSSS$ of the form \eqref{Stilde}. This means that one has
\vspace{-.1cm}
\begin{equation}\label{confull}
\HHHH(\f+F_\f(\f,\psi),\AAA(\psi;\f))
=
A_0\HHHH(\f,\psi) ,
\qquad\qquad
\FFFF(\f,\HHHH(\f,\psi))=F(\f,\psi) .
\vspace{-.1cm}
\end{equation}
It is possible to write a formal solution to \eqref{confull} 
(see also \cite{BKLeb} for a similar argument). Applying the conjugation $\HHHHH$
to the system \eqref{eq:7.4} allows us to reduce it to a system of the form of \eqref{eq:1.1}.
To apply the results of the present paper one then needs to prove that the resulting system has the geometric
and regularity properties needed to satisfy Hypotheses \ref{hyp:1}--\ref{hyp:3}. The above construction, assuming it is successful,
allows one to extend the results on synchronization to the systems as in \eqref{eq:7.4}.
Then, one has to show that the averaging principle proved for the system \eqref{eq:1.1} implies, 
thanks to the existence of $\HHHHH$,
an averaging principle for \eqref{eq:7.4},
more precisely that, writing $F_\f(\f,\psi)=\rho\,f_\f(\f,\psi)$ and starting the evolution at $\f_0$, with
$\psi$ distributed according to $m_0$, the long time evolution generated by \eqref{eq:7.4} for small $\rho$ is described by
the flow $\PPPPhi_t$ defined by
\vspace{-.2cm}
\begin{equation}\label{flowfull}
\left\{
\begin{aligned}
&\frac{\der}{\der t}\PPPPhi_{t}(\f_0)= \ffff(\PPPPhi_{t}(\f_0)) , 
\qquad\qquad \ffff(\f) :=\int_{\TTT^2}f(\f,\psi) \, m_\f(\der\psi), \\
 & \PPPPhi_{0}(\f)=\f _0, \phantom{\sum}
\end{aligned}\right.
\vspace{-.2cm}
\end{equation}
%
where the new measure $m_\f$ can be computed from $\HHHH$
and, as heuristic arguments suggest, it is expected to
be the SRB measure of the Anosov diffeomorphism $\AAA(\cdot;\f)$.




A third possible -- and harder -- generalization to investigate, already in the skew
product case, is obtained by weakening the hypotheses on the dissipative nature
of the map $\SSS$, 
for instance by allowing more general interactions where
the rate of contraction of the neutral variable is non-zero only in average,
as in \cite{DL3,DL2}, where expanding maps are considered instead of Anosov automorphisms.
For such a kind of interactions, \emph{a fortiori} also in the case of systems of the form \eqref{eq:7.4}
new phenomena with respect to those found in the present paper
are expected to occur, such as non-uniform hyperbolicity, positive Lyapunov exponents in the central direction,
non-absolutely integrable central direction and metastability.
Very interesting, from a physical point of view, is the conservative case,
where $\langle f(\cdot,\f) \rangle$ is assumed to vanish for all $\f$.
In such a situation, on the basis of the results available for related models~\cite{De,DL,KM,Ly},
one expects the scaling to be $k=\lfloor t/\rho^2 \rfloor$
and to lead in the limit to a stochastic differential equation,
instead of an ordinary differential equation as in the dissipative case.


Finally, it would be of interest to find a more detailed description of the fluctuations
of the process $X_t$ in \eqref{stpr} around the flow $\Phi_t$ in \eqref{flow}.
By comparison with the results available in the literature~\cite{Ki3,Do,DL,DLPV,DL3,DL2},
we expect the stochastic process
$\Delta_t := 
(X_t-\Phi_t)/\sqrt{\rho}$
to converge in distribution, as $\rho\to 0^+$, to the solution of the stochastic differential equation
\vspace{-.2cm}
\begin{equation} \label{SDE}
\begin{cases}
\displaystyle{ \der \Delta_t=\partial_\f\overline f(\Phi_t(\f_0))\Delta_t \der t +\sigma(\Phi_t(\f_0)) \, \der B(t) } , \phantom{\sum} & \\
 \Delta_0=0 , \phantom{\sum}  &
\end{cases}
\vspace{-.2cm}
\end{equation}
where $B(t)$ is a standard Brownian motion and
$(\sigma(\f))^2 := \bigl\langle\sum_{n=-\infty}^{\infty} \tilde f(\f,\cdot) \, \tilde f(\f, A_0^n \cdot)\bigr\rangle$,
with $\tilde f$ defined according to \eqref{eq:1.222}.
Such a result is known to hold with $\sigma>0$ in the case in which the chaotic system is
an expanding map, if one requires the function $\f \mapsto F(\f,\psi,)$ not to be cohomologous to a
constant function for any $\psi$ \cite{DL3,DL2}. 
Since for conservative interactions the scaling
limit is expected to lead to a stochastic differential equation, 
studying how an equation like \eqref{SDE} emerges from \eqref{eq:1.1} in the scaling regime can be seen
as a precursory step before dealing with the more demanding scaling regime for conservative systems.
The fact that  $\Delta_t$ converges to the solution of equation \eqref{SDE} implies that
$W/\sqrt\rho$, with $W$ (see Subsection~\ref{invariantmanifold}) seen as a random
variable on $\TTT^2$, converges in distribution to a normal random variable with
mean 0 and standard deviation $\sigma(0)$, that is
\vspace{-.2cm}
\begin{equation}\label{CLT}
 \lim_{\rho\to 0^+}m_0 \left( \{\psi \in \TTT^2 :  W(\psi)\le \sqrt\rho z \}
\right)
=\frac{1}{\sqrt{2\pi\sigma(0)^2}}\int_{-\infty}^{z} \der y \, e^{-\frac{y^2}
{2\sigma(0)^2}} .
\vspace{-.1cm}
\end{equation}
To start with, as a consistency check we show how to derive \eqref{CLT}.
First of all, using the notation in Remark~\ref{betadelta}, with $F=\rho f$ and
$\mu(\psi): = 1-\nu(\psi)$, 
set
\vspace{-.2cm}
\[
W_0  := \sum_{i=1}^{\io} \biggl( \, \prod_{j=1}^{i-1} \mu \circ A_0^{-j} \biggr)  \beta \circ A_0^{-i } , \qquad
W_{00}  : = \sum_{i=1}^{\io} \overline\mu^{i-1} \be\circ A_0^{-i } .
\vspace{-.2cm}
\]
We prove in Subsection~\ref{subsec:proof3} that $\aver{|W-W_0|}=O(\rho)$, which
implies that $(W-W_0)/\sqrt\rho$ converges in probability to 0,
so that we just need to prove that $W_0/\sqrt\rho$, in the limit, has the correct normal distribution.
Then, a partial extension of Proposition~\ref{prop:decay} to multi-time correlation functions -- whose proof we omit --  shows that
%
$\average{\left(W_0-W_{00}\right)^2}=O(\rho^2)$.
%
On the other hand, the central limit theorem for Anosov system \cite{Ch} implies
that $W_{00}/\sqrt\rho$ tends to the correct normal limit as $\rho\to 0^+$.
We expect a similar strategy to work for the derivation of \eqref{SDE}, by
proceeding along the lines of Subsection~\ref{proof3-2}.

\subsection{Content of the paper and strategy of the proof} \label{subsec:content}

The rest of the paper, 
devoted to the proof of the results stated in Subsections \ref{subsec:synch} and \ref{subsec:scaling},
%
%
%
%
is organised as follows.

Section~\ref{sec:2} contains the proofs of the Theorems~\ref{thm:1} to
\ref{thm:mixing}, together with the derivation of the properties of both the
invariant manifold and the conjugation that are needed for dealing with the scaling regime.
The existence of the invariant manifold is formulated as a
fixed-point problem in a suitable Banach space (Theorem~\ref{thm:1}).
Thereafter, the conjugation is shown
to admit a series representation which is proved to converge to a
function which satisfies the properties stated in Theorem~\ref{thm:2}. This
result implies the existence of the inverse conjugation,
which satisfies similar properties (Corollary~\ref{l}),
and the mixing properties of the system while evolving toward the invariant manifold (Theorem~\ref{thm:mixing}).
Technically, it turns out to be useful to use the coordinates $(\theta,\psi)$, with $\theta:=\f-W(\psi)$
and $W(\psi)$ as in Subsection~\ref{invariantmanifold},
in terms of which the dynamics is described by 
a map $\SSS_1$ that we call the \emph{translated map} and the attracting invariant manifold is the flat torus $\TTT^2$.

In Section~\ref{sec:5} we discuss the averaging problem in the scaling regime:
setting $F(\f,\theta)=\rho f(\f,\theta)$, the dynamics of $\SSS^k(\f,\psi)$
is studied for $k=\lfloor t/\rho \rfloor $, with $t$ fixed and $\rho\to0^+$.
After deriving 
the deviation laws for the invariant
manifold (Theorem~\ref{thm:3}) and the conjugation (Theorem~\ref{thm:exthm3}), 
we study the deviations of the dynamics from that of the averaged system (Theorem~\ref{thm:4}).
The analysis is based on delicate correlation inequalities, which exploit the hyperbolicity
of $A_0$ and make use of the map being weakly dissipative and the ensuing fact that any trajectory after
a while comes close enough to the attracting invariant manifold. A main issue --
and a major source of technical intricacies -- is that the correlations
inequalities are related to the regularity of the involved functions. In
general, we have to deal with averages of the form $\aver{g_+ g_-\circ\SSS_1^n}$, 
with $g_-\in\calmB_{\al}^-(\Omega,\RRR)$, for which we can expect decay properties
analogous to those of Proposition~\ref{prop:4.3}. However, the invariant
manifold and, hence, the map $\SSS_1$ are only $\al$-H\"older continuous, with
$\al=O(\rho)$, so that a nai\"ve generalization of Proposition~\ref{prop:4.3}
would provide unavailing bounds. If we expand the average $\aver{g_+
g_-\circ\SSS_1^n}$ to second order in $W$, only the first order contributions
require a careful analysis, since the leading terms are regular and the second
order terms can be dealt with by using directly the bounds on the variance provided by
Theorem~\ref{thm:3}. Instead, in order to use the bounds on the average in Theorem~\ref{thm:3}
we need to study in detail the first order contributions.
To this aim, we isolate the contributions which do not depend on the dynamics on the torus
and show that the remaining contributions admit better dimensional bounds.
To implement the scheme outlined above, we introduce a regularized
version $\SSS_2$ of the translated map, that we call the \emph{auxiliary map},
for which we can apply the correlation inequalities, and,
in Subsection~\ref{674} and Appendix~\ref{app:tech2},
we compare the translated map with the auxiliary map through a series of technical lemmas
which aim at extracting and studying the linear dependence on the function $W$.
To do that, we need $\SSS_2$ to regularize $\SSS_1$ and, at the same time, still satisfy Hypotheses
\ref{hyp:1} to \ref{hyp:3}: achieving both goals leads to contributions which,
albeit linear in $W$ so that they be dealt with as outlined before,
unfortunately contain an extra factor $\rho^{-1}$. However such contributions
involve sums of terms in which there appear differences of functions $W$ and the
sums can be rearranged in such a way that the difference is shifted to more
regular functions: this allows us to regain a further factor $\rho$ so as to
compensate the factor $\rho^{-1}$. To implement the idea described above
we perform iterated expansions which make the analysis rather intricate:
in fact, Subsection~\ref{674} and Appendix~\ref{app:tech2} constitute the most
technical part of the paper.

In Section~\ref{last}, we prove Theorems~\ref{thm:6} and \ref{thm:prob}
on the asymptotic behavior of the stochastic process associated to the
dynamics. Again, the crucial issue is that in average the deviations are small,
and hence the evolution of the system is essentially determined by the averaged map.

Appendix~\ref{sec:4}, besides reviewing some basic facts about Anosov automorphisms,
contains the results on the decay of correlations for the
evolution generated by $A_0$ on $\TTT^2$. Such results form the main toolkit we
use in Section~\ref{sec:5} to obtain more general correlation inequalities.
Appendices~\ref{app:tech}, \ref{appC} and mainly -- as said above -- \ref{app:tech2} 
contain mostly the proofs of the more technical results presented in Section~\ref{sec:5}.

\zerarcounters 
\section{Mapping to a simpler model}
\label{sec:2}

In this section we prove Theorems~\ref{thm:1} and \ref{thm:2} by explicitly solving \eqref{inva} and \eqref{eq:1.11}.
The first one follows by relying on Banach fixed-point theorem,
while the second one exploits the dynamics being uniformly contracting around the invariant manifold.
Next, the two results together are showed to yield Theorem~\ref{thm:mixing}.

\subsection{The invariant manifold:~proof of Theorem~\ref{thm:1}} \label{subsec:proofW}

From \eqref{inva} we get
\begin{equation} \label{eq:2.3}
W(A_0\psi)=G(W(\psi),\psi)= W(\psi)+ F(W(\psi),\psi) .
\end{equation}
To show that a solution of \eqref{eq:2.3} exists we define the map
\begin{equation}\label{EEE}
 \EEE[W](\psi) :=G(W(A_0^{-1}\psi),A_0^{-1}\psi) ,
\end{equation}
so that the invariant manifold is the solution of the fixed point 
equation $\EEE[W]=W$.
If we define the set
%
$B :=\{W\in\gotB(\TTT^2,\RRR) : S_m\leq W\leq S_M\}$,
%
then $\EEE(B)\subset B$ by Hypotheses~\ref{hyp:1} and \ref{hyp:2}. Moreover
\[
 \|\EEE[W_1]-\EEE[W_2]\|_\infty = 
 \sup_{\psi\in\TTT^2} \left| G(W_1(A_0^{-1}\psi),A_0^{-1}\psi) - G(W_2(A_0^{-1}\psi),A_0^{-1}\psi) \right|
\le (1-\Gamma)\|W_1-W_2\|_{\infty} ,
\]
that is $\EEE$ is a contraction on $B$, and thus, by the Banach fixed-point theorem, there is a unique $W\in B$ that satisfies \eqref{eq:2.3}.
We pass to discuss the regularity of $W$. Since $F$ is $\alpha_0$-H\"older continuous, we get
\begin{equation}\label{eq:Hold+} 
\begin{aligned}
 |\EEE[W]|_{\al_+}^{+}
& \le \lambda^{-\al_+}\sup_{\psi \in \TTT^2} \sup_{x \in \RRR}|x|^{-\al_+} \Big( |G(W(\psi+x v_+),\psi+x v_+)-G(W(\psi),\psi +x v_+ )|\\
&
+ |G(W(\psi),\psi+ x v_+)-G(W(\psi),\psi) | \Bigr)
\le \lambda^{-\al_+}\bigl((1-\Gamma)|W|_{\al_+}^{+}+|F|_{\al_+}\bigr) ,
\end{aligned}
\end{equation}
for $\al_+\in (0,\alpha_0]$, and similarly we obtain, for $\al_-\in(0,\alpha_0]$,
\begin{equation}\label{eq:Hold-}
|\EEE[W]|_{\al_-}^{-}\leq\lambda^{\al_-} \bigl((1-\Gamma)|W|_{\al_-}^{-}+|F|_{\al_-}\bigr) .
\end{equation}
This implies that the set
$B_{\al_-,\al_+} := \{ W\in \gotB_{\al_-,\al_+}(\TTT^2,\RRR): |W|_{\al_+}^{+}\le \de_+ |F|_{\al_+} \hbox{ and } |W|_{\al_-}^{+}\le \de_- |F|_{\al_-}\}$,
with $\de_{\pm}:=(\la^{\pm \al_{\pm}}\!-\!(1\!-\!\Gamma))^{-1}$,
%
%
%
is invariant under $\EEE$.
Thus, we need to show that $\EEE$ is a contraction on $B_{\al_-,\al_+}$ for suitable $\al_-$ and $\al_+$,
in order to apply once more the Banach Fixed-Point Theorem. To this end,
observe that, for any $W_1,W_2\in B_{\al_-,\al_+}$ we bound
$|\EEE[W_1] - \EEE[W_2]|_{\al_+}^{+} $ by
\vspace{-.2cm}
\begin{equation*}
\begin{aligned}
& \lambda^{-\al_+}\sup_{\psi\in \TTT^2} \sup_{x \in \RRR}|x|^{-\al_+} \Bigl( \,
\Bigl| G( W_1( \psi),\psi) -G(W_2(\psi),\psi)
- G(W_1(\psi+xv_+),\psi) +G(W_2(\psi+xv_+),\psi) \Bigr| \phantom{\sup_x}  \\
& + 
\Bigl| G(W_1(\psi \!+\! xv_+),\psi) -G(W_2(\psi \!+\! xv_+),\psi)
- G(W_1(\psi \!+\! xv_+),\psi \!+\! xv_+)+G(W_2(\psi \!+\! xv_+),\psi \!+\! xv_+ ) \Bigr| \, \Bigr) ,
\end{aligned}
\end{equation*}
where, using twice a standard interpolation between $G(W_1(\psi_1),\psi_2)$ and
$G(W_2(\psi_1),\psi_2)$ in the second line and a 4 point interpolation in the first line we get
\begin{equation} \label{eq:diff+}
|\EEE[W_1] \!-\! \EEE[W_2]|_{\al_+}^+
\! \le \!
\lambda^{-\al_+} \Bigl( (1\!-\!\Gamma)|W_1 \!-\! W_2|_{\al_+}^{+}
\!+\! \bigl(\| \partial_\f^2 F\|_\infty\max_{i=1,2} |W_i|_{\al_+}^{+} \!+\!
|\partial_\f F|_{\al_+}^{+} \bigr) \|W_1\!-\!W_2\|_\infty \Bigr) .
\end{equation}
Analogously, we obtain
\begin{equation} \label{eq:diff-} 
|\EEE[W_1] \!-\! \EEE[W_2]|_{\al_-}^{-}
\!\le\! \lambda^{\al_-} \Bigl( (1 \!-\! \Gamma)|W_1\!-\!W_2|_{\al_-}^{-}
\!+\!  \bigl( \|\partial_\f^2 F\|_\infty\max_{i=1,2}
|W_i|_{\al_-}^{-}
\!+\! |\partial_\f F|_{\al_-}^{-} \bigr)\|W_1 \!-\! W_2\|_\infty \Bigr) .
\end{equation}
Let now fix $\al_-,\al_+\in(0,\al_0]$ so as to satisfy $\lambda^{\al_-}(1-\Gamma) < 1$ and
%
\begin{equation} 
\al_+ \lambda^{-\al_+} \! \biggl( \frac{\| \partial_\f^2 F\|_\infty|F|_{\al_+}}{\lambda^{\al_+}-(1\!-\!\Gamma)}
+ | \partial_\f F|_{\al_+}^{+}\biggr) + \al_- \lambda^{\al_-} \! \biggl(  
\frac{ \lambda^{\al_-} \| \partial_\f^2 F\|_\infty
|F|_{\al_-}}{1 - \lambda^{\al_-}(1\!-\!\Gamma)}+
|\partial_\f F|_{\al_-}^{-}\biggr) \! < \Gamma .
\label{eq:al+1} 
\end{equation}
For such $\al_-$ and $\al_+$, combining the bounds \eqref{eq:Hold+}, \eqref{eq:Hold-}, \eqref{eq:diff+} and \eqref{eq:diff-},
the map $\EEE$ turns out to be a contraction on $B_{\al_-,\al_+}$.
This concludes the proof of Theorem~\ref{thm:1}.

Recalling that $S_m<0<S_M$, by Remark \ref{barf0}, the discussion above implies that the sequence $\EEE^n[0]$ converges
to $W$ in $\gotB(\TTT^2,\RRR)$. Moreover the inequalities above are satisfied for $\al_-,\al_+$
small enough; in particular, if $\rho$ is small and $\norm F\norm_{\al_0,2}=O(\rho)$, then the condition $\lambda^{\al_-}(1-\Gamma) < 1$
requires $\al_-=O(\rho)$, which inserted into \eqref{eq:al+1}  allows $\al_+$ to be $O(1)$ in $\rho$.



In the scaling regime, where $F$ and  its derivatives are proportional to $\rho$,
the best we can say about the invariant manifold is that
$W \in B \cap B_{\al_-,\al_+}$
and hence $|W| \le \max\{|S_m|,S_M\}$, $\al_+=O(1)$ and $\al_-=O(\rho)$,
so that we have $\|W\|_{\io}=O(1)$, 
$|W|_{\al_-}^{-}=O(1)$ and  $|W|_{\al_+}^{+}=O(\rho)$.
This proves Corollary~\ref{lem:1}.

\begin{rmk} \label{rmk:2.1}
\emph{
If $\psi_0$ is a fixed point of $A_0$, that is $A_0\psi_0=\psi_0$, from \eqref{eq:2.3} we get 
$W(\psi_0)=S(\psi_0)$, so that, in the scaling regime, $W(\psi_0)$ does  not depend on $\rho$.
In a similar way, if $\psi_k$ is a periodic point of  period $k$, that is $A_0^k\psi_k=\psi_k$, 
then $W(\psi_k)$ is  found between $\min_i S(A_0^i \psi_k)$ and $\max_i S(A_0^i\psi_k)$. This 
shows that, in the scaling regime, in general $\lim_{\rho\to 0^+}W(\psi)\not=0$.
}
\end{rmk}

\subsection{The conjugation:~proof of Theorem~\ref{thm:2}}
\label{subsec:proof2}

\subsubsection{The translated map}

We construct the conjugation $\HHH$ by
linearizing the dynamics around the invariant manifold.
We write
$\f=\theta + W(\psi)$ so that, in terms of the variables $(\theta,\psi)$, the
dynamics is described by the map
\begin{equation}\label{eq:S1}
\SSS_1(\theta,\psi)=((\SSS_1)_\theta(\theta,\psi),(\SSS_1)_\psi(\theta,\psi)):=(G_1(\theta,\psi), A_0\psi ) , \qquad
G_1(\theta,\psi):=\theta+F_1(\theta,\psi) ,
\end{equation}
with 
\begin{equation}\label{eq:F1}
F_1(\theta,\psi) := F(\theta+W(\psi),\psi)-F(W(\psi),\psi) = 
F(\theta+W(\psi),\psi) + W(\psi) - W(A_0\psi) . 
\end{equation}
We call $\SSS_1$ the \emph{translated map}. Defining
%
$\Omega_1  := \{ (\theta,\psi) \in \TTT \times \TTT^2 \,:\,  \phi_m - W(\psi) \le \theta \le  \phi_M - W(\psi) \}$,
%
we have that $\SSS_1$ is injective from $\Omega_1$ into itself.

\begin{rmk}\label{rmk:twosteps}
\emph{
One easily checks that $\SSS_1(0,\psi)=(0,A_0\psi)$, so that, expressed in terms
of $\theta$, the invariant manifold reduces to
$\ovl\calW=\{(0,\psi):\psi\in\TTT^2\}$ (see Remark \ref{Wbar}).
}
\end{rmk}

\begin{rmk}\label{rmk:SnS1n}
\emph{
The iterations of $\SSS$ and $\SSS_1$ are such that
%
$(\SSS^n)_\f(\f,\psi)=(\SSS_1^n)_\theta(\f-W(\psi),\psi)+W(A_0^n\psi)$
%
and
$(\SSS_1^n)_\theta(\theta,\psi)=(\SSS^n)_\f(\theta+W(\psi),\psi)-W(A_0^n\psi)$,
while $(\SSS^n)_\psi(\f,\psi)=(\SSS_1^n)_\psi(\f-W(\psi),\psi)=A_0^n\psi$.
}
\end{rmk}

In Subsections~\ref{subsec:proofconj} and \ref{subsec:proof2-2} we study the conjugation relation
\vspace{-.1cm}
\begin{equation}\label{eq:Htilde}
\HHH_1 \circ \SSS_1 = \SSS_0\circ \HHH_1 .
\end{equation}
where $\HHH_1:\Omega_1\to\Omega_0$ is of the form
$\HHH_1(\theta,\psi)=(\HH_1(\theta,\psi),\psi)$. We can then write
\begin{equation} \label{HHH1}
\HHH(\f,\psi)  = (\HH_1(\f-W(\psi),\psi),\psi) ,
\vspace{-.1cm}
\end{equation}
with $\HHH$ satisfying the conjugation relation \eqref{eq:1.11}. Thus, if the conjugation $\HHH_1$ exists and is invertible,
the conjugation $\HHH$ exists and is invertible as well, and vice versa.
Moreover, $\HHH$ and $\HHH_1$ have the same image $\Omega_0$   defined in Theorem~\ref{thm:2}.
In analogy with \eqref{HHH1}, we write $\HHH_1^{-1}(\eta,\psi) = (\LL_1(\h,\psi),\psi)$.

Considering also Remark \ref{ambi},
we look for functions $\HH_1\!:\Omega_1  \to \RRR$ and $\LL_1:\Omega_0\to\RRR$ of the form
\vspace{-.1cm}
\begin{equation} \label{HHH1h1}
\HH_1(\theta,\psi) =\theta+\theta^2h_1(\theta,\psi), \qquad
\LL_1(\h,\psi) = \eta + \eta^2 l_1(\eta,\psi) .
\vspace{-.1cm}
\end{equation}
If we show that functions 
of the form \eqref{HHH1h1} exist,
with $h_1 \in \calmB_{\als,1}(\Omega_1,\RRR)$ and $l_1 \in\calmB_{\als,1}(\Omega_0,\RRR)$,
for a suitable $\als\in(0,\al_0)$, then we can write the conjugation $\HHH$ and its inverse as in \eqref{hl},
with
\vspace{-.1cm}
\begin{equation}\label{hh1-ll1}
h(\f,\psi) = h_1(\f-W(\psi),\psi) , \qquad l(\h,\psi)= l_1(\h,\psi).
\vspace{-.1cm}
\end{equation}
%

\begin{rmk} \label{rmk:strip}
\emph{
As a consequence of Theorem~\ref{thm:1}, there exists a closed interval
$\Theta:=[\theta_-,\theta_+]$,
such that $\Theta \times \TTT^2 \subset \Omega_1$ and hence
$\{(\f,\psi)\in \TTT\times \TTT^2 \,:\, \f -W(\psi) \in \Theta\} \subset
\Omega$.}
\end{rmk}

\subsubsection{Dynamics toward the invariant manifold} \label{subsec:proofconj}

From \eqref{eq:Htilde} we see that $\HH_1$ satisfies the equation
$\ka(\psi) \HH_1(\theta,\psi)= \HH_1(\SSS_1(\theta,\psi))$.
%
Using \eqref{HHH1h1}, we find
$h_1(\theta,\psi) = 
(\ka(\psi)\theta^2)^{-1} \bigl(G_1(\theta,\psi) + G_1(\theta,\psi)^2 h_1(\SSS_1(\theta,\psi)) - \ka(\psi) \, \theta \bigr)$,
so that, setting 
\vspace{-.2cm}
\begin{subequations}\label{g1g2}
\begin{align}
p_1(\theta,\psi):=&\frac{(G_1(\theta,\psi))^2}{\theta^2\ka(\psi)}=
\frac{1}{\ka(\psi)} \left( 1 + \int_0^1 \der t \, \partial_\f 
F(t\theta+W(\psi),\psi) \right)^2 ,
\label{g1g2a} \\
q_1(\theta,\psi):=&\frac{G_1(\theta,\psi)-\theta\ka(\psi)}{\theta^2\ka(\psi)} = 
\frac{1}{\ka(\psi)}\int_0^1 \der t \, (1-t) \, \partial_\f^2 
F(t\theta+W(\psi),\psi) \, , 
\label{g1g2b}
\end{align}
\end{subequations}
we obtain
%
$h_1(\theta,\psi)=q_1(\theta,\psi)+p_1(\theta,\psi) \, h_1(\SSS_1(\theta,\psi))$,
%
whose solution can be formally written
as
\vspace{-.1cm}
\begin{equation} \label{lll}
 h_1(\theta,\psi)=\sum_{n=1}^\infty p_1^{(n)}(\theta,\psi) \,
q_1(\SSS_1^n(\theta,\psi)) , \qquad
%
p^{(n)}_1(\theta,\psi) := \prod_{i=0}^{n-1} p_1(\SSS_1^i(\theta,\psi)) .
\vspace{-.2cm}
\end{equation}
where, according to our conventions (see Remark \ref{convention}),we have $p^{(0)}_1(\theta,\psi)=1$.

\begin{rmk} \label{zero}
\emph{
Both $p_1(\theta,\psi)$ and $q_1(\theta,\psi)$ are well defined 
at $\theta=0$. Using \eqref{eq:2.3} and the definition 
of $\ka(\psi)$ after \eqref{eq:sss0}, we find $\ka(\psi) \, q_1(0,\psi) = 
\partial_\f^2 F(0,\psi)$ and $p_1(0,\psi)=\ka(\psi)$. 
}
\end{rmk}

We start by studying the regularity properties of the functions $p_1$ and $q_1$. The 
following result is an easy consequence of the representation in \eqref{g1g2} 
together with Lemma~\ref{lem:easy}.

\begin{lemma}\label{lempq}
Assume $\SSS$ in \eqref{eq:1.1} to satisfy Hypotheses
\ref{hyp:1}--\ref{hyp:3}.
For any $\al\in (0,\min\{\al_-,\al_+\})$, with $\al_-$ and $\al_+$ as in Theorem~\ref{thm:1},
and any $\Gamma'\in(0,\Gamma)$,
one has $p_1,q_1\in \calmB_{\al,5}(\Omega_1,\RRR)$ and
$p_1(\theta,\psi)\leq 1-\Gamma'$
as long as $|\theta|\leq \theta_1$, with
\vspace{-.1cm}
\begin{equation} \label{theta1}
\theta_1 := \frac{\Gamma-\Gamma'}{2\|\ka^{-1} \|_{\io} (1+\|\partial_\f F
\|_{\io}) \|\partial_\f^2 F\|_{\io}} .
\vspace{.1cm}
\end{equation}
\end{lemma}

\noindent\emph{Proof.}
Observe that $p_1(0,\psi)=1+\partial_\f F(W(\psi),\psi) \le 1 - \Gamma$. Thus,
using that, for any $\theta_1>0$ and all $\theta\in[-\theta_1,\theta_1]$, we have
$|p_1(\theta,\psi) - p_1(0,\psi)|\le \|\partial_\theta p_1\|_\io \theta_1$,
if we bound $\partial_\theta p_1$ using \eqref{g1g2a} and fix $\theta_1$ as in
\eqref{theta1}, the bound on $p_1(\theta,\psi)$ for $|\theta|\le\theta_1$
follows immediately. The other bounds too are easily obtained by estimating the
derivatives of $p_1$ and $q_1$.
\qed

\begin{rmk} \label{rmktolempq}
\emph{
In the scaling regime, where $\Gamma=\rho\,\ga$,  one has
$\norm p_1\norm_{\al,5}=1+O(\rho)$ and
$\norm q_1\norm_{\al,5}=O(\rho)$,
while $\norm \partial_\theta p_1\norm_{0,4}=O(\rho)$ and $|p_1|_{\al}=O(\rho)$.
Moreover, if in Lemma~\ref{lempq} one takes $\Gamma'=\rho\,\ga'$
with $\ga'\in(0,\ga)$, then \eqref{theta1} implies that $\theta_1=O(1)$.
}
\end{rmk}

Next, we study the regularity of the iterates of $\SSS_1$.

\begin{lemma}\label{lemSSSn}
Assume $\SSS$ in \eqref{eq:1.1} to satisfy Hypotheses
\ref{hyp:1}--\ref{hyp:3}.
For any $\al$ $\in (0,\min\{\al_-,\al_+\})$, with $\al_-$ and $\al_+$ as in Theorem~\ref{thm:1},
and any $\Gamma' \in (0,\Gamma)$, and all $n\geq 0$, one has $(\SSS^n_1)_\theta \! \in 	\! \calmB_{\al,2}(\Omega_1,\RRR)$ and
\vspace{-.1cm}
\[
\norm \partial_\theta  (\SSS^n_1)_\theta\norm_ {0,1}  \leq D_{0}  \left(1-\Gamma'\right)^{n} ,
\qquad
| (\SSS^n_1)_\theta |_ {\al}  \leq D_{0} \lambda^{\al n} ,
\qquad
| \partial_\theta (\SSS^n_1)_\theta |_ {\al}  \leq D_{0} \lambda^{\al n} ,
\]
with the constant $D_{0}$ depending on $F$ and $\Gamma'$ but not on $n$.
\end{lemma}

\proof
For a fixed $\Gamma'\in(0,\Gamma)$, let $r=r(\Gamma')$ and $N_{r}$ be defined as in Lemma
\ref{lem:easy}, and observe that $N_{r}=O((\Gamma')^{-1})$ in such a case. We have
$\|\partial_\theta G_1\|_\infty\leq 1+\|\partial_\f F\|_\infty$, while 
for $n \ge N_{r}$ we may bound
$\|(\partial_\theta G_1)\circ\SSS^{n}_1\|_\infty\le
\|(1+\partial_\f F)\circ\SSS^{n}\|_\infty\le (1-\Gamma')$.
Thus, noting that $(\SSS^n_1)_\theta=G_1\circ\SSS^{n-1} _1$, we get
\vspace{-.2cm}
\begin{equation} \label{boundSSS1}
\|\partial_\theta(\SSS^n_1)_\theta\|_\infty
\le \prod_{i=0}^{n-1}\|(\partial_\theta G_1)\circ\SSS^{i}_1\|_\infty\le
c_1 \left(1-\Gamma'\right)^{n} ,
\vspace{-.1cm}
\end{equation}
with $c_1 \!:=\! ( \, (1\!+\!\|\partial_\f F\|_\infty )/ (1 \!-\! \Gamma' ) )^{N_{r}}$,
%
%
and 
%
$|G_1\circ\SSS^{n-1}_1|_{\alpha} \!\le\! \|(\partial_\theta G_1)\circ \SSS^{n-1}_1\|_\infty |G_1\circ\SSS^{n-2}_1|_{\alpha} \!+\!  \lambda^{\al (n-1)}  |G_1|_\al$,
%
where $|G_1|_\al \le \| \partial_\f F \|_\io |W|_\al + |F|_\al$,
and hence, iterating, we get
\vspace{-.2cm}
\begin{equation}\label{boundSSS1al}
\null\hspace{-.2cm}
|(\SSS^n_1)_\theta |_{\alpha} 
\le |G_1|_\al\sum_{i=1}^{n} \lambda^{\al(n-i)}\prod_{j=1}^{i-1} \| (\partial_\theta G_1)\circ \SSS^{n-j}_1\|_\infty
\le C_{0}|G_1|_\al \, \lambda^{\al n}\sum_{i=1}^{n}
\left(1 \!-\! \Gamma'\right)^{i-1}
\lambda^{-\al i}\leq
c_2\lambda^{\al n} ,
\vspace{-.1cm}
\end{equation}
with 
%
$c_2 := c_1 |G_1|_\al \, (\Gamma')^{-1}$.
%
Exploiting \eqref{boundSSS1}, we find that
\vspace{-.2cm}
\begin{equation} \label{boundSSS111}
\| \partial_\theta^2 (\SSS^n_1)_\theta \|_\io 
\le \sum_{i=0}^{n-1} 
\left\| (\partial_\theta^2 G_1) \circ \SSS_1^i \right\|_{\io}
\left\| \partial_\theta (\SSS^i_1)_\theta \right\|_\io
\prod_{\substack{ j=0 \\ j \neq i} }^{n-1} \| (\partial_\theta G_1)\circ \SSS^{j}_1 \|_\io \le c_3 \left(1-\Gamma'\right)^{n} ,
\vspace{-.2cm}
\end{equation}
%
%
%
with
%
$c_3 := c_1 \|\partial_\f^2 F\|_{\io} \, (\Gamma')^{-1} $.
%
Finally, noting that
\vspace{-.2cm}
\[
\left| \partial_\theta (\SSS^n_1)_\theta \right|_{\al} \le \sum_{i=0}^{n-1} 
\Bigl(
\left\| \partial_\theta^2 G_1\right\|_\io \left| (\SSS^i_1)_\theta \right|_{\al} +
\lambda^{\al i }  |\partial_\theta G_1|_\al
\Bigr) 
\prod_{\substack{ j=1 \\ j \neq i} }^{n-1} \| (\partial_\theta G_1)\circ \SSS^{j}_1 \|_\io ,
\vspace{-.2cm}
\]
and proceeding as done to get the bound \eqref{boundSSS1al}, we obtain
\vspace{-.1cm}
\begin{equation} \label{boundSSS111111}
| \partial_\theta (\SSS^n_1)_\theta |_ {\al}  \le c_4 \lambda^{\al n} ,
\vspace{-.1cm}
\end{equation}
with
%
$c_4 := \bigl( c_2 \|\partial_\f^2 F\|_{\io} + |\partial_\theta G_1|_\al \bigr) \, (\Gamma')^{-1}$.
%
Collecting together the bounds \eqref{boundSSS1}--\eqref{boundSSS111111}, the assertion follows
follow with $D_{0}=\max\{c_1,c_2,c_3,c_4\}$.
\qed

\begin{rmk} \label{rmktolemSSSn}
\emph{
In the scaling regime, where $\Gamma=\rho\,\ga$, if we choose $\Gamma'=\rho\ga'$, with $\ga'\in(0,\ga)$,
we find that the constant $D_{0}$ is $O(1)$ in $\rho$. Indeed both $\norm F\norm_{0,2}$ and $|F|_{\al}$, and hence $|G_1|_\al$ as well,
are $O(\rho)$, while 
one has $N_{r}=O(\rho^{-1})$ by \eqref{Nr}.
}
\end{rmk}

\begin{rmk} \label{dopo}
\emph{
Since  $(\SSS^n_1)_\theta(0,\psi)=0$ for all $n\in\ZZZ$, we bound also
$|(\SSS^n_1)_\theta(\theta,\psi)| \le \|\partial_\theta (\SSS^n_1)_\theta \|_\io |\theta|$ and hence
$\| (\SSS^n_1)_\theta\|_\io \le c_1  ( \max\{|\phi_m|,\phi_M\} + \max\{|S_m|,S_M\} ) (1-\Gamma')^{n}$.
%
}
\end{rmk}

\subsubsection{Existence and regularity of the conjugation} \label{subsec:proof2-2}

We have all the ingredients to complete the proof of Theorem~\ref{thm:2}.
Fix $\Gamma'\in(0,\Gamma)$ and set $r=r(\Gamma')$, 
$S_r :=\max\{S_M,|S_m|\}+r $ and $M_r := \max\{2 \, (\Gamma')^{-1} \log (S_r/\theta_1),0\}$,
with the notation of Lemma~\ref{lem:easy} and $\theta_1$ as in \eqref{theta1}.
If $(\theta+W(\psi),\psi)\in \Lambda_{r}$ and $n>M_r$, then  
$|(\SSS_{1}^{n}(\theta,\psi))_\theta|\leq \theta_1$ and hence, thanks to the property \ref{prop-Nr} in Lemma~\ref{lem:easy},
$|(\SSS_{1}^{n}(\theta,\psi))_\theta|\leq \theta_1$ for all $(\theta,\psi)\in\Omega_1$ and for 
all $n \ge M_r':=M_r+N_{r}$. Thus, for $n \ge M_r'$,  we obtain
%
$ \| p_1^{(n)} \|_\infty
\le \| p_1 \|_\infty^{M_r'}
(1-\Gamma')^{n-M_r'} $
%
which, inserted in \eqref{lll}, implies
\vspace{-.1cm}
\begin{equation} \label{boundh}
\|h_1\|_\infty\leq 
D_1 \sum_{n=0}^\io 
\left(1- \Gamma'\right)^n\| 
q_1 \|_\infty\leq D_1 (\Gamma')^{-1} \| q_1\|_\infty , \qquad \qquad
D_1 := \left( \frac{\| p_1\|_\infty}{1- \Gamma'}\right)^{M_r'} .
\vspace{-.1cm}
\end{equation}
Furthermore, using Lemma~\ref{lemSSSn}, we see that 
\vspace{-.2cm}
\begin{equation}\label{boundgnp}
\bigl\|\partial_\theta p_1^{(n)}\bigr\|_\infty
\leq 
D_{0} D_1\|\partial_\theta p_1\|_\infty
\left(1-\Gamma'\right)^{n-1}
\sum_{i=0}^{n-1}  
\left(1-\Gamma'\right)^i 
\leq D_{0} D_1 (\Gamma')^{-1} \| \partial_\theta p_1 \|_\infty
\left(1- \Gamma'\right)^{n}\, ,
\vspace{-.1cm}
\end{equation}
with $D_{0}$ as in Lemma~\ref{lemSSSn} and $D_1$ as in \eqref{boundh}, so that, summing over $n$, we get
\vspace{-.1cm}
\begin{equation}\label{boundhder}
\|\partial_\theta h_1 \|_\infty\leq \sum_{n=1}^\infty \left(
\bigl\|\partial_\theta p_1^{(n)}\bigr\|_\infty\|q_1\|_\infty+\bigl\|p_1^{(n)}\bigr\|_\infty
\|\partial_\theta  (q_1\circ\SSS_1^n)\|_\infty\right)
\leq D_2 \, (\Gamma')^{-1} \norm q\norm_{0,1}\,,
\vspace{-.1cm}
\end{equation}
for some constant $D_2$.
Proceeding along the same lines, 
we obtain also (see Appendix~\ref{B0} for details)
%
\begin{equation} \label{boundhder-bis}
\| \partial_\theta^2 h_1 \|_{\infty} \leq D_3 \, (\Gamma')^{-1} \norm q\norm_{0,2}\,,
\end{equation}
for some other constant $D_3$.
Finally, for any $\al\in(0,\min\{\al_-,\al_+\}]$, we get, again relying on Lemma~\ref{lemSSSn}, 
\vspace{-.2cm}
\begin{equation*}  
\bigl| p^{(n)}_1\bigr|_\al
\leq D_{0} D_1 (|p_1|_\al+\|\partial_\theta p_1\|_\infty)
\left(1-\Gamma'\right)^{n-1}
\frac{\lambda^{\al n}-1}{\lambda^{\al}-1}\leq D_4 \, \lambda^{\al n}
\left(1-\Gamma'\right)^n \,,
\end{equation*}
for a suitable constant $D_4$  proportional to $D_1$. 
To sum over $n$ we take $\al=\als$ in Lemma~\ref{lemSSSn}, with $\als$ such that
$\lambda^{\als}(1-\Gamma')<1-\Gamma''$, with $\Gamma''<\Gamma'$, so as to obtain
\vspace{-.1cm}
\begin{equation}\label{boundhderal}
| h_1|_{\als} \leq \sum_{n=1}^\infty \left( \bigl| p_1^{(n)}\bigr|_{\als} \|q\|_\infty+\bigl\|p_1^{(n)}\bigr\|_\infty
|q_1\circ\SSS_1^n|_{\als}\right) \le
D_4\,(\Gamma'')^{-1} \norm q_1\norm_{\als,1} ,
\vspace{-.1cm}
\end{equation}
for a suitable constant $D_4$ proportional to $D_1$.
Once more we bound $| \partial_\theta h_1 |_{\als}$
by reasoning in a similar way (again we refer to Appendix~\ref{B0} for details) and find
\vspace{-.1cm}
\begin{equation}\label{boundhderal-bis}
| \partial_\theta h_1 |_{\als} \leq D_5\,(\Gamma'')^{-1} \norm q_1\norm_{\als,1} ,
\end{equation}
with $D_5$ proportional to $D_2$.
The bounds \eqref{boundh} and \eqref{boundhder}--\eqref{boundhderal-bis}
ensure  that $\|h_1\|_{\als,2}$ is bounded.
Therefore, recalling the first relation in \eqref{hh1-ll1}, Theorem~\ref{thm:2}
is proved, with $\Omega_0= \HHH_1(\Omega_1)=\HHH(\Omega)$.

\begin{rmk} \label{HinLambda}
\emph{
The argument above shows that it is enough to prove the existence of the conjugation
inside $\Lambda$.
Indeed, once the conjugation has been defined in $\Lambda$, it can be extended to the whole $\Omega$ by using
that all trajectories fall inside a neighborhood of the attracting
invariant manifold in a finite time.
}
\end{rmk}


With the notation of Lemma~\ref{lem:easy}, for $r=r(\Gamma')$ we have $N_{r}=O((\Gamma')^{-1})$. 
Thus, from Remark \ref{rmktolemSSSn} it is easy to see that in the scaling regime
the constant $D_{0}$ in Lemma~\ref{lemSSSn} is $O(1)$ in $\rho$.
We can take $\Gamma'=\rho\gamma'$ and $\Gamma''=\rho\gamma''$, with $0<\gamma''<\gamma'<\gamma$.
Using that $\theta_1=O(1)$ in $\rho$ (see Remark \ref{rmktolempq}),
so that $M_r=O(\rho^{-1})$ and hence $M_{r}'=O((\Gamma')^{-1})$ as well, and requiring
$\als$ in Theorem~\ref{thm:2} to be such that
$\lambda^{\als}(1-\rho\,\gamma')<(1-\rho\gamma'')$, so that $\als=O(\rho)$,
we easily check that $D_1$ in \eqref{boundh} and, as a consequence,
the constants $D_2$, $D_3$ and $D_4$ as well are all $O(1)$ in $\rho$.
This implies that both $\norm h \norm_{0,1}$ and $|h|_{\al_*}$ are $O(1)$ in $\rho$.
Collecting together the bounds above immediately implies Corollary \ref{lem:2}.

\subsubsection{The inverse conjugation} 
\label{subsec:proofcor}

Since $\HH_1(\theta,\psi)=\theta+\theta^2h_1(\theta,\psi)$ and $h_1\in \calmB_{\als,1}({\Omega_1},\RRR)$ for a suitable $\als>0$,
there exists $\theta_{*} \le \theta_1$ such that
$\partial_\theta\HH_1(\theta,\psi)>1/2$ for $|\theta|<\theta_{*}$.
We have
$\HH_1(\theta,\psi)= (\ka^{(n)}(\psi))^{-1}\HH_1(\SSS_1^{n}(\theta,\psi))$
for all $(\theta,\psi)\in\Omega_1$ and all $n\in\NNN$.
If we reason like deriving 
the bound on $ \| p_1^{(n)} \|_\infty$ in Subsection~\ref{subsec:proof2-2}
we see that there exists 
$M_2$ such that $|(\SSS_1^{M_2}(\theta,\psi))_\theta|\leq\theta_{*}$. Thus, 
by Lemma~\ref{lem:easy} and Hypothesis~\ref{hyp:1},
we get, for all $(\theta,\psi) \in \Omega_1$,
\vspace{-.2cm}
\begin{equation} \label{chi}
\partial_\theta\HH_1(\theta,\psi)\geq  
\|\ka\|_\infty^{-M_2}\inf_{|\theta| \le \theta^{\star}}
\partial_\theta\HH_1(\theta, \psi)\left( \inf_{(\theta,\psi) \in \Omega_1}
\partial_\theta G_1(\theta,\psi)\right)^{M_2} =: \tau_1 >0 .
\vspace{-.1cm}
\end{equation}
%
By the inverse function theorem, there is
$l_1\!\in\!\calmB_{0,1}(\HHH_1(\Omega_1,\RRR))$ such that
$(\HHH_1^{-1})_\eta (\eta,\psi)\!=\!\eta+\eta^2l_1(\eta,\psi)$.
Furthermore, for any $\psi,\psi' \in\TTT^2$, we can write
$\HH_1 (\HH_1^{-1}(\eta,\psi'),\psi')- \HH_1 (\HH_1^{-1}(\eta,\psi),\psi)=0$,
so that
$|\HH_1 (\HH_1^{-1} (\eta,\psi'),\psi')-
\HH_1 (\HH_1^{-1} (\eta,\psi),\psi')|
=
| \HH_1 (\HH_1^{-1}(\eta,\psi),\psi')-
\HH_1(\HH_1^{-1}(\eta,\psi),\psi)|$,
which, divided by $|\psi'-\psi|^{\als}$, gives
%
$\partial_\theta\HH_1(\theta,\psi) 
\, |\HH_1^{-1}|_{\als}\leq |\HH_1|_{\als}$ for all $(\theta,\psi)\in\Omega_1$,
%
so that we obtain $|\HH_1^{-1}|_{\als} \leq \tau_1^{-1} |\HH_1|_{\als}$.
Therefore, 
also $l_1 \in \calmB_{\als,1}(\HHH_1(\Omega_1),\RRR)$
and, by the second relation of \eqref{hh1-ll1}, Corollorary \ref{l} follows.

\begin{rmk} \label{chibar}
\emph{
By reasoning as in the derivation of \eqref{chi}, we find that there exists $\bar\tau>0$ such that
$\partial_\theta \ovl \HH(\theta) \ge  \bar\tau$ for all $\theta \in \calU$. This will be used later on (see Subsection~\ref{devinverse}).
}
\end{rmk}

\subsection{Physical measure:~proof of Theorem~\ref{thm:mixing}} \label{proofthmpm}

Let $\calO_1$ and $\calO_2$ be two observables in $\calmB_{\als,1}(\Omega,\RRR)$.
By Remark \ref{dopo}, we have
\vspace{-.1cm}
\[
| \calO_2(\SSS^n(\f,\psi))-\calO_2(W(A_0^n\psi),A_0^n\psi)|\le C(1-\Gamma')^n\norm\calO_2\norm_{0,1} ,
\vspace{-.1cm}
\]
for a suitable constant $C$.
Thus, if $\nu_W$ and $\nu_0$ are defined as in Subsection~\ref{2.3.3}, we find
\vspace{-.1cm}
\[
\begin{aligned}
&\left|\nu_0\left(\calO_1\,\calO_2\circ\SSS^n\right)-\nu_0\left(
\calO_1\right)\nu_W\left(\calO_2\right)\right| \\
&\quad \le \int 
d\f\left|\aver*{\calO_1(\f,\cdot)\calO_2(W(A_0^n\cdot),A_0^n\cdot)}-
\aver{\calO_1(\f,\cdot)}\aver{\calO_2(W(\cdot),\cdot)}\right|+
C(1-\Gamma')^n\norm\calO_2\norm_{0,1}\|\calO_1\|_\infty .
\end{aligned}
\vspace{-.2cm}
\]
Since $W\in \gotB_{\al_-}^-(\TTT^2,\RRR)$,
relying on Remark \ref{unia}, we obtain the bound in Theorem~\ref{thm:mixing}.

\zerarcounters 
\section{Averaging and deviations} \label{sec:5}

The study of the convergence of the dynamics of $\SSS$ to the deterministic dynamics $\Phi_{n\rho}$ is
structured in several steps:  we start with the first and second moments of
$W$ (Subsections~\ref{corrineq} and \ref{subsec:proof3});
then we consider the moments of the functions $h$ and $l$ and their derivatives (Subsections~\ref{fluk} to \ref{devinverse});
eventually we draw the conclusions about the deviations of the dynamics
from the averaged system (Subsection~\ref{proof4}).

\begin{rmk} \label{constantC}
\emph{
Throughout this section, as well as in the appendices we refer to for the
proofs (actually from Appendix~\ref{app:tech} on), we assume $\rho$ to be such
that Hypotheses~\ref{hyp:1} to \ref{hyp:3} hold
(see Remark \ref{Psi0ge1}) and
we call $C$ any constant independent of $\rho$ whose numerical value is not relevant.
}
\end{rmk}

\subsection{The extended map} \label{extension}

Even though the set $\Omega$ is positively invariant for the map $\SSS$ in \eqref{eq:1.1bis},
it is useful to extend the map $\SSS|_{\Omega}$ outside $\Omega$.
To do this, we proceed as follows:
\vspace{-.2cm}
\begin{itemize}
\itemsep-0.1em
\item let $\chi_{\rm ext}\!:\RRR\to\RRR$ be a $C^\infty$ function such that
$\chi_{\rm ext}(x)=1$ for $x\le0$ and $\chi_{\rm ext}(x)=0$ for $x\ge1$ while
$\partial_x\chi_{\rm ext}(x)\leq 0$ for every $x\in\RRR$;
\item set $s_M:=\min\{1,\nu_M/2\}$,
where $\nu_M :=\inf_{\psi\in\TTT^2} |f(\f_M,\psi)|$;
\item define $f_{\rm ext} \in\calmB_{\al_0,6}(\RRR\times\TTT^2,\RRR)$,
by setting $f_{\rm ext}(\f,\psi)=f(\f,\psi)$ for $\f_m\leq \f\leq \f_M$
while
\vspace{-.2cm}
\[
f_{\rm ext}(\f,\psi)=\biggl(\sum_{i=0}^6 
\partial_\f^i f(\f_M,\psi)(\f-\f_M)^i \biggr)
\chi_{\rm ext}\left(\frac{\f-\f_M}{s_M}\right)- \frac{\nu_M}2 \left(1-
\chi_{\rm ext}\left(\frac{\f-\f_M}{s_M}\right)\right),
\vspace{-.3cm}
\]
for $\f> \f_M$, and an analogous expression for $\f<\f_m$.
\vspace{-.1cm}
\end{itemize}
One has 
$\norm f_{\rm ext}\norm_{\alpha_0,6}\leq s_M^{-6} \norm\chi_{\rm ext}\norm_{0,6}
\norm F\norm_{\alpha_0,6}$
and 
$\inf_{(\f,\psi)\in\RRR \times\TTT^2\backslash\Lambda_r} |f_{\rm ext}(\f,\psi)|\geq\frac12\inf_{(\f,\psi)\in\Omega\backslash\Lambda_r} |f(\f,\psi)|$
for any $r>0$.
Moreover the map
%
$\SSS_{\rm ext}(\f,\psi):=(\f+\rho f_{\rm ext}(\f,\psi),A_0\psi)$
%
is defined on $\RRR\times\TTT^2$, coincides with $\SSS(\f,\psi)$ for
$(\f,\psi)\in\Omega$ and, restricted to any $\Omega_{\rm ext}=\calU_{\rm
ext}\times \TTT^2$, with $\calU_{\rm ext}\supseteq \calU$ a closed interval,
satisfies Hypotheses~\ref{hyp:1}--\ref{hyp:3} with $\Omega_{\rm ext}$ instead of  $\Omega$.
Extending $\SSS(\f,\psi)$ on $\Omega_{\rm ext}$ naturally defines a map 
$\SSS_{\rm 1, ext}(\theta,\psi) := \SSS_{\rm ext}(\theta+W(\psi),\psi)-W(A_0\psi) $ 
on
$\Omega_{1,{\rm ext}} :=\{ (\theta,\psi)\in \RRR\times\TTT^2  : (\theta + W(\psi),\psi) \in \Omega_{\rm ext} \}$.
In the following discussion we need to choose $\Omega_{\rm ext}$ in such a way
that $\Omega\subset \Omega_{1,{\rm ext}}$ (see Subsection~\ref{proof3-2}).
The conjugation $\HH$ as well can be extended to a function $\HH_{\rm ext} \in
\calmB_{\al_0,1}(\Omega_{\rm ext},\RRR)$,
by reasoning as in Subsection~\ref{subsec:proof2} with
$\SSS_{\rm ext}$ instead of $\SSS$ everywhere.

\begin{rmk} \label{whyext}
\emph{
The reason why we need to extend $\SSS|_\Omega$ outside $\Omega$ to a map
$\SSS_{\rm ext}$, potentially different from the original $\SSS$ on
$(\TTT\times\TTT^2) \backslash
\Omega$, is that we aim to compare $\SSS$ with other maps 
which, albeit being constructed starting from $\SSS$,
not only may fail to admit $\Omega$ as an invariant set (see
Remark \ref{rmk:EEE0}), but also are not necessarily defined in the whole $\Omega$.
In order not to discuss separately the dynamics near the boundary of $\Omega$, it turns out to be easier to extend the maps to a larger domain
$\Omega_{\rm ext}$ in such a way that they satisfy automatically the same properties as $\SSS$.
We stress here that all the functions appearing in Theorems~\ref{thm:3} to \ref{thm:prob} depend
only on $\SSS|_\Omega$ and, when they are estimated, the errors introduced as an effect of
the arbitrariness of the extensions are under control (see Remark \ref{errext}).
}
\end{rmk}


\subsection{A correlation inequality} \label{corrineq}

This subsection is dedicated to a generalization of Proposition~\ref{prop:4.3} that
plays a central role in the proof of Theorem~\ref{thm:3} and Lemma~\ref{lem:flulin}. 

For any two given sets of functions $\vv_0,\ldots,\vv_{n-1}$
and
$\ww_0,\ldots,\ww_{n-1}$ on $\TTT^2$,
set $\uu_i=\vv_i + \rho \, \ww_i$ for $i=0,\ldots,n-1$. Define, for $k=0,\ldots,n$
and $i=0,\ldots,n-k$, 
%
\vspace{-.2cm}
\begin{equation} \label{produ}
\uu^{(k)}_i(\psi) := 
\prod_{j=0}^{k-1} \uu_{i+j}(A_0^{j}\psi) ,
\qquad
\uu^{(k)}(\psi):=\uu^{(k)}_0(\psi) ,
\qquad
\uu^{(-k)}(\psi) := \prod_{j=1}^{k} \uu_{k-j}(A_0^{-j}\psi) = \uu^{(k)} \circ  A_0^{-k} ,
\vspace{-.2cm}
\end{equation}
%
%
%
with $\uu_i{\vphantom{-}}^{(0)}(\psi) 
=1$ according to Remark \ref{convention}, 
and define similarly $\vv^{(k)}_i(\psi)$, $\vv^{(k)}(\psi)$ and $\vv^{(-k)}(\psi)$.

We write
$\uu^{(k)}_i \!\!-\! \vv^{(k)}_i = 
\rho \, \uu^{(k-1)}_i \, \ww_{i+k-1} \circ A_0^{k-1} + ( \uu^{(k-1)}_i \!\!-\! \vv^{(k-1)}_i ) 
\vv_{i+k-1} \circ A_0^{k-1} $,
so that, iterating, we get 
%
\vspace{-.2cm}
\begin{equation} \label{eq:mnsum}
\begin{aligned}
\uu^{(k)}_i = \vv^{(k)}_i & + \rho \sum_{j=0}^{k-1}  \uu^{(j)}_i \, \ww_{i+j} \circ A_0^j \, \vv^{(k-j-1)}_{i+j+1} \circ A_0^{j+1}
=  \vv^{(k)}_i  + \rho \sum_{j=0}^{k-1}  \vv^{(j)}_i \, \ww_{i+j} \circ A_0^j \, \vv^{(k-j-1)}_{i+j+1} \circ A_0^{j+1} \\
& + \rho^2 \sum_{j=1}^{k-1} \sum_{j'=0}^{j-1} \uu^{(j')}_i \, \ww_{i+j'}\circ A_0^{j'} \, \vv^{(j-j'-1)}_{i+j'+1} \circ A_0^{j'+1} \,
\ww_{i+j} \circ A_0^{j} \, \vv^{(k-j-1)}_{i+j+1} \circ A_0^{j+1} .
\end{aligned}
\vspace{-.1cm}
\end{equation}
where the second equality follows applying the first equality to the
factor $\uu^{(j)}_i$ in the first line; 
according to the convention in Remark \ref{convention},
the sum in the second line of \eqref{eq:mnsum} vanishes for $k=0$, while the sum in the last line vanishes for $k=0,1$. 
We can also proceed ``in the opposite direction'' to get
\vspace{-.1cm}
\begin{equation}\label{eq:summn}
\uu^{(k)}_i = \vv^{(k)}_i + \rho \sum_{j=0}^{k-1}  \vv^{(j)}_i \, \ww_{i+j} \circ A_0^j \, \uu^{(k-j-1)}_{i+j+1} \circ A_0^{j+1} ,
\vspace{-.1cm}
\end{equation}
where we avoid writing the equivalent of the second expansion in \eqref{eq:mnsum} since we will not need it.

The following result plays a key role in the forthcoming
analysis. The proof is based on \eqref{eq:mnsum}. 

\begin{prop} \label{prop:decay}
Let $\uu_0,\ldots,\uu_{n-1}$ be any functions in
$\gotB_{\al}(\TTT^2,\RRR)$, with $\alpha\in (0,1]$, such that
\vspace{-.1cm}
\begin{enumerate}
\itemsep-0.2em
\item $0<\aver{\uu_i} \le \|\uu_i\|_{\io} \le 1-\rho\,\ga$ for all $i=0,\ldots,n-1$,
\item $\tilde \uu_i=O(\rho)$ for all $i=0,\ldots,n-1$,
\vspace{-.1cm}
\end{enumerate}
and set $\aver{\uu}^{(n)}:=\aver{\uu_0} \aver{\uu_1} \ldots \aver{\uu_{n-1}}$.
Given $g_+\in\gotB^+_{\al}(\TTT^2,\RRR)$ and $g_-\in\gotB^-_{\al}(\TTT^2,\RRR)$, one has
\vspace{-.1cm}
\begin{equation*}  
\begin{aligned}
& \Bigl| \aver{ g_+\,\uu^{(n)}\circ A_0\,g_- \circ  A_0^{n+1} } -\aver{\uu}^{(n)}\aver{g_+}\aver{g_-} \Bigr| \\
& \quad \le C(1\!-\!\rho\,\ga)^n\Bigl( (1\!+\!\al n) \lambda^{-\al n} \|\tilde g_+\|_{\al}^+\|\tilde g_-\|_{\al}^-
+ \rho \bigl( \| \tilde g_+\|_{\al}^+\| g_-\|_{\al}^- + \|g_+\|_{\al}^+\| \tilde g_-\|_{\al}^- \bigr)
+ \rho^2 n |\aver{g_+}||\aver{g_-}| \Bigr)  .
\end{aligned}
\vspace{.1cm}
\end{equation*}
\end{prop}

\proof For any function $\uu_i$, with $i=0,\ldots,n-1$,
let $\vv_i:=\aver{\uu_i}$ and $\rho\,\ww_i:=\tilde\uu_i$
and introduce the notation $\aver{\uu}^{(k)}_i
:=\aver{\uu_i}\ldots\aver{\uu_{k-1}}$,
so that $\aver{\uu}^{(k)}=\aver{\uu}^{(k)}_0$.
Then we write
%
\begin{equation} \label{eq:1+2+3}
 \aver{g_+\,\uu^{(n)}\circ A_0\, g_-\circ A_0^{n+1}}=
\aver{g_+ \, \uu^{(n)}\circ A_0\, \tilde g_-\circ A_0^{n+1}}+
\aver{\tilde g_+\,\uu^{(n)}\circ A_0}\aver{g_-}+
\aver{g_+}\aver{\uu^{(n)}}\aver{g_-} ,
\end{equation}
so that, using the first line of \eqref{eq:mnsum}, 
with $\uu^{(n)}\circ A_0$ instead of $\uu^{(n)}$,
we rewrite the first term in \eqref{eq:1+2+3} as
\vspace{-.1cm}
\[
\aver{ g_+ \, \uu^{(n)}\circ A_0 \, \tilde g_-\circ
A_0^{n+1}} = \aver{\uu}^n\aver{ g_+ \, \tilde g_-\circ
A_0^{n+1}} 
+ \rho \sum_{k=0}^{n-1} \aver{\uu}^{n-k-1}_{k+1} \aver{ g_+ \, \uu^{(k)}\circ
A_0 \,
\tilde\uu_{k} \circ A_0^{k+1} \, \tilde g_- \circ A_0^{n+1} }  .
\vspace{-.1cm}
\]

Using \eqref{eq:1.6a} and \eqref{def:lbar}, we get
\[
\begin{aligned}
 \bigl| g_+ \circ A_0^{-(k+1)}\uu^{(k)} \!\circ\!  A_0^{-k} \, \tilde \uu_k
\bigr|_{\al}^{+}
& \leq \| g_+ \circ A_0^{-(k+1)} \, \tilde \uu_k \bigr\|_{\io} \bigl| \uu^{(k)}
\circ  A_0^{-k} \bigr|_{\al}^{+} +
\bigl| g_+ \circ A_0^{-(k+1)} \, \tilde \uu_k \bigr|_{\al}^+ \| \uu^{(k)}
\|_{\io}  \\
& \le C (1-\rho\,\ga)^{k}\|g_+\circ A_0^{-(k+1)} \, \tilde \uu_k\|_\al^+ ,
\end{aligned}
\]
so that, from \eqref{eq:1.6p} and Proposition~\ref{prop:4.3}, we obtain in
\eqref{eq:1+2+3}
\begin{equation} \label{eq:1}
\bigl|\aver{ g_+ \, \uu^{(n)}\circ A_0 \, \tilde g_- \circ
A_0^{n+1}}\bigr|
\leq
C(1-\rho\,\ga)^n\lambda^{-\al n} (1+\al n) \|\tilde g_+\|_{\al}^+\|\tilde
g_-\|_{\al}^-+
C\rho(1-\rho\,\ga)^n \| g_+\|_{\al}^+\|\tilde g_-\|_{\al}^-  \, .
\end{equation}

Analogously to \eqref{eq:1}, using \eqref{eq:summn}, with $\uu^{(n)}\circ A_0$
instead of $\uu^{(n)}$, and Proposition~\ref{prop:4.3}, we get
\vspace{-.1cm}
\begin{equation} \label{eq:2}
\begin{aligned}
\bigl|\aver{ \tilde g_+ \, \uu^{(n)}\circ A_0}\bigr|
 \le \rho \sum_{k=0}^{n-1} \aver\uu^{k}_0  | \aver{ \tilde g_+ \, \tilde \uu_k
\circ
A_0^k \, \uu^{(n-k-1)} \circ A_0^{k+1}} \,  |
 \leq C \rho (1-\rho\,\ga)^n\|\tilde g_+ \|_{\al}^+\, .
\end{aligned}
\vspace{-.1cm}
\end{equation}

Finally, for $n\ge 2$, since $\aver{\tilde \uu_k}=0$ for all $k=0,\ldots,n-1$,
using the third line of \eqref{eq:mnsum},
once more with $\uu^{(n)}\circ A_0$ instead of $\uu^{(n)}$, we have
\vspace{-.1cm}
\[
\begin{aligned}
\bigl| \aver{ \, \uu^{(n)}}-\aver{\uu}^n \bigr|
& = \Bigl| \rho^2 \sum_{k=1}^{n-1} \sum_{j=0}^{k-1} \aver{\uu}^{k-j-1}_{j+1}
\aver{\uu}^{n-k-1}_{k+1}
\aver{ \uu^{(j)} \circ A_0^{-j} \,\tilde \uu_j \, \tilde\uu_k \circ A_0^{k-j} } \Bigr| \\
%
%
& \le C \rho^2 \sum_{k=1}^{n-1} \sum_{j=0}^{n-1}
j^2 \la^{-\al j}  (1-\rho\,\ga)^{n}  \| \tilde \uu_j\|_{\al}^+ \| \tilde \uu_k
\|_{\al}^-
\le  C\rho^2n(1-\rho\,\ga)^{n-2} \| \tilde \uu_j\|_{\al}^+ \| \tilde \uu_k \|_{\al}^- ,
\end{aligned}
\vspace{-.1cm}
\]
where we have used Proposition~\ref{prop:4.3} and \eqref{produ}, so that we obtain
\begin{equation} \label{eq:3}
\bigl|\aver{ \, \uu^{(n)}}-\aver{\uu}^n_0 \bigr|
\leq C\rho^2n(1-\rho\,\ga)^n\, .
\end{equation}
Collecting all contributions \eqref{eq:1}--\eqref{eq:3} gives the thesis. \qed

\begin{rmk} \label{tildeg}
\emph{
The factor $\rho \bigl( \| \tilde g_+\|_{\al}^+\| g_-\|_{\al}^- + \|g_+\|_{\al}^+\| \tilde g_-\|_{\al}^- \bigr)$ 
appearing in the estimate of Proposition \ref{prop:decay}, in principle, might be replaced more pragmatically with
$2 \rho \, \| g_+\|_{\al}^+\| g_-\|_{\al}^-$. However, it may be useful in some cases. In particular, for $g_+=g_-=1$, we obtain
$| \aver{ \uu^{(n)}} -\aver{\uu}^{(n)}| \leq C\rho^2n(1-\rho\,\ga)^n$ and hence
$| \aver{ \uu^{(n)}} -\aver{\uu}^{(n)}|\leq C\rho(1-\rho\,\ga')^n$ for any $\gamma'\in(0,\gamma)$,
with the constant $C$ depending on $\gamma'$.
}
\end{rmk}

\subsection{Oscillations of the invariant manifold:~proof of Theorem~\ref{thm:3}} 
\label{subsec:proof3}

Following Remark~\ref{betadelta}, we set $b(\psi) :=f(0,\psi)$ and $v(\psi) :=\partial_\f f(0,\psi)$, with $\| v \|_{\io} \le C$ by Remark \ref{Psi0ge1},
and write
\vspace{-.1cm}
\begin{equation}\label{defd}
f(\f,\psi)=b(\psi) + v(\psi) \, \f + d(\f,\psi)\f^2 ,
\end{equation}
which implicitly defines the function  $d(\f,\psi)$. We also set
%
$\mu(\psi):=1 + \rho v(\psi)=1+\rho\partial_\f f(0,\psi)$
%
and, for $i\geq 0$, according to \eqref{produ}, 
\vspace{-.2cm}
\begin{equation}\label{prodm}
\mu^{(i)}(\psi) = \prod_{j=0}^{i-1} \mu(A_0^{j}\psi) ,
\qquad\qquad
\mu^{(-i)}(\psi) = \prod_{j=1}^{i}\mu(A_0^{-j}\psi) = \mu^{(i)}(A^{-i}\psi) .
\vspace{-.1cm}
\end{equation}

Decomposing $\mu(\psi)=\aver{\mu}+\tilde \mu(\psi)$ according to \eqref{eq:1.222}, we have
$\aver\mu=\ol\mu$, with $\ol\mu$ as in \eqref{Sbarlin}, and
$\tilde \mu(\psi)= \rho \tilde v(\psi)$. We get $\aver\mu\leq \|\mu\|_{\io} \le 1-\rho\,\ga$, since
$\partial_\f f(\f,\psi) \le -\gamma$ for all $\f\in[S_m,S_M]$ and $\bar\f=0\in [S_m,S_M]$
(see the definition of $\ga$ after \eqref{eq:1.1bis} and Remark \ref{barf0}) and $|\mu|_{\al_0}^+ = O(\rho)$. 

In Section~\ref{subsec:proofW} we proved that $W=\lim_{n\to\infty} \EEE^{n}[0]$ uniformly on $\TTT^2$,
%
with $\EEE$ defined in \eqref{EEE}. Thus, to prove \eqref{Wflu}, we show that, for all $n\in \NNN$,
one has
$\big| \langle \EEE^{n}[0]\rangle \bigr| \le C\rho$
and
$\aver{(\EEE^{n}[0])^2}\le C\rho$,
for some constant $C$ independent of $n$.
To this end, analogously to \eqref{EEE}, we set
\vspace{-.1cm}
\begin{equation*}
\EEE_0[W](\psi) :=G_0(W(A_0^{-1}\psi),A_0^{-1}\psi) ,
\qquad
%
G_0(\f,\psi) :=\f+\rho f_0(\f,\psi) , \qquad
f_0(\f,\psi) := b(\psi) + v(\psi)\f 
\vspace{-.1cm}
\end{equation*}
and observe that, for any $h:\TTT^2\to\RRR$,
\vspace{-.2cm}
\begin{equation}\label{eq:EEE0}
\EEE_0^n[h]=\rho\sum_{i=1}^n\mu^{(-i+1)}b\circ A_0^{-i }  +\mu^{(-n)}  h\circ A_0^{-n} .
\vspace{-.1cm}
\end{equation}

\begin{rmk} \label{rmk:W0}
\emph{
Since $\EEE_0$ is a contraction on the space of bounded continuous 
functions from $\TTT^2$ to $\RRR$, the fixed point equation 
$\EEE_0[W]=W$ admits a unique solution $W_0$ and $\EEE_0^n[0]$ 
converges uniformly to $W_0$. We can write
\vspace{-.2cm}
\begin{equation}\label{eq:W0exp}
 W_0=\rho\sum_{i=1}^n\mu^{(-i+1)}b\circ A_0^{-i }  +\mu^{(-n)}  W_0\circ A_0^{-n}= \rho\sum_{i=1}^{\io} \mu^{(-i+1)}b\circ A_0^{-i }\, ,
\vspace{-.1cm}
\end{equation}
from which we get $\|W_0\|_\infty\leq \gamma^{-1}\|b\|_\infty$,
while $|W_0|_{\al_0}^+=O(\rho)$ by \eqref{gAn1}.
}
\end{rmk}

To prove Theorem~\ref{thm:3} we show, first, by explicit computation,
that $|\aver{W_0}|=O(\rho)$ and $\aver{W_0^2}=O(\rho)$, then that
$\aver{|W-W_0|}=O(\rho)$,  by comparing the iterates of $\EEE_0$ and of $\EEE$ and using their contractivity. 

From the last definition in \eqref{produ} we get
$\langle\mu^{(-i+1)}b\circ A_0^{-i}\rangle=\langle b\, \mu^{(i-1)}\circ 
A_0\rangle$. Since, by definition, $\langle b\rangle=0$, using Proposition~\ref{prop:decay}, 
with $n=i-1$, $\uu_k=\mu$ for $k=0,\ldots,n-1$, $g_+=b$ and $g_-=1$, we find
\vspace{-.2cm}
\begin{equation*} \nonumber 
\bigl |\langle\EEE_0^n[0]\rangle \bigr| \le C\rho^2\sum_{i=0}^{n-1} (1-\rho\,\ga)^i \|b\|_{\alpha_0}  \le C\rho\, .
\vspace{-.3cm}
\end{equation*}
Moreover we have
\[
\vspace{-.2cm}
(\EEE_0^n[0])^2
= \rho^2\sum_{i=1}^{n} (\mu^{(-i+1)})^2 (b \circ A_0^{-i})^2 \\
+ 2\rho^2\sum_{ 1 \le i < j \le n}
(\mu^{(-i+1)})^2 \, b \circ A_0^{-i} \, \mu \circ A_0^{-i} \, 
\mu^{-(j-i-1)}\circ A_0^{-i} \, b\circ A_0^{-j} ,
\vspace{-.1cm}
\]
where
%
\[ 
\bigl\langle (\mu^{(-i+1)})^2 \, b \circ A_0^{-i} \, \mu  \circ A_0^{-i} \, 
\mu^{(-j+i+1)} \circ A_0^{-i} \, b \circ A_0^{-j} \big\rangle
= \bigl\langle b \, \mu^{(j-i-1)} \circ A_0
(\mu^{(i-1)})^2 \circ A_0^{j-i+1} \, b \circ A_0^{j-i} \, \mu \circ A_0^{j-i} \bigr\rangle ,
\vspace{-.1cm}
\]
so that, first using \eqref{eq:1.6p} and \eqref{gAn1} to estimate
\vspace{-.1cm}
\[
\bigl| (\mu^{(i-1)})^2 \circ A_0 \, b \, \mu \bigr|_{\alpha_0}^-
\le \bigl| (\mu^{(i-1)})^2 \circ A_0 \bigr|_{\alpha_0}^+ \bigl\| b \, \mu \bigr\|_{\io}  
+ \bigl\| (\mu^{(i-1)})^2 \bigr\|_{\io}  \bigl| b \, \mu \bigr|_{\alpha_0}^+ 
\le C(1-\rho\,\ga)^{2(i-1)} ,
\vspace{-.1cm}
\]
then applying once more Proposition~\ref{prop:decay},
with $n=j-i-1$, $\uu_k=\mu$ for $k=0,\ldots,n-1$, 
$g_+=b$ and $g_-=(\mu^{(i-1)})^2 \circ A_0 \, b \, \mu$,
we get
%
\begin{equation*} 
 \begin{aligned}
 \langle\EEE_0^n[0]^2\rangle
&  \le C \rho \| b\|_{\io}^2  + C\rho^3 \!\!\!\!\!\! \sum_{ 1 \le i < j \le n} \!\!\!\! (1 \!-\!\rho \ga)^{j-i} 
\|b\|_{\al_0} \| (\mu^{(i-1)})^2 \circ A_0 \, b \, \mu \|_{\alpha_0}^-
\le C \rho + C \rho^3\!\!\!\!\!  \sum_{1 \le i < j \le n}^n \!\!\!\!\!
(1\!-\!\rho\ga)^{j+i}\leq 
C\rho .
\end{aligned}
\vspace{-.1cm}
\end{equation*}
Thus, if $W_0$ is defined as in Remark \ref{rmk:W0}, we obtain
\vspace{-.1cm}
\begin{equation} \label{W0}
\langle W_0\rangle = \lim_{n\to\infty} \langle \EEE_0^n[0] \rangle = O(\rho) ,
\qquad
\langle W_0^2\rangle = \lim_{n\to\io}  \langle\EEE_0^n[0]^2\rangle = O(\rho) .
\vspace{-.1cm}
\end{equation}
%

For any $n \ge 0$, we can write
\vspace{-.2cm}
\begin{equation} \label{telescopio}
\EEE^n[0] \!-\! \EEE_0^n[0]
= \sum_{i=0}^{n-1} 
\EEE^{n-i}[\EEE_0^{i}[0]] \!-\! \EEE^{n-i-1}[\EEE_0^{i+1}[0]] 
= \sum_{i=0}^{n-1} 
\EEE^{n-i-1}[ \EEE[\EEE_0^i[0]]] \!-\! \EEE^{n-i-1}[\EEE_0[\EEE_0^i[0]] ] \, .
\vspace{-.1cm}
\end{equation}

\begin{rmk} \label{rmk:EEE0}
\emph{
It may happen that $\EEE_0^i[0]\not\subset \Omega$ for some $i$.
In such a case, we observe that there exists an interval $\calU' :=[\phi_m',\phi_M']
\supset \calU$ such that $\Omega':=\calU'\times \TTT^2$
and $\EEE_0^n[0](\psi) \in \Omega'$ for all $n\ge 0$ and all $\psi\in\TTT^2$.
Thus for the r.h.s.~of \eqref{telescopio} to be well defined, we
compute $\EEE$ by replacing the map $\SSS$ with the extension $\SSS_{\rm ext}$
introduced in Subsection~\ref{extension} such that $\Omega_{\rm ext}\supset
\Omega'$
(see also Remark \ref{whyext}).}
\end{rmk}

By taking $\SSS_{\rm ext}$ according to Remark \ref{rmk:EEE0} and reasoning as in the proof of Lemma
\ref{lem:easy}, we find that for any $r>0$ there exists 
%
%
$N'_r \le \max\{ \phi_M'-S_M,S_m-\phi_m' \}/(\rho \inf_{\Omega'\backslash\Lambda_r} |f_{\rm ext}(\f,\psi)|)$
such that $\SSS_{\rm ext}^{N_r'}(\Omega') \subset \Lambda_r$,
and, given $W_1,W_2:\!\TTT^2\to[\phi'_m,\phi'_M]$,  we have $\EEE^{N'_r}[W_i]\in
[S_m-r,S_M+r]$ for $i=1,2$, while
$|\EEE^{N'_r}[W_1]-\EEE^{N'_r}[W_2]|\leq (1+\rho \|\partial_\f f\|_\io )^{N'_r}|W_1-W_2|$.
Thus, for any $\ga'\in(0,\ga)$, with the notation in property \ref{prop-rG} of Lemma~\ref{lem:easy},
we can choose $r=r(\rho\,\ga')$ and obtain $N_r'=O(1/\rho\,\ga')$ and,
for any $W_1,W_2\in [S_m-r,S_M+r]$ and any $k\ge0$, we may bound
$|\EEE^{k}[W_1]-\EEE^{k}[W_2]|\leq  (1-\rho\,\ga')^{k}|W_1-W_2|$.
Summing up, \eqref{telescopio} gives
\vspace{-.2cm}
\begin{equation*}  
\begin{aligned}
& |\EEE^n[0](\psi) \!-\! \EEE_0^n[0](\psi)|
\!\le\! \left(\frac{1+\rho \| \partial_\f f\|_{\io} }{1 - \rho\,\ga'} \right)^{\! N_r'} \!\!
\rho \! \sum_{i=0}^{n-1} 
\left(1\!-\!\rho\,\ga' \right)^{n-i-1} \!
\left|d(\EEE_0^{i}[0](A_0^{-1}\psi),A_0^{-1}\psi) \right|
(\EEE_0^i[0](A_0^{-1}\psi))^2 ,
\end{aligned}
\vspace{-.1cm}
\end{equation*}
with $d(\f,\psi)$ implicitly defined in \eqref{defd}.
Integrating over $\psi$ and using the bound on $\langle\EEE_0^n[0]^2\rangle$,
we obtain
\vspace{-.2cm}
\[
\langle | \EEE^n[0]-\EEE_0^n[0] | \rangle \le C\rho\sum_{i=0}^{n-1}
\left(1-\rho\,\ga' \right)^i \|d\|_\io \aver{\EEE_0^i[0]^2} \leq C\rho ,
\]
which yields
\begin{equation}\label{W-W0}
\aver*{|W-W_0|}\leq C\rho ,
\vspace{.1cm}
\end{equation}
which, together with \eqref{W0}, gives
%
$|\langle W\rangle|\le |\langle W-W_0\rangle|+ |\langle W_0\rangle|\le \langle | W-W_0| \rangle+ |\langle W_0\rangle|\leq C\rho$,
%
that is the first bound in \eqref{Wflu}.
Finally, we have 
$|\aver{W^2-W_0^2}|\leq\aver*{|W-W_0|\;|W+W_0|}\leq C\rho$,
because of \eqref{W-W0},
and hence, thanks to \eqref{W0},
%
$\langle W^2\rangle \leq |\langle W^2-W_0^2\rangle| +  \langle W_0^2\rangle \leq C\rho$,
%
that is the second bound in \eqref{Wflu}.

\subsection{Fluctuations of the linearized dynamics:~proof of Lemma~\ref{lem:flulin}} \label{fluk}

Having studied the oscillations of the invariant manifold we now focus our
attention on the linearized dynamics.
Using that $\ka(\psi)=1+\rho \, \partial_\f f(W(\psi),\psi)$ and $\mu(\psi)=1+\rho \, \partial_\f f(0,\psi)=1-\rho\,v(\psi)$, we write
\begin{equation} \label{xi}
\ka(\psi) = \mu(\psi)+\rho \, \xi(\psi) , \qquad
\xi(\psi) := 
d_1(\psi) \, W(\psi)+ d_2(\psi) \, W^2(\psi) , \qquad d_1(\psi) := \partial_\f^2 f(0,\psi) ,
\end{equation}
which implicitly defines the function $d_2\in \gotB^*_{\al_-,\al_+}\!(\TTT^2,\RRR)$.

From Proposition~\ref{prop:decay}, with $g_+=g_-=1$ and either $\uu_i=\mu$ or $\uu_i=\mu^2$ for all $i$,
and Remark \ref{tildeg},  it follows that
$\bigl|\aver{\mu^{(n)} - \ovl\mu^{n}} \bigr|\leq C \rho^2 n (1-\rho\gamma)^n
$ and $
\average{(\mu^{(n)} - \ovl\mu^{n})^2}\leq C \rho^2 n
(1-\rho\gamma)^{2n}$,
and hence, in order to prove \eqref{linear}, it is
enough to show that
\vspace{-.2cm}
\begin{equation} \label{linearbis}
\bigl|\aver{\kappa^{(n)} - \mu^{(n)}}\bigr|\leq C \rho
(1-\rho\gamma)^n\sum_{k=0}^2\rho^kn^k ,\qquad
\average{(\kappa^{(n)} - \mu^{(n)})^2} \leq C \rho
(1-\rho\gamma)^{2n}\sum_{k=0}^2\rho^kn^k .
\end{equation}
%

\begin{rmk} \label{rmkA}
\emph{
The proof of \eqref{linearbis} represents a first instance of the strategy
outlined in Section~\ref{subsec:content}. In order to bypass the low
regularity of $\kappa$ due to its dependence on the invariant
manifold, we expand it up to the second order in $W$ so as to use Theorem
\ref{thm:3} to estimate the average of the quadratic contributions, the
bound \eqref{W-W0} to reduce the analysis of the average of the linear
contributions to the more manageable function $W_0$ and  the
explicit expression for $W_0$ in \eqref{eq:W0exp} to obtain the desired
estimates.
}
\end{rmk}

Thus, we proceed with the proof of \eqref{linearbis}. By Remark \eqref{eq:mnsum} we can write
\vspace{-.3cm}
\begin{equation} \label{eq:first}
\begin{aligned} 
& \kappa^{(n)} - \mu^{(n)} = \rho\sum_{j=0}^{n-1} \kappa^{(j)}\,\xi\circ A_0^j\,\mu^{(n-j-1)}\circ A^{j+1} \\
& \; =  \rho\sum_{j=0}^{n-1} \mu^{(j)}\,\xi\circ A_0^j\,\mu^{(n-j-1)}\circ A^{j+1}
+ \rho^2\sum_{j=1}^{n-1}\sum_{k=0}^{j-1}
\ka^{(k)}\,\xi\circ A_0^k\, \mu^{(j-k-1)}\circ 
A_0^{k+1}\,\xi\circ A_0^{j}\, \mu^{(n-j-1)} \circ A_0^{j+1} .
 \end{aligned}
\vspace{-.2cm}
\end{equation}
Since
$| \aver{ \ka^{(k)} \xi\circ A_0^k  \mu^{(j-k-1)} \circ A_0^{k+1}\xi \circ A_0^{j} \mu^{(n-j-1)} \circ A_0^{j+1}} | 
\le (1-\rho\,\ga)^{n-2}\aver{| \xi \circ A_0^k \, \xi \circ A_0^j|}
\leq (1-\rho\,\ga)^{n-2}\aver{\xi^2}$
and $\aver{\xi^2}\leq C\rho$, by \eqref{xi} and \eqref{Wflu}, eventually we get, for $n  \ge 2$,
%
\[
\biggl|\sum_{j=0}^{n}\sum_{k=0}^{j-1}\aver*{\ka^{(k)}\,\xi\circ A_0^k\, 
\mu^{(j-k-1)}\circ 
A_0^{k+1}\,\xi\circ A_0^{j}\, \mu^{(n-j-1)} A_0^{j+1}}\biggr|\leq C\rho \, n^2 (1-\rho\,\ga)^n .
\vspace{-.1cm}
\]
On the other hand, since we have, again by \eqref{Wflu},
%
\[
\bigl| \bigl\langle \mu^{(j)}\,(d_2W^2)\circ A_0^j\, \mu^{(n-j-1)}\circ A^{j+1} \bigr\rangle
\bigl| \leq(1-\rho\,\ga)^{n-1}\|  d_2\|_\io\aver{W^2}\leq  C\rho (1-\rho\,\ga)^n .
\]
we need to estimate
$\aver{ \mu^{(j)}\,(d_1W)\circ A_0^j\, \mu^{(n-j-1)}\circ A_0^{j+1}} \!=\!
\aver{ \mu^{(j)} \circ A_0^{-j} \, d_1W \, \mu^{(n-j-1)}\circ A_0}.$
By Proposition~\ref{prop:decay},
with $\al\!=\!\al_+$, $n\!-\!j\!-\!1$ instead of $n$, $\uu_i\!=\!\mu$ for $i\!=\!0,\ldots,n\!-\!j\!-\!2$,  $g_+= \mu^{(j)} \circ A_0^{-j} \, d_1W$ and $g_-\!=\!1$, we find
%
\[
\bigl| \bigl\langle \mu^{(j)}\,(d_1W)\circ A_0^j\, \mu^{(n-j-1)}\circ A_0^{j+1} \bigr\rangle \bigr| 
\le C(1-\rho\,\ga)^{n-j}\bigl| \bigl\langle \mu^{(j) \bigl\rangle \circ A_0^{-j}\,d_1W}\bigr| + C\rho \,(1-\rho\,\ga)^{n}\, (1+\rho\,(n-j)) ,
\]
where we have used also that $\|W\|_{\al_+}^+ \le C$ and $\|\mu^{(j)}\circ A_0^{-j}\|_{\al_0}^+ \le C$, 
because of Theorem~\ref{thm:2} and \eqref{gAn1}, respectively. 
Then, by \eqref{W-W0} and Remark \ref{rmk:W0}, we find
\vspace{-.1cm}
\[
\begin{aligned}
\null\!\!\!\!
&  \left| \aver*{ \mu^{(j)} \circ A_0^{-j} d_1W } \right| = \left|\aver*{ \mu^{(j)}\,(d_1W)\circ A_0^{j}}\right|
\leq \left|\aver*{ 
\mu^{(j)}\,(d_1W_0)\circ A_0^{j}}\right|+C\rho(1-\rho\,\ga)^{j} \\ 
\null\!\!\!\!
& \quad \le \left|\aver*{ 
W_0\,(\mu^{(j)})^2\,d_1\circ A_0^{j}}\right|+\rho \! \sum_{k=1}^j\left|\aver*{ 
\mu^{(j-k)}b\circ A_0^{j-k}(\mu^{(k-1)})^2\circ A_0^{j-k} \mu\circ A_0^{j-1} d_1\circ  A_0^{j}}\right| +C\rho(1-\rho\,\ga)^{j} ,
\end{aligned}
\vspace{-.2cm}
\]
where, since  $\aver{W_0}=O(\rho)$, we have
$|\aver{ W_0 \, (\mu^{(j)})^2 \,d_1\circ A_0^{j}} |\le C \, (1-\rho\,\ga)^{2j} \, (1+\al_0 j)\la^{-\al_0 j}+\rho+\rho^3 j )$,
by Proposition~\ref{prop:decay}, with $n=j$, $\uu_i=\mu^2$ for $i=0,\ldots,j-1$,  $g_+=W_0 \circ A_0^{-1}$ and $g_-=d_1$,
while
\vspace{-.1cm}
\[
\begin{aligned}
& \left|\aver*{ \mu^{(j-k)}b\circ A_0^{j-k}(\mu^{(k-1)})^2\circ A_0^{j-k} \mu \circ A_0^{j-1}\,d_1\circ A_0^{j} }\right| \\
& \qquad \qquad
\le (1-\rho\,\ga)^{j-k}\left|\aver*{ b(\mu^{(k-1)})^2\, \mu \circ A_0^{k-1} \, d_1\circ A_0^{k}}\right|+C\rho(1-\rho\,\ga)^{j+k} \, ( 1 + \rho\,(j-k))  \\
& \qquad \qquad
\le C(1-\rho\,\ga)^{j-k} \bigl( (1+\al_0 k)\lambda^{-\alpha_0 k} + \rho \bigr) \|b\|_{\al_0}\|d_1\|_{\al_0}+C\rho(1-\rho\,\ga)^{j+k} ,
\end{aligned}
\vspace{-.1cm}
\]
where we have used twice Proposition~\ref{prop:decay}, first
with $n=j-k$, $\uu_i=\mu$ for $i=0,\ldots,j-k-1$, $g_+=1$ and $g_-=b\,(\mu^{(k-1)})^2\mu\circ A_0^{k-1} d_1 \circ A_0^k$,
then with $n=k$, $\uu_i=\mu^2$ for $i=0,\ldots,k-1$ and $\uu_{k}=\mu$, $g_+=b$ and $g_-=d_1$.
Inserting all the bounds into \eqref{eq:first}, we obtain the first of \eqref{linearbis}.

For the second of \eqref{linear}, we use the first line of \eqref{eq:first} to write
\vspace{-.2cm}
\[
\begin{aligned}
\aver*{ \!\bigl( \kappa^{(n)} \!-\! \mu^{(n)}\bigr)^{\!2} }
& 
\!\le\!
\rho^2 \!\! \sum_{i,j=1}^{n-1}\!
\left|\aver*{\kappa^{(i)}\,\xi\circ A_0^i\,\mu^{(n\!-\!i\!-\!1)}\circ A^{i+1} \kappa^{(j)}\,\xi\circ A_0^j\,\mu^{(n\!-\!j\!-\!1)}\circ  A^{j+1}}\right|
\le C \, n^2 \rho^2 \left( 1 \!-\! \rho\,\ga \right)^{2(n-1)} \!\aver*{\xi^2} .
\end{aligned}
\vspace{-.2cm}
\]
Thus, the second bound in \eqref{linearbis} follows using again the second bound in \eqref{Wflu}.

\subsection{Iterated products}\label{subsec:iterated}

We introduce here some notation that will be used widely throughout the rest of the paper.
Let $\SSS$ be any map on $\calU\times\TTT^2$ of the form
\vspace{-.2cm}
\begin{equation} \label{SSS}
\SSS(\f,\psi) = (\SSS_\f(\f,\psi),\SSS_\psi(\f,\psi)) :=(G(\f,\psi),A_0\psi)\, ,
\vspace{-.1cm}
\end{equation}
and let $\pp_0,\ldots,\pp_{n-1}$ be functions defined in
$\calU\times\TTT^2$.
Define, for $k=0,\ldots,n$ and $i=0,\ldots,n-k$,
\vspace{-.2cm}
\begin{equation} \label{defpn}
\pp^{(k)}_{[i]} (\SSS;\f,\psi) 
:= \prod_{j=0}^{k-1} \pp_{i+j}(\SSS^j(\f,\psi)) , \qquad \qquad \pp^{(k)}(\SSS;\f,\psi) := \pp^{(k)}_{[0]} (\SSS;\f,\psi) ,
\vspace{-.2cm}
\end{equation}
and set $\pp^{(-1)}_{[i]} (\SSS;\f,\psi)=\pp^{(0)}_{[i]} (\SSS;\f,\psi)=1$ according to the convention established in Remark \ref{convention}.

\begin{rmk} \label{SnoS}
\emph{
The map $\SSS$ in \eqref{SSS} is not necessarily the map \eqref{eq:1.1} which defines our model. In particular, in what follows,
we shall use the notation \eqref{defpn} for several maps, including the translated map $\SSS_1$ and the auxiliary map
$\SSS_2$ which will be introduced in Subsection~\ref{aux}.
}
\end{rmk}

If the function $\SSS_\f(\f,\psi)$ does not  depend on $\psi$ and the functions
$\pp_0,\ldots,\pp_{n-1}$ are independent of $\psi$ as well, instead of
\eqref{defpn} we may consider
%
\begin{equation} \label{defpnnopsi}
\pp^{(k)}_{[i]}(\SSS;\f) := \prod_{j=0}^{k-1} \pp_{i+j}(G^j(\f)) , 
\qquad \qquad \pp^{(k)}(\SSS;\f):=\pp^{(k)}_{[0]}(\SSS;\f) ,
\vspace{-.2cm}
\end{equation}
with $G(\f):=\SSS_\f(\f,\psi)$ and $G^j$ denoting is the composition of $G$ with
itself $j$ times. In particular, if $
\ovl\SSS(\f,\psi)
=(\ovl G(\f),A_0\psi)$ is the averaged map \eqref{Sbar}
and $\pp_0,\ldots,\pp_{n-1}$ are functions on $\calU\times\TTT^2$, we have
\vspace{-.2cm}
\begin{equation} \label{defpnaver}
\aver{\pp}^{(k)}_{[i]} (\ovl\SSS;\f)  = \prod_{j=0}^{k-1} \aver{\pp_{i+j}} (\ovl G^j(\f)) , 
\qquad \qquad \aver{\pp}^{(k)}(\ovl\SSS;\f) :=\aver{\pp}^{(k)}_{[0]}(\ovl\SSS;\f) .
\vspace{-.2cm}
\end{equation}
%

\subsection{Conjugation of the averaged dynamics:~proof of Lemma
\ref{lem:Hbar}}
\label{subsec:barH}

Proceeding as in Subsection~\ref{subsec:proofconj} we see that the function $\overline h\!:\calU \to \calU_0$
introduced in \eqref{def:lbar} satisfies the equation
%
$\overline h(\f)=\overline q(\f) + \overline p(\f) \, \overline h(\overline G(\f))$,
%
with $\overline G(\f)=\ovl \SSS_\f(\f,\psi)$ defined in \eqref{Sbar} and
\vspace{-.1cm}
\begin{subequations} \label{pbarqbar}
\begin{align}
\overline p(\f) & := \frac{(\overline G(\f))^2}{\f^2 \, \ovl\mu} 
= \frac{1}{1+\rho \, \partial_\f \overline 
f(0)}\left(\frac{\f+\rho \overline f(\f)}{\f}\right)^{\!\!2} ,
\label{pbarqbara} \\
\overline q(\f) & := \frac{\overline G(\f) - \f \, \ovl\mu}{\f^2 \, \ovl\mu} 
= \frac{\rho}{1+\rho \, \partial_\f \overline f (0)} 
\frac{\overline f(\f)- \partial_\f \overline f (0)\f}{\f^2}.
\label{pbarqbarb}
\end{align}
\end{subequations}
Thus we can write (see \eqref{defpnaver} for the notation)
\vspace{-.2cm}
\begin{equation} \label{hbar}
\overline h(\f)=\sum_{n=1}^\io \overline p^{(n)}(\f) \, \overline q({\overline G}^n(\f)), \qquad\qquad
%
%
\overline p^{(n)}(\f)=\ovl p^{(n)}(\ovl \SSS;\f) =\prod_{i=0}^{n-1} \overline p({\overline G}^i(\f)) .
\vspace{-.2cm}
\end{equation}
%

Analogously to Lemma~\ref{lempq} the following result holds.

\begin{lemma} \label{thetabar}
In $\calU$ one has $\norm \overline p\norm_{0,5}=1+O(\rho)$,
$\norm \overline q\norm_{0,5}=O(\rho)$ and $\norm \partial_\theta\overline p\norm_{0,4}=O(\rho)$.
Moreover, for any $\ga'\in(0,\ga)$,
there exists $\overline \theta=O(1)$ such that $\overline p(\f)\leq 1-\rho\,\ga'$ for $|\f|\leq \ovl\theta$.
\end{lemma}

\begin{rmk} \label{rmkB}
\emph{
By reasoning as in the proof of Lemma~\ref{lemSSSn}
(and hence, actually, Lemma \ref{lem:easy}),
we find that, for any $\ga'\in(0,\ga)$, there exist
$r>0$ and $\ol N=O(\rho^{-1})$ such that $\ol{\SSS}^{n}(\calU) \subset [S_m-r,S_M+r]$ and
$(1+\rho\partial_\f\ovl f) \circ\ol{\SSS}^{n}\leq (1-\rho\,\ga')$ for all $n \ge \ol N$.}
\end{rmk}

The following result is proved in Appendix~\ref{proof:pn}.

\begin{lemma} \label{lem:pn}
Given $\ga'\in(0,\ga)$ and $\ovl\theta$ as in Lemma~\ref{thetabar},
consider any functions $\pp_0,\ldots,\pp_{n-1}$ in $\calmB_{0,3}(\Omega,\RRR)$,
independent of $\psi$, such that
\vspace{-.2cm}
\begin{enumerate}
\itemsep0em
\item $\|\pp_i - 1 \|_{0,3}  = O(\rho)$ for all $i=0,\ldots,n-1$,
\item
$|\pp_i(\f)| \leq 1-\rho\,\ga'$ for any
$|\f|\le \bar\theta$ and for all $i=0,\ldots,n-1$,
\vspace{-.1cm}
\end{enumerate}
and define $\pp^{(n)}(\f)=\pp^{(n)}(\ovl \SSS;\f)$ 
as in \eqref{defpnnopsi} with $\SSS=\ovl\SSS$.
One has 
\begin{equation} \label{deriderideripna}
\norm \pp^{(n)}\norm_{0,3} \leq  C
(1-\rho \,\ga')^n , \qquad \qquad \|\partial_\f^k \ovl G^n\|_\infty \leq C  (1-{\rho}\,\ga')^n\, ,  \qquad k=1,2,3 .
\end{equation}
where the constant $C$ does not depend on $n$.
%
%
\end{lemma}

\begin{rmk} \label{ex-point3}
\emph{
An immediate consequence of the functions $\pp_0,\ldots,\pp_{n-1}$ satisfying condition 1 in Lemma~\ref{lem:pn}
is that $\norm\partial_\f \pp_i \norm_{0,2}=O(\rho)$ for all $i=0,\ldots,n-1$.
}
\end{rmk}

\begin{rmk} \label{ex-point3bis}
\emph{
In the following we have to consider cases in which condition 2 of Lemma~\ref{lem:pn} holds
for all $i=0,\ldots,n-1$ except at most $n_*$ values, for some $n_*<n$ independent of $n$.
However, such cases are easily reduced to Lemma~\ref{lem:pn}. Indeed, if $\pp_{i_1},\ldots,\pp_{i_{n_*}}$ are the functions
which do not satisfy condition 2, setting
$I_*=\{ i_1,\ldots i_{n_*} \}$
and
$K_* = \max\{ \|\pp_i\|_{\io} : i \in I_* \}$,
we may define
\vspace{-.1cm}
\[
\tilde \pp_i(\f) = 
\begin{cases}
(1-\rho\,\ga')^{-1} K_*^{-1} \pp_i(\f) , & i \in I_* , \\
\pp_i(\f) , & i \notin I_* ,
\end{cases}
\vspace{-.1cm}
\]
so that $|\tilde\pp_i(\f) | \le (1-\rho\,\ga')^{-1}$ for $i=1,\ldots,n-1$ and hence, since $\|\tilde\pp_i-1\|_{0,3}=O(\rho)$ if $\|\pp_i-1\|_{0,3}=O(\rho)$,
the functions $\tilde \pp_0,\ldots,\tilde\pp_{n-1}$  satisfy the hypotheses of Lemma~\ref{lem:pn}. Therefore, we can write
$\pp_i(\f) = (1-\rho\,\ga') \, K_* \tilde \pp_i(\f)$ for $i \in I_*$ and incorporate the factor $((1-\rho\,\ga')\,K_*)^{n_*}$
into the constant $C$ in \eqref{deriderideripna}.
}
\end{rmk}

The bounds for $\ovl h$ now follow easily considering \eqref{hbar},
reasoning like in Subsection~\ref{subsec:proof2-2} and using the bound \eqref{deriderideripna}
with $\pp_i=\ovl p$ for $i=0,\ldots,n-1$.
Invertibility of $\ovl \HHH$ follows by the same argument used in the proof of Corollary~\ref{l} (see Subsection~\ref{subsec:proofcor}), and
the bounds for the function $\bar l$ are easily obtained using the inverse function theorem.

\subsection{The averaged and the continuous time system:~proof of Lemma~\ref{lem:Phik}}
\label{subsec:proofPhi}

Let $\Phi_t(\f)$ be the solution of \eqref{flow}. Observe first that
\vspace{-.2cm}
\begin{equation} \label{Phirho}
\Phi_{\rho}(\f)=\f + \rho \overline
f(\f) + \rho^2 \int_0^1(1-t) \, \partial_\f \overline f(\Phi_{t \rho}(\f)) \,  \overline f(\Phi_{t \rho}(\f)) \, \der t
\vspace{-.1cm}
\end{equation}
and that for any $\ga''\in(0,\ga)$ there exists a constant $C$ independent of $\rho$ such that, for all $n\ge0$,
\vspace{-.1cm}
\begin{equation} \label{Phinrho}
| \Phi_{n\rho}(\f) | \leq C(1-\rho\,\ga'')^n , \qquad\qquad |(\ovl\SSS^{n})_\f(\f,\psi)| \leq C(1-\rho\,\ga'')^n .
\vspace{-.1cm}
\end{equation}
Then, using \eqref{Sbar} and \eqref{Phirho}, we obtain, for some $\f_n$  between
$(\overline \SSS^{n})_\f(\f,\psi)$ and $\Phi_{n\rho}(\f)$,
\[
 \left| (\overline \SSS^{n+1})_\f(\f,\psi)-\Phi_{(n+1)\rho}(\f) \right| \leq
(1+\rho \partial_\f \ovl f(\f_n))  |(\overline\SSS^{n})_\f(\f,\psi)-\Phi_{n\rho}(\f)|+C\rho^2(1-\rho\,\ga'')^n
\]
where we have used \eqref{Phinrho} and the fact that $\ovl f(0)=0$ in order to
bound  $|\overline f(\Phi_{t \rho}(\f))| \le C |\Phi_{t \rho}(\f))|$.
Iterating we get, for suitable $\f_1,\ldots,\f_{n-1}$,
\vspace{-.2cm}
\[
|(\overline \SSS^n)_\f(\f,\psi)-\Phi_{n\rho}(\f)|\le
C\rho^2 \, \sum_{i=0}^{n-1} (1-\rho\,\ga'')^{n-1-i} \!\!\! \prod_{j=n-i}^{n}
\!\! \left(1+ \rho \, \partial_\f \ovl f(\f_j)\right) .
\vspace{-.1cm}
\]
Let $\ol{N}$ be defined as in Remark \ref{rmkB}. By the first of \eqref{Phinrho},
there exists $\ol M \ge N$ such that both $(\overline \SSS^{k})_\f(\f,\psi)$ and $\Phi_{k\rho}(\f)$
-- and hence $\f_k$ as well -- are in $[S_m-r,S_M+r]$ for $k \ge M$, with $r$ such that
$| 1+\rho \, \partial_\f \ovl f(\f) | \le C (1-\rho \ga'')$ for all $\f \in [S_m-r,S_M+r]$.
Thus, we get
\vspace{-.2cm}
\[
\rho^2 \sum_{i=0}^{n-1} (1-\rho\,\ga'')^{n-1-i} \!\! \prod_{j=n-i}^{n}
\!\! (1+\rho \, \partial_\f \ovl f(\f_j) )  \le C \rho
\left( \frac{ 1+\rho \| \partial_\f f \|_{\io}}{1-\rho\,\ga''}  \right)^{\ol M} n  \rho (1-\rho\,\ga'')^{n} ,
\vspace{-.2cm}
\]
so that the follows by taking $\ga'' \in (0,\ga)$ and
$\ga'\in(0,\ga'')$ such that $n\rho(1-\rho\,\ga'')^n \le (1-\rho\,\ga')^n$.
\qed

\subsection{Deviations of the conjugation:~proof of Theorem~\ref{thm:exthm3}} 
\label{proof3-2}

To compare $\SSS^n$ with $\ovl\SSS^n$, according to
\eqref{S1n-Sbarn} we need to control the deviations of $h$ from $\ovl h$ and of $l$ with respect to $\bar l$.
Since $\overline G(0)=0$, the map $\overline \SSS$ admits the same invariant manifold
$\overline \calW:=\{(0,\psi)\,|\, \psi\in\TTT^2\}$ as $\SSS_1$ (see Remarks \ref{Wbar} and \ref{rmk:twosteps}).
Thus, in order to complete the program outlined at the beginning of Subsection
\ref{averaged}, it is more convenient to compare first $h_1$ with $\ovl h$ and show that, in average, they are close
to each other, and then show that the same happens when comparing $h$ with $h_1$.

The domain $\Omega_1$ of $\SSS_1(\f,\psi)$ and $h_1(\f,\psi)$
has the form considered in Remark \ref{moregeneral}, with $a_+(\psi)=\phi_M-W(\psi)$
and $a_-(\psi)=\phi_m-W(\psi)$. For $\aver{h_1(\f,\cdot)}$ to make sense we
need $h_1(\f,\psi)$ to be defined for every $\psi$, but this happens only for
$\f\in \Theta$ with $\Theta$ strictly contained in $\calU$ (see Remark
\ref{rmk:strip}). Since we want to compare $\ovl h(\f)$ with
$\aver{h(\f,\cdot)}$ for all $\f\in\calU$, we need to extend $h_1(\f,\psi)$ to
the whole  set $\Omega$. One way to accomplish this is to extend both $F$ and
$\HH$, as described in Subsection~\ref{extension}, to functions $F_{\rm ext}$ and $\HH_{\rm ext}$
defined on a larger domain $\Omega_{\rm ext}$ 
such that the extended map $\SSS_{\rm 1, ext}(\theta,\psi)$
is defined for all $(\theta,\psi)\in\Omega_{\rm ext}$. To this end we set
%
$\Omega_{\rm ext} := \calU_{\rm ext} \times \TTT^2$,
with
$\calU_{\rm ext}=[\phi_m+\min_{\psi\in\TTT^2} W(\psi), \phi_M+\max_{\psi\in \TTT^2} W(\psi)]$,
%
so that we have $\Omega_{\rm ext} \supset \{(\f,\psi) \in \RRR\times \TTT^2 : \phi_m +
\min\{0,W(\psi)\} \le \f \le \phi_M + \max\{0,W(\psi)\} \}$.
Then, when considering $h(\f,\psi) - \ovl h(\f)$, we may write, for $\f\in\calU$,
%
\begin{equation} \label{h1-hbar}
\begin{aligned} 
h(\f,\psi) & = h_{\rm ext}(\f,\psi) = h_{\rm ext}(\f+W(\psi),\psi) - \bigl(  h_{\rm ext}(\f+W(\psi),\psi) - h_{\rm ext}(\f,\psi) \bigr) \\
& = h_{1,\rm ext}(\f,\psi) - W(\psi) \, \partial_\f h_{\rm ext}(\f,\psi) -
(W(\psi))^2 \!\! \int_0^1 \!\! \der t \, (1-t) \, \partial_\f^2 h_{\rm ext}(\f -t W(\psi),\psi) 
\end{aligned}
\vspace{-.1cm}
\end{equation}
and start by studying the average of $h_{1,\rm ext}(\f,\psi) - \ovl h(\f)$. The
next step is to show, thanks to Theorem~\ref{thm:3},
that the average of the other terms appearing in \eqref{h1-hbar} produce corrections of order $\rho$.

\begin{rmk} \label{errext}
\emph{
Although we use the extended maps along the proof of Theorem~\ref{thm:4},
the final result does not depend on the extension we take,
because the difference between $\Omega_1$ and $\Omega$
has measure of order $\rho$ and hence any extended map we may consider produces
corrections which are at most of order $\rho$. A different choice could be $h_1(\f,\psi)=0$ for $(\f,\psi)\in \Omega\setminus\Omega_1$;
the advantage of taking $h_{1,{\rm ext}}$ as in Subsection \ref{extension}
is that it has the same regularity of the original $h_1$. 
}
\end{rmk}

\begin{rmk} \label{labelext}
\emph{
Throughout the rest of the section, we work with the extended functions, but we
drop the subscript `${\rm ext}$' not to overwhelm the notation. Since we first
compare $\ovl h$ with $h_1$, to avoid confusion, we call $\theta$ the first
variable not only of $\SSS_1$ and $h_1$, but also of $\ovl h$ and $\ovl\SSS$.
From the above discussion, it follows that, for all
$\psi\in\TTT^2$, the range of the
variable $\theta$ contains the whole interval $\calU$. Only at the end, when
comparing $h$ with $\bar h$, we compute $\ovl h$ and $h_1$ at $\theta=\f$.
}
\end{rmk}

The rest of the subsection is mostly devoted to the proof of the following proposition and some of its implications.

\begin{prop} \label{prop:h1hbar}
Let $h_1$ and $\ovl h$ be as in \eqref{HHH1h1} and in \eqref{def:lbar}, respectively.
For all $\theta\in\calU$, one has
\vspace{-.1cm}
\begin{subequations} \label{h1hbar}
\begin{align}
\bigl| \aver*{ h_1(\theta,\cdot) - \ovl h(\theta) } \bigr| &\leq C\rho , \qquad&
\bigl| \aver*{ ( h_1(\theta,\cdot) -  \ovl h(\theta))^2 } \bigr| &\leq C\rho ,
\label{h1hbara}\\
\bigl| \aver*{\partial_\theta h_1(\theta,\cdot) - \partial_\theta \ovl h(\theta) } \bigr| & \leq C\rho , \qquad &
\bigl| \aver*{ (\partial_\theta h_1(\theta,\cdot) - \partial_\theta \ovl h(\theta))^2 } \bigr| &\leq C\rho .
\label{h1hbarb}
\end{align}
\end{subequations}
\end{prop}

After proving Proposition~\ref{prop:h1hbar}, to complete the proof of Theorem~\ref{thm:exthm3} we need to reexpress $h_1$
in terms of $h$. This is done in the last Subsection~\ref{675}. 

\subsubsection{The auxiliary map} \label{aux}

Let $r_2$ be such that $\theta(f(\theta,\psi)-f(0,\psi))< 0$ for
$(\theta,\psi)\in\Lambda_{2r_2}$ and let $\chi:\calU\to\RRR$ be a $C^\infty$ function
such that $\chi(\theta)=1$ for $\theta\in [S_m-r_2,S_M+r_2]$ 
and $\chi(\theta)=0$ for $\theta\not\in [S_m-2r_2,S_M+2r_2]$. 
We introduce the \emph{auxiliary map}
\begin{equation} \label{eq:S2}
\SSS_2(\theta,\psi) := (G_2(\theta,\psi),A_0\psi) ,  \qquad G_2(\theta,\psi) := \theta+\rho f_2(\theta,\psi),
\qquad
%
%
f_2(\theta,\psi):=f(\theta,\psi)-\chi(\theta)f(0,\psi) .
\end{equation}
%

\begin{rmk}\label{rmk:S3}
\emph{
Instead of $\SSS_2$, one might like to consider the simpler map
%
$\SSS_3(\theta,\psi) := (G_3(\theta,\psi),A_0\psi)$, 
with $G_3(\theta,\psi) \!:=\! \theta+\rho f_3(\theta,\psi)$ 
and $f_3(\theta,\psi) \!:=\! f(\theta,\psi)-f(0,\psi)$.
%
However, 
it could happen that $f_3(\theta,\psi)\!=\!0$ also for some $\theta\not=0$,
so that $\SSS_3$ would not satisfy Hypothesis \ref{hyp:2} and $\ovl\calW$
might fail to be a global attractor for $(\Omega,\SSS_3)$; hence, in general, $\SSS_3$ could not be conjugated with $\SSS_0$.
On the other hand, if one is willing to restrict the map $\SSS$
to a smaller set inside $\Omega$, say the set $\Lambda_{2r_2}$ defined above, then one can define
$f_2(\theta,\psi)=f(\theta,\psi)-f(0,\psi)$, without introducing the function $\chi$,
and the corresponding map $\SSS_2$ satisfies all Hypotheses~\ref{hyp:1}--\ref{hyp:3}.
The same goal might be achieved by assuming stronger hypotheses on the map $\SSS$,
say by requiring it to be uniformly contracting along the direction of the slow variable on the whole $\Omega$,
but the conditions so introduced would be too restrictive and unnecessary for the result to hold.
}
\end{rmk}

If we set $F_1(\theta,\psi)=\rho f_1(\theta,\psi)$ in \eqref{eq:F1}, with
$f_1(\theta,\psi) := f(\theta+W(\psi),\psi)-f(W(\psi),\psi)$, we can write $f_1(\theta)=f_2(\theta,\psi)+ \zeta(\theta,\psi)$, with
%
\vspace{-.1cm}
\begin{subequations} \label{eq:zetac0}
\begin{align}
\zeta(\theta,\psi) & := c_0(\theta) \, \gotf(\psi)  + c_1(\theta,\psi) \, W(\psi) + c_2(\theta,\psi)\,W(\psi)^2 , 
\label{eq:zeta} \\
c_0(\theta) & := \chi(\theta) - 1 , \phantom{ \int_0 \der t } 
\label{eq:c0} \\
c_1(\theta,\psi) & := \left( \partial_\theta  f(\theta,\psi) - \chi(\theta) \, \partial_\theta f(0,\psi) \right) ,
\label{eq:c1} \\
c_2(\theta,\psi) & := \int_0^1 \der t \, (1-t) \left( \partial_\theta^2
f(\theta+tW(\psi),\psi) - \chi(\theta) \, \partial_\theta^2  f(tW(\psi),\psi) \right),
\label{eq:c2} \\
\gotf(\psi) & :=f(W(\psi),\psi) = \rho^{-1}( W(A_0\psi)-W(\psi)) . \phantom{\sum_i}
\label{fWW} 
\end{align}
\end{subequations}
%
%
%
%
\vspace{-.6cm}

\noindent where the identity \eqref{fWW} derives from \eqref{eq:2.3}, with $F=\rho f$, and is used at length in the following.

\begin{rmk}\label{rmk:exS2id}
\emph{
The definition of the function $\chi$ ensures that $\SSS_2$ satisfies Hypotheses~\ref{hyp:1}--\ref{hyp:3},
provided $\rho$ is such that
%
$\rho \, \partial_\theta f_2(\theta,\psi) > -1$ for all $ (\theta,\psi) \in \Omega$.
%
Thus, if we wish to use the results in Subsections~\ref{subsec:synch} and
\ref{subsec:proof2} with $\SSS_2$ in place of $\SSS$ we need to
restrict $\rho$ to an interval $(0,\rho_0)$ with $\rho_0>0$ possibly smaller
than $\rho_*$,
as defined in Remark \ref{Psi0ge1}. This is not a problem since we are mainly
interested in the regime in which $\rho$ tends to zero. Moreover, for any
fixed $\rho_0\in(0,\rho_*)$, the bounds in Theorem~\ref{thm:exthm3} become
trivial for $\rho\in[\rho_0,\rho_*)$ by taking, if needed, larger values for the
involved constants $C$ (see also Remark \ref{regSS2}).
}
\end{rmk}

\begin{rmk}\label{rmk:S2id}
\emph{
By construction $\aver {f_2(\theta,\cdot)}=\aver
{f(\theta,\cdot)}$ for $\theta\in\calU$ and $\partial_\theta
f_2(0,\psi)=\partial_\theta f(0,\psi)$. On the other hand, the fact that $f_1(\theta)=f_2(\theta,\psi)+ \zeta(\theta,\psi)$
shows
that $\SSS_2$ can be seen as a regularization of $\SSS_1$. In particular, one has
$\SSS_2(0,\psi)=(0,A_0\psi)$, so that also the invariant manifold of $\SSS_2$ is
given by $\overline\calW=\{(0,\psi) : \psi\in\TTT^2\}$, and it is thus the
same as that of $\ovl\SSS$ and $\SSS_1$ (see Remarks \ref{Wbar} and \ref{rmk:twosteps}).
}
\end{rmk}

From Theorem~\ref{thm:2}, with $\SSS_2$ instead of $\SSS$, and Remark \ref{rmk:S2id}
it follows that there exists a set $\Omega_2 \subset \RRR \times \TTT^2$ and a
map $\HHH_2\!:\Omega \to \Omega_2$ of the
form
\vspace{-.1cm}
\begin{equation}
\HHH_2(\theta,\psi):=(\calH_2(\theta,\psi),\psi), \qquad
 \HH_2(\theta,\psi)
:=\theta+\theta^2 h_2(\theta,\psi) ,
\label{H2}
\vspace{-.1cm}
\end{equation}
which conjugates $\SSS_2$ to its linearization
$(\mu(\psi)\theta,A_0\psi)$, i.e.~such that
$\mu(\psi)\HH_2(\theta,\psi)=\HH_2(\SSS_2(\theta,\psi))$,
with $\mu(\psi)$ defined after \eqref{defd}.
Introducing the functions
%
\vspace{-.2cm}
\begin{subequations} \label{p2q2}
\begin{align}
p_2(\theta,\psi)
& := \frac{1}{1+\rho \partial_\theta f_2(0,\psi)}\left(\frac{\theta+\rho
f_2(\theta,\psi)}{\theta}\right)^2,
\label{p2q2a} \\
q_2(\theta,\psi)
& :=\frac{\rho}{1+\rho \, \partial_\theta f_2(0,\psi)}
\frac{f_2(\theta,\psi)-\partial_\theta f_2(0,\psi)\theta}{\theta^2} ,
\label{p2q2b}
\end{align}
\vspace{-.1cm}
\end{subequations}
%
%
%
we get
\vspace{-.2cm}
\begin{equation}
h_2(\theta,\psi) :=\sum_{n=0}^\io p^{(n)}_{2}(\theta,\psi) \,
q_2(\SSS_2^n(\theta,\psi)) , \qquad
p^{(n)}_{2}(\theta,\psi)  := \prod_{i=0}^{n-1} p_2(\SSS_2^i(\theta,\psi)) .
\label{p2n}
\vspace{-.2cm}
\end{equation}
%


In order to study the average of $h_1(\theta,\psi) - \ovl h (\theta)$ and of its derivative, we split
\begin{subequations} \label{h1h2hbarnoderder}
\begin{align}
h_1(\theta,\psi) - \ovl h (\theta) & = \left( h_1 (\theta,\psi) - h_2 (\theta,\psi) \right) + \left(  h_2 (\theta,\psi) - \ovl h (\theta)\right) ,
\label{h1h2hbar} \\
\partial_\theta h_1 (\theta,\psi) - \partial_\theta \ovl h (\theta)& = \left( \partial_\theta h_1 (\theta,\psi) - \partial_\theta h_2 (\theta,\psi) \right) +
\left( \partial_\theta h_2(\theta,\psi) - \partial_\theta \ovl h (\theta) \right) ,
\label{h1h2hbarder1}
\end{align}
\end{subequations}
and study separately the two contributions in each of \eqref{h1h2hbar} and \eqref{h1h2hbarder1}. This is the content of
Propositions~\ref{prop:h2hbar}, \ref{prop:fluhder1},  \ref{prop:hh2}, and \ref{prop:fluhder2} below,
which combined immediately imply Proposition~\ref{prop:h1hbar}.
For both maps $\SSS_1$ and $\SSS_2$, the $\theta$-component  vanishes at $\theta=0$.
However, while $\SSS_1$ depends on $W$ and hence inherits the low regularity of the
invariant manifold, $\SSS_2$ has the same regularity as $\SSS$.
Therefore, through the splitting \eqref{h1h2hbarnoderder}, we control first the deviations of $h_2$ from $\ovl h$
(see Subsections~\ref{672} and \ref{673}) using the regularity of $\SSS_2$ and the
fact that $\SSS$ and $\SSS_2$ share the same averaged map $\ovl\SSS$; next we
show that the deviations of $h_1$ from $h_2$ are small thanks to \eqref{eq:zeta} and Theorem~\ref{thm:3}
(see Subsection~\ref{674}).

\begin{rmk} \label{regSS2}
\emph{
For $\SSS_2$ to satisfy Hypotheses~\ref{hyp:1}--\ref{hyp:3} we need to restrict
the maximum value allowed for $\rho$ to $\rho_0$ since the condition
$\rho \, \partial_\theta f_2(\theta,\psi) > -1$ is more stringent than the condition
$\rho \, \partial_\f f(\f,\psi) > -1$ (see Remark \ref{rmk:exS2id}). 
However, since the bounds in Theorem~\ref{thm:exthm3} are trivially satisfied for any
fixed $\rho$, by possibly taking a larger constant $C_7$,
we may and do take for granted that the bounds hold for $\rho \ge \rho_0$.
Thus, we confine ourselves to consider $\rho\in(0,\rho_0)$ and assume that both
$\SSS$ and $\SSS_2$ satisfy Hypotheses~\ref{hyp:1}--\ref{hyp:3}.
}
\end{rmk}

\subsubsection{A new correlation inequality} \label{672}

To fulfill the program outlined at the end of the previous section, we start by comparing $h_2$ with $\ovl h$. To this aim,
by using the expansion \eqref{p2n}, we find useful to compare first $\aver{p_2{\vphantom{p}}^{(n)} q_2\circ\SSS_2^n}$ 
with $\aver{p_2}^{(n)} \aver{q_2}\circ\ovl G{\vphantom{G}}^n$ (see \eqref{h2ah2} below).
This comparison is similar to the comparison in Proposition~\ref{prop:decay},
the main difference being that the analogues of $g_+$, $\uu$ and $g_-$ now depend also on $\theta$.
Thus we need a correlation inequality that generalizes the previous one to this case.

The following preliminary result, proved in Appendix~\ref{proof:pna}, shows
that bounds analogous to those in Lemma~\ref{lem:pn}, which hold for functions
depending only on the slow variable, extend to functions
depending also on the fast variables, as far as the latter dependence is regular
enough.

\begin{lemma}\label{pna} 
Let $\pp_0,\ldots,\pp_{n-1}$ be any functions in $\calmB_{\al_0,3}(\Omega,\RRR)$ such that,
for some $\ga' \in (0,\ga)$,
\vspace{-.2cm}
\begin{enumerate}
\itemsep0em
\item $\norm\pp_i - 1 \norm_{\al_0,3}= O(\rho)$ for all $i=0,\ldots,n-1$, \label{1}
\item $|\pp_i(\theta)| \leq 1-\rho\,\ga'$ \label{2}
for $|\theta|\leq \theta'$ for some $\theta'$ independent of $\rho$ and
for all $i=0,\ldots,n-1$,
\vspace{-.2cm}
\end{enumerate}
and set $\pp^{(n)}(\theta,\psi) := \pp^{(n)}(\SSS_2;\theta,\psi)$, with $\pp^{(n)}(\SSS_2;\theta,\psi)$ defined according to \eqref{defpn}.
One has
\vspace{-.2cm}
\begin{equation} \label{bounds-pna}
\norm \pp^{(n)}\norm_{\al_0,3}^- \leq C \, (1-\rho\,\ga')^{n } , \qquad
%
%
%
\norm \partial_\theta (\SSS^n_2)_\theta \norm_{\al_0,4}^-
\leq C \, (1-\rho\,\ga')^n ,
\vspace{-.2cm}
\end{equation}
where the constant $C$ does not depend on $n$.
\end{lemma}

\begin{rmk} \label{ex3}
\emph{
Condition \ref{1} in Lemma~\ref{pna} implies that
$|\pp_i|_{\al_0}$, 
$\norm\partial_\theta \pp_i\norm_{\al_0,2}$
and
$\| \pp_i \!-\!\aver{\pp_i}\|_\infty$ are $O(\rho)$.
}
\end{rmk}

\begin{rmk} \label{ex-point3ter}
\emph{
A comment analogous to Remark \ref{ex-point3bis} applies also to Lemma~\ref{pna} and the forthcoming Proposition~\ref{prop:pCWu}:
if we assume that property 2 in Lemma~\ref{pna} holds for all $i=0,\ldots,n-1$
except at most $n_*$ values, with $n_* <n$ independent of $n$,
then both results still hold.
}
\end{rmk}

We are now ready to state the new correlation inequality, which is proved in Appendix~\ref{app:NCI}.

\begin{prop}\label{prop:decaynonltot}
Let $\pp_0,\ldots,\pp_{n-1}$ be any functions in $\calmB_{\al_0,3}(\Omega,\RRR)$ satisfying,
for some $\gamma'\in(0,\gamma)$, properties \ref{1} and \ref{2} in
Lemma~\ref{pna}.
For any $g_+\in \calmB_{\al_0}^+(\Omega,\RRR)$ and $g_-\in\calmB_{\al_0,3}^-(\Omega,\RRR)$  one has
\vspace{-.1cm}
\[
\begin{aligned}
& \left|\aver*{ g_+ \pp^{(n)} g_-\circ \SSS_2^n}-\aver {g_+}\aver 
\pp^{(n)}\aver{g_-}\circ \ovl G^n\right| \\
& \quad \leq 
C \left(1-\rho\,\ga'\right)^n 
\Bigl( (1+\al_0 n)\lambda^{-\al_0 n}\|  \tilde g_+\|^+_{\al_0}\|\tilde  g_-\|^-_{\al_0} 
+  \rho \| g_+\|^+_{\al_0} \norm g_-\norm^-_{\al_0,2} 
+ \rho^2 n \|\aver {g_+}\|_\infty \norm g_-\norm_{\al_0,3}^- \Bigr) ,
\end{aligned}
\vspace{-.2cm}
\]
where $\pp^{(n)}(\theta,\psi):=\pp^{(n)}(\SSS_2;\theta,\psi)$ and $\aver{\pp}^{(n)}(\theta) : =\aver{\pp}^{(n)}(\ovl \SSS;\theta)$,
with $\pp^{(n)}(\SSS_2;\theta,\psi)$ and $\aver{\pp}^{(n)}(\ovl \SSS;\theta)$ defined according to \eqref{bounds-pna} and
\eqref{defpnaver}, respectively, and $C$ is a constant independent of $n$.
\end{prop}

\begin{rmk} \label{rmkC}
\emph{
Proposition~\ref{prop:decaynonltot} can be seen as a generalization of
Proposition~\ref{prop:decay} to functions which also depend smoothly on the slow
variable. This will be exploited in Subsection~\ref{673} to compare the averaged
map with the auxiliary map, by using that all the involved functions are
regular -- i.e.~at least $\al_0$-H\"older continuous for some $\al_0$
independent of $\rho$ -- in the fast variable. The next step, to be achieved in
Subsection~\ref{674}, is to compare the auxiliary map with the translated
map, where the dependence on the fast variable is only $\als$-H\"older
continuous, with $\als=O(\rho)$.
}
\end{rmk}
\subsubsection{Deviations of the conjugation of the auxiliary map} \label{673}

As discussed in Remark \ref{regSS2}, we assume, without loss of generality, that $\rho\leq\rho_0$ and hence
that  $\SSS_2$ satisfies Hypotheses~\ref{hyp:1}--\ref{hyp:3}.
Recall also that we are working with the extension of $\SSS$, although not explicitly indicated (see Remark \ref{labelext});
thus, all functions appearing in what follows refer to such an extension.

We start by comparing $h_2$ with $\ovl h$. By relying on the expansions \eqref{p2n} and \eqref{hbar}, we may write
\vspace{-.1cm}
\begin{equation} \label{h2mhbar}
\null\hspace{-.2cm}
\aver{p_2}^{\!(n)}\aver{q_2}\circ \ovl G^n \!\!\!- \ovl p^{(n)} \, \ovl q\circ \ovl G^n
\!\!=\! \aver{q_2}\circ \ovl G^n
\sum_{i=0}^{n-1} \aver{p_2}^{\!(i)} \! \bigl( \aver{p_2} \!-\! \ovl p \bigr)  \circ \ovl G^{i} \, \ovl p^{(n-i-2)}\circ \ovl G^{i+1}
\!\!\!+\! \ovl p^{(n)} \bigl( \aver{q_2}\circ \ovl G^n \!\!\!- \ovl q\circ \ovl G^n \bigr) .
\vspace{-.1cm}
\end{equation}

We can now prove the following result.

\begin{prop} \label{prop:h2hbar}
Let $h_2$ and $\ovl h$ be defined according to \eqref{H2} and \eqref{def:lbar},
respectively. For all $\theta\in\calU$, one has
\vspace{-.1cm}
\begin{equation} \label{fluh2}
\bigl| \aver*{h_2(\theta,\cdot)-\ovl h(\theta) } \bigr|  \leq C\rho , \qquad\qquad
 \aver*{(h_2 (\theta,\cdot)-\ovl h(\theta))^2}\leq C\rho\, .
\end{equation}
\end{prop}

\noindent\emph{Proof}. 
Observe that $p_2(\theta,\psi)=1+O(\rho)$, so that (compare with Remark \ref{ex3})
\vspace{-.1cm}
\begin{equation} \label{deriv}
\|p_2-\aver{p_2} \|_{\infty} \leq C\rho, \qquad
|p_2|_{\al_0}\leq C\rho, \qquad
\norm\partial_\theta p_2\norm_{\al_0,2}\leq C\rho\, . 
\vspace{-.1cm}
\end{equation}
Since $p_2(0,\psi)=1+\rho \partial_\theta f_2(0,\psi)\leq 
1-\rho$, for any $\rho'\in(1,\rho)$ there exists $\theta_2$ such that $p_2(\theta,\psi)\leq 1-\rho'$ for 
$|\theta|\leq\theta_2$. Moreover, we easily check that
%
\begin{equation} \label{q2}
\norm q_2\norm_{\al_0,3}\leq C\rho ,
\end{equation}
so that we can apply Proposition~\ref{prop:decaynonltot} and obtain
%
\begin{equation} \label{h2ah2}
\bigl| \aver{p_2^{(n)} \! q_2 \circ \SSS_2^n}-\aver{p_2}^{(n)} \aver{q_2}\circ \ovl G^n\bigr| \le
C(1-\rho\,\ga')^n \rho^2 \, (1+\rho n) ,
\end{equation}
where $\aver{p_2}^{(n)}(\theta)$ is given by \eqref{defpnaver}, with $\pp_i=p_2$ $\forall i=0,\ldots,n-1$. Since
\vspace{-.1cm}
\begin{equation} \label{p2pbar}
\norm \aver{p_2}-\overline p \norm_{0,1} \leq C\rho^2  ,
\qquad
\norm  \aver{q_2}-\overline q \norm_{0,1} \leq C\rho^2 ,
\vspace{-.1cm}
\end{equation}
we have also, by \eqref{h2mhbar},
\begin{equation}\label{h2mhbar-bound}
\begin{aligned}
\bigl| \aver{p_2}^{(n)}\aver{q_2}\circ \ovl G^n- \ovl p^{(n)} \, \ovl q\circ \ovl G^n \bigr|
\le C(1 \!-\! \rho\,\ga')^n \left( n \rho^2  \| q_2 \|_{\io} + 
\|\aver{ q_2 } \!-\! \bar q \|_{\io} \right)
\le C (1 \!-\! \rho\,\ga')^n \rho^2 (1\!+\!\rho n) ,
\end{aligned}
\end{equation}
so that, combining \eqref{h2ah2} and \eqref{h2mhbar-bound}, we obtain
$| \aver{p_2^{(n)} \! q_2 \circ \SSS_2^n}-\ovl p^{(n)} \, \ovl q \circ \ovl G^n | \le C(1-\rho\,\ga')^n \rho^2(1+\rho n)$.
Summing over $n$ gives the first bound in \eqref{fluh2}. 

For the second bound, we start considering
\vspace{-.1cm}
\[
\begin{aligned}
&\bigl(p_2^{(n_1)} \! q_2 \circ \SSS_2^{n_1} - \aver{p_2}^{(n_1)}\aver{q_2}\circ \ovl G^{n_1}\bigr)
\bigl(p_2^{(n_2)} \! q_2  \circ \SSS_2^{n_2} - \aver{p_2}^{(n_2)}\aver{q_2}\circ \ovl G^{n_2}\bigr)\\
& \;
=p_2^{(n_1)} \! q_2 \circ \SSS_2^{n_1} \, p_2^{(n_2)} \! q_2 \circ \SSS_2^{n_2} 
- \aver{p_2}^{(n_1)}\aver{p_2}^{(n_2)}
\aver{q_2}\circ \ovl G^{n_1}\aver{q_2}\circ\ovl G^{n_2}\\
& \;
- \bigl(p_2^{(n_1)} \! q_2 \SSS_2^{n_1} \!-\! \aver{p_2}^{\!(n_1)}\aver{q_2}\circ \ovl G^{n_1}\bigr)
\aver{p_2}^{\!(n_2)}\aver{q_2}\circ \ovl G^{n_2}
-  \aver{p_2}^{\!(n_1)}\aver{q_2}\circ \ovl G^{n_1} \!
\bigl(p_2^{(n_2)} \! q_2 \SSS_2^{n_1} \!-\!  \aver{p_2}^{\!(n_2)}\aver{q_2}\circ \ovl G^{n_2}\bigr) ,
\end{aligned}
\]
and observe that, thanks to \eqref{h2ah2},
the averages of both contributions in the last two lines
are bounded by $C\rho^3(1+\rho (n_1+n_2))(1-\rho\,\ga')^{n_1+n_2}$.
As to the contribution in the second line, assuming $n_1 \le n_2$, we can write
\vspace{-.2cm}
\[
 p_2^{(n_1)} (\theta,\psi) \, q_2(\SSS^{n_1}_2(\theta,\psi)) \,
p_2^{(n_2)} (\theta,\psi) \, q_2(\SSS^{n_2}_2(\theta,\psi))= \left(
\prod_{i=0}^{n_1-1}
p_3(\SSS_2^i(\theta,\psi)) \right) q_{3,n_2-n_1}(\SSS_2^{n_1}(\theta,\psi)) ,
\vspace{-.2cm}
\]
with 
$p_3(\theta,\psi) \!:=\! \left( p_2(\theta,\psi) \right)^2$ and 
$q_{3,n}(\theta,\psi) \!:=\! q_2(\theta,\psi) p^{(n)}_2(\theta,\psi)\, q_2(\SSS^{n}_2(\theta,\psi))$.
Proposition~\ref{prop:decaynonltot} gives
\[
\bigl| p_2^{(n_1)} \! q_2\circ \SSS^{n_1}_2 \, p_2^{(n_2)} \! q_2
\circ \SSS^{n_2}_2 -\aver{p_3}^{(n_1)} \aver{ q_{3,n_2-n_1}} \circ \ovl
G^{n_1} \bigr|
\leq C \, (1-\rho\,\ga')^{n_1} \rho \, (1 + n_1\rho) \norm q_{3,n_2-n_1}\norm_{\al_0,3}^- \, ,
\]
where we have $\norm q_{3,n_2-n_1}\norm_{\al_0,3}^-  \leq C\rho^2(1-\rho\,\ga')^{n_2-n_1}$,
as a consequence of Lemma~\ref{pna}, of the bound \eqref{q2} and of the
inequality \eqref{eq:normalk}. Thus we are left with studying
\vspace{-.1cm}
\[ 
\begin{aligned}
& \aver{p_3}^{(n_1)}
 \aver{ q_{3,n_2-n_1}} \circ \, \ovl G^{n_1} - \aver{p_2}^{(n_1)}\aver{p_2}^{(n_2)}
\aver{q_2}\circ \, \ovl G^{n_1}\aver{q_2}\circ\ovl G^{n_2} \\
& \qquad \qquad
= \aver{p_3}^{(n_1)} \bigl(
\aver{q_{3,n_2-n_1}} \circ \,\ovl G^{n_1} -\aver{q_2} \circ \ovl G^{n_1}\aver{p_2}^{(n_2-n_1)} \circ \ovl G^{n_1} \aver{q_2}\circ \ovl G^{n_2} \bigr) \\
& \qquad \qquad
+ \bigl( \aver{p_3}^{(n_1)}  \aver{p_2}^{(n_2 - n_1)} \circ \ovl G^{n_1} - \aver{p_2}^{(n_1)} \aver{p_2}^{(n_2)} \bigr) 
\aver{q_2} \circ \ovl G^{n_1} \aver{q_2} \circ \ovl G^{n_2} .
\end{aligned}
\vspace{-.1cm}
\]
Using again Proposition~\ref{prop:decaynonltot} and \eqref{q2}, we obtain
\vspace{-.2cm}
\[
\begin{aligned}
& \bigl| \aver{q_{3,n_2-n_1}}-\aver{q_2}\aver{p_2}^{(n_2-n_1)}\aver{q_2}\circ \ovl G^{n_2-n_1} \bigr| \\
& \qquad \qquad \leq 
C\rho^2(1-\rho\,\ga')^{n_2-n_1} \bigl( (1+\al_0(n_2-n_1)) \lambda^{-\al_0(n_2-n_1)}+ \rho +\rho^2 (n_2-n_1) \bigr) ,
\end{aligned} 
\vspace{-.1cm}
\]
while the first bound in \eqref{deriv} yields $|\aver{p_3}-\aver{p_2}^2|\leq C\rho^2$, which in turn gives
\[
\vspace{-.2cm}
\begin{aligned}
& \bigl| \aver{p_3}^{(n_1)} 
\aver{p_2}^{(n_2 - n_1)} \circ \ovl G^{n_1} 
 - \aver{p_2}^{(n_1)} \aver{p_2}^{(n_2)} \bigr|
= \bigl| \bigl( \aver{p_3}^{(n_1)} - \aver{p_2}^{(n_1)} \aver{p_2}^{(n_1)}  \bigr) \aver{p_2}^{(n_2-n_1)} \circ \ovl G^{n_1} \bigr| \\
& \qquad = \big| \sum_{i=0}^{n_1} \aver{p_3}^{(i)} \bigl( \aver{p_3} - \aver{p_2}^2 \bigr) \circ \ovl G^i \, \aver{p_2}^{(n_2-n_1)} \circ \ovl G^{n_1} \bigr|
\le C \, (1 - \rho\,\ga')^{n_1+n_2} n_1 \rho^2 ,
\end{aligned}
\vspace{-.1cm}
\]
Combining all the estimates together and proceeding like in \eqref{h2mhbar}, we get, respectively, 
\vspace{-.2cm}
\begin{equation*}  
\biggl\langle \biggl( \,  \sum_{n=0}^{\io} \left( p^{(n)}_{2} \, q_2 \circ \SSS_2^n - \aver{p_2}^{(n)}\aver{q_2}\circ \ovl  G^{n} \right) \biggr)^{\!2} \biggr\rangle \leq C\rho ,
\qquad
\biggl(  \, \sum_{n=0}^{\io} \left( \aver{p_2}^{(n)}\aver{q_2}\circ \ovl G^{n}  - \ovl p^{(n)} \, \ovl q\circ \ovl G^n   \right) \biggr)^{\!\!2}  \leq C\rho .
\vspace{-.2cm}
\end{equation*}
which, together, provides the second bound in \eqref{fluh2}.
\qed

\medskip

The following result extends the analysis above to the first derivatives of $h_2$ and $\ovl h$;
the proof, based on the same ideas used for Proposition~\ref{prop:h2hbar}
up to technical intricacies, is deferred to Appendix~\ref{proof:fluhder1}.

\begin{prop} \label{prop:fluhder1}
\!Let $h_2$ and $\ovl h$ be defined as in \eqref{p2n} and
\eqref{hbar}, respectively.~For all $\theta \!\in\! \calU$, one has
\vspace{-.2cm}
\begin{equation} \label{fluhder1}
\bigl| \aver*{ \partial_\theta h_2(\theta,\cdot) - \partial_\theta \ovl h(\theta)} \bigr| \leq C\rho , 
\qquad \aver*{(\partial_\theta h_2(\theta,\cdot) - \partial_\theta \ovl h (\theta) )^2} \leq C\rho\, .
\vspace{.1cm}
\end{equation}
\end{prop}

\begin{rmk} \label{pre-regularity2}
\emph{
To prove Proposition~\ref{prop:fluhder1} we need $f\in \calmB_{\al_0,5}$ while to prove
Theorems~\ref{thm:1} to \ref{thm:3} it would be enough to assume $f\in
\calmB_{\al_0,2}$. The full
regularity assumed in our hypotheses is required to prove the forthcoming
Proposition~\ref{prop:fluhder2} (see Remark \ref{regularity2}).
}
\end{rmk}

\subsubsection{Comparison between the translated map and the auxiliary map} \label{674}

We start by introducing some notation. 
If, for $\pp_1 , \pp_2\in\calmB_{0,3}(\Omega,\RRR)$, we define
%
\vspace{-.3cm}
\begin{equation} \label{p1np2ndef}
\pp_1^{(n)} := 
\prod_{i=0}^{n-1} \pp_1 \circ \SSS_1^i , \qquad
\pp_2^{(n)}  := 
\prod_{i=0}^{n-1} \pp_2 \circ \SSS_2^i  ,
\vspace{-.2cm}
\end{equation}
reasoning as in Subsection~\ref{fluk} we may write
\vspace{-.3cm}
\begin{subequations} \label{p1n-p2n+h1-h2}
\begin{align}
\pp_1^{(n)} - \pp_2^{(n)}
& = \sum_{k=0}^{n-1} \pp_1^{(k)}
\bigl( \pp_1\circ\SSS_1^k  - \pp_2\circ\SSS_2^k \bigr)
\, \pp_2^{(n-k-1)} \circ \SSS_2^{k+1} ,
\label{p1n-p2n} \\
\pp_1^{(n)} \qq_1 \circ \SSS_1^n - \pp_2^{(n)} \, \qq_2 \circ \SSS_2 ^n
& = \bigl( \pp_1^{(n)} - \pp_2^{(n)} \bigr) \, \qq_2\circ\SSS_2^n
+ \pp_1^{(n)} \, \bigl( \qq_1\circ\SSS_1^n-\qq_2\circ\SSS_2^n \bigr) \, ,
\label{h1-h2}
\end{align}
\vspace{-.2cm}
\end{subequations}
and, in a similar way, we find 
%
\begin{subequations} \label{pnS1pnS2+p1nS1p1nS2}
\begin{align}
\pp_2^{(n)} \circ \SSS_1^i - \pp_2^{(n)} \circ \SSS_2 ^i
& = \sum_{k=0}^{n-1} \pp_2^{(k)}\circ\SSS_1^i
\bigl( \pp_2\circ\SSS_2^k \circ \SSS_1^i  - \pp_2\circ\SSS_2^k \circ \SSS_2^i
\bigr)
\, \pp_2^{(n-k-1)} \circ \SSS_2^{k+1+i}
\label{pnS1pnS2} \\
\pp_1^{(n)} \circ \SSS_1^i - \pp_2^{(n)} \circ \SSS_2 ^i
& = \sum_{k=0}^{n-1} \pp_1^{(k)}\circ\SSS_1^i
\bigl( \pp_1\circ\SSS_1^{k+i}  - \pp_2\circ\SSS_2^{k+i} \bigr)
\, \pp_2^{(n-k-1)} \circ \SSS_2^{k+1+i} ,
\label{p1nS1p1nS2}
\end{align}
\end{subequations}
with the latter reducing to \eqref{p1n-p2n} for $i=0$.

Now, we come back to Proposition~\ref{prop:h1hbar}. To complete the proof
we are left to study the contributions  $h_1 - h_2$ and $\partial_\theta h_1 -
\partial_\theta h_2$ in \eqref{h1h2hbarnoderder}.
In the light of \eqref{lll} and of the the expansion \eqref{p1n-p2n}, we may write
\vspace{-.2cm}
\begin{equation} \label{h1-h2true}
\begin{aligned}
 & h_1 - h_2 = \sum_{n=0}^\io \bigl( p_1^{(n)} \, q_1 \circ \SSS_1^n -p_2^{(n)} q_2 \circ \SSS_2 ^n \bigr) =
\sum_{n=0}^\io p_2^{(n)} \bigl( q_1 \circ \SSS_1^n - q_2 \circ \SSS_2 ^n \bigr) \\
& \quad + \sum_{n=0}^\io \bigl( p_1^{(n)} \!-\! p_2^{(n)} \bigr) \, q_2 \circ \SSS_2 ^n 
+ \sum_{n=0}^\io\sum_{k=0}^{n-1}
p_1^{(k)} \, (p_1\circ\SSS_1^k \!-\! p_2\circ\SSS_2^k) \, p_2^{(n-k-1)} \circ \SSS_2^{k+1} (q_1\circ\SSS_1^n - q_2\circ\SSS_2^n ) .
\end{aligned}
\vspace{-.1cm}
\end{equation}

We estimate $h_1-h_2$ by studying the differences that appear in \eqref{h1-h2true}.
First, we prove a series a technical lemmas based on the structure of the difference $f_1-f_2$,
and come back to \eqref{h1-h2true} in Lemma~\ref{lem:intermediate-1} below.
To start with, we write $\zeta$ as in \eqref{eq:zeta} and note
%
that $\norm \zeta\norm_{0,2}\leq C$,
while, setting $\zeta_0=c_1(\theta,\psi) \, W(\psi) + c_2(\theta,\psi)\,W(\psi)^2$,
we have $\aver{\zeta_0^2}\leq C\rho$ and
$\aver{(\partial_\theta\zeta_0)^2}\leq C\rho$ by \eqref{Wflu} in Theorem~\ref{thm:3}.

\begin{rmk} \label{rmk:last}
\emph{
A key observation in the argument used below and in the related appendices is the following.
According to 
\eqref{eq:zeta},
the function $\zeta$ can be written as sum of three terms. While the last two terms,
which depend linearly or quadratically on $W$, are controlled by Theorem~\ref{thm:3},
the first one depends on $W$ through the function $\gotf$.
One can write $\gotf$ in terms of the difference $W\circ A_0-W$ (and hence linearly in $W$) as in \eqref{fWW},
but, in doing so, a factor $\rho$ is lost. However, in order to compare $\SSS_1$ with $\SSS_2$,
one deals with sums over $i$ of contributions of the form $\Xi_i \circ \SSS_2^ i (W\circ A_0^{i+1} -W \circ A_0^i)$,
with $\Xi_i$ more regular than $W$
(see for instance 
the proof of Lemma~\ref{lem:intermediate-1} below).
Thus, one can rearrange the sums and obtain summands of the form
$(\Xi_{i+1}\circ\SSS_2^{i-1} - \Xi_i\circ\SSS_2^i) W\circ A_0^i$ (see Lemmas~\ref{lem:new} and \ref{lem:new1}),
where the differences $\Xi_{i+1}\circ\SSS_2^{i-1} - \Xi_i\circ\SSS_2^i$
allow to gain a compensating factor $\rho$. 
}
\end{rmk}

The following result plays a crucial role in the forthcoming discussion. The proof,
based on the idea illustrated in Remark \ref{rmk:last}, is given in Appendix~\ref{proof:S1-S2}
(which we refer to for more details).

\begin{lemma}\label{S1-S2}
For any $\pp$ in $\calmB_{\al_0,3}(\Omega,\RRR)$ one has
\vspace{-.3cm}
\begin{equation} \label{pS1-pS2}
\pp\circ \SSS_1^n-\pp\circ \SSS_2^n=
\Delta_{\pp,W,n} := \rho\sum_{i=0}^{n} \gotC_{\pp,n,n-i} \circ
\SSS_2^{i}\; W\circ A_0^i+  \rho \, \gotR_{\pp,n} ,
\vspace{-.2cm}
\end{equation}
for suitable functions $\gotC_{\pp,n,0},\ldots,\gotC_{\pp,n,n} \in \calmB_{\al_0,2}(\Omega,\RRR)$ and
$\gotR_{\pp,n} \in\calmB_{0,2}(\Omega,\RRR)$ such that
%
\vspace{-.1cm}
\begin{subequations} \label{sumCkR}
\begin{align}
& \norm \gotC_{\pp,n,k}\norm_{\al_0,2}^-  \le C\, (1-\rho\,\ga')^k \, \norm\partial_\theta \pp\norm_{\al_0,2} ,
\qquad k=1,\ldots,n-1, 
\label{Ck} \\
& \sum_{k=0}^{n} \norm \gotC_{\pp,n,k}\norm_{\al_0,2}^- \le C \, \rho^{-1} \norm\partial_\theta \pp\norm_{\al_0,2} ,
\label{sumCk} \\
& \rho \| \gotR_{\pp,n}\|_\infty + \aver{| \gotR_{\pp,n}|} \le C \, \norm\partial_\theta \pp\norm_{\al_0,2} .
\label{R}
\end{align}
\end{subequations}
\end{lemma}

The two next results are immediate consequences of Lemma~\ref{S1-S2}.

\begin{lemma} \label{p1S1p2S2}
For any $\pp_2 \in \calmB_{\al_0,3}(\Omega,\RRR)$ and any $\pp_1\in \calmB_{0,3}(\Omega,\RRR)$ such that
\vspace{-.1cm}
\begin{equation}\label{p1p2}
(\pp_1-\pp_2)(\theta,\psi) =
\rho \, \ccc_1(\theta,\psi)W(\psi)+\rho \, \ccc_2(\theta,\psi) + \rho \, \ccc_3(\theta) \, \gotf (\psi) , 
\vspace{-.1cm}
\end{equation}
with $\ccc_1\in\calmB_{\al_0,2}(\Omega,\RRR)$, $\ccc_2 
\in\calmB_{0,2}(\Omega,\RRR)$, $\ccc_3\in C^1(\calU,\RRR)$ and $\gotf$ as
in \eqref{fWW},
%
%
one has
\vspace{-.1cm}
\begin{equation} \label{diffp1p2}
\pp_1\circ \SSS_1^n-\pp_2\circ \SSS_2^n =
\Delta_{\pp_1,\pp_2,W,n} + \rho \, \ccc_3 \circ \SSS_1^n \, \gotf \circ A_0^n , 
\vspace{-.3cm}
\end{equation}
with
\vspace{-.4cm}
\begin{subequations} \label{CpCpRW}
\begin{align}
\Delta_{\pp_1,\pp_2,W,n} &:= 
\rho \sum_{i=0}^{n}  \gotC_{\pp_1,\pp_2,n,n-i} \circ\SSS_2^i\, W\circ A_0^i + \rho \, \gotR_{\pp_1,\pp_2,n} 
\label{bbbbb} \\
\gotC_{\pp_1,\pp_2,n,k} & := \gotC_{\pp_2,n,k} + \de_{k,0} \ccc_1 , \qquad k=0,\ldots,n , 
\label{CpCp} \\
\gotR_{\pp_1,\pp_2,n} & := \gotR_{\pp_2,n} + \ccc_2 \circ \SSS_1^n + 
( \ccc_1 \circ \SSS_1^n - \ccc_1 \circ \SSS_2^n ) \, W \circ A_0^n . \phantom{\sum^{n}}
\label{CpCpR} 
\end{align}
\end{subequations}
\end{lemma}

\vspace{-.3cm}

\proof
Write
$\pp_1\circ \SSS_1^n-\pp_2\circ \SSS_2^n  =  \pp_2\circ \SSS_1^n-\pp_2\circ \SSS_2^n + (\pp_1-\pp_2)\circ \SSS_1^n$
and 
$\ccc_1 \circ \SSS_1^n = \ccc_1 \circ \SSS_1^n - \ccc_1 \circ \SSS_2^n +  \ccc_1 \circ \SSS_2^n$,
and use Lemma~\ref{S1-S2} to deal with the contribution $\pp_2\circ \SSS_1^n-\pp_2\circ \SSS_2^n$.
\qed

\begin{coro} \label{p1S1p2S2bis}
Let $\pp_1,\pp_2 \in \calmB_{\al_0,3}(\Omega,\RRR)$ be such that 
$\norm \partial_\theta \pp_2 \norm_{\al_0,2} \le C \rho$ 
and 
\eqref{p1p2} holds, with
%
$\norm \ccc_1 \norm_{\al_0,2}^- \le C$,
and
$\rho \| \ccc_2 \|_\infty  + \aver{|\ccc_2|} \le C \rho$.
The functions \eqref{CpCpRW} satisfy the bounds
\vspace{-.2cm}
\begin{equation*} 
\sum_{k=0}^n \norm \gotC_{\pp_1,\pp_2,n,k}\norm_{\al_0,2}  \leq C  ,\qquad\qquad
\rho  \| \gotR_{\pp_1,\pp_2,n}\|_\infty + \aver{|\gotR_{\pp_1,\pp_2,n}|}\leq C\rho .
\vspace{-.1cm}
\end{equation*}
\end{coro}

\proof
The assertion follows immediately from the bounds \eqref{sumCkR} and from Theorem~\ref{thm:3}.
\qed

\begin{rmk} \label{rmk:diffp1p2}
\emph{
Taking either $\pp_1=p_1$ and $\pp_2=p_2$ or $\pp_1=q_1$ and $\pp_2=q_2$,
with $p_1$ and $q_1$ as in \eqref{g1g2} and $p_2$ and $q_2$ as in \eqref{p2q2},
the hypotheses of Lemma~\ref{p1S1p2S2} are verified, with the functions $\ccc_1$ and $\ccc_2$ satisfying the
estimates in Corollary \ref{p1S1p2S2bis} in both cases. A straightforward computation gives
$\ccc_3(\theta) = \gota_1(\theta):=-2\theta^{-1} c_0(\theta)$
and $\ccc_3(\theta) = \gota_2(\theta):=-\theta^{-2} c_0(\theta)$, respectively, so that
\vspace{-.1cm}
\begin{equation} \label{p1p2q1q2}
p_1\circ \SSS_1^n \!-\! p_2\circ \SSS_2^n
\!=\! \Delta_{p_1,p_2,W,n}
+ \rho \, \gota_1 \circ \SSS_1^n \, \gotf \circ A_0^n , 
\qquad
q_1\circ \SSS_1^n \!-\! q_2\circ \SSS_2^n
\!=\! \Delta_{q_1,q_2,W,n}
+ \rho \, \gota_2 \circ \SSS_1^n \, \gotf \circ A_0^n , 
\vspace{-.1cm}
\end{equation}
with $\gota_1, \gota_2 \in C^\io(\calU,\RRR)$, $\| \gota_1 \|_{\io} \le C$ and $\| \gota_2 \|_{\io} \le C$. Moreover, Theorem~\ref{thm:3},
for any $n\ge 0$ one has
\vspace{-.1cm}
\begin{equation} \label{shorten-norm}
\norm \Delta_{p_1,p_2,W,n} \norm_{\al_0,2}^-  \le C \rho , \qquad
\norm \Delta_{q_1,q_2,W,n} \norm_{\al_0,2}^-  \le C \rho .
\vspace{-.1cm}
\end{equation}
%
If one restricts $\SSS$ to $\Lambda_{2r_2}$ (see Remark \ref{rmk:S3}), $c_0$ is replaced with $0$ and both functions $\gota_1$ and $\gota_2$ vanish, so that
the terms with $\gotc_3$ in \eqref{p1p2} and \eqref{diffp1p2} disappear.
In particular, the coming Lemmas~\ref{lem:new} and \ref{lem:new1}
are not needed, and, in the discussion of the remaining results,
all terms involving the functions $\gota_1$ and $\gota_2$ vanish,
with a substantial simplification of the proofs.
}
\end{rmk}


\begin{rmk} \label{rmk:exshorten}
\emph{
Bounds analogous to \eqref{shorten-norm} hold for $\Delta_{p_2,W,n}$ and $\Delta_{q_2,W,n}$. Indeed, for any $n\ge 0$, one has
%
$\norm \Delta_{p_2,W,n} \norm_{\al_0,2}^- \le C \rho$
and
$\norm \Delta_{q_2,W,n} \norm_{\al_0,2}^- \le C \rho$.
Moreover, by Theorem~\ref{thm:3}, for any $n,k\ge 0$ one has
\vspace{-.2cm}
\begin{subequations} \label{shorten-inf}
\begin{align}
& \average{ \bigl| \Delta_{p_2,W,n} \, W \circ A_0^k \bigr| } \le C \rho^2 , \phantom{\sum} 
& \qquad \average{ \bigl| \Delta_{q_2,W,n} \, W \circ A_0^k \bigr| } \le C \rho^2 ,
\label{shorten-infa} \\
& \average{\bigl| \Delta_{p_1,p_2,W,n} \, W \circ A_0^k \bigr| } \le C \rho^2 , \qquad
& \average{ \bigl| \Delta_{p_1,p_2,W,n} \, \Delta_{q_1,q_2,W,k}  \bigr| } \le C\rho^3 .
\label{shorten-infb}
\end{align}
\end{subequations}
}
\end{rmk}

Corollary~\ref{p1S1p2S2bis} allows us to deal with the first two contributions in the r.h.s.~of \eqref{diffp1p2}.
In order to deal with the last contribution we need also the following two results.

\begin{lemma} \label{lem:new}
For $\pp_1,\pp_2 \in\calmB_{0,3}(\Omega,\RRR)$, such that \eqref{p1p2} is satisfied and $\| \pp_1 -1 \|_{\io} \le C\rho$,
let $\pp_1^{(n)}$ and $\pp_2^{(n)}$ be defined as in \eqref{p1np2ndef}.
For any function $\gota \!: \calU \to \RRR$ of class $C^1$ and any $n\ge 1$ and
$j\ge 0$, one has
\vspace{-.1cm}
\begin{equation} \label{new}
\begin{aligned}
& \sum_{k=0}^{n-1} \pp_1^{(k)} \circ \SSS_1^j \, \gota \circ \SSS_1^{k+j} \pp_2^{(n-1-k)} \circ \SSS_2^{k+1+j} \bigl( W \circ A_0^{k+1+j} - W \circ A_0^{k+j} \bigr) \\
& \qquad \qquad =
\sum_{k=0}^{n} \pp_1^{(k-1)}\circ\SSS_1^j \, \gotD_{\gota,n,k,j} \circ \SSS_1^{k+j} \, \pp_2^{(n-1-k)} \circ \SSS_2^{k+1+j} \, W \circ A_0^{k+j} ,
\end{aligned}
\vspace{-.2cm}
\end{equation}
with
\begin{equation} \label{new-def}
\gotD_{\gota,n,k,j} :=
\begin{cases}
-\gota , & \quad k= 0 , \phantom{\displaystyle{\sum}} \\
\gota \circ \SSS_1^{-1} \, \pp_2 \circ \SSS_2^{k+j} \circ \SSS_1^{-(k+j)} - \pp_1 \circ \SSS_1^{-1} \, \gota , & \quad k=1,\ldots,n-1 , \\
\gota \circ \SSS_1^{-1} , & \quad k=n , \phantom{\displaystyle{\sum}}
\end{cases}
\end{equation}
such that
\begin{equation} \label{new-bound}
\| \gotD_{\gota,n,k,j} \|_{\io} \le C \rho \, \norm \gota \norm_{0,1}, \qquad k=1,\ldots,n-1 .
\end{equation}
\end{lemma}

\proof
The identity \eqref{new} is easily checked. To bound $\gotD_{\gota,n,k,j}$ for $k=1,\ldots,n-1$, write
\[
\begin{aligned}
& \gota \circ \SSS_1^{-1} \, \pp_2 \circ \SSS_2^{k+j} \circ \SSS_1^{-(k+j)} - \gota \, \pp_1 \circ \SSS_1^{-1} \\
& \qquad  = 
\bigl( \gota \circ \SSS_1^{-1} - \gota \bigr) \pp_2 \circ \SSS_2^{k+j} \circ \SSS_1^{-(k+j)} +
\gota \bigl( \pp_2\circ \SSS_2^{k+j}  - \pp_1 \circ \SSS_1^{k+j} \bigr) \circ \SSS_1^{-(k+j)} 
+  \gota \bigl( \pp_1  - \pp_1 \circ \SSS_1^{-1} \bigr) ,
\end{aligned}
\]
and use that
\[
\begin{aligned}
\bigl\| \bigr( \gota \circ \SSS_1^{-1} - \gota \bigr) \pp_2 \bigr\|_{\io} & \le
\| \partial_\theta \gota \|_{\io} \| (\SSS_1)_\theta - \uno \|_\io \le C \rho \, \| \partial_\theta \gota \|_{\io} , \\
\bigl\| \pp_2 \circ \SSS_2^{k+j} - \pp_1 \circ \SSS_1^{k+j} \bigr\|_{\io} & \le
\rho \sum_{i=0}^{k+j} \| \gotC_{\pp_1,\pp_2,k+j,k+j-i} \|_{\io} +  \rho \| \gotR_{\pp_1,\pp_2,k+j} \|_{\io} 
+ \rho \| \ccc_3 \|_{\io} \| \gotf \|_{\io}  \le C \rho , \\
\bigl\| \pp_1 - \pp_1 \circ \SSS_2^{-1} \bigr\|_{\io} & \le
\| ( 1 + O(\rho) ) - (1 + O(\rho))  \|_{\io} \le C \rho ,
\end{aligned}
\]
with the second inequality following from \eqref{diffp1p2} in Lemma~\ref{p1S1p2S2} and the bounds in Corollary~\ref{p1S1p2S2bis}.
\qed

\begin{rmk} \label{Dnk0}
\emph{
The coefficients $\gotD_{\gota,n,k,j}$ with $k=0,\ldots,n-1$ in \eqref{new-def} do not depend on $n$, in the sense that 
$\gotD_{\gota,n,k,j} := \gotD_{\gota,n',k,j}$ for all $k < \min\{n,n'\}$. Thus, we may define, for future convenience,
\vspace{-.1cm}
\[
\gotD_{\gota,k,j} := \gota \circ \SSS_1{\vphantom{s}}^{-1} \, \pp_2 \circ \SSS_2{\vphantom{s}}^{k+j} \circ
\SSS_1{\vphantom{s}}^{-(k+j)} - \pp_1 \circ \SSS_1{\vphantom{s}}^{-1} \gota .
\]
}
\end{rmk}

\begin{lemma} \label{lem:new1}
Let $\pp_1,\pp_2 \in \calmB_{0,3}(\Omega,\RRR)$ be as in Lemma~\ref{lem:new}, and let
$\pp_1^{(n)}$ and $\pp^{(n)}_2$ be defined as in \eqref{p1np2ndef}.
For any function $\gota \!: \calU \to \RRR$ of class $C^3$ one has
%
\begin{equation} \label{new1}
\sum_{n=0}^{\io} \pp_1^{(n)} \, \gota \circ \SSS_1^n \bigl( W \circ A_0^{n+1} - W \circ A_0^n \bigr) =
\sum_{n=0}^{\io} \pp_1^{(n-1)} \, \gotD_{0,\gota,n} \circ \SSS_1^{n} \, W \circ A_0^n ,
\end{equation}
with
\begin{equation} \label{new-def1}
\gotD_{0,\gota,n} :=
\begin{cases}
-\gota , & \quad n= 0 , \phantom{\displaystyle{\sum}} \\
\gota \circ \SSS_1^{-1} - \pp_1 \circ \SSS_1^{-1} \, \gota, & \quad n \ge 1 ,
\end{cases}
\end{equation}
and, similarly, for $r=1,2$,
\begin{equation} \label{new2}
\sum_{n=0}^{\io} \pp_2^{(n)} \, \gota \circ \SSS_r^n \bigl( W \circ A_0^{n+1} - W \circ A_0^n \bigr) =
\sum_{n=0}^{\io} \pp_2^{(n-1)} \, \gotD_{r,\gota,n} \circ \SSS_r^{n} \, W \circ A_0^n ,
\vspace{-.2cm}
\end{equation}
with
\begin{equation} \label{new-def2}
\gotD_{r,\gota,n} :=
\begin{cases}
-\gota , & \quad n= 0 , \phantom{\displaystyle{\sum}} \\
\gota \circ \SSS_r^{-1} - \pp_2 \circ \SSS_2^{n-1} \circ \SSS_r^{-n} \, \gota, & \quad n \ge 1 .
\end{cases}
\end{equation}
Furthermore,
if $\rho$ is such that $\SSS_2$ satisfies Hypotheses~\ref{hyp:1}--\ref{hyp:3},
one has 
\vspace{-.2cm}
\begin{subequations} \label{new-bound12+2}
\begin{align}
\| \gotD_{r,\gota,n} \|_{\io} & \le C \rho \, \norm \gota \norm_{0,1} , \qquad n \ge 1 , \quad r=0,1,2 , \phantom{\Big(}
\label{new-bound12} \\
\norm \gotD_{2,\gota,n} \norm_{0,2} & \le C \rho \, \norm \gota \norm_{0,3} , \qquad n \ge 1 . \phantom{\Big(}
\label{new-bound2}
\end{align}
\end{subequations}
\end{lemma}

\proof
After checking \eqref{new1} and \eqref{new2} by direct computation,
the bounds \eqref{new-bound12} are easily obtained by writing
%
$\gota \circ \SSS_1^{-1} - \pp_2 \circ \SSS_2^{n-1} \circ \SSS_1^{-n} \, \gota =
\gota \circ \SSS_1^{-1} - \pp_1 \circ \SSS_1^{-1} \, \gota +
\left( \pp_1 \circ \SSS_1^{n-1} - \pp_2 \circ \SSS_2^{n-1} \right) \circ \SSS_1^{-n}$
in \eqref{new-def2}
%
and using that
$\| \gota \circ \SSS_r^{-1} - \pp_r \circ \SSS_r^{-1} \, \gota \|_{\io} 
\| \gota \circ \SSS_r^{-1} - \gota \|_{\io} + \| (1 - \pp_r \circ \SSS_r^{-1} ) \|_\io 
\| \gota \|_{\io} \le C\rho \, \norm  \gota \norm_{0,1}$
for $r=1,2$,
and
$\| \pp_1 \circ \SSS_1^{n-1} - \pp_2 \circ \SSS_2^{n-1} \|_\io
\le C\rho$,
with the second bound holding by Lemma~\ref{p1S1p2S2}.

The bound \eqref{new-bound2} is obtained by writing
$\gota \circ \SSS_2^{-1} - \pp_2 \circ \SSS_2^{-1} \, \gota=
\bigl( \gota \circ \SSS_2^{-1} - \gota \bigr) + \bigl(  1 - \pp_2 \circ \SSS_2^{-1} \bigr) \gota$
and using the bounds of Lemma~\ref{pna}. \qed


\medskip

%
%
%
%

Also the next result, essentially based on Proposition \ref{prop:decaynonltot}, is used at length in what follows.
 
\begin{prop} \label{prop:pCWu}
Let $\pp_0,\ldots,\pp_{k}\in\calmB_{\al_0,3}(\Omega,\RRR)$ satisfy
the properties 1 and 2 in Lemma~\ref{pna},
and let $\pp^{(k)}(\theta,\psi):=\pp^{(k)}_0(\SSS_2;\theta,\psi)$ be defined according to \eqref{defpn}.
For any $\gotC\in \calmB_{\al_0,2}(\Omega,\RRR)$ and $\uu\in
\calmB_{\al_0,3}(\Omega,\RRR)$,
and for any $\gamma'\in(0,\gamma)$ and $i=0,\ldots,k$, one has
\vspace{-.1cm}
\[
\Bigl| \Average{\pp^{(k)} \gotC \circ \SSS_2^i \, W \circ A_0^i \, \uu \circ \SSS_2^{k+1} }  \Bigr|
\le
C (1-\rho\,\ga')^{k} \left((1+\al_0 i) \la^{-\al_0 i} + \rho + i \rho^2 + i^2 \rho^3 \right) 
\norm \gotC  \norm_{\al_0,2}^- \norm \uu \norm_{\al_0,3}^- .
\vspace{-.1cm}
\]
\end{prop}

\proof
For $i=0,\ldots,k$,
let $\pp^{(k)}_{[i]}(\theta,\psi):=\pp^{(k)}_{[i]}(\SSS_2;\theta,\psi)$ be as in \eqref{defpn} with $\SSS=\SSS_2$.
We have
\vspace{-.3cm}
\[
\begin{aligned}
\null \hspace{.2cm} \pp^{(k)} \gotC \circ\SSS_2^i\, W \circ A_0^i \, \uu\circ \SSS_2^{k+1} & 
=  \pp^{(k)} \, \gotC \circ\SSS_2^i\, 
(W-W_0 ) \circ A_0^i  \, \uu \circ \SSS_2^{k+1}
+ \pp^{(i)} \, \pp^{(k-i)}_{[i]} \circ \SSS_2^i \, 
\gotC \circ\SSS_2^i\, \mu^{(i)} \, W_0 \,
\uu \circ \SSS_2^{k+1} \phantom{\sum^i } \\
& + \rho  \sum_{j=1}^{i} 
\pp^{(i-j)} \, \pp^{(k-i+j)}_{[i-j]} \circ \SSS_2^{i-j} \, \gotC \circ \SSS_2^i\, 
\mu^{(j-1)} \circ A_0^{i-(j-1)} \, b \circ A_0^{i-j} \,
\uu \circ  \SSS_2^{k+1} \, , 
\end{aligned}
\vspace{-.2cm}
\]
by \eqref{produ} and Remark \ref{rmk:W0}, so that its average can be bounded as
\[
\begin{aligned}
& \Bigl| \Average{ \pp^{(k)} \gotC  \circ\SSS_2^i\, W \circ A_0^i \, \uu \circ \SSS_2^{k+1} } \Bigr| 
 \le
C  (1-\rho\,\ga')^{k}  \rho \, \| \gotC  \|_{\io} \| \uu \|_{\io} \\
& \qquad +C  (1-\rho\,\ga')^{k} \Bigl( \bigl( (1+\al_0 i) \la^{-\al_0 i} + \rho + i \rho^2 \bigr)) +  
\rho  \sum_{j=1}^{i} \left( \rho + (i-j) \rho^2 \right) \Bigr) \norm \gotC  \norm_{\al_0,3}^- \norm \uu \norm_{\al_0,3}^- 
\\ & \qquad + C \rho \sum_{j=1}^{i}  \Bigl| \aver{\pp}^{(i-j)} \aver*{b \,\pp_{i-j}  (\pp \, \mu )^{(j-1)}_{[i-j+1]}\circ\SSS_2 \, 
( \pp^{(k-i)}_{[i]} \, \gotC) \circ \SSS_2^{j} \, \uu \circ \SSS_2^{k+1-i+j} } \Bigr|  ,
\end{aligned}
\vspace{-.1cm}
\]
where we have used \eqref{W-W0} to obtain the first line and
applied Proposition~\ref{prop:decaynonltot} twice: first with $g_+\!=\!W_0$, $g_-\!=\!\pp^{(k-i)}_{[i]} \gotC \, \uu \circ \SSS_2^{k+1-i} $
and $\pp^{(n)}$ replaced with $(\pp\mu)^{(i)}$ 
to obtain the second line; then, after writing
%
$\pp_{[i-j]}^{(k-i+j)} \!\!=\! \pp_{i-j} \pp_{[i-j+1]}^{(j-1)} \circ \SSS_2\pp_{[i]}^{(k-i)}\circ\SSS_2^j $,
with $g_+\!=\!1$, $g_-\!=\! b \,\pp_{i-j} (\pp \, \mu)^{(j-1)}_{[i-j+1]}\circ\SSS_2  \, ( \pp^{(k-i)}_{[i]} \, \gotC ) \circ \SSS_2^{j} \,
\uu \circ \SSS_2^{k+1-i+j} $ and $\pp^{(n)}$ replaced with $\pp^{(i-j)}$, to obtain the last two lines.

Using once more Proposition~\ref{prop:decaynonltot}, with
$g_+=b$, $g_-= ( \pp^{(k-i)}_{[i]} \, \gotC )\, \uu\circ \SSS_2^{k+1-i}$
and $\pp^{(n)}$ replaced with $\pp_{i-j}  (\pp \, \mu )^{(j-1)}_{[i-j+1]}\circ\SSS_2$,
in the last line we bound
\vspace{-.1cm}
\[
\begin{aligned}
& \Bigl| \Bigl\langle b \,\pp_{i-j}  (\pp \, \mu )^{(j-1)}_{[i-j+1]}\circ\SSS_2 \, ( \pp^{(k-i)}_{[i]} \, \gotC) \circ \SSS_2^{j} \,
\uu \circ \SSS_2^{k+1-i+j} \Bigr\rangle \Bigr| \\
& \qquad \le C (1-\rho\,\ga')^{k-(i-j)} \left((1+\al_0 j)\la^{-\al_0 j} + \rho + j
\rho^2 \right) \norm \gotC \norm_{\al_0,3}^- \norm \uu \norm_{\al_0,3}^-  .
\end{aligned}
\vspace{-.1cm}
\]
Collecting all the bounds together, we obtain the assertion.
\qed

\medskip

We can now come back to the study of $h_1-h_2$ as represented in
\eqref{h1-h2true}. We study separately the averages of the three contributions
in \eqref{h1-h2true}, starting from the last one.

\begin{lemma} \label{lem:intermediate-1}
For $p_1$ and $q_1$ as in \eqref{g1g2}, and
$p_2$ and $q_2$ as in \eqref{p2q2},
let $p_1^{(n)}$ and $p_2^{(n)}$ be defined as in \eqref{lll} 
and in \eqref{p2n}, respectively. 
One has
\vspace{-.1cm}
\[
\left| \sum_{n=0}^\io\sum_{k=0}^{n-1}
\aver*{ p_1^{(k)}
(p_1\circ\SSS_1^k \!-\! p_2\circ\SSS_2^k) \, p_2^{(n-k-1)} \circ \SSS_2^{k+1}
(q_1\circ\SSS_1^n \!-\! q_2\circ\SSS_2^n)}\right| \le C \rho .
\vspace{-.1cm}
\]
\end{lemma}

\proof
Using \eqref{fWW} and the notation in \eqref{pS1-pS2} and in \eqref{diffp1p2}, we get
\vspace{-.1cm}
\begin{equation} \label{canc1}
\begin{aligned}
& p_1^{(k)} (p_1\circ\SSS_1^k \!-\! p_2\circ\SSS_2^k) \, p_2^{(n-k-1)} \circ \SSS_2^{k+1}
(q_1\circ\SSS_1^n \!-\! q_2\circ\SSS_2^n) 
\\ & \qquad =
p_1^{(k)} \Delta_{p_1,p_2,W,k}
 \, p_2^{(n-k-1)} \circ \SSS_2^{k+1} \Delta_{q_1,q_2,W,n} 
 \\& \qquad 
 + 
p_1^{(k)} \, \gota_1\circ \SSS_1^k \, ( W\circ A_0^{k+1}-W\circ A_0^{k})\,  p_2^{(n-k-1)} \circ \SSS_2^{k+1}  \Delta_{q_1,q_2,W,n}
\\ & \qquad 
+ 
p_1^{(k)}  \Delta_{p_1,p_2,W,k} \, p_2^{(n-k-1)} \circ \SSS_2^{k+1} 
\, \gota_2 \circ \SSS_1^n \, ( W\circ A_0^{n+1}-W\circ A_0^{n} ) \\ 
& \qquad +
p_1^{(k)} \, \gota_1\circ \SSS_1^k \, ( W\circ A_0^{k+1}-W\circ A_0^{k} ) \, p_2^{(n-k-1)} \circ \SSS_2^{k+1} 
\, \gota_2 \circ \SSS_1^n \, ( W\circ A_0^{n+1}-W\circ A_0^{n} ) . 
\end{aligned}
\vspace{-.1cm}
\end{equation}
The bounds \eqref{shorten-inf} give immediately
\vspace{-.1cm}
\begin{equation*} 
\left| \aver*{p_1^{(k)} \Delta_{p_1,p_2,W,k} \, p_2^{(n-k-1)} \circ \SSS_2^{k+1} \Delta_{q_1,q_2,W,n} } \right|
 \le C \rho^3 \, n \, (1-\rho\,\ga')^{n} .
\vspace{-.1cm}
\end{equation*}
Furthermore, by applying Lemma~\ref{lem:new} using the bounds \eqref{shorten-inf} and \eqref{new-bound}, we obtain
%
\begin{equation*}  
\begin{aligned}
&
\left| \sum_{k=0}^{n-1}  \aver*{
p_1^{(k)} \, \gota_1\circ \SSS_1^k \, ( W\circ A_0^{k+1}-W\circ A_0^{k})\,  p_2^{(n-k-1)} \circ \SSS_2^{k+1}  \Delta_{q_1,q_2,W,n} } \right| \\
& \qquad = \left| \sum_{k=0}^{n}
\aver*{p_1^{(k-1)} \, \gotD_{\gota_1,n,k,0} \circ \SSS_1^k \, W \circ A_0^k \, p_2^{(n-k-1)} \circ \SSS_2^{k+1}  \,
\Delta_{q_1,q_2,W,n}  } \right|
%
%
\le C \rho^2 \, n \, (1-\rho\,\ga')^n .
\end{aligned}
\end{equation*}
Analogously, using Lemma~\ref{lem:new1} and the expansion \eqref{pnS1pnS2} 
together with the notation in \eqref{diffp1p2}, to write
%
\begin{equation*} 
\begin{aligned}
\null\hspace{-.3truecm}
& \biggl| \sum_{n=0}^{\io} \sum_{k=0}^{n-1} \aver*{ p_1^{(k)}  \Delta_{p_1,p_2,W,k} \, p_2^{(n-k-1)} \circ \SSS_2^{k+1} 
\, \gota_2 \circ \SSS_1^n \, ( W\circ A_0^{n+1}-W\circ A_0^{n} ) } \biggr| \\
\null\hspace{-.3truecm}
& =
\biggl| \sum_{k=0}^{\io} \Bigl\langle p_1^{(k)}  \Delta_{p_1,p_2,W,k} 
\sum_{n=0}^{\io} \bigl( 
p_2^{(n-1)} \gotD_{1,\gota_2,n} \circ  \SSS_1^n W\circ A_0^{n} \bigr) \circ \SSS_1^{k+1} \Bigr\rangle
-  \sum_{k=0}^{\io} \Bigl\langle p_1^{(k)}  \Delta_{p_1,p_2,W,k} \Bigr\rangle  \\
\null\hspace{-.3truecm}
&   - \! \sum_{k=0}^{\io} \Bigl\langle p_1^{(k)}  \! \Delta_{p_1,p_2,W,k} 
\! \sum_{n=0}^{\io} \sum_{i=0}^{n-1} p_2^{(i)} \circ \SSS_1^{k+1}
\Delta_{p_2\circ\SSS_2^i,W,k+1} p_2^{(n-1-i)} \circ \SSS_2^{k+2+i} \rho \gota_2 \circ \SSS_1^{n+k+1} \gotf \circ A_0^{n+k+1} \Bigr\rangle \biggr| 
 \le C \rho , \\
\end{aligned}
\end{equation*}
%
where we have used \eqref{shorten-inf} and \eqref{new-bound12}.
%
%
Finally, using Lemma~\ref{lem:new} and Remark \ref{Dnk0},
hence Lemma~\ref{lem:new1} and thence the expansions \eqref{pnS1pnS2+p1nS1p1nS2}, we obtain
\[
\begin{aligned}
& \sum_{n=0}^{\io} \sum_{k=0}^{n-1}
\Bigl\langle p_1^{(k)} \, \gota_1\circ \SSS_1^k \, ( W\circ A_0^{k+1}-W\circ A_0^{k} ) \, p_2^{(n-k-1)} \circ \SSS_2^{k+1} 
\, \gota_2 \circ \SSS_1^n \, ( W\circ A_0^{n+1}-W\circ A_0^{n} ) \Bigr\rangle \\
& \; = \!
\sum_{k=0}^{\io} \Bigl\langle p_1^{(k-1)} \, \gotD_{\gota_1,k,0} \circ \SSS_1^{k} \, W \circ A_0^k \,
\sum_{n=0}^{\io}  \bigl( 
p_2^{(n-1)} \, \gotD_{1,\gota_2,n} \circ  \SSS_1^n \, W\circ A_0^{n} \bigr) \circ \SSS_1^{k+1} \Bigr\rangle \\
& \; - \!
\! \sum_{k=0}^{\io} \Bigl\langle p_1^{(k-1)} \gotD_{\gota_1,k,0} \circ \SSS_1^{k} W \circ A_0^k \!\sum_{n=0}^{\io} \! \sum_{i=0}^{n-1}
p_2^{(i)}\! \circ \SSS_1^{k+1} \! \Delta_{p_2\circ\SSS_2^i,W,k+1} 
p_2^{(n-1-i)} \! \circ \SSS_2^{k+2+i} \! \rho \gota_2 \circ \SSS_1^{n+k+1} \gotf \circ A_0^{n+k+1} \Bigr\rangle \\
& \; + \! 
\frac{1}{2} \sum_{n=0}^{\io} \Bigl\langle p_1^{(n-1)} \, \Delta_{1,\gota_1 \gota_2 \circ \SSS_1,n} \circ \SSS_1^{n} 
\bigl( W\circ A_0^{n+1} \bigr)^2 \Bigr\rangle -
\frac{1}{2} \sum_{n=0}^{\io} \Bigl\langle p_1^{(n)} \, \big( \gota_1 \, \gota_2 \circ \SSS_1 \bigr) \circ \SSS_1^{n} \bigl( \rho \, \gotf \circ A_0^{n+1} \bigr)^2 \Bigr\rangle  , \\
\end{aligned}
\]
where the two contributions in the last line have been obtained writing
%
%
%
\begin{equation} \label{Wf=WW}
\begin{aligned}
& \sum_{n=1}^{\io} p_1^{(n-1)} \, \gota_1 \circ \SSS_1^{n-1} \, W \circ A_0^n \rho \, \gota_2 \circ \SSS_1^{n} \gotf \circ A_0^n , \\
& \qquad = \frac{1}{2} \sum_{n=0}^{\io} p_1^{(n)} \, \big( \gota_1 \, \gota_2 \circ \SSS_1 \bigr) \circ \SSS_1^{n}
\Bigl( ( W\circ A_0^{n+2})^2 - (W\circ A_0^{n+1})^2  - \bigl( \rho \,
\gotf \circ A_0^{n+1} \bigr)^2 \Bigr) 
\end{aligned}
\vspace{-.2cm}
\end{equation}
%
and then proceeding as in the proof of Lemma~\ref{lem:new}.
Thus, relying once more on \eqref{shorten-inf}, \eqref{new-bound} and \eqref{new-bound12}, we get the assertion.
\qed

\medskip

The first application of Proposition~\ref{prop:pCWu} is to estimate the other two contributions in \eqref{h1-h2true}.
This leads to the two following lemmas, whose proof makes also use of the
argument given in Remark \ref{rmk:last}
(in particular of Lemmas~\ref{S1-S2}, \ref{lem:new} and \ref{lem:new1}, which
are based on the latter).

\begin{lemma} \label{lem:intermediate-2}
Let $p_2^{(n)}$ be defined as in \eqref{p2n}, with $q_1$ as in \eqref{g1g2} and $p_2,q_2$ as in \eqref{p2q2}.
One has
\vspace{-.1cm}
\[
\biggl| \sum_{n=0}^\io \aver*{ p_2^{(n)} \bigl( q_1 \circ \SSS_1^n - q_2 \circ \SSS_2 ^n \bigr) }\biggr| \le C \rho .
\vspace{-.2cm}
\]
\end{lemma}

\proof
Using Lemma~\ref{p1S1p2S2} and Remark \ref{rmk:diffp1p2}, we write
\vspace{-.1cm}
\begin{equation} \label{firstline}
\begin{aligned}
& \null \hspace{-.3cm}
\biggl|\sum_{n=0}^\infty\aver*{ p_2^{(n)} \left(q_1\circ\SSS_1^n- q_2\circ\SSS_2^n \right) } \biggr|
\le \biggl| \rho \sum_{n=0}^\io \sum_{k=0}^n \aver*{ p_2^{(n)} \gotC_{q_1,q_2,n,n-k} \circ \SSS_2^k \, W \circ A_0^k } \biggr|
\!+\! \biggl| \rho \sum_{n=0}^\io \aver*{ p_2^{(n)} R_{q_1,q_2,n} } \biggr| \\
& \qquad
+  \biggl| \sum_{n=0}^\io \aver*{ p_2^{(n)} \gota_2 \circ \SSS_2^n \bigl( W \circ A_0^{n+1} - W \circ A_0^n \bigr) } \biggr| 
+  \biggl| \sum_{n=0}^\io \aver*{ p_2^{(n)} \, \rho \bigl( \gota_2 \circ \SSS_1^n - \gota_2 \circ \SSS_2^n \bigr) \gotf \circ A_0^n } \biggr| .
\end{aligned}
\vspace{-.1cm}
\end{equation}
We use Proposition~\ref{prop:pCWu}, with $k$ and $i$ replaced with $n$ and $k$, respectively,
$\pp_j=p_2$ $\forall j=1,\ldots,k-1$,
$\gotC=\gotC_{q_1,q_2,n,n-k}$ and $\uu=1$, and the estimates \eqref{sumCkR}
in Lemma~\ref{S1-S2} to bound the second line in \eqref{firstline}.
In the first contribution of the third line, using Lemma~\ref{lem:new1},
we write, for $n\ge 1$,
\vspace{-.1cm}
\begin{equation} \label{firstcontribution}
\biggl| \sum_{n=0}^\io \aver*{ p_2^{(n)} \gota_2 \circ \SSS_2^n \bigl( W \circ A_0^{n+1} - W \circ A_0^n \bigr) } \biggr|
= \biggl| \sum_{n=0}^\io \aver*{ p_2^{(n-1)} \gotD_{2,\gota_2,n} \circ \SSS_2^n \, W \circ A_0^{n} } \biggr| \\
\le C \rho ,
\vspace{-.1cm}
\end{equation}
where the last bound follows from Proposition~\ref{prop:pCWu}, with $k=n-1$, $i=n$, $\pp_j=p_2$ for $j=0,\ldots,n-2$
(see Remark \ref{ex-point3ter}), $\gotC= \gotD_{2,\gota_2,n}$,
so that $\norm \gotC \norm_{\al_0,2}^- = \norm  \gotD_{2,\gota_2,n}\norm_{\al_0,2}^- \le C\rho$ for $n\ge 1$,
and $\uu=1$. To deal with the second contribution in the third line of \eqref{firstline}, setting
\vspace{-.2cm}
\begin{equation} \label{f0gn}
\gotf_0(\psi): = f(0,\psi) , \qquad
\gotf \circ A_0^n - \gotf_0 \circ A_0^n = \gotg_n \, W \circ A_0^n ,
\qquad \gotg_n(\psi) := \int_0^1 \, \partial_{\theta} \gotf (t W(A_0^n\psi),A_0\psi)  ,
\vspace{-.2cm}
\end{equation}
and using Lemma~\ref{S1-S2}, we rewrite
\vspace{-.1cm}
\[
\bigl( \gota_2 \circ \SSS_1^n - \gota_2 \circ \SSS_2^n \bigr) \gotf \circ A_0^n 
= \biggl( \rho \sum_{k=0}^{n}\gotC_{\gota_2,n,n-k} \circ \SSS_2^k \, W \circ A_0^k + \rho R_{\gota_2,n} \biggr) 
\bigl( \gotf_0 \circ A_0^n + \gotg_n \, W \circ A_0^n \bigr)  ,
\vspace{-.1cm}
\]
so that, by exploiting Proposition~\ref{prop:pCWu}, with $k$ and $i$ replaced with $n$ and $k$, respectively, and with $\uu=\gotf_0\circ A_0^{-1}$,
Theorem~\ref{thm:3} and the bounds of Lemma~\ref{S1-S2}, we obtain,
for any $\gamma'\in(0,\gamma)$
\vspace{-.1cm}
\[
\begin{aligned}
& \left| \sum_{n=0}^\io  \sum_{k=0}^{n} \rho^2 \aver*{ p_2^{(n)} 
\gotC_{\gota_2,n,n-k} \circ \SSS_2^k \, W \circ A_0^k \, \gotf_0 \circ A_0^n } \right| 
= \left| \sum_{k=0}^\io  \sum_{n=0}^{\io} \rho^2 \aver*{ p_2^{(n+k)} 
\gotC_{\gota_2,n+k,n} \circ \SSS_2^k \, W \circ A_0^k \, \gotf_0 \circ A_0^{n+k} } \right| \\
& \qquad \qquad
\le \sum_{k=0}^\io \rho^2 (1-\rho\,\ga')^{k} \bigl( (1+ \al_0 k) \, \lambda^{- \al_0 k} + \rho + k\rho^2 + k\rho^3 \bigr)
\sum_{n=0}^{\io} \norm \gotC_{\gota_2,n+k,n} \norm_{\al_0,2}^- \le C \rho ,
\end{aligned}
\]
and 
\[
\begin{aligned}
& \biggl| \sum_{n=0}^\io \rho^2 \aver*{ p_2^{(n)} 
\, R_{\gota_2,n} \, \gotf_0 \circ A_0^ n } \biggr|
\le C \sum_{n=0}^\io \rho^2 (1-\rho\,\ga')^{n} \bigl\langle \bigl| \gotR_{\gota_2,n} \bigr| \bigr\rangle \le C \rho , \\
& \biggl| \sum_{n=0}^\io \rho^2 \biggl\langle
p_2^{(n)} \gotg_n \biggl( \sum_{k=0}^{n}\gotC_{\gota_2,n,n-k} \circ \SSS_2^k \, W \circ A_0^k 
+ R_{\gota_2,n} \biggr) W \circ A_0^n \biggr\rangle \biggr| \le C\rho ,
\end{aligned}
\]
which imply the desired bound.
\qed

\begin{lemma} \label{lem:intermediate-3}
Let $p_1^{(n)}$ and $p_2^{(n)}$ be defined as in \eqref{lll} 
and in \eqref{p2n}, respectively, with $p_1$ and $q_1$ as in \eqref{g1g2}, and
$p_2$ and $q_2$ as in \eqref{p2q2}. One has
\vspace{-.1cm}
\[
\biggl| \sum_{n=0}^\io \aver*{ \bigl( p_1^{(n)} - p_2^{(n)} \bigr) \, q_2 \circ \SSS_2 ^n } \biggr| \le C \rho .
\vspace{-.2cm}
\]
\end{lemma}

\proof
We expand $p_1^{(n)} - p_2^{(n)}$ as in \eqref{p1n-p2n}, hence split
$p_1^{(k)} = p_2^{(k)} + (p_1^{(k)} - p_2^{(k)}) $ and thence expand again $p_1^{(k)} -p_2^{(k)}$, so as to obtain
\vspace{-.2cm}
\begin{equation} \label{lastbound}
\begin{aligned}
\null\hspace{-.3cm} & 
\sum_{n=0}^\io \aver*{ \bigl( p_1^{(n)} \! - \!  p_2^{(n)} \bigr) \, q_2 \circ \SSS_2 ^n } 
= \sum_{n=0}^{\io} \sum_{k=0}^{n-1} p_2^{(k)} \aver*{ \bigl( p_1\circ\SSS_1^k- p_2\circ\SSS_2^k \bigr) p_2^{(n-k-1)} \circ \SSS_2^{k+1} \, q_2\circ\SSS_2^n } \\
\null\hspace{-.3cm} & \; +  \sum_{n=0}^{\io} \sum_{k=0}^{n-1}\sum_{j=0}^{k-1} 
\aver*{
p_1^{(j)} \bigl( p_1\circ\SSS_1^j  \!-\! p_2\circ\SSS_2^j \bigr) 
p_2^{(k-1-j)} \circ \SSS_2^{j+1} \, \bigl( p_1\circ\SSS_1^k \!-\! p_2\circ\SSS_2^k \bigr)
p_2^{(n-k-1)} \circ \SSS_2^{k+1} \, q_2\circ\SSS_2^n} ,
\end{aligned}
\vspace{-.2cm}
\end{equation}

To bound the second line of \eqref{lastbound}, we write
%
$p_1^{(j)} \bigl( p_1\circ\SSS_1^j  - p_2\circ\SSS_2^j \bigr) p_2^{(k-1-j)} \circ \SSS_2^{j+1} \, \bigl( p_1\circ\SSS_1^k- p_2\circ\SSS_2^k \bigr)$
as in \eqref{canc1}, with $k$, $j$, $p_1$ and $p_2$ instead of $n$, $k$, $q_1$ and $q_2$, respectively,
and we reason as in the proof of Lemma~\ref{lem:intermediate-1},
by relying on Lemmas~\ref{lem:new} and \ref{lem:new1}, and using
\eqref{Wf=WW}. Thus, the second line becomes
\vspace{-.1cm}
\[
\begin{aligned}
& \sum_{n=0}^{\io} \sum_{k=0}^{n-1} \sum_{j=0}^{k-1} 
\aver*{
p_1^{(j)} \Delta_{p_1,p_2,W,j} \, p_2^{(k-1-j)} \circ \SSS_2^{j+1} \, \Delta_{p_1,p_2,W,k} \, p_2^{(n-k-1)} \circ \SSS_2^{k+1} \, q_2\circ\SSS_2^n } \\
& \;+ 
\sum_{n=0}^\io \sum_{k=0}^{n-1} \sum_{j=0}^{k-1}
\aver*{
p_1^{(j-1)} \, \gotD_{\gota_1,j,0} \circ \SSS_1^j \, W \circ A_0^j \, p_2^{(k-j-1)} \circ \SSS_2^{j+1}  \,
\Delta_{p_1,p_2,W,k} \, p_2^{(n-k-1)} \circ \SSS_2^{k+1} \, q_2\circ\SSS_2^n } \\
& \; + 
\sum_{n=0}^\io \sum_{k=0}^{n-1}
\aver*{ p_1^{(k-1)} \, \gota_1 \circ \SSS_1^{k-1} \, W \circ A_0^k \, 
\Delta_{p_1,p_2,W,k} \, p_2^{(n-k-1)} \circ \SSS_2^{k+1} \, q_2\circ\SSS_2^n } \\
& \; +
\sum_{j=0}^\io
\Bigl\langle
p_1^{(j)} \, \Delta_{p_1,p_2,W,j} 
\sum_{k=0}^{\io} 
\bigl( p_2^{(k-1)} \, \gotD_{1,\gota_2,k} \circ\SSS_1^k \, W \circ A_0^k \bigr) \circ \SSS_1^{j+1}
\calQ_{k+1}
\Bigr\rangle \\
& \; -
\sum_{j=0}^\io
\Bigl\langle
p_1^{(j)} \, \Delta_{p_1,p_2,W,j} 
\sum_{k=0}^{\io} \sum_{i=0}^{k} p_2^{(i)} \circ \SSS_1^{j+1}
\Delta_{p_2\circ\SSS_2^i,W,j+1}
\calP_{j,k,i} 
( \gota_2 
\rho \, \gotf ) \circ \SSS_1^{k+j+1}
\calQ_{k+1} \Bigr\rangle \\
& \; +
\sum_{j=0}^{\io} 
\Bigl\langle
p_1^{(j-1)} \, \gotD_{\gota_1,j,0} \circ \SSS_1^{j} \, W \circ A_0^j 
\sum_{k=0}^{\io} \bigl( p_2^{(k-1)} \gotD_{1,\gota_2,k}  \circ \SSS_1^{k} \, W\circ A_0^{k} \bigr) \circ \SSS_1^{j+1}
\calQ_{k+1}\circ\SSS_2^{j+1} 
\Bigr\rangle \\
& \; -
\sum_{j=0}^{\io} 
\Bigl\langle 
p_1^{(j-1)} \, \gotD_{\gota_1,j,0} \circ \SSS_1^{j} \, W \circ A_0^j \, \sum_{k=0}^{\io} \sum_{i=0}^k p_2^{(i)} \circ \SSS_1^{j+1} 
\Delta_{p_2\circ\SSS_2^i,W,j+1}
\calP_{j,k,i} 
(\gota_2 
\rho \,  \gotf ) \circ \SSS_1^{k+j+1} 
\calQ_{k+1}\circ\SSS_2^{j+1} 
\Bigr\rangle
 \\ & \; 
+
\frac{1}{2} \sum_{k=0}^{\io} 
\Bigl\langle
p_1^{(k-1)} \gotD_{\gota_1\gota_2\circ\SSS_1^{-1},k,0} \circ \SSS_1^{k} \bigl( W \circ A_0^{k+1} \bigr)^{\!2}
\calQ_{k+1}
\Bigr\rangle
- 
\frac{1}{2} \sum_{k=0}^{\io} 
\Bigl\langle p_1^{(k)} \! \bigl( \gota_1\gota_2\circ\SSS_1^{-1} \bigr) \circ \SSS_1^{k} \bigl( \rho \, \gotf \circ A_0^{k+1} \bigr)^{\!2} 
\calQ_{k+1}
\Bigr\rangle , \\
\end{aligned}
\]
with $\calQ_k:= \sum_{n=0}^{\io}\bigl( p_2^{(n)} \, q_2\circ\SSS_2^n \bigr) \circ \SSS_2^{k}$
and $\calP_{j,k,i}:=p_2^{(k-1-i)} \circ \SSS_2^{j+i+2}$,
so that all contributions are found to be bounded as $C \rho$. 

The contribution in the first line of the r.h.s.~of \eqref{lastbound} is dealt with by writing $p_1\circ\SSS_1^k- p_2\circ\SSS_2^k$
once more according to Remark \ref{rmk:diffp1p2}, so that we obtain
\vspace{-.2cm}
\begin{equation} \label{lastboundlast}
\begin{aligned}
& \sum_{n=0}^{\io} \sum_{k=0}^{n-1} 
\aver*{p_2^{(k)} \left( p_1\circ\SSS_1^k- p_2\circ\SSS_2^k \right) p_2^{(n-k-1)} \circ \SSS_2^{k+1} \, q_2\circ\SSS_2^n} \\
& \qquad =
\sum_{n=0}^{\io} \sum_{k=0}^{n-1} 
\biggl\langle p_2^{(k)} 
\biggl( \rho \sum_{i=0}^{k}  \gotC_{p_1,p_2,k,k-i} \circ\SSS_2^i\, W\circ A_0^i + \rho \, \gotR_{p_1,p_2,k} \biggr)
p_2^{(n-k-1)} \circ \SSS_2^{k+1} \, q_2\circ\SSS_2^n \biggr\rangle \\
& \qquad +
\sum_{n=0}^{\io} \sum_{k=0}^{n} 
\aver*{p_2^{(k)} \,  \rho \, \gota_2 \circ \SSS_1^k \, \gotf \circ A_0^{k} \, p_2^{(n-k-1)} \circ \SSS_2^{k+1} \, q_2\circ\SSS_2^n }.
\end{aligned}
\end{equation}
In \eqref{lastboundlast} the second line
is bounded by Proposition~\ref{prop:pCWu}, with $\pp_i=p_2$ $\forall i=1,\ldots,k-1$,
$\gotC= \gotC_{p_1,p_2,k,k-i}$ and $\uu=p_2^{(n-k-1)} \, q_2 \circ \SSS^{n-k-1}$,
while in order to bound the third line we proceed as follows. After splitting
$\gota_2 \circ \SSS_1^k=\gota_2 \circ \SSS_2^k+(\gota_2 \circ \SSS_1^k-\gota_2 \circ \SSS_2^k)$,
we reason as done for the last line of \eqref{firstline}:
the contribution with $\gota_2 \circ \SSS_2^k$ is dealt with as \eqref{firstcontribution},
the only difference being that, when applying Proposition~\ref{prop:pCWu}, 
one sets $\uu=p_2^{(n-k-1)} \circ \SSS_2\, q_2\circ\SSS_2^{n-k}$;
in the second contribution we expand $\gota_2 \circ \SSS_1^k - \gota_2 \circ \SSS_2^k$ as in \eqref{pS1-pS2}
and write $\gotf\circ A_0^k$ as in \eqref{f0gn}, so as to obtain the three contributions
\vspace{-.1cm}
\[
\begin{aligned}
& \sum_{i=0}^{\io} \sum_{k=0}^{\io} \sum_{n=0}^{\io} 
\aver*{\rho \, 
p_2^{(k+i)} \,  \rho \, \gotC_{\gota_2,k+i,k} \circ \SSS_2^i \, W\circ A_0^i \, 
\gotf_0 \circ A_0^{k+1} \,
p_2^{(n-1)} \circ \SSS_2^{k+i+1} \, q_2\circ\SSS_2^{n+k+i} } , \\
& \sum_{n=0}^{\io} \sum_{k=0}^{n} 
\aver*{ \rho \,
p_2^{(k)} \,  \rho \, \gotR_{\gota_2,k} \, \gotf_0 \circ A_0^{k+1} \,
p_2^{(n-k-1)} \circ \SSS_2^{k+1} \, q_2\circ\SSS_2^{n} } , \\
& \sum_{n=0}^{\io} \sum_{k=0}^{n} 
\biggl\langle \rho \, 
p_2^{(k)} \biggl( \rho \sum_{i=0}^{k} \gotC_{\gota_2,k,k-i} \circ \SSS_2^i \, W\circ A_0^i + \rho \, \gotR_{\gota_2,k} \biggr)
\gotg_k \, W\circ A_0^k \, p_2^{(n-k-1)} \circ \SSS_2^{k+1} \, q_2\circ\SSS_2^{ni} \biggr\rangle , \\
\end{aligned}
\vspace{-.1cm}
\]
which are all bounded proportionally to $\rho$.
\qed

\medskip

The following result puts together the bounds obtained above and extends the analysis to the square of $h_1-h_2$.
Together with the forthcoming Proposition~\ref{prop:fluhder2}, it completes the first step in order to prove Proposition~\ref{prop:fluhder1},
as outlined at the beginning of the present subsection.

\begin{prop} \label{prop:hh2}
Let $h_2$ and $h_1$ be as in \eqref{p2n} and in \eqref{HHH1},
respectively. For all $\theta\!\in\!\calU$ one has
\vspace{-.1cm}
\begin{equation} \label{fluh}
 |\aver*{h_1(\theta,\cdot)-h_2(\theta,\cdot)} |\leq C\rho , \qquad \aver*{(h_1(\theta,\cdot)-h_2(\theta,\cdot))^2} \leq C\rho\, .
\end{equation}
\end{prop}

\proof
%
%
The first bound in \eqref{fluh} follows immediately from \eqref{h1-h2true}
and from the estimates in Lemma~\ref{lem:intermediate-1}, \ref{lem:intermediate-2} and \ref{lem:intermediate-3}.
To obtain the second bound in \eqref{fluh}, we expand $(h_1 - h_2)^2$ as
\vspace{-.2cm}
\[
\begin{aligned}
& \sum_{n_1,n_2=0}^{\io}
\!\! \Bigl(
p_1^{(n_1)} \! \bigl( q_1\circ \SSS_1^{n_1} \!\!-\! q_2\circ \SSS_2^{n_1} \bigr)
p_1^{(n_2)} \! \bigl( q_1\circ \SSS_1^{n_2} \!\!-\! q_2\circ \SSS_2^{n_2} \bigr) 
\!+\! 
p_1^{(n_1)} \! \bigl( q_1\circ \SSS_1^{n_1}\! \!- q_2\circ \SSS_2^{n_1} \bigr) \!
\bigl( p_1^{(n_2)} \!\!-\! p_2^{(n_2)} \bigr) q_2\circ \SSS_2^{n_2}
\\
& + 
\bigl( p_1^{(n_1)} - p_2^{(n_1)} \bigr) \, q_2\circ \SSS_2^{n_1}
\,  p_1^{(n_2)} \, \bigl( q_1\circ \SSS_1^{n_2} \!-\! q_2\circ \SSS_2^{n_2} \bigr) 
+ 
\bigl( p_1^{(n_1)} \!-\! p_2^{(n_1)} \bigr)\,  q_2\circ \SSS_2^{n_1} \,
\bigl( p_1^{(n_2)} \!-\! p_2^{(n_2)} \bigr) q_2\circ \SSS_2^{n_2} \Bigr) ,
\end{aligned}
\vspace{-.2cm}
\]
where, writing, for $i=1,2$,
$q_1\circ \SSS_1^{n_i} - q_2\circ \SSS_2^{n_i}$ according to \eqref{p1p2q1q2}
and $p_1^{(n_i)} - p_2^{(n_i)}$
using first \eqref{p1n-p2n} and then once more \eqref{p1p2q1q2},
we obtain a sum of contributions 
which can be dealt with as the contributions in \eqref{canc1}.
Then, using also that $\norm q_2\norm_{\al_0,3}\leq C\rho$, the second bound follows.
\qed

\medskip

Finally, the following results shows that bounds analogous to those of
Proposition~\ref{prop:hh2} extend to
the derivatives of the two functions $h_1$ and $h_2$. The proof, which follows the same lines of the proof
of Proposition~\ref{prop:hh2}, is given in Appendix~\ref{proof:fluhder2}.

\begin{prop} \label{prop:fluhder2}
Let $h_1$ and $h_2$ be as in \eqref{HHH1} and in \eqref{p2n},
respectively. For all $\theta\in\calU$, one has
\begin{equation} \label{fluhder2}
|\aver*{ \partial_\theta h_1(\theta,\cdot) - \partial_\theta h_2(\theta,\cdot)} | \leq C\rho , \qquad 
\aver*{(\partial_\theta h_1(\theta,\cdot) - \partial_\theta
h_2(\theta,\cdot))^2} \leq C\rho\, .
\vspace{.1cm}
\end{equation}
\end{prop}

\begin{rmk} \label{regularity2}
\emph{
By looking at the proof of Proposition~\ref{prop:fluhder2}, 
we see that $F=\rho f$ has to be required to belong to $\calmB_{\al_0,5}$,
because we need to apply Proposition~\ref{prop:pCWu}.
The further condition 
that $F$ be in $\calmB_{\al_0,6}$ is needed in order to control the deviations of the
second derivative of the conjugation, which in turn is used to estimate the deviations of the
first derivative of the inverse conjugation (see Remark \ref{regularity4}).
}
\end{rmk}

\subsubsection{Deviations of the conjugation:~proof of the bounds (\ref{bounds-thm3a}) and (\ref{bounds-thm3b})} \label{675} 

We now come back to $h(\f,\psi) - \ovl h(\f)$, with $h(\f,\psi)$  written as in \eqref{h1-hbar}
(taking in mind Remark \ref{labelext} for the notation), and estimate the whole average. 
After rewriting 
\hspace{-.2cm}
\[
\aver{ h(\f,\cdot) - \ovl h(\f) } =
\average{h_1(\f,\cdot) - \ovl h(\f)}  + \average{W(\cdot) \, \partial_\f
h_1(\f,\cdot)
} + \!
\int_0^1 \!\! \der t \, (1-t)
\, \average{ (W(\cdot))^2 \, \partial_\f^2 h_1(\f + t W(\cdot),\cdot) } ,
\hspace{-.2cm}
\]
we bound $ |\aver{ h_1(\f,\cdot ) - \ovl h(\f)}| \le C \rho$
by Proposition~\ref{prop:h1hbar}, and
$\aver{ (W(\cdot))^2 } \le C\rho$ by Theorem~\ref{thm:3}. Writing
$\average{W(\cdot) \, \partial_\f h_1(\f,\cdot)}=\aver*{W(\cdot) \,
\partial_\f ( h_1(\f,\cdot)-\ovl h(\f))}+\aver W \partial_\f\ovl h(\f)$,
Theorem~\ref{thm:3} gives $|\aver W \partial_\f\ovl h(\f)|\leq C\rho$ while we
use first the Cauchy-Schwarz inequality, then
Proposition~\ref{prop:h1hbar}, together again with Theorem~\ref{thm:3},  to obtain
%
$|\average{W(\cdot) \, \partial_\f ( h_1(\f,\cdot)- \ovl h(\f))} |
\le
(\average{(W(\cdot))^2} 
\langle (
\partial_\f h_1(\f,\cdot)-
\partial_\f \ovl h(\f) 
)^2 \rangle
)^\frac{1}{2} \le C \rho$.
%
This concludes the proof of the first bound in \eqref{bounds-thm3a}.

The second bound in \eqref{bounds-thm3a} is easily obtained by writing
\vspace{-.1cm}
\[
\begin{aligned}
\bigl( h(\f,\psi) - \ovl h(\f) \bigr)^2 
&  = \biggl( h_1(\f,\psi) - \ovl  h(\f) - W(\psi) \int_0^1 \der t \, \partial_\f h(\f-tW(\psi),\psi) \biggr)^2 \\
%
& \le  2 \left( \left( h_1(\f,\psi) - \ovl  h(\f) \right)^2 + (W(\psi))^2 \left\| \partial_\f h \right\|^2 \right) \\
& \le  4 \left( h_1(\f,\psi) -  h_2(\f,\psi) \right)^2 +
4 \left( h_2(\f,\psi) - \ovl  h(\f) \right)^2 + 2 (W(\psi))^2 \left\| \partial_\f h \right\|^2 , \phantom{\biggr)^2}
\end{aligned}
\]
which in turn is bounded using once more Proposition~\ref{prop:h1hbar} and Theorem~\ref{thm:3}.


Now we pass to the bounds \eqref{bounds-thm3}.
In order to study the average of $\partial_\f h(\f,\psi) - \partial_\f\ovl
h(\f)$, we write
\begin{equation} \label{bohbohboh}
\begin{aligned}
\partial_\f h(\f,\psi) 
& = \partial_\f h_1(\f,\psi) -  W(\psi) \int_0^1 \der t \, \partial_\f^2 h (\f - t W(\psi),\psi) 
= \partial_\f h_1(\f,\psi) \\
& 
- W(\psi) \left( \partial_\f^2 h_1 (\f,\psi) \!-\! \partial_\f^2 \ovl h (\f) \right)
-  W(\psi) \, \partial_\f^2 \ovl h(\f) + (W(\psi))^2 \!\! \int_0^1 \!\! \der t \,
\partial_\f^3 h(\f -  t W(\psi),\psi) .
\end{aligned}
\end{equation}
Therefore, using the second expansion in \eqref{bohbohboh}, we can bound
\begin{equation} \label{h-hbar-der}
\begin{aligned}
& \left| \aver*{\partial_\f h(\f,\cdot) - \partial_\f \ovl h(\f) } \right|
\le \left| \aver*{\partial_\f h_1(\f,\cdot) - \partial_\f \ovl h(\f) } \right| \\
& \qquad +  
\left| \aver*{W(\cdot) \left( \partial_\f^2 h_1 (\f,\cdot) - \partial_\f^2 \ovl h (\f) \right)} \right|
+ \aver*{W(\psi)} \partial_\f^2 \ovl h(\f) + \aver*{ (W(\psi))^2} \|
\partial_\f^3 h \|_{\io} .
\end{aligned}
\end{equation}
The term in the first line and the last two terms in the second line are bounded by \eqref{h1hbarb} 
and by Theorem~\ref{thm:3}, respectively.
To bound the first term in the second line of \eqref{h-hbar-der}
we need the following result, which is proved in Appendix~\ref{proof:fluhderder}.

\begin{lemma} \label{lem:fluhderder}
Let $h_1$ and $\ovl h$ be defined
as in \eqref{HHH1} and in \eqref{hbar}, respectively.
For all $\theta\in\calU$ one has
\[
\bigl|\aver*{ (\partial_\theta^2 h_1(\theta,\cdot) - \partial_\theta^2 \ovl h(\theta))^2 } \bigr| \le C\rho .
\]
\end{lemma}

\begin{rmk} \label{regularity4} 
\emph{
As mentioned in Remark \ref{regularity2}, the
condition $F\in\calmB_{\alpha_0,6}$ is required to obtain the bounds in Lemma
\ref{lem:fluhderder}.
A bound like $|\aver*{ \partial_\theta^2 h_1(\theta,\cdot) - \partial_\theta^2 \ovl h(\theta) } | \leq C\rho$
could be obtained as well, but actually we need only the bound on the squared deviations because
the latter will be needed in order to estimate the deviations of first derivative of the inverse conjugation.
}
\end{rmk}

By the Cauchy-Schwarz inequality and Lemma \ref{lem:fluhderder}, we get
$ \left| \aver*{W(\cdot) \left( \partial_\f^2 h_1 (\f,\cdot) - \partial_\f^2 \ovl h (\f) \right)} \right| \le C\rho$,
which inserted into \eqref{h-hbar-der} yields the first bound in \eqref{bounds-thm3b}.

Finally, we pass to the second bound in \eqref{bounds-thm3b}.
By using the first expansion in \eqref{bohbohboh}, 
we may bound
$ | \partial_\f h(\f,\psi) - \partial_\f \ovl h(\f)  \le
| \partial_\f h_1(\f,\psi) - \partial_\f \ovl h(\f) | + |W(\psi)| \,
\| \partial_\f^2 h \|_{\io} $,
which allows us to estimate
$ \langle ( \partial_\f h(\f,\cdot) - \partial_\f \ovl h(\f) )^2 \rangle \le
2 \langle ( \partial_\f h_1(\f,\cdot) - \partial_\f \ovl h(\f) )^2 \rangle
+ 2 \| \partial_\f^2 h \|_{\io}^2 \aver{(W(\cdot))^2} $.
This immediately implies the second bound in \eqref{bounds-thm3b} by
\eqref{fluhder1} and by Theorem~\ref{thm:3}.

\subsection{Deviations of the inverse conjugation} 
\label{devinverse}

To deal with the inverse conjugation, we exploit the following trivial identity.
\begin{equation} \label{identity} 
\HH_1 (\HH_1^{-1}(\h,\psi),\psi) = \ovl\HH (\ovl\HH^{-1}(\h)) = \h ,
\end{equation}
which holds for all $(\h,\psi) \in \Omega_{0} = \calU_0\times \TTT^2 :=\overline \HHH(\Omega)$,
provided $\Omega_{\rm ext}$ is such that
$\HHH(\Omega_{\rm ext})\supset \overline \HHH(\Omega)$ and
$\Omega_{1,\rm ext}\supset \Omega$
(recall that we are working with the extended maps).

\begin{lemma} \label{lem:l1}
Let $l_1$ and $\ovl l$ be defined as in \eqref{HHH1h1} and in \eqref{def:lbar}, respectively.
For all $\h\in\calU_0$, one has
\begin{equation} \label{fluhl}
|\aver*{l_1(\h,\cdot)- \ovl l(\h) } | \leq C\rho , \qquad \aver*{(l_1(\h,\cdot)-\ovl l(\h))^2} \leq C\rho\, .
\end{equation}
\end{lemma}

\proof 
By using \eqref{identity} we get
\vspace{-.1cm}

\begin{equation} \label{formula}
\HH_1 (\HH_1^{-1}(\h,\psi),\psi) - \HH_1 (\ovl\HH^{-1}(\h),\psi) = \ovl\HH
(\ovl\HH^{-1}(\h) )-\HH_1 (\ovl\HH^{-1}(\h),\psi).
\vspace{-.1cm}
\end{equation}
If we write the l.h.s. in \eqref{formula} as
\vspace{-.1cm}
\[
\bigl( \HH_1^{-1}(\h,\psi) -\ovl\HH^{-1}(\h) \bigr) 
\int_0^1 \der t \, \partial_\theta \HH_1 (\ovl\HH^{-1}(\h) + t (\HH_1^{-1}(\h,\psi) -\ovl\HH^{-1}(\h) ) ,\psi)
\vspace{-.1cm}
\]
and use the lower bound \eqref{chi}, we obtain from \eqref{formula} 
\vspace{-.1cm}
\begin{equation} \label{hinv1}
\ga_1^2 \average{\bigl( \HH_1^{-1}  -\ovl\HH^{-1} \bigr)^2}(\eta) \le
\average{ \left( \HH_1  - \ovl\HH \right)^2 }\bigl( \ovl\HH^{-1}(\eta) \bigr) .\
\vspace{-.1cm}
\end{equation}
Analogously, if we write l.h.s.~of \eqref{formula} as
\vspace{-.1cm}
\[
\begin{aligned}
& \bigl( \HH_1^{-1}(\h,\psi) -\ovl\HH^{-1}(\h) \bigr) \partial_\theta \ovl \HH (\ovl\HH^{-1}(\h))
+ \bigl( \HH_1^{-1}(\h,\psi) -\ovl\HH^{-1}(\h) \bigr) 
\left( \partial_\theta \HH_1 (\ovl\HH^{-1}(\h) ,\psi)  - \partial_\theta \ovl \HH (\ovl\HH^{-1}(\h))  \right) \\
& \qquad + \bigl( \HH_1^{-1}(\h,\psi) -\ovl\HH^{-1}(\h) \bigr)^2
\int_0^1 (1-t) \, \der t \, \partial_\theta^2 \HH_1 (\ovl\HH^{-1}(\h) + t (\HH_1^{-1}(\h,\psi) -\ovl\HH^{-1}(\h) ) ,\psi) ,
\end{aligned}
\vspace{-.1cm}
\]
then we find
\vspace{-.1cm}
\begin{equation} \label{hinv2}
\begin{aligned}
& \ovl\ga \bigl| \average{ \HH_1^{-1} -\ovl\HH^{-1} }(\eta) \bigr|
\le \partial_\theta \ovl \HH (\ovl\HH^{-1}(\h)) \bigl| \average{\HH_1^{-1} - \ovl\HH^{-1} }(\eta)  \bigr|
\le \bigl| \average{ \HH_1  - \ovl\HH  } \bigl(\ovl \HH^{-1}(\eta)\bigr) \bigr|
\\ & \quad
+ \| \partial^2_\theta \HH_1\|_{\io}  \average{ \bigl( \HH_1^{-1} -\ovl\HH^{-1} \bigr)^2 }(\eta)
+ \Bigl( \average{ \bigl( \HH_1^{-1} -\ovl\HH^{-1} \bigr)^2}(\eta)
\average{ \bigl( \partial_\theta \HH_1  - \partial_\theta \ovl \HH \bigr)^2 } \bigl(\ovl \HH^{-1}(\eta)\bigr)\Bigr)^{1/2}  ,
\end{aligned}
\vspace{-.1cm}
\end{equation}
with $\ovl \ga=O(1)$ as in Remark \ref{chibar}. 
Thus, using that
%
$\HH_1 (\theta,\psi) -\ovl \HH(\theta) = \theta^2 \bigl( h_1(\theta,\psi) - \ovl h(\theta) \bigr)$ and 
$\HH_1^{-1}(\h,\psi) -\ovl \HH{\vphantom{\HH}}^{-1}(\h) = \h^2 \bigl( l_1(\h,\psi) - \ovl l(\h) \bigr)$
and that, for $\theta=\ovl\HH{\vphantom{\HH}}^{-1}(\h)$, there exists a constant $d_0$ such that $d_0^{-1} \le \theta/\eta \le d_0$ 
(see Section~\ref{subsec:proofcor}), the bound \eqref{hinv1}, together with the second bound in \eqref{h1hbara},
gives the second bound in \eqref{fluhl},
which, in turn, inserted into \eqref{hinv2}, together with the first bound in \eqref{h1hbara} and the second bound in \eqref{h1hbarb},
yields the first bound in \eqref{fluhl}.
\qed


\begin{lemma} \label{lem:l1der}
Let $l_1$ and $\ovl l$ be defined as in \eqref{HHH1h1} and in \eqref{def:lbar}, respectively. 
For all $\h \in \calU_0$, one has
\vspace{-.1cm}
\begin{equation} \label{fluhlder}
|\aver*{\partial_\h l_1 (\h,\cdot) - \partial_\h \ovl l (\h) } | \leq C\rho , \qquad \aver*{(\partial_\h l_1(\h,\cdot)- \partial_\h \ovl l(\h) )^2} \leq C\rho\, .
\vspace{-.1cm}
\end{equation}
\end{lemma}

\proof 
Differentiating \eqref{identity} with respect to $\h$, we obtain
\vspace{-.1cm}
\begin{equation} \label{formula2}
\partial_\theta \HH_1 (\HH_1^{-1}(\h,\psi),\psi) \, \partial_\h \HH_1^{-1}(\h,\psi) =
\partial_\theta \ovl\HH (\ovl\HH^{-1}(\h) ) \, \partial_\h \ovl \HH^{-1}(\h) = 1 ,
\vspace{-.2cm}
\end{equation}
that implies
%
\begin{equation} \label{useful}
\partial_\h \HH_1^{-1}(\h,\psi) - \partial_\h \ovl \HH^{-1}(\h) =
- \frac{\partial_\theta \HH_1 (\HH_1^{-1}(\h,\psi),\psi) - \partial_\theta \ovl\HH (\ovl\HH^{-1}(\h) )}
{\partial_\theta \HH_1 (\HH_1^{-1}(\h,\psi),\psi) \, \partial_\theta \ovl\HH (\ovl\HH^{-1}(\h) )} .
\vspace{-.1cm}
\end{equation}
Thus, we find that square $\bigl( \partial_\h \HH_1^{-1}(\h,\psi) \!-\! \partial_\h \ovl \HH^{-1}(\h) \bigr)^2$ is bounded by
\vspace{-.1cm}
\[
\frac{2}{\tau_1^2\ovl\tau^2}
\bigl( \| \partial_\theta^2 \HH_1 \|_{\io}^2
\bigl( \HH_1^{-1}(\h,\psi) \!-\! \ovl\HH^{-1}(\h) \bigr)^{2} +
\bigl( \partial_\theta \HH_1 (\ovl\HH^{-1}(\h),\psi) \!-\! \partial_\theta \ovl\HH (\ovl\HH^{-1}(\h)) \bigr)^{2} \bigr) ,
\vspace{-.1cm}
\]
where $\tau_1$ and $\ovl \tau$ are on lower bounds, respectively, on $\partial_\theta \HH_1$ and $\partial_\theta \ovl\HH$
(see Subsection~\ref{subsec:proofcor}), so that
\vspace{-.1cm}
\[
\average{ \bigl( \partial_\h l_1  - \partial_\h \ovl l  \bigr)^2 } \le
\frac{2}{\tau_1^2\ovl\tau^2}
\left( \| \partial_\theta^2 \HH_1 \|_{\io}^2 \average{ \bigl( l_1 - \ovl l \bigr)^{2} } +
\average{ \bigl( \partial_\theta h_1 - \partial_\theta \ovl h \bigr)^{2} } \right) ,
\vspace{-.1cm}
\]
which, together with \eqref{h1hbarb} and \eqref{fluhl}, implies the second of \eqref{fluhlder}.
Rewriting \eqref{useful} as
%
\[
- \frac{\partial_\theta \HH_1 (\HH_1^{-1}(\h,\psi),\psi) - \partial_\theta \ovl\HH (\ovl\HH^{-1}(\h) )}
{\bigl(\partial_\theta \ovl\HH (\ovl\HH^{-1}(\h) )\bigr)^2} 
+  \frac{\bigl( \partial_\theta \HH_1 (\HH_1^{-1}(\h,\psi),\psi) - \partial_\theta \ovl\HH (\ovl\HH^{-1}(\h)) \bigr)^2}
{\partial_\theta \HH_1 (\HH_1^{-1}(\h,\psi),\psi) \, \bigl(\partial_\theta \ovl\HH (\ovl\HH^{-1}(\h))\bigr)^2} ,
\vspace{-.2cm}
\]
and expanding
\vspace{-.1cm}
\[
\begin{aligned}
& \partial_\theta \HH_1 (\HH_1^{-1}(\h,\psi),\psi) = \partial_\theta \HH_1 (\ovl \HH^{-1}(\h),\psi) 
+ \partial_\theta^2 \ovl \HH (\ovl \HH^{-1}(\h)) 
\, \left( (\HH_1^{-1}(\h,\psi) - \ovl \HH^{-1}(\h) \right) \\
& \qquad\qquad  +
\left( \partial_\theta^2 \HH_1(\ovl \HH^{-1}(\h),\psi) 
- \partial_\theta^2 \ovl \HH (\ovl \HH^{-1}(\h)) \right)
\, \left( (\HH_1^{-1}(\h,\psi) - \ovl \HH^{-1}(\h) \right) \\
& \qquad \qquad  +
\bigl( (\HH_1^{-1}(\h,\psi) - \ovl \HH^{-1}(\h) \bigr)^2
\int_0^1 \der t \, (1-t) \partial_\theta^3 \HH_1 \bigl( \ovl \HH^{-1}(\h),\psi) + t \bigl( (\HH_1^{-1}(\h,\psi) - \ovl \HH^{-1}(\h) \bigr) \bigr) ,
\end{aligned}
\vspace{-.2cm}
\]
we obtain eventually
\vspace{-.2cm}
\[
\begin{aligned}
\bigl| \average{ \partial_\h l_1  - \partial_\h \ovl l (\h) } \bigr| 
& \le \frac{1}{\ovl\tau^2}
\Bigl( 
\bigl| \average{ \partial_\theta h_1  - \partial_\theta \ovl h } \bigr|
+ \| \partial_\theta^2 \ovl \HH \|_{\io} \bigl| \aver{ l_1  - \ovl l  } \bigr|  +
\bigl( \average{ ( \partial_\theta^2 h_1  - \partial_\theta^2 \ovl h )^2}
\aver{ ( l_1  - \ovl l  )^2 } \bigr)^{1/2} \\
& +  \| \partial_\theta^2 \HH_1 \|_\io \bigl| \aver{ ( l_1  - \ovl l  )^2 } \bigr| \Bigr) + \frac{2}{\tau_1^2 \ovl\tau} \Bigl(
\| \partial_\theta^3 \HH_1 \|_\io \bigl| \aver{ ( l_1  - \ovl l )^2 } \bigr| +
\bigl| \average{ ( \partial_\theta h_1  - \partial_\theta \ovl h )^2 } \bigr| \Bigr) ,
\vspace{-.1cm}
\end{aligned}
\]
so that the first of \eqref{fluhlder} follows from the estimates of Propositions~\ref{prop:h1hbar}
and Lemmas~\ref{lem:fluhderder} and \ref{lem:l1der}. 
\qed

\medskip

Recalling that $l=l_1$, by \eqref{hh1-ll1}, Lemmas~\ref{lem:l1} and \ref{lem:l1der} immediately imply the following result.

\begin{prop} \label{prop:dascrivere}
Let $l$ be defined as in \eqref{hl}.
There 
is a constant $C$ such that, for all $\h\in\calU_0$,
\begin{subequations} \label{bounds-thm3cd}
\begin{align}
\left| \aver*{ l(\h,\cdot) - \overline l (\h) } \right| & \le  C\rho ,
& \aver*{( l(\h,\cdot)- \overline h(\h))^2}  & \le C\rho
\label{bounds-thm3c} \\
\left| \aver*{\partial_\h l(\h,\cdot) -\partial_\h \overline l (\h) } \right| & \le  C\rho ,
& \aver*{( \partial_\h l(\h,\cdot)- \partial_\h \overline l(\h))^2}  & \le C\rho  \, .
\label{bounds-thm3d}
\end{align}
\end{subequations}
\end{prop}

\subsection{Fluctuations of the dynamics:~proof of 
Theorem~\ref{thm:4}}
\label{proof4}

We start  by proving the following result for the extension of the translated map $\SSS_1$.

\begin{lemma} \label{lemma:fluc1}
For any $\gamma'\in(0,\gamma)$
there exists a constant $C$ such
that for all $\theta\in\calU$ and all $n\in \NNN$
 \[
\bigl|  \average{ (\SSS_{1}^n)_\theta(\theta,\cdot)- (\overline  \SSS^n)_\theta 
(\theta)} \bigr| \leq C \rho(1-\rho\,\ga')^{n} , \qquad
\bigl\langle \bigl( (\SSS_{1}^n)_\theta(\theta,\cdot)- (\overline  \SSS^n )_\theta 
(\theta)\bigr)^2 \bigr\rangle \leq C \rho \, (1-\rho\,\ga')^{2n} . 
 \]
\end{lemma}

\prova From \eqref{eq:Hbar2}, \eqref{eq:Htilde} 
and \eqref{HHH1h1}, we find
\begin{equation} \label{*}
\begin{aligned}
&( \SSS_1^n)_\theta (\theta,\psi) - ( \overline \SSS^n)_\theta (\theta)=
\HH_1^{-1} (\SSS_0^n(\HH_1(\theta,\psi),\psi)) - \ovl \HH^{-1} (\ovl\SSS_0^n(\ovl\HH(\theta),\psi)) \\
& \qquad \qquad
= 
\ka^{(n)}(\psi) \, \HH_{1} (\theta,\psi) -  \ovl \mu^n  \ovl \HH (\theta)
+ 
\Bigl( \bigl(\ka^{(n)}(\psi) \, \HH_{1} (\theta,\psi) \bigr)^2 -  \bigl(\ovl \mu^{n} \, \ovl \HH (\theta) \bigr)^2 \Bigr) \,
\ovl l \bigl( \overline \mu^n\; \overline\HH(\theta) \bigr) \Bigr)  \\
& \qquad\qquad
+
\Bigl( \bigl(\ka^{(n)}(\psi) \, \HH_{1} (\theta,\psi) \bigr)^2 -  \bigl(\ovl \mu^{n} \, \ovl \HH (\theta) \bigr)^2 \Bigr)
\Bigl( l_1 \bigl(\ka^{(n)}(\psi) \HH_{1} (\theta,\psi), A_0^n\psi\bigr) - \ovl l \bigl( \overline \mu^n\; \overline\HH(\theta) \bigr) \Bigr)  \\ 
& \qquad \qquad
+ \bigl( \ovl\m^{n} \, \ovl \HH(\theta) \bigr)^2  \Bigl(  l_1 \bigl(\ka^{(n)}(\psi) \HH_{1} (\theta,\psi), A_0^n\psi\bigr)-
\ovl l \bigl( \overline \mu^n\; \overline\HH(\theta) \bigr) \Bigr) .
\end{aligned}
\end{equation}
Shorten for notational simplicity
\vspace{-.1cm}
\[
\begin{aligned}
\AAA_n (\theta,\psi) & := \ka^{(n)}(\psi) \HH_{1}(\theta,\psi) - \overline \mu^n\; \overline\HH , \phantom{\int} 
\hspace{1cm}
\la_1(\theta,\psi)  := l_1 \bigl(\ka^{(n)}(\psi) \HH_{1} (\theta,\psi), A_0^n\psi\bigr) , \phantom{\int}  \\
\bar\la_1(\theta,\psi) & := l_1 \bigl( \overline \mu^n\; \overline\HH(\theta) , A_0^n\psi\bigr) \bigr) , \phantom{\int} 
\hspace{1.8cm}
\la_2(\theta,\psi)   := \partial_\h l_1 \bigl(\ka^{(n)}(\psi) \HH_{1} (\theta,\psi), A_0^n\psi\bigr) , \phantom{\int}  \\
\bar \la(\theta) & := \ovl l \bigl( \overline \mu^n\; \overline\HH(\theta) \bigr) \bigr) , \phantom{\int}  
\hspace{2.8cm}
\bar \la_2(\theta) := \partial_\h \ovl l \bigl( \overline \mu^n\; \overline\HH(\theta) \bigr) \bigr) , \phantom{\int}  \\
I_n(\theta,\psi)  & := \!\!\! \int_0^1 \!\!\! \der t \, (1\!-\!t) \,
\partial_\h^2  l_1 \bigl( \overline \mu^n\; \overline\HH(\theta) + t
\bigl(\ka^{(n)}(\psi) \HH_{1} (\theta,\psi) \!-\! \overline \mu^n\; \overline\HH(\theta) \bigr) , A_0^n\psi \bigr) ,
\vspace{-.1cm}
\end{aligned}
\]
and observe that $(\ka^{(n)} \HH_{1})^2 -  (\ovl \mu^{n} \, \ovl \HH)^2 = \AAA_n^2 + 2 \, \ovl \mu^n \, \ovl \HH \, \AAA_n$ and
\begin{equation} \label{**}
\begin{aligned}
\AAA_n & =
\bigl( \ka^{(n)} - \overline \mu^n \bigr) \bigl( \HH_1 -  \overline\HH \bigr) +
\bigl( \ka^{(n)} - \overline \mu^n \bigr) \overline\HH +
\overline\mu^n \bigl( \HH_1 -  \overline\HH \bigr)  \\
& = \theta^2 \bigl( \ka^{(n)} - \overline \mu^n \bigr) \bigl( h_1 -  \overline h \bigr) +
\bigl( \ka^{(n)} - \overline \mu^n \bigr) \overline\HH +
\theta^2 \overline\mu^n \bigl( h_1 -  \overline h \bigr) , \\
\la_1 - \bar \la & = \la_2 \AAA_n + I_n \, \AAA_n^2 + \bigl( \bar\la_1 - \bar \la \bigr)
\\ & 
= \ovl\lambda_2 \, \AAA_n + \left( \lambda_2 - \ovl \lambda_2 \right) \AAA_n 
+ I_n \AAA_n^2 + \left( \ovl \lambda_1 - \ovl \lambda \right) .
\end{aligned}
\end{equation}
Then, we rewrite \eqref{*} as
\[
( \SSS_1^n)_\theta  - ( \overline \SSS^n)_\theta  = \AAA_n +
\bigl( \AAA_n^2 + 2  \, \ovl \mu^{n} \, \ovl \HH \, \AAA_n \bigr) \ovl \la +
\bigl( \AAA_n^2 + 2  \, \ovl \mu^{n} \, \ovl \HH \, \AAA_n \bigr) \bigl( \la_1 - \ovl \la \bigr) +
\bigl( \ovl \mu^{n} \, \ovl \HH \bigr)^2 \bigl( \la_1 - \bar \la \bigr) ,
\]
with $\AAA_n$ and $\la_1 - \ovl \la$ as in the second and fourth line of \eqref{**}, respectively.
Thus, using Lemma~\ref{lem:flulin}, Propositions~\ref{prop:h1hbar} and \ref{lem:l1der}, and the Cauchy-Schwarz 
inequality, the bound \eqref{mediaSSS} follows. 

To obtain the second bound in \eqref{boundSSS} we write \eqref{*} as
\[
( \SSS_1^n)_\theta  - ( \overline \SSS^n)_\theta  = \AAA_n +
\bigl( \AAA_n^2 + 2  \, \ovl \mu^{n} \, \ovl \HH \, \AAA_n \bigr) \la_1 +
\bigl( \ovl \mu^{n} \, \ovl \HH \bigr)^2 \bigl( \la_1 - \bar \la \bigr) ,
\]
with $\AAA_n$ and $\la_1 - \ovl \la$ as in the second and third line of \eqref{**}, respectively,
so that, bounding
\[
\begin{aligned}
\bigl( ( \SSS_1^n)_\theta  - ( \overline \SSS^n)_\theta \bigr)^2 
& \le 3 \left( \AAA_n^2 +  2 \, \AAA_n^2
\bigl( \AAA_n^2 + 4 \, ( \ovl \mu^{n} \, \ovl \HH )^2 \bigr) \la_1^2 +
\bigl( \ovl \mu^{n} \, \ovl \HH \bigr)^4 \bigl( \la_1 - \bar \la \bigr)^2 \right) , \\
\bigl( \la_1 - \bar \la \bigr)^2 & \le 3 \left( \la_2^2 \AAA_n^2 + I_n^2 \, \AAA_n^4 + \bigl( \bar\la_1 - \bar \la \bigr)^2 \right) , \\
\AAA_n^2 & \le 2 \left( \bigl( \ka^{(n)} - \overline \mu^n \bigr)^2 \overline\HH +
\theta^4 \overline \mu^{2n} \bigl( h_{1} - \overline h  \bigr)^2 \right),
\end{aligned}
\]
we obtain that $\aver{(\,( \SSS_1^n)_\theta (\theta,\cdot) - ( \overline \SSS{\vphantom{S}}^n)_\theta (\theta)\, )^2}$ is bounded by
\[
C \Bigl( \bigl\langle \bigl( \ka^{(n)}(\cdot) - \overline \mu^n \bigr)^2 \bigr\rangle
+ \ovl\mu^{2n} \bigl\langle \bigl( h_1(\theta,\cdot) - \bar h (\theta) \bigr)^2 \bigr\rangle
+ \ovl\mu^{4n} \bigl\langle \bigl( l_1 (\ovl\mu^n\ovl \HH(\theta),\cdot) - \bar l (\ovl\mu^n\ovl \HH(\theta)\bigr)^2 \bigr\rangle \Bigr) .
\]
Then, using Lemma~\ref{lem:flulin} and Propositions~\ref{prop:h1hbar} and \ref{lem:l1der} gives immediately
the bound \eqref{varSSS}.
\qed

\medskip

Finally, we come back to the map $\SSS$. In order to compare
the full dynamics  generated by $\SSS$ with that generated by  $\overline \SSS$, 
we proceed along the same lines as the proof of Lemma~\ref{lemma:fluc1}. 
Thus, we decompose $(\SSS^n \big)_\f (\f,\psi) - ( \overline \SSS^n)_\f (\f)$
as done for $(\SSS_1^n \big)_\theta (\theta,\psi) - ( \overline \SSS^n)_\theta(\theta)$,
relying on \eqref{eq:1.11} and \eqref{hl} instead of \eqref{eq:Htilde} and \eqref{HHH1h1},
and obtain
\[
\begin{aligned}
& ( \SSS^n)_\f (\f,\psi) - ( \overline \SSS^n)_\f(\f) =
\HH^{-1} (\SSS_0^n(\HH(\theta,\psi),\psi)) - \ovl \HH^{-1} (\ovl\SSS_0^n(\ovl\HH(\theta),\psi)) \\
& \quad
= W(A_0^n\psi) + \ka^{(n)}(\psi) \, \HH (\theta,\psi) - \ovl \mu^{n} \, \ovl \HH (\theta)
+ \Bigl( \ka^{(n)}(\psi) \, \HH (\theta,\psi) -  \ovl \mu^{n} \, \ovl \HH (\theta) \Bigr)^2 \,
\ovl l \bigl( \overline \mu^n\; \overline\HH(\theta) \bigr) \Bigr) \\
& \quad
+ \Bigl( \ka^{(n)}(\psi) \, \HH (\theta,\psi) -  \ovl \mu^{n} \, \ovl \HH (\theta) \Bigr)^2 \,
\Bigl( l \bigl( \ka^{(n)}(\psi) \, \HH(\theta,\psi),\psi \bigr) - \ovl l \big( \overline \mu^n\; \overline\HH(\theta) \bigr) \Bigr) \\
& \quad
+  2  \, \ovl \mu^{n} \, \ovl \HH (\theta) \, \Bigl( \ka^{(n)}(\psi) \, \HH (\theta,\psi) -  \ovl \mu^{n} \, \ovl \HH (\theta) \Bigr) \,
\ovl l \bigl( \overline \mu^n\; \overline\HH(\theta) \bigr) \Bigr)  \\
& \quad
+ 2  \, \ovl \mu^{n} \, \ovl \HH (\theta) \, \Bigl( \ka^{(n)}(\psi) \, \HH (\theta,\psi) -  \ovl \mu^{n} \, \ovl \HH (\theta) \Bigr) \,
\Bigl( l \bigl( \ka^{(n)}(\psi) \, \HH(\theta,\psi),\psi \bigr) - \ovl l \big( \overline \mu^n\; \overline\HH(\theta) \bigr) \Bigr) \\
& \quad
+ \bigl( \ovl\m^{n} \, \ovl \HH(\theta) \bigr)^2  \Bigl(  
\partial_\h \ovl l \bigl( \overline \mu^n\; \overline\HH(\theta) \bigr) \,
\bigl(\ka^{(n)}(\psi) \HH (\theta,\psi) - \overline \mu^n\; \overline\HH(\theta) \bigr) \\
&  \quad
+ \bigl( \partial_\h l \bigl( \overline \mu^n \ovl \HH (\theta), A_0^n\psi\bigr) \bigr)
- \partial_\h \ovl l \bigl( \overline \mu^n\; \overline\HH(\theta) \bigr) \,
\bigl(\ka^{(n)}(\psi) \HH  (\theta,\psi)  - \overline \mu^n\; \overline\HH(\theta) \bigr) \\
& \quad
+  \bigl(\ka^{(n)}(\psi) \HH  (\theta,\psi) - \overline \mu^n\; \overline\HH(\theta) \bigr)^2 
\!\! \int_0^1 \!\!\! \der t \, (1\!-\!t) \,
\partial_\h^2  l \bigl( \overline \mu^n\; \overline\HH(\theta) + t
\bigl(\ka^{(n)}(\psi) \HH (\theta,\psi) \!-\! \overline \mu^n\; \overline\HH(\theta) \bigr) , A_0^n\psi \bigr) \\
& \quad
+  l \bigl( \overline \mu^n\; \overline\HH(\theta) , A_0^n\psi\bigr)-
\ovl l \bigl( \overline \mu^n\; \overline\HH(\theta) \bigr) \Bigr) ,
\end{aligned}
\]
where we expand 
\[
\begin{aligned}
& \ka^{(n)}(\psi) \, \HH (\theta,\psi) - \ovl \mu^{n} \, \ovl \HH (\theta) \\
& \quad =
\theta^2 \bigl( \ka^{(n)}(\psi) - \overline \mu^n \bigr) \, \bigl( h (\theta,\psi) - \overline h (\theta) \bigr)
+ \bigl( \ka^{(n)}(\psi) -  \overline \mu^n \bigr) \, \overline\HH(\theta) +
\theta^2 \overline \mu^n \bigl( h (\theta,\psi), A_0^n\psi\bigr) - \overline h (\theta) \bigr) .
\end{aligned}
\]
Then, using the bounds of Lemma~\ref{lem:flulin}, Theorem~\ref{thm:3}, Proposition~\ref{prop:dascrivere}
and the Cauchy-Schwarz inequality, we obtain the bound \eqref{mediaSSS} of Theorem~\ref{thm:4}.
Again, the bound \eqref{varSSS} is obtained in a similar way.

\zerarcounters 
\section{Convergence in probability} \label{last}

In this section we collect the previous results to prove Theorems~\ref{thm:6} and \ref{thm:prob},
so as to provide the probabilistic description of the results discussed in Subsection~\ref{subsec:probability}.
Below, as in Subsections~\ref{proof3-2} to \ref{proof4}, for notational simplicity, $\SSS_1$ denotes the extended map $\SSS_{1,{\rm ext}}$.

\subsection{The probability of deviations:~proof of Theorem~\ref{thm:6}}
\label{proof:prob}

From Theorem~\ref{thm:1} and a direct application of Chebyshev inequality 
we obtain that, for any $\delta>0$,
\vspace{-.2cm}
\begin{equation*}  
 m_0 \Bigl( \Bigl\{\psi \in\TTT^2 : |W(\psi)|>\delta \Bigr\} \Bigr)\leq \frac{C
\rho}{\delta^2} ,
\vspace{-.1cm}
\end{equation*}
for a suitable positive constant $C$ independent of $n$. 
Therefore, the invariant manifold $\calW$ for $\SSS$ converges in probability to $\overline \calW=\{(0,\psi) : \psi\in \TTT^2\}$.

Similarly, from Lemma~\ref{lemma:fluc1},
we obtain that, if $\ga'\in(0,\ga)$, then, for fixed $\theta$ and $n$, and for any $\de>0$,
\vspace{-.3cm}
\begin{equation} \label{1timep}
m_0\Bigl(\Bigl\{\psi \in \TTT^2 : (1-\rho\,\ga')^{-n} | (\SSS_1^n)_\theta(\theta,\psi) - (\overline\SSS^n)_\theta(\theta,\psi)  |
> \delta  \Bigr\}\Bigr)\leq  \frac{c_1(\ga') \,\rho}{\delta^2}  ,
\vspace{-.1cm}
\end{equation}
for a suitable positive constant $c_1(\ga')$
depending on $\ga'$ but neither on $\theta$ nor $n$.

The following result provides a uniform version of the estimate \eqref{1timep} 
and shows that, for most values of $\psi\in\TTT^2$, the quantity
$(1-\rho\,\ga')^{-n} | (\SSS_1^n)_\theta(\theta,\psi)-(\overline\SSS^n)_\theta(\theta,\psi) |$ remains small for all $n\ge 0$.

\begin{lemma}\label{lem:prob} 
Consider the dynamical system described by $\SSS$ in \eqref{eq:1.1bis} satisfying Hypotheses~\ref{hyp:1}--\ref{hyp:3}.
Let $\SSS_1$ be defined as in \eqref{eq:S1}. 
For $\rho$ small enough and any $\ga'\in(0,\ga)$
there exists a constant $C$ such that, for all $\theta\in\calU$, 
\vspace{-.4cm}
\[
m_0\Bigl(\Bigl\{\psi \in \TTT^2 : \sup_{n\ge 0} \, (1-\rho\,\ga')^{-n} | \,
(\SSS_1^n)_\theta (\theta,\psi)-(\overline\SSS^n)_\theta(\theta,\psi) |>\delta \Bigr\}\Bigr)\leq  \frac{C\rho}{\delta^3} .
\vspace{-.2cm}
\]
\end{lemma}

\prova
We have
$| (\SSS_1)_\theta (\theta,\psi)-(\overline\SSS)_\theta(\theta',\psi)|
\geq (1-\rho)|\theta-\theta'|-
\rho| f_1(\theta,\psi)-\bar f(\theta)|\geq (1-\rho)|\theta-\theta'| - C\rho|\theta| $
for all $\theta, \theta'\in\calU$, where
we have used that $\norm f \norm_{\al_0,6}=1$ and $f_1(0,\psi)=\ovl f(0)=0$, as it follows from the definition of $f_1$ in Subsection \ref{aux}
and from Remark \ref{barf0}.
Iterating we obtain, for any $\theta, \theta'\in\calU$ and for any $k\ge 0$,
 $|(\SSS_1^k)_\theta(\theta,\psi)-(\overline\SSS_\theta^k(\theta',\psi)|\ge   (1-\rho)^k|\theta-\theta'| - Ck(1-\rho\,\ga')^k\rho|\theta|$,
where we have used that $(1-\rho)< (1-\rho\,\ga')$, since $\ga'<\ga<1$ (see Remark \ref{Psi0ge1}), and that, by Remark \ref{rmktolemSSSn},
$|(\SSS_1^k)_\theta(\theta,\psi)| \leq  C(1-\rho\,\ga')^k|\theta|$. Assuming that
$(1-\rho\,\ga')^{-n} | 
(\SSS^n_1)_\theta(\theta,\psi)-(\overline\SSS^n)_\theta(\theta,\psi)| > \delta$
for some $n>0$, we find that 
\vspace{-.3cm}
\[
 (1-\rho\,\ga')^{-(n+k)} | 
(\SSS^{n+k}_1)_\theta(\theta,\psi)-(\overline\SSS^{n+k})_\theta(\theta,\psi)|
\geq \left(\frac{1-\rho}{1-\rho\,\ga'}\right)^k \delta  -  Ck\rho\, .
\vspace{-.2cm}
\]
For $\rho$ small enough, we can now find 
$M\in\NNN$ of the form $M =M_0\de/\rho$, with $M_0$ independent of $\rho$ and $\de$, such that
$Ck\rho\le \de/4$ and $((1-\rho)/(1-\rho\,\ga'))^k\ge 1/2$.
for all $k\leq M$.
This means that for $k\leq M$ we have
$(1-\rho\,\ga')^{-(n+k)} | (\SSS^{n+k}_1)_\theta(\theta,\psi)-(\overline\SSS{\vphantom{S}}^{n+k})_\theta(\theta,\psi)| \ge \de/4$.
We thus obtain that
\vspace{-.2cm}
\begin{equation} \label{sumsup}
\begin{aligned}
& \sum_{n=0}^{\infty}  m_0\left(\left\{\psi \in \TTT^2 :  (1-\rho\,\ga')^{-n} | 
(\SSS^n_1)_\theta(\theta,\psi)-(\overline\SSS^n)_\theta(\theta,\psi)|> \frac{\delta}{4} \right\}\right)  \\
& \qquad \ge \frac{M_0\de}{\rho} \, m_0\Bigl(\Bigl\{\psi \in \TTT^2 : \sup_{n\ge 0} \,  (1-\rho\,\ga')^{-n} | 
(\SSS^n_1)_\theta(\theta,\psi)-(\overline\SSS^n)_\theta(\theta,\psi)|> \delta  \Bigr\}\Bigr) .
\end{aligned}
\vspace{-.1cm}
\end{equation}
On the other hand, for any $\ga',\ga'' \in (0,\ga)$ we find, by using \eqref{1timep},
\[
m_0 \left( \left\{\psi \in \TTT^2 :  (1-\rho\,\ga'')^{-n} | (\SSS^n_1)_\theta(\theta,\psi)-(\overline\SSS^n)_\theta(\theta,\psi)| > 
\frac{\delta}{4}  \right\}\right) \le \left(\frac{1-\rho\,\ga''}{1-\rho\,\ga'}\right)^{2n} \frac{16 \, c_1(\ga')\,\rho}{\delta^2} ,
\vspace{-.1cm}
\]
and in the same way, by inverting the roles of $\ga'$ and $\ga''$,
%
\[
 m_0 \left(\left\{\psi \in\TTT^2 :  (1-\rho\,\ga')^{-n} | (\SSS^n_1)_\theta(\theta,\psi)-(\overline\SSS^n)_\theta(\theta,\psi)|
 > \frac{\delta}{4} \right\}\right) \le \left(\frac{1-\rho\,\ga'}{1-\rho\,\ga''}\right)^{2n}  \frac{16 \, c_1(\ga'')\,\rho}{\delta^2} ,
\vspace{-.1cm}
\]
so that, fixing $\ga'\in(0,\ga)$ and taking $\ga''=\ga'/2$, we obtain
%
\begin{equation} \label{sum}
\sum_{n=0}^{\infty} m_0\left(\left\{\psi \in \TTT^2 :  (1-\rho\,\ga')^{-n} | (\SSS^n_1)_\theta(\theta,\psi)-(\overline\SSS^n)_\theta(\theta,\psi)|
> \frac{\delta}{4} \right\}\right) \le \frac{c_2(\ga')}{\delta^2} ,
\vspace{-.1cm}
\end{equation}
%
%
with
\vspace{-.2cm}
\[
c_2(\ga')  = 16 \, c_1(\ga'/2)\,\rho \sum_{n=0}^{\io}  \left(\frac{1-\rho\,\ga'}{1-\rho \, \ga'/2}\right)^{2n} .
\vspace{-.1cm}
\]
Inserting \eqref{sum} into \eqref{sumsup} delivers the thesis. \qed

\medskip


To deal with $\SSS$, we follow the same argument used for $\SSS_1$.
First of all, from Theorem~\ref{thm:4}
and Chebyshev inequality, if $\ga'\in(0,\ga)$, then, for fixed $\theta$ and $n$, and for any $\de>0$,
\vspace{-.2cm}
\[
m_0\Bigl(\Bigl\{\psi \in \TTT^2 : (1-\rho\,\ga')^{-n} | (\SSS^n)_\f(\f,\psi) - (\overline\SSS^n)_\theta(\f,\psi)  - W(A_0^n\psi)|
>\delta  \Bigr\}\Bigr)\leq 
\frac{c_3(\ga')\,\rho}{\delta^2}  ,
\vspace{-.2cm}
\]
with the constant $c_3(\ga')$
independent of both $\theta$ and $n$. Next, observe that we can write
\vspace{-.1cm}
\[
\begin{aligned}
& \bigl| \SSS_\f(\f,\psi) - \overline\SSS_\f(\f',\psi)  - W(A_0\psi) \bigr| = \bigl| \SSS_\f(\f,\psi) - \overline\SSS_\f(\f',\psi)  - W(\psi) - \rho \, f(W(\psi),\psi) \bigr|
\phantom{\Big)} \\
& \quad =
\bigl| \f - W(\psi) - \f' + \rho \bigl(  f(\f,\psi) - \rho \, f(W(\psi),\psi) - \rho \, \ovl f(\f-W(\psi) \bigr) + \rho \bigl( \ovl f(\f-W(\psi)) - \ovl f(\f') \bigr) \bigr| \phantom{\Big)} \\
& \quad \ge (1 - \rho) \bigl| \f - W(\psi) - \f' \bigr| - C \rho \bigl| \f - W(\psi) , \phantom{\Big)}
\end{aligned}
\vspace{-.2cm}
\]
so that, iterating and using that
$| (\SSS^i)_\f (\f,\psi) \!-\! W(A_0^i\psi)| = | (\SSS_1^i)_\theta(\theta,\f)| \le C(1\!-\!\rho\,\ga')^i |\theta|$,
with $\theta=\f\!-\!W(\psi)$, we find, once again thanks to Lemma~\ref{lemSSSn},
\vspace{-.2cm}
\[
\bigl| (\SSS^k)_\f(\f,\psi) - (\overline\SSS^k)_\f(\f',\psi)  - W(A_0^n\psi) \bigr| \ge
(1 - \rho)^k \bigl| \f - W(\psi) - \f' \bigr| - C k (1-\rho\,\ga')^k \rho \bigl| \f - W(\psi) \bigr| .
\vspace{-.1cm}
\]
Therefore, assuming that 
$(1-\rho\,\ga')^{-n} |  (\SSS^n)_\theta(\theta,\psi)-(\overline\SSS^n)_\theta(\theta,\psi)-W(A_0^n\psi)| > \de$
for some $n>0$ and proceeding as done for $\SSS_1$ leads to 
\[
\begin{aligned}
\frac{c_4(\ga')}{\de^2}
& \ge \sum_{n=0}^{\infty}  m_0\left(\left\{\psi \in \TTT^2 :  (1-\rho\,\ga')^{-n} | 
(\SSS^n)_\f(\f,\psi)-(\overline\SSS^n)_\f(\f,\psi) - W(A_0^n\psi)|> \frac{\delta}{4} \right\}\right)  \\
& \ge \frac{M_0\de}{\rho} \, m_0\Bigl(\Bigl\{\psi \in \TTT^2 : \sup_{n\ge 0} \,  (1-\rho\,\ga')^{-n} | 
(\SSS^n)_\f(\f,\psi)-(\overline\SSS^n)_\f(\f,\psi)-W(A_0^n\psi)|> \delta  \Bigr\}\Bigr) .
\end{aligned}
\]
for a suitable constant $c_4(\ga')$. This completes the proof of Theorem
\ref{thm:6}.

\subsection{Continuous time:~proof of Theorem~\ref{thm:prob}}
\label{proof:probcont}

Relying on Remark \ref{rmk:SnS1n}, we get
\[
\begin{aligned}
\XXXX_t(\psi) & = (\SSS_1^{\lfloor t/\rho 
\rfloor})_\theta(\f_0-W(\psi),\psi)
\\ & 
+ (t/\rho-\lfloor t/\rho 
\rfloor)\left((\SSS_1^{\lfloor t/\rho \rfloor+1})_\theta(\f_0-W(\psi),\psi)-
(\SSS_1^{\lfloor t/\rho \rfloor})_\theta(\f_0-W(\psi),\psi)\right)\,.
\end{aligned}
\]
By Lemma~\ref{lemSSSn}, we obtain the bound
%
$\left| (\SSS_1^k)_\theta(\f_0-W(\psi),\psi) - (\SSS_1^k)_\theta(\f_0,\psi) \right| \leq C(1-\rho\,\ga')^k|W(\psi)|$,
%
which, combined with Lemmas~\ref{lemma:fluc1} and \ref{lem:Phik}, allows us to estimate, for $\xi\in(0,\ga)$,
\[
 \sup_{t\ge 0} e^{\xi t} |\XXXX_t(\psi)-\Phi_t(\f_0)|\leq \sup_{n\ge 0} \,
 (1-\rho\,\xi)^{-n}  \left| (\SSS^n_1)_\theta(\theta,\psi)-(\overline\SSS^n)_\theta(\theta,\psi) \right|+ C \left( |W(\psi)|+\rho \right) ,
\vspace{-.1cm}
\]
so that, for every $\delta>0$, we have
\[
 \lim_{\rho\to 0^+} \, m_0 \Bigl( \Bigl\{ \psi \in \TTT^2 : \sup_{t\ge 0} \, e^{\xi t}|\XXXX_t(\psi)-\Phi_t(\f_0)|\geq \delta \Bigr\} \Bigr) = 0 ,
\vspace{-.1cm}
\]
because of \eqref{0time} and Lemma~\ref{lem:prob}.

\medskip

\noindent
{\it Acknowledgements}.
F.B.~has been partially supported by the NSF grant DMS-1907643.
G.G.~has been partially supported by the research project PRIN 20223J85K3$\_$002
``Mathematical Interacting Quantum Fields" of the Italian Ministry of Education and Research (MIUR).

%
%
%

\appendix

\zerarcounters 
\section{Decay of correlations for H\"older continuous functions} \label{sec:4} 

\subsection{Symbolic dynamics} \label{app:corre}

Consider a Markov partition ${\mathcal Q}=\{Q_1,Q_2,\ldots,Q_s\}$ for the 
Anosov automorphism $A_0$ in \eqref{eq:1.1} and call $T$ the corresponding compatibility matrix.
Let $a$ denote the mixing time, i.e.~the minimum value $a\in\NNN$ such that all the entries of 
$T^{a+1}$ are different from $0$ (since every Anosov automorphism is transitive, $a$ is finite),
and let $\lambda_i$ be the eigenvalues of $T$ ordered according to their modulus.
Then $\lambda_1$, which is real and simple by the Perron-Frobenius Theorem,
equals $\lambda=|\lambda_+|$, with $\lambda_+$ being the
eigenvalue of $A_0$ with the largest modulus \cite[Ch.~6]{GBG}.

Set $\NN=\{1,2,\ldots,s\}$. We say that a sequence $\ul{\sigma}\in\NN^\ZZZ$ is 
$T$-compatible if $T_{\sigma_{k},\sigma_{k+1}}=1$ for all $k\in\ZZZ$; let
$\NN^\ZZZ_T$ denote the set of $T$-compatible sequences. We use the symbol 
$\ul{\sigma}$ also to denote subsequences, i.e.~elements of $\NN^{[n,m]}$, with
finite $n$ and $m$.
Given two sequences $\ul{\sigma}$ and $\ul{\sigma}'$, we set
$\nu(\ul{\sigma},\ul{\sigma}'):=\max\{k \in \NNN : \sigma_i=\sigma'_i, |i|<k\}$.  On 
$\NN^\ZZZ_T$ we consider the topology generated by the distance 
$d(\ul{\sigma},\ul{\sigma}')=\lambda^{-\nu(\ul{\sigma},\ul{\sigma}')}$.
Given $\ul{\sigma}\in \NN^\ZZZ_T$, let $\psi=\psi(\ul{\sigma})\in\TTT^2$ be the 
unique point whose symbolic representation is $\ul{\sigma}$; then we have
$A_0\psi(\ul{\sigma}) = \psi(\tau\ul{\sigma})$,
where $\tau$ is the left translation, that is $(\tau\ul{\sigma})_i=\sigma_{i+1}$, and
$|\psi(\ul{\sigma})-\psi(\ul{\sigma}')|\leq D_s \, d(\ul{\sigma},\ul{\sigma}')$,
where $D_s$ is the maximum diameter of the sets $Q_1,\ldots,Q_s$.
We call $C_{\ul{\sigma}}^J$ the  $T$-compatible cylinder
with base $J = (j_1,\ldots,j_q) \subset\ZZZ$ and specification $\underline{\sigma} 
=(\sigma_{j_1},\ldots,\sigma_{j_q})\in \NN^q_T$, i.e.~the set of sequences 
$\ul{\sigma'}\in\NN^\ZZZ_T$ such that $\sigma_{j_k}'=\sigma_{j_k}$ for $k=1,\ldots,q$.
Let $\calF^{\leq k}$ be the $\sigma$-algebra generated by the cylinder
sets  $C_{\ul{\sigma}}^{J}$ with $J\subset (-\infty,k]$ and $\calF^{\geq k}$ be 
the $\sigma$-algebra generated by the cylinders $C_{\ul{\sigma}}^{J}$ with $J\subset [k,\infty)$. 
The following well-known result follow easily from the mixing properties of $T$.

\begin{lemma}\label{lem:dec}
Given $k,k' \in\ZZZ$, with $k'>k$, let $g_-$ be a bounded $\calF^{\leq k}$-measurable function
and $g_-$ be a bounded $\calF^{\geq k'}$-measurable function.
One has
\vspace{-.2cm}
\begin{equation*} 
\left|\aver{g_-g_+}-\aver {g_-}\aver {g_+}\right|\leq K_1 \lambda_0^{- (k'-k)}\|g_-\|_\infty\|g_+\|_\infty\, .
\vspace{-.1cm}
\end{equation*}
with $\lambda_0:=\lambda/|\lambda_2|$ and $K_1$ a suitable constant.
\end{lemma}
\begin{rmk} \label{rmk:4.2}
\emph{
The eigenvalues of the compatibility matrix of any Anosov automorphism on
$\TTT^2$ are $\lambda$, $\lambda^{-1}$, together with $0$'s and roots of unity
\cite{Sn}. Therefore we have $\lambda_0 \ge \lambda$ in Lemma~\ref{lem:dec}.
}
\end{rmk}

\begin{rmk} 
\emph{
If $g_-(\ul{\sigma})$ is $\calF^{\leq k}$-measurable, then $g_-(\ul{\sigma})$  depends only on $\sigma_i$ with $i\leq k$,
that is $g_-(\ul{\sigma})=g_-(\ul{\sigma}')$ if $\sigma_i=\sigma'_i$ for every $i\leq k$. 
Similarly, if $g_+(\ul{\sigma})$ is $\calF^{\geq k}$, then $g_+(\ul{\sigma})$ depends only on $\sigma_i$ with $i\geq k'$,
that is $g_+(\ul{\sigma})=g_+(\ul{\sigma}')$ if $\sigma_i=\sigma'_i$ for every  $i\geq k'$.}
\end{rmk}


Let the sequences $\ul{\sigma}, \ul{\sigma}' \in \NN^\ZZZ_T$ be such that $\sigma_i=\sigma'_i$ for 
$i\leq n$. If $n>0$ then $d(\ul{\sigma},\ul{\sigma}')\leq\lambda^{-n}$ and, for 
$N\geq 0$, $d(\tau^{-N}\ul{\sigma},\tau^{-N}\ul{\sigma}')= 
\lambda^{-N}d(\ul{\sigma},\ul{\sigma }')$. Thus $\psi(\ul{\sigma}')$ is on the 
unstable manifold of $\psi(\ul{\sigma})$ and, since the unstable manifold is a 
straight line, we have $\psi(\ul{\sigma}')=\psi(\ul{\sigma})+x \, v_+$, with $|x|\leq D_s\lambda^{-n}$.
If $n\leq 0$ we still have that  $d(\tau^{-N}(\ul{\sigma}),\tau^{-N}(\ul{\sigma}')) \to0$ as $N\to\io$, so that 
even in this case we have $\psi(\ul{\sigma}')= \psi(\ul{\sigma})+x v_+$ for some 
$x\in\RRR$. Observe that $\psi(\tau^{2n}\ul{\sigma}')= 
\psi(\tau^{2n}\ul{\sigma})+\lambda^{2n} x \,v_+$ while, since  
$(\tau^{2n}\ul{\sigma}')_i =(\tau^{2n}\ul{\sigma})_i$ for $i\leq -n$, we have 
$\|\psi(\tau^{2n}\ul{\sigma}')-\psi(\tau^{2n}\ul{\sigma})\|\leq D_s  \lambda^{-n}$,
so that, by the previous argument, we find again that $|x| \leq D_s\lambda^{-n}$.
In the same way one shows that, if $\ul{\sigma}$ and $\ul{\sigma}'$ are such 
that $\sigma_i=\sigma'_i$ for $i\geq n$, then 
$\psi(\ul{\sigma}')=\psi(\ul{\sigma})+x v_-$, with $|x| \leq D_s\lambda^{n}$.

For $\ul{\sigma},\ul{\sigma}'\in\NN_T^\ZZZ$ let 
$\ul{\sigma}_{(-\infty,n]}\vee\ul{\sigma}_{(n,\infty)}'$ be the 
sequence that agrees with $\ul{\sigma}$ on $(-\infty,n]$ and with  
$\ul{\sigma}'$ outside such an interval and call
$m(\der\ul{\sigma}'_{(n,\infty)}|\ul{\sigma}_{(-\infty,n]})$ the conditional 
probability measure on $\ul{\sigma}'_{(n,\infty)}$ given  
$\ul{\sigma}_{(-\infty,n]}$ \cite[prop. 5.3.2]{GBG}. Calling 
$\NN(\ul{\sigma})=\{\ul{\sigma}'\,|\, 
\ul{\sigma}_{(-\infty,n]}\vee\ul{\sigma}_{(n,\infty)}'\in\NN_T^\ZZZ\}$ we have 
$m(\NN(\ul{\sigma})|\ul{\sigma}_{(-\infty,n]})=1$.

Given $g\in\gotB_{\al_-,\al_+}(\TTT^2,\RRR)$ write $\hat g(\ul{\sigma}):=g(\psi(\ul{\sigma}))$ and observe that 
$\hat g(\ul{\sigma}_{(-\infty,n]}\vee\ul{\sigma}'_{(n,\infty)})$ is almost everywhere well defined with respect to 
$m(\der\ul{\sigma}'_{(n,\infty)}|\ul{\sigma}_{(-\infty,n]})$. We can thus define 
%
\begin{subequations} \label{nonserve}
\begin{align}
\hat g^{(\leq n)}(\ul{\sigma}) & \! := \!\! \int \hat  
g(\ul{\sigma}_{(-\infty,n]} \!\!\vee\! \ul{\sigma}'_{(n,\infty)}) 
m(\der\ul{\sigma}'_{(n,\infty)}|\ul{\sigma}_{(-\infty,n]}) ,
\label{nonservea} \\
 \hat g^{(\geq n)}(\ul{\sigma}) & \! := \!\! \int \hat  
g(\ul{\sigma}_{(-\infty,n)}' \!\!\vee\! \ul{\sigma}_{[n,\infty)}) 
m(\der\ul{\sigma}_{(-\infty,n)}'|\ul{\sigma}_{[n,\infty)}) 
\label{nonserveb}
\end{align}
\end{subequations}
By construction $\hat g^{(\leq n)}(\ul{\sigma})$
 is $\calF^{(\leq n)}$-measurable and
$\langle \hat g^{(\geq n)}\rangle=\langle \hat g\rangle$, where, with a slight abuse of notation,
we use $\langle\cdot\rangle$ also for the average with  respect to $m$.  Analogous
considerations hold for $\hat g^{(\leq n)}(\ul{\sigma})$ as well.
Finally, we define
\vspace{-.1cm}
\begin{equation*}  
\hat g^{(n,+)}(\ul{\sigma}) := \hat  g^{(\leq \lfloor 
n/\al_+\rfloor)}(\ul{\sigma})-g^{(\leq  \lfloor (n-1)/\al_+\rfloor)}(\ul{\sigma}) , \qquad
\hat g^{(n,-)}(\ul{\sigma}) := \hat  g^{(\geq \lfloor n/\al_-\rfloor)}(\ul{\sigma})-g^{(\geq \lfloor (n+1)/\al_-\rfloor))}(\ul{\sigma}) .
\vspace{-.1cm}
\end{equation*}
which allows us to decompose
\vspace{-.1cm}
\[
 \hat g(\ul\sigma)  =\hat g^{(\leq 0)}(\ul\sigma)+\sum_{k=1}^{\infty}  \hat g^{(k,+)}(\ul\sigma) ,
\vspace{-.2cm}
\]
with $\hat g^{(k,+)}(\ul\sigma)$ depending only on $\ul\sigma_{(-\infty,\lfloor k/\al_+\rfloor)}$.
%

\begin{lemma} \label{lem:holder}
Let $g$ be $(\al_-,\al_+)$-H\"older continuous on $\TTT^2$.
One has
\begin{subequations} \label{eq:decay}
\begin{align}
& |\hat g^{(\leq n)}(\ul{\sigma}) -g(\ul{\sigma}) |\leq 
K_2 \lambda^{-\al_+n}|g|_{\al_+}^{+}\, , 
\qquad
|\hat g^{(\geq n)}(\ul{\sigma}) - g(\ul{\sigma})  |\leq
K_2 \lambda^{\al_-n}|g|_{\al_-}^{-} ,
\label{eq:decay2} \\
& |\hat g^{(n,+)}(\ul{\sigma})|\leq K_2 \lambda^{-n}|g|_{\al_+}^{+} ,
\qquad\qquad\qquad
 |\hat g^{(n,-)}(\ul{\sigma})|\leq K_2 \lambda^{n}|g|_{\al_-}^{-} ,
\label{eq:decay1}
\end{align}
\end{subequations}
for some positive constant $K_2$.
\end{lemma}

\proof
 If $\ul\sigma$ and $\ul \sigma'$ are $T$-compatible 
sequences and $\ul{\sigma}_{(-\infty,n]}\vee\ul{\sigma}'_{(n,\infty)}$ is 
$T$-compatible as well, then
$(\ul\sigma)_i=(\ul{\sigma}_{(-\infty,n]}\vee\ul{\sigma}'_{(n,\infty)})_i$ for 
$i\leq n$ and from the discussion at 
the beginning of this subsection we get 
$\psi(\ul{\sigma}_{(-\infty,n]}\vee\ul{\sigma}'_{(n,\infty)})=
\psi(\sigma)+xv_+$ with $|x|\leq D_s \lambda^{-n}$, so that
$|\hat g(\ul{\sigma}_{(-\infty,n]}\vee\ul{\sigma}'_{(n,\infty)})-\hat  g(\ul\sigma)|\leq  D_s \lambda^{-n\al_+} |g|_{\al_+}^{+} $.
Integrating over $m(\der\ul{\sigma}'_{(n,\infty)}|\ul{\sigma}_{(-\infty,n]})$ 
gives the first bound in \eqref{eq:decay2}. The second bound is derived in a similar way.

Let now $\ul \omega$ be a $T$-compatible sequence and, for 
any pair of symbols $\sigma,\sigma'\in\{1,\ldots,s\}$, let 
$\pi=\pi(\sigma,\sigma')$ be a $T$-compatible sequence of length $a$ such that 
$T_{\sigma\pi_1}= T_{\pi_a\sigma'}=1$. For every $T$-compatible 
sequence $\ul\sigma$, define the $T$-compatible sequence $\ul\omega_n(\ul\sigma) 
:=\ul{\sigma}_{(-\infty,n]}\vee \pi(\sigma_n,\omega_{n+a})\vee 
\ul\omega_{[n+a,\infty)}$. Reasoning as above and using that
$\lfloor (n-1)/\al_+\rfloor\al_+ \ge n-1-\al_+$, we get
\begin{equation} \label{prima}
\begin{aligned}
& |\hat g(\ul{\sigma}_{(-\infty,\lfloor (n-1)/\al_+\rfloor]}\vee\ul{\sigma}'_{(\lfloor (n-1)/\al_+\rfloor,\infty)}) - 
\hat g(\ul \omega_{\lfloor(n-1)/\al_+\rfloor}(\ul\sigma))|
\le D_s \lambda^{\al_+}\lambda^{-(n-1)}|g|_{\al_+}^{+} 
\end{aligned}
\end{equation}
and, analogously,
\vspace{-.1cm}
\begin{equation} \label{seconda}
 |\hat g(\ul{\sigma}_{(-\infty,\lfloor n/\al_+\rfloor]}\vee\ul{\sigma}'_{(\lfloor 
n/\al_+\rfloor],\infty)})-\hat g(\ul \omega_{\lfloor(n-1)/\al_+\rfloor}(\ul\sigma))| \leq
D_s \lambda^{\al_+}\lambda^{-(n-1)} |g|_{\al_+}^{+} .
\end{equation}
The first bound in \eqref{eq:decay1} follows by
integrating \eqref{prima} over $m(\der\ul{\sigma}'_{(\lfloor(n-1)/\al_+\rfloor,\infty)}| \ul{\sigma}_{(-\infty,\lfloor(n-1)/\al_+\rfloor]})$ and 
\eqref{seconda} over $m(\der\ul{\sigma}'_{(\lfloor n/\al_+\rfloor,\infty)}|\ul{\sigma}_{(-\infty,\lfloor n/\al_+\rfloor]})$, so as to obtain
\[
\begin{aligned}
|\hat g^{(\leq \lfloor(n-1)/\al_+\rfloor)}(\ul{\sigma})-\hat g(\ul 
\omega_{\lfloor(n-1)/\al_+\rfloor}(\ul\sigma))|&\leq
D_s \lambda^{\al_+}\lambda^{-(n-1)} |g|_{\al_+}^{+} ,\\
|\hat g^{(\leq \lfloor 
n/\al_+\rfloor)}(\ul{\sigma})-\hat g(\ul 
\omega_{\lfloor(n-1)/\al_+\rfloor}(\ul\sigma))|&\leq
D_s \lambda^{\al_+}\lambda^{-(n-1)} |g|_{\al_+}^{+} ,
\end{aligned}
\vspace{-.1cm}
\]
which gives with $K_2=2D_s\lambda^{1+\alpha_+}$. In a similar way we obtain the second bound.
\qed

\subsection{A correlation inequality:~proof of Proposition~\ref{prop:4.3}} \label{app:corre1}

Since $\aver*{ g_+ \, g_-\circ A_0^{n} }- \aver g_+\aver g_-  =\aver*{ \tilde g_+ \, \tilde g_-\circ A_0^{n} }$,
with the notation in \eqref{eq:1.222},
we can assume that $\aver{g_+}=\aver{g_-}=0$. Then, calling $\tilde n :=\lfloor  \al n\rfloor$, we write,
with obvious meaning of the symbols,
\vspace{-.1cm}
\[
\begin{aligned}
 \hat g_+(\ul\sigma) & =\hat g_+^{(\leq 0)}(\ul\sigma)+\sum_{k=1}^{\tilde n} 
\hat g_+^{(k,+)}(\ul\sigma)+ (\hat g_+(\ul\sigma)-\hat g_+^{(\leq 
\lfloor\tilde n/\al\rfloor)}(\ul\sigma)) 
=: \sum_{k=0}^{\tilde n+1}\check g_+^{(k)}(\ul\sigma) , \\
\hat g_-(\ul\sigma) & =
\hat g_-^{(\geq 0)}(\ul\sigma)+\sum_{k=1}^{\tilde n} \hat g_-^{(-k,-)}(\ul\sigma)+ (\hat g_-(\ul\sigma)-\hat g_-^{(\geq 
-\lfloor\tilde n/\al\rfloor}(\ul\sigma)) =: \sum_{k=0}^{\tilde n+1}\check g_-^{(k)}(\ul\sigma)  ,
\end{aligned}
\]
where we have
$\|\check g_+^{(k)}\|_\infty\leq K_2 \lambda^{- k}\|g_+\|_{\al}^+$ and
$\|\check g_-^{(k)}\|_\infty\leq K_2\lambda^{-k}\|g_-\|_{\al}^-$ by Lemma~\ref{lem:holder}.
Since $(\tau^n\ul{\sigma})_{(-k,\infty)}\!=\!\ul{\sigma}_{(n-k,\infty)}$,
so that $\check g^{(k)}_-(\tau^n\ul\sigma)$ depends only on $\ul{\sigma}_{(n-k,\infty)}$, we have
\vspace{-.1cm}
\begin{equation*}
\begin{aligned}
| \aver{\check g_+^{(k)} \, \check  g_-^{(k')} \circ \tau^n } |
& \le  K_1 K_2 \|g_+\|_{\al}^+\|g_-\|_{\al}^- \lambda^{- (k+k') }\lambda_0^{-(n-\lfloor k/\al\rfloor-\lfloor k'/\al\rfloor) } ,
\qquad & \hbox{if $k+k'<\tilde n$}, \\
%
| \aver{\check g_+^{(k)} \, \check  g_-^{(k')} \circ \tau^n}|
& \le K_2  \|g_+\|_{\al}^+\|g_-\|_{\al}^-\lambda^{-(k+k)} , 
\qquad & \hbox{if $k+k'\geq \tilde n$} ,
\end{aligned}
\end{equation*}
by Lemmas~\ref{lem:dec} and \ref{lem:holder}.
Summing over $k$ and $k'$ and using that
$(n-\lfloor k/\al\rfloor-\lfloor k'/\al\rfloor)\geq (\tilde n-k-k')/\al-3$, so that 
$(k+k')+(n-\lfloor k/\al\rfloor-\lfloor 
k'/\al\rfloor)\geq \tilde n/\al-(k+k')(1-1/\al)-3\geq \tilde 
n +(1-\al)\al^{-1}(\tilde n-k-k')-3$,
we obtain
\vspace{-.3cm}
\[
\begin{aligned}
| \aver*{ g_+ \, g_-\circ A_0^{n} }|
& \le K_2 \lambda^{-\tilde 
n}\|g_+\|_{\al}^+\|g_-\|_{\al}^-  \Bigg(
\sum_{\substack{k,k'=0,\ldots,\tilde n+1 \\ k+k'<\tilde n}} 
K_1 \lambda^3  \lambda^{-(\tilde n-k-k')\frac{1-\al}{\al}} +
\sum_{\substack{k,k'=0,\ldots,\tilde n+1 \\ k+k' \ge \tilde n}} 
\lambda^{\tilde n-k-k'}  \Biggr) \\
& \le 
\max\{1,K_1\} K_2 \lambda^3\|g_+\|_{\al}^+\|g_-\|_{\al}^- \lambda^{-\tilde n}\left(\tilde 
n\sum_{q=0}^{\tilde n}  \lambda^{-q\frac{1-\al}{\al}}+\sum_{q=0}^{\tilde n + 2}  (\tilde n + q + 1)\lambda^{-q}\right) ,
\end{aligned}
\vspace{-.1cm}
\]
from which the estimate in Proposition~\ref{prop:4.3}
follows observing that $\lambda^{-\tilde n}\leq \lambda^{\al}\lambda^{-\al n} \le \lambda \, \lambda^{-\al n}$.

\zerarcounters 
\section{Bounds on the norms of the iterated products } \label{app:tech}

Let $\pp^{(n)} (\f,\psi)$ be defined as in \eqref{defpn}. Set
%
$\PP_{ij}(\f,\psi) := (\partial_\f^i \pp_j) (\SSS^j (\f,\psi)) =
(\partial_\f^i \pp_j \circ \SSS^j )(\f,\psi)$ and
$\SSSS_{ij}(\f,\psi) := \partial_\f^i (\SSS^j)_\f (\f,\psi)$.
%
The following lemma is easily checked by direct computation.

\begin{lemma}\label{3deri}
Let $\SSS$ be a map on $\calU\times\TTT^2$ of the form \eqref{SSS}, 
with $\SSS_\f(\f,\psi)$ of class $C^3$ in $\f$, 
and let $\pp_0,\ldots,\pp_{n-1}$ be any functions defined in $\calU\times\TTT^2$ and of class $C^3$ in the first 
variable $\f$.
%
%
One has
\vspace{-.1cm}
\begin{subequations} \label{deripn-mammamia}
\begin{align}
\partial_\f \pp^{(n)}
& = \sum_{i=0}^{n-1} \PP_{1i} \SSSS_{1i} 
\prod_{\substack{j=0 \\ j\not=i}}^{n-1} \PP_{0j} 
\label{deripn1} \\
 \partial^2_\f \pp^{(n)}
 & = \sum_{i=0}^{n-1} 
\Bigl( 
\PP_{2i} \SSSS_{1i}^2 + \PP_{1i} \SSSS_{2i}
\Bigr)
\prod_{\substack{ j=0 \\ j\not=i}}^{n-1} \PP_{0j} 
%
+ \sum_{\substack{ i,j = 0 \\ i\not=j}}^{n-1} 
\PP_{1i}\SSSS_{1i} \PP_{1j} \SSSS_{1j}
\prod_{\substack{k=0 \\ k \not=i,j}}^{n-1} \PP_{0k} 
\label{deripn2} \\ 
\partial^3_\f \pp^{(n)} 
& = \sum_{i=0}^{n-1} 
\Bigl( 
\PP_{3i} \SSSS_{1i}^3 + 3 \PP_{2i} \SSSS_{1i} \SSSS_{2i} + \PP_{1i} \SSSS_{3i} 
%
\Bigr) 
 \prod_{\substack{ j=0 \\ j\not=i}}^{n-1} \PP_{0j} 
\nonumber  \\ & 
+ 2 \sum_{\substack{ i,j = 0 \\ i\not=j}}^{n-1}  
\Bigl( \PP_{2i} \SSSS_{1i}^2 
%
+ \PP_{1i} \SSSS_{2i} 
\Bigr)
 \left( \partial_\f \pp_j \circ \SSS^j \right) \PP_{1j} \SSSS_{1j} 
 \prod_{\substack{k=0 \\ k \not=i,j}}^{n-1} \PP_{0k} 
 \label{deripn3} \\ &
+ \sum_{\substack{ i,j,k = 0 \\ i\not=j \neq k \neq i}}^{n-1} \PP_{1i} 
%
\SSSS_{1i} \PP_{1j} \SSSS_{1j} \PP_{1k} \SSSS_{1k}
\prod_{\substack{h=0 \\ h \not=i,j,k}}^{n-1} \PP_{0h} . 
\nonumber
\end{align}
\end{subequations}
\end{lemma}

\begin{rmk} \label{rmk:zero2}
\emph{
If both $\SSS_\f(\f,\psi)$ and the functions $\pp_0,\ldots,\pp_{n-1}$ do not 
depend on $\psi$, then setting $\pp^{(n)}(\f) = \pp^{(n)}(\SSS;\f)$,
equations \eqref{deripn-mammamia} still hold, with $\SSS(\f,\psi)$ and 
$\pp_i(\SSS^i(\f,\psi))$ replaced, respectively, with $G(\f)$ and $\pp_i(G^i(\f))$,
and with $\PP_{ij}(\f,\psi)$ and $\SSSS_{ij}(\f,\psi)$ defined accordingly.
}
\end{rmk}

\subsection{A first application: completion of the proof of Theorem~\ref{thm:2} } \label{B0}

Differentiating twice the function $h_1$ in \eqref{lll} gives
%
\begin{equation*}
\|\partial_\theta^2 h_1 \|_\infty\leq \sum_{n=1}^\infty \left(
\bigl\|\partial_\theta^2 p_1^{(n)}\bigr\|_\infty\|q_1\|_\infty
+ \bigl\|  \partial_\theta p_1^{(n)}\bigr\|_\infty \|\partial_\theta  (q_1\circ\SSS_1^n)\|_\infty
+ \bigl\|p_1^{(n)}\bigr\|_\infty \|\partial_\theta^2  (q_1\circ\SSS_1^n)\|_\infty \right) ,
\vspace{-.1cm}
\end{equation*}
where $\bigl\|  \partial_\theta p_1^{(n)}\bigr\|_\infty$ can be bounded as in \eqref{boundgnp}.
Thus, noting that
\vspace{-.1cm}
\[
\partial_\theta  (q_1\circ\SSS_1^n) = (\partial_\theta q_1 \circ\SSS_1^n ) \, \partial_\theta \SSS_1^n ,
\qquad
\partial_\theta^2  (q_1\circ\SSS_1^n) = (\partial_\theta^2 q_1 \circ\SSS_1^n ) \, (\partial_\theta \SSS_1^n)^2 +
(\partial_\theta q_1\circ\SSS_1^n )\, \partial_\theta^2 \SSS_1^n ,
\vspace{-.1cm}
\]
and writing $\partial_\theta^2 p_1^{(n)}$ according to \eqref{deripn2}, we can use Lemma~\ref{lemSSSn}
to obtain the bound \eqref{boundhder-bis}, for a suitable constant $D_3$.
Finally, since
\vspace{-.3cm}
\[
\partial_\theta h_1 = \sum_{n=1}^\infty \left( \bigl(\partial_\theta p_1^{(n)} \bigr) q_1 +p_1^{(n)} \partial_\theta  (q_1\circ\SSS_1^n) \right) ,
\vspace{-.2cm}
\]
we find
\vspace{-.2cm}
\begin{equation*}
| \partial_\theta h_1|_{\als} 
\le \sum_{n=1}^\infty \Bigl( \bigl| \partial_\theta p_1^{(n)}\bigr|_{\als} \|q_1\|_\infty
+\bigl\| \partial_\theta p_1^{(n)}\bigr\|_\infty |q_1\circ\SSS_1^n|_{\als} 
+ \bigl| p_1^{(n)}\bigr|_{\als} \|\partial_\theta q\|_\infty
+\bigl\| p_1^{(n)}\bigr\|_\infty | \partial_\theta q_1\circ\SSS_1^n|_{\als}\Bigr) ,
\vspace{-.1cm}
\end{equation*}
where the factors which are not differentiated with respect to $\theta$ are bounded as to obtain \eqref{boundhderal},
while the differentiated factors are bounded once more by relaying on
\eqref{deripn1} and the bounds in Lemma~\ref{lemSSSn}.

\subsection{Products independent of the fast variable:~proof of Lemma~\ref{lem:pn}} \label{proof:pn}

Let $\overline N=O(\rho^{-1})$ be defined as in Remark \ref{rmkB}.
Then, using that
$|\overline G^n(\f)|\leq \ovl\theta$ for $n\geq \overline N$
and that
\vspace{-.2cm}
\begin{equation*}  
 \partial_\f \ovl G^n(\f)=\prod_{i=0}^{n-1} \partial_\f \ovl  G
(\ovl G^i(\f)) ,
\vspace{-.2cm}
\end{equation*}
we get
%
\begin{equation} \label{deriGn}
\|\partial_\f \ovl G^n\|_\infty\leq C (1 - \rho\,\ga')^n\, .
\end{equation}
%

\begin{rmk} \label{constantC1}
\emph{
As observed in Remark \ref{constantC}, from here on -- including the following appendices -- we call $C$ 
any constant whose exact value is not important.
}
\end{rmk}

Then, combining \eqref{deripn1} and \eqref{defpnnopsi} together with the bound \eqref{deriGn},
we obtain
%
\vspace{-.1cm}
\begin{equation*}
\biggl| \prod_{i=0}^{n-1} \pp_i(\ovl G^i(\f)\biggr|
\le C (1 \!-\! \rho\ga')^{n} ,
\qquad
\biggl|\partial_\f\biggl( \; \prod_{i=0}^{n-1} \pp_i(\ovl G^i(\f))\biggr)\biggr|
\leq C\rho
(1 \!- \!\rho\ga')^{n-1} 
\sum_{i=0}^{n-1}
(1\! - \!\rho\ga')^i
\leq C
(1 \!-\! \rho\ga')^{n} ,
\vspace{-.1cm}
\end{equation*}
so that $\norm \pp^{(n)}\norm_{0,0} = \| \pp^{(n)} \|_{\io} \leq C (1 - \rho\,\ga')^n$ and
%
%
\begin{equation} \label{deripn}
\norm \pp^{(n)}\norm_{0,1}\leq C (1 - \rho\,\ga')^n , 
\qquad
\|\partial_\f^2\ovl  G^n\|_\infty\leq C (1 - \rho\,\ga')^n ,
\end{equation}
where the second one is a special case of the first one, with $\pp_i=\partial_\f \ovl G$ $\forall i=0,\ldots,n-1$.
Inserting the bounds \eqref{deripn} in \eqref{deripn2} (again see \eqref{defpnnopsi} for notations)
and reasoning in a similar way, we get
\begin{equation} \label{derideripn}
\norm \pp^{(n)}\norm_{0,2}\leq C(1 - \rho\,\ga')^n , 
\qquad
\|\partial_\f^3\ovl G^n\|_\infty\leq C(1 - \rho\,\ga')^n .
\end{equation}
Using \eqref{derideripn} in \eqref{deripn3} gives \eqref{deriderideripna}, while
from \eqref{deriGn}, \eqref{deripn} and \eqref{derideripn}
we get the second bound in \eqref{deriderideripna}.

\subsection{Products depending on the fast variable:  proof of Lemma~\ref{pna}} \label{proof:pna}

Assume the hypotheses of Lemma~\ref{pna} to be satisfied.
Reasoning as in the proof of Lemma~\ref{lem:pn}, we find
%
$\norm \pp^{(n)}\norm_{0,3}\leq C (1-\rho\,\ga')^n$.
Since $\partial_\theta(\SSS^n_2)_\theta=(\partial_\theta G_2)^{(n)}$,
where $(\partial_\theta G_2)^{(n)}$ is given by \eqref{defpn}
with $\pp_i=\partial_\theta G_2$ for all $i=0,\ldots,n-1$, and $(\SSS_2)_\theta(0,\psi)=0$, we also have
\begin{equation} \label{S2123}
\|\partial^k_\theta(\SSS^n_2)_\theta\|_\io  \leq C  (1 - \rho\,\ga')^n
\qquad 
k=0,1,2,3 .
\end{equation}
On the other hand, writing
\[
\begin{aligned}
 G_2(\SSS_2^n(\theta,\psi))&-G_2(\SSS_2^n(\theta,\psi'))=
  G_2((\SSS_2^n)_\theta(\theta,\psi),A_0^n\psi)-G_2((\SSS_2^n)_\theta(\theta,\psi'),A_0^n\psi)\\
&+(\SSS_2^n)_\theta(\theta,\psi')\int_0^1 \left( \partial_\theta G_2(t(\SSS_2^n)_\theta(\theta,\psi'),A_0^n\psi)-
\partial_\theta G_2(t(\SSS_2^n)_\theta(\theta,\psi'),A_0^n\psi') \right)\der t ,
\end{aligned}
\]
we have, for $n \ge 2$,
\[
|(\SSS_2^n)_\theta|_{\al_0}^-
=|G_2\circ\SSS^{n-1}_2|_{\alpha_0}^-\leq 
\|(\partial_\theta G_2)\circ 
\SSS^{n-1}_2\|_\infty
 |G_2\circ\SSS^{n-2}_2|_{\alpha_0}^-+(1-\rho\,\ga')^{n-1}\lambda^{-\al_0 (n-1)}|\partial_\theta G_2|^-_{\al_0} ,
\]
so that, iterating, thanks to the condition assumed on $\rho$, we get
\vspace{-.1cm}
\begin{equation}\label{chenoia0}
|(\SSS_2^n)_\theta|_{\al_0}^- 
\le C\norm G_2\norm_{\al_0,1}(1-\rho\,\ga')^{n-1}\sum_{i=0}^{n-1}
\lambda^{-\al_0(n-1- i)}
\le C (1-\rho\,\ga')^n\, ,
\vspace{-.1cm}
\end{equation}
which holds true also for $n=1$. We can now write, by \eqref{eq:1.6a},
\vspace{-.1cm}
\[
 | \pp^{(n)}|_{\al_0}^-\leq\sum_{i=0}^{n-1}| \pp_i\circ \SSS^{i}_2|_{\al_0}^-
 \prod_{\substack{j=0 \\ j \not=i}}^{n-1} \| \pp_i\circ  \SSS^{j}_2\|_\infty ,
\vspace{-.2cm}
\]
where
$| \pp_i\circ\SSS^{i}_2|_{\al_0}^-\leq \|\partial_\theta \pp_i\|_\infty |(\SSS^{i}_2)_\theta|_{\al_0}^-+\lambda^{-\al_0 i } | \pp_i|_{\al_0}$,
with the first term missing if $i=0$, so that we get
\vspace{-.1cm}
\begin{equation}\label{chenoia}
\begin{aligned}
| \pp^{(n)}|_{\al_0}^-
& \le C (1 - \rho\,\ga')^{n-1}
\left(
\rho\sum_{i=1}^{n-1} (1 - \rho\,\ga')^i + \rho \sum_{i=0}^{n-1}\lambda^{-\al_0 i}
\right)
\le C (1 - \rho\,\ga')^n ,
\end{aligned}
\vspace{-.1cm}
\end{equation}
which implies, taking $\pp_i=\partial_\theta G_2$ $\forall i=0,\ldots,n-1$, also the bound
\vspace{-.1cm}
\begin{equation}\label{chenoia2}
|\partial_\theta   (\SSS^n_2)_\theta|_{\al_0}^-\leq C (1 - \rho\,\ga')^n .
\vspace{-.1cm}
\end{equation}
Analogously, using the expression \eqref{deripn2} for $\partial_\theta^2 \pp^{(n)}(\theta)$ and 
the bound \eqref{S2123}, we get
\vspace{-.1cm}
\[
\begin{aligned}
| \partial_\theta \pp^{(n)}|_{\al_0}^-\leq&
\sum_{i=0}^{n-1}\Bigl( |\partial_\theta  \pp_i \circ  \SSS^{i}_2|_{\al_0}^-\|\partial_\theta (\SSS^{i}_2)_\theta \|_\infty+
\|\partial_\theta  \pp_i \circ \SSS^{i}_2\|_\infty|\partial_\theta(\SSS^{i}_2)_\theta |_{\al_0}^-\Bigr)
\prod_{\substack{ j=0 \\ j\not=i}}^{n-1} \| \pp_j \circ \SSS^{j}_2\|_\infty\\
+&\sum_{ \substack{ i, j=0 \\ i \neq j}}^{n-1}
\| \partial_\theta \pp_i \circ \SSS^{i}_2\|_\infty\|\partial_\theta(\SSS^{i}_2)_\theta \|_\infty
| \pp_j\circ \SSS^{j}_2|_{\al_0}^-
\prod_{\substack{ k=0 \\ k \not=i,j}}^{n-1} \| \pp_k \circ  \SSS^{k}_2\|_\infty ,
\end{aligned}
\vspace{-.1cm}
\]
that, by the same argument used in \eqref{chenoia}, delivers
\vspace{-.1cm}
\begin{equation}\label{chenoia3}
 |\partial_\theta  \pp^{(n)}|_{\al_0}^-\leq C  \, (1 - \rho\,\ga' )^n , 
\qquad\qquad 
|\partial^2_\theta   (\SSS^n_2)_\theta|_{\al_0}^-\leq C \, (1- \rho\,\ga' )^n .
\vspace{-.1cm}
\end{equation}
It is now easy, by using the expression \eqref{deripn3} for $\partial_\theta^3 p^{(n)}(\theta)$
and the bounds \eqref{S2123}--\eqref{chenoia3}, and reasoning once more as in in \eqref{chenoia},
to obtain
\vspace{-.2cm}
\begin{equation} \label{chenoia4}
 |\partial_\theta^2 \pp^{(n)}|_{\al_0}^-\leq C  \, (1-  \rho\,\ga' )^n , 
\vspace{-.1cm}
\end{equation}
so as to complete the proof of the first bound in \eqref{bounds-pna}.

Finally the second bound in \eqref{bounds-pna} follow collecting together the bounds \eqref{S2123}--\eqref{chenoia4}. 

\zerarcounters 
\section{A correlation inequality involving the slow variable} \label{appC}

In this and the following Appendix~\ref{app:tech2}, the maps $\SSS$, $\SSS_1$ and $\SSS_2$ are meant as the extended maps
which are obtained by following the procedure described in Subsection~\ref{extension}, and, analogously,
the domains are meant as the extended domains where the extended maps are defined.

\subsection{Some preliminary rewriting}

Define $\pp^{(k)}_{[i]}:= \pp^{(k)}_{[i]}(\SSS_2;\theta,\psi)$ and $\aver{\pp}^{(k)}_{[i]}:= \aver{\pp}^{(k)}_{[i]}(\ovl \SSS;\theta)$,
for $k=0,\ldots,n-1$ and $i=0,\ldots,n-k$. according to \eqref{defpn} and \eqref{defpnaver}, respectively,
and set $\pp^{(n)}=\pp^{(n)}_{[0]}$ and $\aver{\pp}^{(n)}=\aver{\pp}^{(n)}_{[0]}$. Observe that 
\begin{equation}\label{1diff}
\begin{aligned}
&  \pp^{(n)} (\theta,\psi) 
\, g_- (\SSS_2^n  (\theta,\psi))
-  \aver{\pp}^{(n)} (\theta) 
\,  g_- (\ovl \SSS^n  (\theta,\psi)) 
\\
& \qquad \qquad = \sum_{i=0}^{n-1}\Bigl(
\aver{\pp}^{(i)}(\theta) \, \pp_{[i]}^{(n-i)}(\ovl\SSS^i(\theta,\psi)) \, g_-(\SSS_2^{n-i}  (\ovl\SSS^i(\theta,\psi)))\\
& \qquad \qquad 
-\aver{\pp}^{(i+1)}(\theta) \, \pp_{[i+1]}^{(n-i-1)}( \ovl\SSS^{i+1}(\theta,\psi)) \, g_-( \SSS_2^{n-i-1}  (\ovl\SSS^{i+1}(\theta,\psi)))\Bigr) .
\end{aligned}
\end{equation}
Moreover we have
%
$\pp^{(n-i)}_{[i]}(\ovl\SSS^i(\theta,\psi)) \, g_-(\SSS_2^{n-i}(\ovl\SSS^i(\theta,\psi)))=
\pp^{(n-i)}_{[i]}(\ovl G^i(\theta),A_0^{i}\psi) \, g_-(\SSS_2^{n-i}(\ovl G^i(\theta),A_0^{i}\psi))$,
%
that is $\aver{\pp}^{(i)}(\theta) \, \pp_{[i]}^{(n-i)}(\ovl\SSS^i(\theta,\psi)) \, g_-(\SSS_2^{n-i}(\ovl\SSS^i(\theta,\psi)))$
depends on $\psi$ only through $\psi':=A_0^{i}\psi$;
a similar consideration holds for the second term inside the summation in \eqref{1diff}. Defining
\vspace{-.1cm}
\begin{equation} \label{pngDelta00}
\Delta^{(k)}_i(\theta,\psi) : = \pp^{(k)}_{[i]}(\theta,\psi) \,
g_-(\SSS_2^{k}(\theta,\psi))
- \aver{\pp_i}( \theta) \, \pp^{(k-1)}_{[i+1]}( \ovl G(\theta),A_0\psi) \, g_-(\SSS_2^{k-1}(\ovl G(\theta),A_0\psi)) ,
\vspace{-.1cm}
\end{equation}
we get
\vspace{-.4cm}
\begin{equation} \label{pngDelta}
\pp^{(n)}(\theta,\psi) \, g_-(\SSS_2^n(\theta,\psi))-\aver{\pp}^{(n)}(\theta) \, g_-( \ovl \SSS^n(\theta,\psi))
= \sum_{i=0}^{n-1}\aver{\pp}^{(i)}(\theta) \, \Delta^{(n-i)}_i(\ovl \SSS^{i}(\theta,\psi)) .
\vspace{-.1cm}
\end{equation}
%

\subsection{The new correlation inequality:~proof of Proposition~\ref{prop:decaynonltot}} \label{app:NCI}

We start with a particular case, by assuming the function $g_+$ in Proposition~\ref{prop:decaynonltot}
to have zero average. Eventually we extend the result to any $g_+ \in \calmB_{\al_0}^+(\Omega,\RRR)$.

\begin{lemma}\label{decaynonl}
Assume $\rho$ to be such that the map $\SSS_2$ satisfies Hypotheses~\ref{hyp:1}--\ref{hyp:3}.
Let $\pp_0,\ldots,\pp_{n-1}$ be any functions in $\calmB_{0,3}(\Omega,\RRR)$ such that,
for some $\rho' \in (0,1)$,
\vspace{-.2cm}
\begin{enumerate}
\itemsep0em
\item $\|\pp_i - 1 \|_{\io}= O(\rho)$ for all $i=0,\ldots,n-1$,
\item $|\pp_i(\theta)| \leq 1-\rho\ga'$
for $|\theta|\leq \theta'$ for some $\theta'=O(1)$ in $\rho$ and for all $i=0,\ldots,n-1$, 
\vspace{-.2cm}
\end{enumerate}
and set $\pp^{(n)}(\theta,\psi) = \pp^{(n)}(\SSS_2;\theta,\psi)$, with $\pp^{(n)}(\SSS_2;\theta,\psi)$ as in \eqref{defpn}.
For any $g_+\in \calmB_{\al_0}^+(\Omega,\RRR)$ with $\aver{g_+}(\theta) = 0$
and any $g_-\in\calmB_{\al_0,2}^-(\Omega,\RRR)$ one has
\[
 \left|\aver*{ g_+ \pp^{(n)} g_-\circ \SSS_2^n}\right|\leq C
 (1 - \rho\,\ga')^n
 \left(\lambda^{-\al_0 n}\| g_+\|^+_{\al_0}\|\tilde  g_-\|^-_{\al_0}+
\rho \, \| g_+\|^+_{\al_0} 
\norm g_-\norm^-_{\al_0,2}
\right) ,
\]
where the constant $C$  does not depends on $n$.
\end{lemma}

\proof
In \eqref{pngDelta}, $\Delta^{(n-i)}_i(\ovl \SSS^{i}(\theta,\psi))=\Delta^{(n-i)}_i(\ovl G^{i}(\theta),A_0\psi))$ 
depends on $\psi$ only through $\psi':=A_0^{i}\psi$.
Thus, using Proposition~\ref{prop:4.3}, we bound
\begin{equation} \label{boh} 
\begin{aligned}
& \null\hspace{-.3cm} \left| \aver*{ g_+ \pp^{(n)} g_-\circ \SSS_2^n} \right| 
\le \left| \aver*{ g_+ \pp^{(n)} g_- \circ \SSS_2^n}  -\aver \pp^{(n)}  \aver*{ g_+ g_-\circ \ovl \SSS^n}\right| +
\bigl| \aver \pp^{(n)} \bigr|  \left| \aver*{ g_+ \, g_-\circ \ovl \SSS^n}\right| \\
& \le C\sum_{i=0}^{n-1}  \|\pp^{(i)}\|_\infty (1\!+\! \al_0 i) \, \lambda^{-\al_0 i}\|g_+\|_{\al_0}^+\|\Delta^{(n-i)}_i\|_{\al_0}^-
+ C\|\aver{\pp}^{(n)}\|_\infty (1\!+\!\al_0 n) \, \lambda^{-\al_0 n}\| g_+\|^+_{\al_0}\|\tilde  g_-\|^-_{\al_0} .
\end{aligned}
\end{equation}
Moreover, $\SSS_2(\theta,\psi)=(\ovl G(\theta)+\rho \Delta f(\theta,\psi),A_0\psi)$,
with
%
$\Delta f(\theta,\psi):= \tilde f(\theta,\psi) - \chi(\theta)\,f(0,\psi)$,
%
so that we have
\begin{align*}
\Delta^{(k)}_i(\theta,\psi) & =
\bigl( \pp_i(\theta,\psi) - \aver{\pp_i} (\theta) \bigr) \pp_{[i+1]}^{(k-1)}(\SSS_2(\theta,\psi)) 
\, g_-(\SSS_2^{k}(\theta,\psi)) \\
& +  \aver{\pp_i}( \theta) \Bigl( \pp_{[i+1]}^{(k-1)}( \ovl G(\theta)+\rho\Delta f(\theta,\psi),A_0\psi) -\pp_{[i+1]}^{(k-1)}( \ovl G(\theta),A_0\psi) \Bigr)
g_-(\SSS_2^{k}(\theta,\psi)) \\
& + \aver{\pp_i}( \theta) \, \pp_{[i+1]}^{(k-1)}( \ovl G(\theta),A_0\psi) \bigl(
g_-(\SSS_2^{k-1}(\ovl G(\theta,\psi) + \rho \Delta f(\theta,\psi))  - 
g_-(\SSS_2^{k-1}(\ovl G(\theta),A_0\psi)) \bigr) .
\end{align*}
Thus, writing
\begin{equation} \label{bounds12}
\begin{aligned}
& \pp_{[i+1]}^{(k-1)}( \ovl G(\theta)+\rho\Delta f(\theta,\psi),A_0\psi) - \pp_{[i+1]}^{(k-1)}( \ovl G(\theta),A_0\psi) \\
& \qquad \qquad = \rho\Delta f(\theta,\psi) \int_0^1 \der t \, \partial_\theta 
\pp_{[i+1]}^{(k-1)}( \ovl G(\theta)+ t \rho\Delta f(\theta,\psi),A_0\psi) , \\
%
%
& g_-(\SSS_2^{k-1}(\ovl G(\theta) + \rho \Delta f(\theta,\psi))  -g_-(\SSS_2^{k-1}(\ovl G(\theta),A_0\psi))
\\ & \qquad\qquad 
= \rho\Delta f(\theta,\psi) \int_0^1 \der t 
\left( ( \partial_\theta g_- \circ \SSS_2^{k-1} ) 
( \partial_\theta (\SSS_2^{k-1})_\theta) \right) (\ovl G(\theta)+ t \rho\Delta f(\theta,\psi),A_0\psi) , 
\end{aligned}
\end{equation}
and using that 
$| \partial_\theta g_- \circ \SSS_2^k|_{\al_0}^- \le \| 
\partial^2_\theta g_- \circ \SSS_2^k \|_\io | (\SSS_2^k)_\theta |_{\al_0}^-  + \la^{-\al_0 k} | \partial_\theta g_- |_{\al_0}^-$,
%
the assumptions on the functions $\pp_0,\ldots,\pp_{n-1}$ and the bounds provided in Lemma~\ref{pna} and Remark \ref{ex3}
provide the bound
$\|{\Delta_i^{(k)}} \|_{\al_0}^-\leq C \rho (1 - \rho\,\ga')^{k-1} \norm g_- \norm_{\al_0,2}^-$,
which, inserted into \eqref{boh}, completes the proof. \qed

\begin{rmk} \label{Delta}
\emph{
If $g_-$ is in $\calmB_{\al_0,3}^-(\Omega,\RRR)$,
then, using \eqref{bounds12} and still relying on Lemma~\ref{pna}
in the same way as we obatined the bound on $\|\Delta_i^{(k)}\|_{\al_0}^-$
we may prove that
$\norm\Delta^{(k)}_i \norm_{\al_0,2}^-\leq C \rho (1-\rho\,\ga')^{k-1} \norm g_-\norm_{\al_0,3}^- $.
}
\end{rmk}

\begin{lemma}\label{Omega}
Assume $\rho$ to be such that the map $\SSS_2$ satisfies Hypotheses~\ref{hyp:1}--\ref{hyp:3}.
Let $\pp_0,\ldots,\pp_{n-1}$ be any functions in $\calmB_{0,3}(\Omega,\RRR)$
verifying the hypotheses in Lemma~\ref{decaynonl}, and define $\aver{\pp}^{(n)}$ as in \eqref{defpn} and
$\Xi^{(k)}(\theta,\psi) : =\aver{\pp}^{(k)}(\theta) \, g_-( \ovl \SSS{\vphantom{s}}^{k}(\theta,\psi))$.
For any $g_- \in\calmB_{\al_0,2}^-(\Omega,\RRR)$ one has
\vspace{-.1cm}
\[
\left| \aver*{ \Xi^{(k)} \circ \SSS_2- \Xi^{(k)} \circ \ovl\SSS} \right| \leq
C \rho(1-\rho\,\ga')^k \bigl( (1+\al_0 k) \, \lambda^{-\al_0k}\norm 
g_-\norm_{\al_0,2}^-+\rho\norm g_-\norm_{0,2} \bigr) \, .
\vspace{-.1cm}
\]
\end{lemma}

\proof
Writing
\[
\begin{aligned}
& \Xi^{(k)}(\SSS_2(\theta,\psi))-  \Xi^{(k)}(\ovl\SSS(\theta,\psi)) \\
& \quad =\rho\Delta  f(\theta,\psi) \,
\partial_\theta \Xi^{(k)}(\ovl\SSS(\theta,\psi)) 
+ (\rho \Delta f(\theta,\psi))^2  
 \!\! \int_0^1 \der t \,(1-t) \, \partial^2_\theta \Xi^{(k)}(\ovl G(\theta,\psi)+ t\rho\Delta f(\theta,\psi),A_0\psi) 
\end{aligned}
\vspace{-.2cm}
\]
and observing that
$\Xi^{(k)}(\ovl\SSS(\theta,\psi))=\aver{\pp}^{(k)}(\ovl G(\theta)) \, g_-( \ovl G^{k+1}(\theta),A_0^{k+1}\psi))$,
we get, by using Lemmas~\ref{lem:pn} and \ref{decaynonl},
$|  \aver{\Delta f\, \partial_\theta \Xi^{(k)}\circ\ovl\SSS} | 
\leq C (1+\al_0 k ) \, \lambda^{-\al_0k}\|\Delta f\|_{\al_0}(1-\rho\,\ga')^k\norm g_-\norm_{\al_0,2}^-$,
while, using \eqref{eq:normalk}, we find
 $\|\partial^2_\theta \Xi^{(k)}\|_{\infty}\leq C  \norm\aver{\pp}^{(k)}\norm_{0,2} \,
 \norm g_-\circ\ovl\SSS{\vphantom{s}}^{(k)}\norm_{0,2}\leq C (1-\rho\,\ga')^k\norm g_-\norm_{0,2}$, again by Lemma~\ref{lem:pn}.
Combining the above estimates we obtain the desired bound.
\qed

\medskip

Now we prove Proposition~\ref{prop:decaynonltot}.
Observe that
$\aver{ g_+ \pp^{(n)} g_-\circ \SSS_2^n} \!=\!\aver*{ g_+}\aver{\pp^{(n)} g_-\circ  \SSS_2^n}\!+\!\aver*{\tilde g_+ \pp^{(n)} g_-\circ \SSS_2^n}$.
The second term is bounded using Lemma~\ref{decaynonl}.
If we write the first term using \eqref{pngDelta}, we need to estimate $\aver{\Delta_i{\vphantom{s}}^{(n-i)}}$. We write, for any $k\ge 1$,
%
\begin{equation} \label{boundDelta2}
\begin{aligned}
\Delta_i^{(k)} & = (\pp_i -\aver{\pp_i})\Bigl(\pp^{(k-1)}_{[i+1]} \circ \SSS_2  \, 
g_- \circ \SSS_2^{k} - \pp^{(k-1)}_{[i+1]} \circ \ovl\SSS \, g_- \circ \SSS_2^{k-1} \circ \ovl\SSS \Bigr) \\
& +(\pp_i -\aver{\pp_i}) \, \pp^{(k-1)}_{[i+1]} \circ \ovl\SSS \, g_- \circ \SSS_2^{k-1}\circ \ovl\SSS
+\aver{\pp_i} \Bigl(\pp^{(k-1)}_{[i+1]} \circ \SSS_2 \, g_- \circ \SSS_2^{k}-\pp^{(k-1)}_{[i+1]} \circ \ovl\SSS \, g_- \circ \SSS_2^{k-1}\circ \ovl\SSS \Bigr)  .
\end{aligned}
\vspace{-.1cm}
\end{equation}
We have $\|(\pp_i(\theta,\psi)-\aver{\pp_i}(\theta))\|_{\io} \leq C\rho$, while, 
reasoning like when studying \eqref{bounds12}, we get
\[
\| \pp^{(k-1)}_{[i+1]}
(\SSS_2(\theta,\psi)) 	\, g_-(\SSS_2^{k}(\theta,\psi))-\pp^{(k-1)}_{[i+1]}
(\ovl\SSS( \theta,\psi)) \, g_-(\SSS_2^{k-1}(\ovl\SSS(\theta,\psi)))\|_\infty
\leq  C\rho(1-\rho\,\ga')^k\norm g_-\norm_{0,1} ,
\]
so that the average of the first contribution in the r.h.s.~of \eqref{boundDelta2} is bounded by 
$C\rho^2(1-\rho\,\ga')^k\norm g_-\norm_{0,1}$.
A similar bound for the second contribution 
is obtained by using \eqref{bounds12}, as in the proof of Lemma~\ref{decaynonl},
with the gain of a further factor $\rho$ because of the factor $\pp_i-\aver{\pp_i}$.
As to the last contribution, we use \eqref{pngDelta} twice and write
\vspace{-.2cm}
\begin{equation*}  
\begin{aligned}
\pp^{(k-1)}_{[i+1]} \circ \ovl\SSS \, g_- \circ \SSS_2^{k-1} \circ \ovl\SSS 
& = \sum_{j=0}^{k-2} \Bigl( \aver{\pp}^{(j)}_{[i+1]} \circ G_2
\Delta^{(k-1-j)}_j \circ \ovl \SSS^{j} \circ \SSS_2 -\aver{\pp}^{(j)}_{[i+1]}\circ \ovl  G \, \Delta^{(k-1-j)}_j \circ \ovl  \SSS^{j+1} \Bigr)  \\ & 
+ \aver{\pp}^{(k-1)}_{[i+1]} \circ G_2 \, g_- \circ \ovl  \SSS^{k-1} \circ \SSS_2 - \aver{\pp}^{(k-1)}_{[i+1]}\circ \ovl G \, g_- \circ \ovl \SSS^{k}  .
\end{aligned}
\end{equation*}
Thus, Lemma~\ref{Omega}, applied to the last line, gives
\[
\bigl| \aver{
\aver{\pp}^{(k-1)}_{[i+1]} \!\circ G_2 \, g_- \circ \ovl\SSS^{k-1}\!\!\!\circ \SSS_2 - \aver{\pp}^{(k-1)}_{[i+1]} \circ \ovl G \, g_- \circ \ovl \SSS^{k} } \bigr| 
\le C \rho(1-\rho\,\ga')^k \bigl( (1 + \al_0 k) \lambda^{-\al_0k}\norm 
g_-\norm_{\al_0,2}^-+\rho\norm g_-\norm_{0,2} \bigr) ,
\]
and, applied to the second line, with $g_-=\Delta^{(k-1-j)}_j$, gives
\[
\begin{aligned}
& \bigl| \aver{\aver{\pp}^{(j)}_{[i+1]}\circ G_2\,\Delta^{(k-1-j)}_j\circ\ovl\SSS^{j}\circ\SSS_2-
\aver{\pp}^{(j)}_{[i+1]}\circ\ovl G\,\Delta^{(k-1-j)}_j \circ \ovl \SSS^{j+1}} \bigr| \\
& \qquad \qquad \qquad \leq C\rho(1-\rho\,\ga')^j \bigl( (1+\al_0 j) \lambda^{-\al_0j}
\norm \Delta^{(k-1-j)}_j \norm_{\al_0,2}^-+\rho\norm\Delta^{(k-1-j)}_j \norm_{0,2}^- \bigr)\, ,
\end{aligned}
\]
where $\norm \Delta^{(k-1-j)}_j\norm_{\al_0,2}^-$
and hence also $\norm\Delta^{(k-1-j)}_j \norm_{0,2}^-$ are bounded as discussed in Remark \ref{Delta}.
This concludes the proof of Proposition~\ref{prop:decaynonltot}. 

\zerarcounters 
\section{Proof of some technical results} \label{app:tech2}

As mentioned at the beginning of Appendix~\ref{appC},  we use $\SSS$, $\SSS_1$ and $\SSS_2$ as a shortened notation for  the
corresponding extended maps. In the light of Remark \ref{regSS2}, we also assume $\rho<\rho_0$.

\subsection{Derivative of the auxiliary map:~proof of Proposition~\ref{prop:fluhder1}} \label{proof:fluhder1}

In order to bound $\aver{\partial_\theta \, (p_2^{(n)}  \, q_2 \circ \SSS_2^n ) - \partial_\theta \, (\ovl p^{(n)} \, \ovl q \circ \ovl G^n )}$,
we split
\vspace{-.1cm}
\begin{subequations} \label{derp2q2}
\begin{align}
\partial_\theta \, (p_2^{(n)}  \, q_2 \circ \SSS_2^n )
& = \partial_\theta p_2^{(n)} \, q_2 \circ \SSS_2^n  + p_2^{(n)} \, \partial_\theta  (q_2 \circ \SSS_2^n ) , 
\label{derp2q2a} \\
\partial_\theta \, (\ovl p^{(n)} \, \ovl q \circ \ovl G^n )
& = \partial_\theta \ovl p^{(n)} \, \ovl q \circ \ovl G^n + \ovl p^{(n)} \, \partial_\theta ( \ovl q \circ \ovl G^n  ) ,
\label{derp2q2b}
\end{align}
\end{subequations}
with both $\partial_\theta p_2^{(n)}$ and $\partial_\theta \ovl p^{(n)}$ written according to \eqref{deripn1},
and we bound separately the two contributions
%
$\aver{ \partial_\theta p_2^{(n)} \, q_2 \circ \SSS_2^n } -  \partial_\theta \ovl p^{(n)} \, \ovl q \circ \ovl G{\vphantom{G}}^n$
and
$\aver{p_2^{(n)} \, \partial_\theta  (q_2 \circ \SSS_2^n ) } - \ovl p^{(n)} \, \partial_\theta ( \ovl q \circ \ovl G{\vphantom{G}}^n )$.
Since $\partial_\theta (\SSS_2)^n_\theta=(\partial_\theta G_2)^{(n)}$ and $\partial_\theta \ovl G^n =(\partial_\theta \ovl G)^{(n)}$, we bound
the first contribution as
\vspace{-.2cm}
\begin{equation} \label{tobebounded}
\begin{aligned}
& \bigl| \aver{ \partial_\theta p_2^{(n)} \, q_2 \circ \SSS_2^n } -  \partial_\theta \ovl p^{(n)} \, \ovl q \circ \ovl G^n  \bigr|
\\ & \;
\le \sum_{k=0}^{n-1} \Bigl| \aver{ (p_2 \, \partial_\theta G_2)^{(k)} 
\bigl( \partial_\theta p_2   \, 
P_{n-1-k} Q_{n-k}
\bigr) \circ \SSS_2^{k} } 
- \aver{p_2 \, \partial_\theta G_2}^{(k)} \aver{\partial_\theta p _2 \, 
P_{n-1-k} Q_{n-k}
} \circ \ovl G^{k} \Bigr|
\\ & \quad
+  \sum_{k=0}^{n-1} \bigl| \aver{ p_2 \, \partial_\theta G_2 }^{(k)}
\bigr|  \bigl|
\bigl( \aver{\partial_\theta p_2  \, 
P_{n-1-k} Q_{n-k}
}\circ \ovl G^k 
- \aver{\partial_\theta p_2 } \aver{
P_{n-1-k}
} \aver{q_2} \circ\ovl G^{n-k}  \bigr) \circ \ovl G^k \bigr|
\\ & \quad
+ \sum_{k=0}^{n-1} 
\bigl| \aver{ p_2 \partial_\theta G_2}^{(k)} \bigl( \aver{\partial_\theta p_2 } \aver{
P_{n-1-k}
}
\aver{q_2} \circ \ovl G^{n-k} \bigr) \circ \ovl G^k  
\!-\! (\ovl p \, \partial_\theta \ovl G)^{(k)} \bigl( \partial_\theta \ovl p \, 
\ovl P_{n-1-k}
\,
\ovl Q_{n-k} 
\bigr) \circ \ovl G^k \big| ,
\end{aligned}
\vspace{-.2cm}
\end{equation}
where we have set $P_n \!:=\! p_2^{(n)}\circ\SSS_2$, $Q_n\!:=\!q_2\circ\SSS_2^n$,
$\ovl P_n\!:=\!\ovl p^{(n)}\circ\ovl G$ and $\ovl Q_n \!:=\! \ovl q\circ\ovl G^n$.
We use Proposition~\ref{prop:decaynonltot},
with $k$ instead of $n$, $g_+=1$, $\pp_i = p_2 \partial_\theta G_2$ for $i=1,\ldots,k$ and $g_-= \partial_\theta p_2  \, p_2^{(n-1-k)} \circ\SSS_2 \,  q_2\circ \SSS^{n-k}$,
to bound the second line of \eqref{tobebounded} with
\vspace{-.2cm}
\begin{equation} \label{bounded1}
\sum_{k=1}^n  C (1-\rho\,\ga')^{k} \bigl( \rho+ k \rho^2 \bigr) \norm \partial_\theta p_2  \, p_2^{(n-1-k)} \circ \SSS_2 \,  q_2\circ \SSS_2^{n-k} \norm_{\al_0,3}^- ,
\vspace{-.1cm}
\end{equation}
%
and with $n-1-k$ instead of $n$, $g_+=\partial_\theta p_2$, $\pp_i= p_2\circ\SSS_2$ for $i=1,\ldots,n-k-1$ and $g_-=q_2$,
to bound the third line of \eqref{tobebounded} with
\vspace{-.2cm}
\begin{equation} \label{bounded2}
\sum_{k=0}^{n-1} 
C (1-\rho\,\ga')^{k} \Bigl( (n-1-k) \la^{-\al_0(n-1-k)} + \bigl( \rho + (n-1-k)\rho^2 \bigr) \norm \partial_\theta p_2 \norm_{\al_0}^+ \norm q_2 \norm_{\al_0,3}^- \Bigr) .
\vspace{-.1cm}
\end{equation}
After estimating
\[
\begin{aligned}
& \bigl| \aver{ (p_2 \, \partial_\theta G_2)^{(k)}} \bigl( \aver{\partial_\theta p_2 } \aver{
P_{n-1-k} 
} \aver{q_2} \circ \ovl G^{n-k}\bigr)\circ\ovl G^k
- (\ovl p \, \partial_\theta \ovl G)^{(k)} \bigl( \partial_\theta \ovl p \, \, 
\ovl P_{n-1-k}
\,
\ovl Q_{n-k}
\bigr) \circ \ovl G^k \bigl| \\
& \qquad \le
\bigl| \aver{ p_2 \, \partial_\theta G_2 }^{(k)} - (\ovl p \, \partial_\theta \ovl G)^{(k)} \bigr|
\bigl| \bigl( \aver{\partial_\theta p_2 } \aver{
P_{n-1-k} 
} \aver{q_2} \circ \ovl G^{n-k} \bigr) \circ \ovl G^k \bigr|
\\ & \qquad \qquad 
+
\bigl| (\ovl p \, \partial_\theta \ovl G)^{(k)} \bigr|
\bigl| ( \aver{\partial_\theta p _2} - \partial_\theta \ovl p \bigr) \circ \ovl G^k \bigr|
\bigl| ( \aver{ 
P_{n-1-k} 
} \aver{q_2} \circ \ovl G^{n-k} \bigr) \circ \ovl G^k\bigr| \phantom{\Bigr(} \\
& \qquad \qquad 
+ 
\bigl| (\ovl p \, \partial_\theta \ovl G)^{(k)} \, \partial_\theta \ovl p \circ \ovl G^k  \bigr|
\bigl| \bigl( \aver{ 
P_{n-1-k} 
} -
\ovl P_{n-1-k}
\bigr) \circ \ovl G^k \bigr| 
\bigr| \aver{q_2} \circ \ovl G^{n}  \bigr| 
\\ & \qquad \qquad
+
\bigl| (\ovl p \, \partial_\theta \ovl G)^{(k)} \, \bigl( \partial_\theta \ovl p \, \,
\ovl P_{n-1-k}
\bigr) \circ \ovl G^k \bigr|
\bigl| \aver{q_2} \circ \ovl G^{n}  -
\ovl Q_{n}
\bigl| , \phantom{\Bigr(}  \\
\end{aligned}
\]
we rely once more Proposition~\ref{prop:decaynonltot} to bound the fourth line of \eqref{tobebounded} with
\begin{equation} \label{bounded3}
\begin{aligned}
& \hspace{-.4cm} \sum_{k=0}^{n-1}  
C (1 \!-\! \rho\ga')^{k} \Bigl( {(\rho + k  \rho^2)} \| \partial_\theta p_2 \|_{\io} \|p_2^{(n-1-k)} \|_{\io} \|q_2 \|_{\io} 
+ \rho \,  \| \aver{\partial_\theta p_2} \!-\! \partial_\theta \ovl p \|_{\io}  \|p_2^{(n-1-k)} \|_{\io} \|q_2 \|_{\io} \\
& \qquad \qquad 
+ {(\rho + (n-k-1)  \, \rho^2 ) } \, (1-\rho\,\ga')^{n-k-1} \|q_2 \|_{\io} +
\rho \, (1- \rho\,\ga')^{n-1-k} \| \aver{q_2} - \ovl q \|_{\io} \Bigr) .
\end{aligned}
\end{equation}
Therefore, thanks to \eqref{deriv} and \eqref{p2pbar}, which yield
\[
\norm\partial_\theta p_2\norm_{\al_0,2}\leq C\rho ,
\qquad \norm \aver{p_2} -\ovl p \norm_{0,1} \le C\rho^2 ,
\qquad \|q_2\|_{\io} \le C \rho  , 
\qquad \| \aver{q_2} - \ovl q\|_{\io} \le C\rho^2 , 
\]
by collecting together the bounds \eqref{bounded1}--\eqref{bounded3} and using \eqref{eq:normalk}, we obtain
\vspace{-.1cm}
\begin{equation} \label{fluh2der1}
\bigl| \aver{ \partial_\theta p_2^{(n)} \, q_2 \circ \SSS_2^n } -  \partial_\theta \ovl p^{(n)} \, \ovl q \circ \ovl G^n  \bigr| 
\le \sum_{k=0}^{n-1} C (1-\rho\,\ga')^n \rho^3  \left( 1 + k \rho + k^2 \rho^2 \right) .
\vspace{-.1cm}
\end{equation}

Next, we consider the second contribution, that we rewrite as
\[
\bigl( \aver{p_2^{(n)} \, \partial_\theta  (q_2 \circ \SSS_2^n )}  - 
\aver{p_2}^{(n)}\aver{ \partial_\theta (q_2 \circ \ovl G^n)}  \bigr) +
\bigl( \aver{p_2}^{(n)}\aver{ \partial_\theta (q_2\circ \ovl G^n }-
 \ovl p^{(n)} \, \partial_\theta (\ovl q\circ \ovl G^n) \bigr) ,
 \]
and, using that
$\partial_\theta  (q_2 \circ \SSS_2^n ) \!=\! (\partial_\theta (\SSS_2)_\theta )^{(n)} (\partial_\theta  q_2) \circ \SSS_2^n$,
$\partial_\theta( q_2 \circ \ovl G^n) \!=\! (\partial_\theta \ovl G)^{(n)}  \partial_\theta q_2 \circ \ovl G^n$ and
$\partial_\theta (\ovl q\circ \ovl G^n) \!=\! (\partial_\theta \ovl G)^{(n)} (\partial_\theta \ovl q) \circ \ovl G^n$,
we apply first Proposition~\ref{prop:decaynonltot}, with $\pp_i=p_2 \, \partial_\theta\SSS_2$, $g_+=1$ and $g_-=q_2$ to bound
\[ 
\bigl| \aver{ ( p_2 \, \partial_\theta (\SSS_2)_\theta )^{(n)} (\partial _\theta q_2) \circ \SSS_2^n} -
\aver{p_2 \, \partial_\theta (\SSS_2)_\theta }^{(n)} \aver {\partial_\theta q_2} \circ \ovl G^n \bigr| \le 
C (1-\rho\,\ga')^{n}{ \left( \rho + \rho^2 n \right)} \norm \partial_\theta q_2 \norm_{\al_0,3} ,
\]
then reason as when dealing with \eqref{h2mhbar} to get
\[
\begin{aligned}
& \left|\aver{p_2 \partial_\theta(\SSS_2)_\theta }^{(n)}\aver{ \partial_\theta q_2}\circ \ovl G^n -
( \ovl p \, \partial_\theta \ovl G)^{(n)} \, (\partial_\theta \ovl q) \circ \ovl G^n \right| \\
& \qquad \le
\aver{\partial_\theta q_2}\circ \ovl G^n
\sum_{i=0}^{n-1} \aver{p_2 \, \partial_\theta (\SSS_2)_\theta }^{(i)} \left( \aver{p_2 \partial_\theta (\SSS_2)_\theta }-\ovl p \, \partial_\theta \ovl G\right)
\circ \ovl G^{i} \, (\ovl p \, \partial_\theta \ovl G)^{(n-i-2)}\circ \ovl G^{i+1}
\\ & \qquad \qquad + 
(\ovl p \, \partial_\theta \ovl G)^{(n)} \left( \aver{\partial_\theta q_2}\circ \ovl G^n- \partial_\theta \ovl q\circ \ovl G^n \right)
\le C(1-\rho\,\ga')^n n \, \| p_2 \partial_\theta (\SSS_2)_\theta  - \ovl p \, \partial_\theta \ovl G \|_{\io} \, \| \partial q_2 \|_{\io} \\
 & \qquad \qquad  
 +  C(1-\rho\,\ga')^n \,  \| \aver{ \partial_\theta q_2 } - \partial_\theta \bar q \|_{\io} 
\le C (1-\rho\,\ga')^n n \rho^2 , \phantom{\sum_{i=0}^{n-1}}
\end{aligned}
\]
so that eventually we obtain
\begin{equation} \label{fluh2der2}
\bigl| \aver{p_2^{(n)} \, \partial_\theta  (q_2 \circ \SSS_2^n ) - \ovl p^{(n)} \, \partial_\theta ( \ovl q \circ \ovl G^n  ) } \bigr| \leq C\rho .
\end{equation}
Thus, summing together \eqref{fluh2der1} and \eqref{fluh2der2},
the first bound in \eqref{fluhder1} follows.

The second bound in \eqref{fluhder1} is obtained in a similar way by reasoning
as in the second part of the proof of Proposition~\ref{prop:h2hbar}
and using \eqref{fluh2der1} and \eqref{fluh2der2} instead of \eqref{h2ah2}.

\subsection{Comparing the translated and auxiliary maps}

\subsubsection{Proof of Lemma~\ref{S1-S2}} \label{proof:S1-S2}


Following the strategy outlined in Remark \ref{rmk:last}, we rearrange
the sums containing terms $W\circ A_0^{i+1} - W\circ A_0^i$
in such a way that the new summands contain differences of more regular functions 
(see Lemmas~\ref{W-W} and \ref{W-Wbis} below).
We define, for $k \ge 0$,
\vspace{-.2cm}
\begin{subequations} \label{D}
\begin{align}
D_{\pp,1,k} (\theta,\psi) 
& := \int_0^1 \der t \, \partial_\theta(\pp\circ\SSS_1^{k}) \,
(G_2(\theta,\psi)+t \, \rho \, \zeta(\theta,\psi),A_0\psi) ,
\label{Dabis} \\
D_{\pp,2,k} (\theta,\psi) 
& :=\int_0^1 \der t \, (1-t) \, \partial^2_\theta( \pp\circ\SSS_1^{k}) \,
(G_2(\theta,\psi)+ t \, \rho \, \zeta(\theta,\psi ), A_0\psi) .
\label{Db} \\
R_{\pp,1,k}(\theta,\psi) & :=  \sum_{i=0}^{k} \left( \partial_\theta
\bigl( ( D_{\pp,1,k-i} \, \zeta ) \circ\SSS_2^{i} \bigr) \right) (\theta,\psi) ,
\label{R1} \\
R_{\pp,2,k}(\theta,\psi) & := \sum_{i=0}^{k} \left( \left( D_{\pp,2,k-i} \, \zeta^2 \right) \circ \SSS_2^{i} \right) (\theta,\psi) ,
\label{R2}
\end{align}
\end{subequations}
and expand, for $n\ge 1$,
\vspace{-.3cm}
\begin{equation} \label{pS1n-pS2n}
\pp\circ\SSS_1^n-\pp\circ\SSS_2^n 
= \rho \sum_{i=0}^{n-1} \bigl( D_{\pp,1,n-1-i} \, \zeta \bigr) \circ\SSS_2^{i} 
= \rho \sum_{i=0}^{n-1}
\bigl(  \partial_\theta(\pp\circ \SSS_1^{n-1-i})\circ \SSS_2 \, \zeta \bigr) \circ \SSS_2^{i} +
\rho^2 R_{\pp,2,n-1} .
\vspace{-.2cm}
\end{equation}
Using the first expansion in \eqref{pS1n-pS2n}, with $n-i-1$ instead of $n$, to write
\vspace{-.2cm}
\[
\partial_\theta(\pp\circ \SSS_1^{n-1-i}) =
\partial_\theta(\pp\circ \SSS_2^{n-1-i}) +  
\rho \sum_{i=0}^{n-1-i} \partial_\theta \bigl( \bigl( D_{\pp,1,n-1-i-j} \, \zeta \bigr) \circ\SSS_2^{j} \bigr) 
\vspace{-.2cm}
\]
in the second expansion, we obtain
\vspace{-.2cm}
\begin{equation} \label{pS1n-pS2n-first}
\pp\circ\SSS_1^n-\pp\circ\SSS_2^n =
\rho \!\sum_{i=0}^{n-1} \bigl(  \partial_\theta(\pp\circ \SSS_2^{n-1-i})\circ \SSS_2 \, \zeta \bigr) \circ \SSS_2^{i} +
\rho^2 \sum_{i=0}^{n-1} \left( R_{\pp,1,n-1-i} \circ \SSS_2 \, \zeta \right) \circ \SSS_2^{i} +\rho^2 R_{\pp,2,n-1} ,
\vspace{-.2cm}
\end{equation}
where, by Lemma~\ref{lempq} and Remark \ref{dopo}, for all $n\ge 1$ we bound
\vspace{-.1cm}
\begin{equation} \label{R2-bounds} 
\|R_{\pp,2,n-1}\|_\infty\leq C , \qquad \aver*{|R_{\pp,2,n-1}|}\leq C .
\vspace{-.1cm}
\end{equation}

We start with two abstract results, easy to check, in the spirit of some lemmas in Subsection \ref{674}.

\begin{lemma} \label{W-W}
Let $\{E_k\}_{k=0}^{n-1}$ be a set of functions $E_k\!:\Omega\to\Omega$, with $n\ge 1$.
Define, for $m=1,\ldots,n$,
\begin{equation} \label{DeltaE}
\Delta E_{m,0} = E_0 \circ \SSS_2^{-1} , \quad
\Delta E_{m,k} = E_{k} \circ \SSS_2^{-1} - E_{k-1} , \, k=1,\ldots,m-1 , \quad 
\Delta E_{m,m} = - E_{m-1} ,
\end{equation}
where $\Delta E_{m,k}$ with $0<k<m$ are meaningful only if $m\ge 2$.
For any $m$ as above one has
\vspace{-.2cm}
\begin{equation} \label{eq:canc0}
\sum_{i=0}^{m-1} E_{m-1-i} \circ \SSS_2^i \, (W \circ A_0^{i+1}-W \circ A_0^{i}) \\
= \sum_{i=0}^{m}
\Delta E_{m,m-i} \circ\SSS_2^{i} \, W \circ A_0^{i} .
\vspace{-.2cm}
\end{equation}
\end{lemma}

\begin{rmk} \label{rmk:DeltaE}
\emph{
If the functions $E_k$ in Lemma~\ref{W-W} are such that
\vspace{-.1cm}
\begin{equation} \label{hypE0}
\norm E_{k} \norm_{\al_0,3}^- \le (1-\rho\,\ga')^k \, \Gamma_0 , 
\quad 
\norm E_k \circ \SSS_2^{-1} - E_{k-1} \norm_{\al_0,3}^- \le \rho \, (1-\rho\,\ga')^k \, \Gamma_0 , 
\qquad k=1,\ldots,n-1 ,
\end{equation}
for some constant $\Gamma_0$, then one has
%
$\sum_{i=0}^{n} \norm \Delta E_{n,n-i} \norm_{\al_0,3}^- \le C \, \Gamma_0$.
%
Analogously, if
\vspace{-.1cm}
\begin{equation} \label{hypE1}
\| E_k \|_{\io} \le (1-\rho\,\ga')^k \, \Gamma_0 , \quad 
\| E_k \circ \SSS_2^{-1} - E_{k-1} \|_{\io} \le \rho \, (1-\rho\,\ga')^k \, \Gamma_0 , \qquad k=1,\ldots,n-1 ,
\vspace{-.1cm}
\end{equation}
one has
%
$\sum_{i=0}^{n} \| \Delta E_{n,n-i} \|_{\io} \le C \, \Gamma_0$.
}
\end{rmk}


\begin{lemma} \label{W-Wbis}
Let $\{E_k\}_{k=0}^{n-1}$ and, for $1\le m\le n$, $\{\Delta E_{m,k}\}_{k=1}^m$ be as in Lemma~\ref{W-W},
and let $\{F_k\}_{k=0}^{n}$ be a set of functions $F_k\!:\Omega\to\Omega$.
For any $m$ as above one has 
\vspace{-.2cm}
\begin{equation*}
\begin{aligned}
& \sum_{i=0}^{m-1} \! E_{m-1-i} \circ \SSS_2^i \, F_i  (W \circ A_0^{i+1} \!-\! W \circ A_0^{i})
\!=\! \sum_{i=0}^{m} \Delta E_{m,m-i} \circ\SSS_2^{i} \, F_i W \circ A_0^{i} +
\sum_{i=1}^{m} \! E_{m-i} \circ\SSS_2^{i-1} \bigl( F_{i-1} \!-\! F_i \bigr) W \circ A_0^{i} .
\end{aligned}
\end{equation*}
\end{lemma}

Next, for $r=0,1,2$, we introduce the functions
\vspace{-.1cm}
\begin{subequations} \label{E0+Q}
\begin{align}
E_{\pp,r,k} :=\partial_\theta (\pp \circ \SSS_2^{k}) \circ \SSS_2 \, c_r , \qquad 
Q_{\pp,r,k} :=  \partial_\theta \bigl( D_{\pp,1,k} \, c_r \bigr) , \qquad
Q_{r,k} :=\partial_\theta (\SSS_2^k)_\theta \circ \SSS_2 \, c_r,  
\label{EzEc}
\end{align}
\end{subequations}
and define $\Delta E_{\pp,0,m,k}$, $\Delta Q_{\pp,r,m,k}$ and $\Delta Q_{r,m,k}$
as in \eqref{DeltaE} with $E_k\!=\!E_{\pp,0,k}$, $E_{k}\!=\!Q_{\pp,r,k}$ and $E_k\!=\!Q_{r,k}$, respectively.
Set also $\Delta_2 Q_{0,m}\!:=\!\Delta Q_{0,m}  \!-\! \Delta Q_{0,m+1}$ and
$\Delta_2 Q_{0,m,k}\!:=\!\Delta Q_{0,m,k} \!-\! \Delta Q_{0,m+1,k+1}$.


Using that $\partial_\theta (\pp\circ\SSS_2^k) = \partial_\theta \pp \circ\SSS_2^k \partial_\theta (\SSS_2^k)_\theta$
and the bounds \eqref{bounds-pna} in Lemma~\ref{pna}, we may bound
\vspace{-.1cm}
\[
\begin{aligned}
& \norm E_{\pp,0,k} \circ\SSS_2^{-1} - E_{\pp,0,k-1} \norm_{\al_0,3} ^-
\le \norm\partial_\theta (\pp \circ \SSS_2^{k})\norm_{\al_0,3}^- \norm c_0\circ \SSS_2^{-1} - c_0\norm_{\al_0,3} \\
& \quad + \norm\partial_\theta (\pp \circ \SSS_2^{k}) \circ\SSS_2^{-1}  - \partial_\theta (\pp \circ \SSS_2^{k-1}) \norm_{\al_0,3}^-
\norm c_0\norm_{\al_0,3}
\le \norm\partial_\theta \pp \norm_{\al_0,3} \norm \partial_\theta (\SSS_2^{k})_\theta \norm_{\al_0,3}^- 
\norm c_0\circ \SSS_2^{-1} - c_0\norm_{\al_0,3} \\
& \quad +
\norm\partial_\theta \pp \norm_{\al_0,3} \norm \partial_\theta (\SSS_2^{k-1})_\theta \norm_{\al_0,3}^-
\norm \partial_\theta (\SSS_2)_\theta - \uno \norm_{\al_0,3}^- \norm c_0\norm_{\al_0,3}
\le C \rho \, (1-\rho\,\ga')^k \, \norm \partial_\theta\pp\norm_{\al_0,3} , \\
%
& \| \partial_\theta ( \SSS_2^{k})_\theta \, c_r \circ \SSS_2^{-1} - 
\partial_\theta ( \SSS_2^{k})_\theta \circ \SSS_2  \, c_r \|_{\io} \\
& \quad \le
\| \partial_\theta ( \SSS_2^{k})_\theta  \|_{\io} 
\| c_r \circ (\SSS_2^{-1})_\theta - c_r  \|_{\io}  +
\| \partial_\theta^2 ( \SSS_2^{k})_\theta  \|_{\io} 
\| (\SSS_2)_\theta - \uno  \|_{\io}  \| \, c_r \|_{\io}
\le C\rho(1-\rho\,\ga')^k , \\
%
%
& \| \partial_\theta \bigl( D_{\pp,1,k} \, c_r \bigr) \circ \SSS_2^{-1} -
\partial_\theta \bigl( D_{\pp,1,k-1} \, c_r \bigr)  \|_{\io}  \\
& \quad \le
\| \partial_\theta \bigl( D_{\pp,1,k} \, c_r \bigr)  \|_{\io}  \| \circ \SSS_2^{-1} \!-\! \uno  \|_{\io} 
+
\| \partial_\theta \bigl( D_{\pp,1,k} \, c_r \bigr) \!-\! \partial_\theta \bigl( D_{\pp,1,k-1} \, c_r \bigr)  \|_{\io}
\le C \rho \, (1\!-\!\rho\,\ga')^{k+j} \norm \partial_\theta\pp\norm_{\al_0,3} ,
\vspace{-.2cm}
\end{aligned}
\]
so that in \eqref{E0+Q} the functions $E_{\pp,0,k}$ satisfy the bounds \eqref{hypE0},
with $\Gamma_0=C\norm \partial_\theta\pp\norm_{\al_0,3}$, while
the functions $Q_{r,k}$ and $Q_{\pp,r,k}$, for $r=0,1,2$,
satisfy the bounds \eqref{hypE1}, with $\Gamma_0=C$ and $\Gamma_0=C \norm \partial_\theta\pp\norm_{\al_0,3}$, respectively.
Thus, for $r=0,1$, $m\ge 2$ and $k=1,\ldots,m-1$, we obtain
\vspace{-.2cm}
\begin{subequations} 
\begin{align}
& \sum_{i=0}^{n} \norm \Delta E_{\pp,0,n,n-i} \norm_{\al_0,3}^- \le C \, \norm \partial_\theta\pp\norm_{\al_0,3} , 
\label{estimate1} \\
%
%
%
& \| \Delta Q_{r,m,k} \|_{\io}  \le C \rho(1-\rho\,\ga')^k
\qquad
\| \Delta Q_{\pp,r,m,k} \|_{\io} \le C \rho (1-\rho\,\ga')^k \, \norm \partial_\theta\pp\norm_{\al_0,3} .
\label{DeltaQ}
\end{align}
\vspace{-.3cm}
\end{subequations}
%


Lets us come back to \eqref{pS1n-pS2n-first}.
Recalling \eqref{eq:zeta} and \eqref{fWW}, we write 
$\left(\partial_\theta(\pp\circ\SSS_2^{n-1-i})\circ \SSS_2 \, \zeta\right)\circ \SSS_2^i$ as
%
\vspace{-.3cm}
\begin{equation} \label{bohboh}
\rho^{-1} E_{\pp,0,n-1-i} \circ \SSS_2^i \, (W \circ A_0^{i+1}-W \circ A_0^{i}) 
+ \sum_{r=1}^2 E_{\pp,r,n-1-i}  \circ \SSS_2^i \, \left( W \circ A_0^i \right)^r ,
\vspace{-.3cm}
\end{equation}
which, thanks to Lemma~\ref{W-W}, allows to write the first sum in \eqref{pS1n-pS2n-first} as
\vspace{-.1cm}
\begin{equation} \label{eq:canc2}
\sum_{i=0}^{n}
\Delta E_{\pp,0,n,n-i} 
\circ\SSS_2^{i} \, W \circ A_0^{i} +
\rho \sum_{i=1}^{n-1} \sum_{r=1}^2 E_{\pp,r,n-1-i}  \circ \SSS_2^i \, \left( W \circ A_0^i \right)^r .
\vspace{-.1cm}
\end{equation}
%
The second sum on the r.h.s.~of \eqref{pS1n-pS2n-first}, in terms of the functions in \eqref{EzEc}, becomes
\vspace{-.2cm}
\begin{equation*}
\rho^2 \! \sum_{i=0}^{n-1} \! \left( R_{\pp,1,n-1-i} \circ \SSS_2 \, \zeta \right) \! \circ \SSS_2^i
= \rho^2 \! \sum_{i=0}^{n-1} \sum_{j=0}^{i} 
\bigl( \partial_\theta \bigl( \! \bigl( D_{\pp,1,n-1-i} \, \zeta \bigr) \circ\SSS_2^{i-j} \bigr) \circ \SSS_2 \, \zeta \bigr) \circ \SSS_2^j =
\rho^2 \mathscr{A}_{n,2} + \rho \mathscr{A}_{n,1} + \mathscr{A}_{n,0} ,
\vspace{-.1cm}
\end{equation*}
with
\vspace{-.4cm}
\begin{equation*} 
\begin{aligned}
\mathscr{A}_2 & := \sum_{i=0}^{n-1} \sum_{j=0}^i \sum_{r,s=1}^2 Q_{\pp,r,n-1-i} \circ \SSS_2^{i+1} Q_{s,i-j} \circ \SSS_2^j \,
\bigl( W \circ A_0^{i+1} \bigr)^{r} \bigl( W \circ A_0^{j} \bigr)^{s} \\
\mathscr{A}_1& := \sum_{i=0}^{n-1} \sum_{j=0}^{i+1} \sum_{r=1}^2 Q_{\pp,r,n-1-i} \circ \SSS_2^{i+1} 
\Delta Q_{0,i+1,i+1-j} \circ \SSS_2^j \, \bigl( W \circ A_0^{i+1} \bigr)^r W \circ A_0^j \\
& + \sum_{j=0}^{n-1} \sum_{i=0}^{n-j} \sum_{s=1}^2 
\bigl( \bigl( \Delta Q_{\pp,0,n-j,n-j-i} \circ \SSS_2^{i+1} Q_{s,i} + Q_{\pp,0,n-j-i} \circ \SSS_2^{i} \Delta_2 Q_{s,i-1} 
\bigr) W \circ A_0^{i+1} W^s \bigr) \circ \SSS_2^j \\ 
\mathscr{A}_0 & := \sum_{j=0}^{n-1} \sum_{i=0}^{n-j} 
\bigl( \bigl( \Delta Q_{\pp,0,n-j,n-j-i} \circ \SSS_2^{i+1} \Delta Q_{0,i+1,i+1} 
+ Q_{\pp,0,n-j-i} \circ \SSS_2^i \Delta_2 Q_{0,i,i}
\bigr) W \circ A_0^{i+1} W \bigr) \circ \SSS_2^{j} \\
& + \sum_{i=0}^{n-1} 
Q_{\pp,0,n-1-i} \circ \SSS_2^{i+1} \Delta Q_{0,i+1,0} \circ \SSS_2^{i+1} \bigl( W \circ A_0^{i+2} - W\circ A_0^{i+1} \bigr) W\circ A_0^{i+1} ,
\end{aligned}
\end{equation*}
where the terms with negative integer labels are meant as 0
and we have used Lemma~\ref{W-W} and Lemma~\ref{W-Wbis} to obtain the first and second line of $\mathscr{A}_1$,
respectively, and both Lemmas~\ref{W-W} and \ref{W-Wbis} to obtain $\mathscr{A}_0$.

\begin{rmk} \label{rmk:remark6}
\emph{
To bound the contributions to $\mathscr{A}_0$ with $\Delta_2 Q_{0,i,i}$ 
we use that, for $m > k\ge0$, we have
$\|\Delta_2 Q_{0,m,k} \|_\io
\!=\! \|\partial_\theta(\SSS_2^{k-1})_\theta \left( \uno \!-\! \partial_\theta(\SSS_2)_\theta \right)
\left( \left( \partial_\theta( \SSS_2)_\theta \!-\! \uno \right) c_0 \circ \SSS_2^{-1} + \left( c_0 \circ \SSS_2^{-1} \!-\! c_0 \right) \right) \|_\io
\!\le\! C\rho^2 (1-\rho\,\ga')^{k}$.
}
\end{rmk}

The bounds \eqref{bounds-pna}, 
the second bound in \eqref{Wflu}, 
the bounds \eqref{DeltaQ} and 
Remark 
\ref{rmk:remark6} yield that
\vspace{-.1cm}
\begin{equation}
\label{estimate2}
\| \rho^2 \mathscr{A}_2 + \rho \mathscr{A}_1 + \mathscr{A}_0 \| \le C \, \rho^{-2} \norm \partial_\theta\pp\norm_{\al_0,3}, 
\qquad
\aver{ | \rho^2 \mathscr{A}_2 + \rho \mathscr{A}_1 + \mathscr{A}_0 | }
\le C \rho^{-1} \norm \partial_\theta\pp\norm_{\al_0,3} .
\vspace{-.1cm}
\end{equation}
In conclusion, if we set $C_{\pp,n,0} = \Delta E_{\pp,1,n,0}$ and
$C_{\pp,n,k} = E_{\pp,1,k-1} + \rho^{-1} \Delta E_{\pp,0,n,k}$ for $k=1,\ldots,n$, i.e.
\vspace{-.1cm}
\begin{equation*} 
\begin{aligned}
& \null\hspace{-.2cm} 
C_{\pp,n,0} :=  \rho^{-1} \partial_\theta \pp \, c_0 \circ \SSS_2^{-1} , \\ 
& \null\hspace{-.2cm} 
C_{\pp, n,k} := \partial_\theta(\pp\circ\SSS_2^{k-1})\circ \SSS_2 \, c_1 
\!+\! \rho^{-1} \bigl( \partial_\theta (\pp \circ\SSS_2^{k}) \, 
c_0 \circ \SSS_2^{-1} \!-\! \partial_\theta (\pp \circ\SSS_2^{k-1}) \circ \SSS_2 \, c_0 \bigr) , 
\quad k=1,\ldots,n \!-\! 1, \\
& \null\hspace{-.2cm} 
C_{\pp,n,n} := \partial_\theta (\pp \circ\SSS_2^{n-1}) \circ \SSS_2 \, \ccc_1
- \rho^{-1} \partial_\theta (\pp \circ\SSS_2^{n-1}) \, c_0 \circ \SSS_2^{-1}
\end{aligned}
\vspace{-.1cm}
\end{equation*}
and define
\vspace{-.1cm}
\begin{equation} \label{Rpn}
\begin{aligned}
\!\!\! R_ {\pp,n} := 
\sum_{i=0}^{n-1} \Bigl( \partial_\theta (\pp \circ \SSS_2^{n-1-i}) \circ \SSS_2 \, c_2 \Bigr) \circ \SSS_2^i \, W^2 \circ A_0^i 
+ \rho\sum_{i=0}^{n-1} \left( R_{\pp,1,n-1-i} \circ \SSS_2 \, \zeta \right) \circ \SSS_2^i + \rho \,
R_{\pp,2,n-1} 
\end{aligned}
\vspace{-.1cm}
\end{equation}
we obtain \eqref{pS1-pS2}, with the bounds \eqref{Ck} and \eqref{sumCk} following from the estimates \eqref{bounds-pna}
and \eqref{estimate1}, while the bound \eqref{R} follows from the bounds \eqref{R2-bounds} and \eqref{estimate2}.


\subsubsection{Proof of Proposition \ref{prop:fluhder2}} \label{proof:fluhder2}

We have
%
%
which, together with \eqref{derp2q2a}, allows us to write the first contribution in \eqref{h1h2hbarder1} as
\vspace{-.2cm}
\begin{equation} \label{derh1-derh2}
\begin{aligned}
\partial_\theta h_1 - \partial_\theta h_2
& = \sum_{n=0}^\io \Bigl( 
( \partial_\theta p_1^{(n)} -  \partial_\theta p_2^{(n)}) \, (q_1 \circ \SSS_1^n -  q_2 \circ \SSS_2^n ) 
 + 
 ( \partial_\theta p_1^{(n)} -  \partial_\theta p_2^{(n)}) \,  q_2 \circ \SSS_2^n
 \\ & + 
\partial_\theta p_2^{(n)} \, (q_1 \circ \SSS_1^n - q_2 \circ \SSS_2^n )
+ 
( p_1^{(n)} - p_2^{(n)} ) \, (\partial_\theta  q_1 \circ \SSS_1^n  - \partial_\theta  q_2 \circ \SSS_2^n ) \\
& + 
( p_1^{(n)} -  p_2^{(n)} ) \, \partial_\theta  q_2 \circ \SSS_2^n  +
p_2^{(n)} \, (\partial_\theta  q_1 \circ \SSS_1^n  \!\!-\! \partial_\theta  q_2 \circ \SSS_2^n ) \Bigr) . \phantom{\sum^\io_n}
\end{aligned}
\vspace{-.4cm}
\end{equation}
%
where we expand
%
%
%
\vspace{-.1cm}
\begin{subequations} \label{omnia}
\begin{align}
& \null\hspace{-0.6cm}
p_1^{(n)} \! - p_2^{(n)}
= \sum_{k=0}^{n-1} p_1^{(k)} \bigl( p_1\circ\SSS_1^k  - p_2\circ\SSS_2^k \bigr) \, p_2^{(n-k-1)} \circ \SSS_2^{k+1} ,
\label{derp1np2n-a} \\
& \null\hspace{-0.6cm} 
\partial_\theta p_1^{(n)} \!\!-\! \partial_\theta p_2^{(n)}
= 
\sum_{k=0}^{n-1}
\sum_{i=0}^{k-1} \pi_1^{(i)} \Delta_{\pi_1,\pi_2,W,i} \pi_2^{(k-i-1)} \circ \SSS_2^{i+1}
\partial_\theta p_2 \circ \SSS_2^k \, p_2^{(n-k-1)} \circ\SSS_2^{k+1}
\nonumber \\
& \null\hspace{0.4cm}
+ \sum_{k=0}^{n-1}
\sum_{i=0}^{k-1} \pi_1^{(i)} \rho  \gota_3 \circ\SSS_1^i \gotf \circ A_0^i  \pi_2^{(k-i-1)} \circ \SSS_2^{i+1}
\partial_\theta p_2 \circ \SSS_2^k \, p_2^{(n-k-1)} \circ\SSS_2^{k+1}
\nonumber \\
& \null\hspace{0.4cm} 
+
\sum_{k=0}^{n-1} \pi_1^{(k)}
\Delta_{\partial_\theta p_1, \partial_\theta p_2,W,k}
p_2^{(n-k-1)} \circ\SSS_2^{k+1} 
+
\sum_{k=0}^{n-1} \pi_1^{(k)}
\rho \gota_4 \circ \SSS_1^k \, \gotf \circ A_0^k
p_2^{(n-k-1)} \circ\SSS_2^{k+1} 
\label{derp1np2n-b} \\
& \null\hspace{0.4cm} 
+
\sum_{k=0}^{n-1} \!\pi_1^{(k)}  \partial_\theta p_1 \circ \SSS_1^k \!\!\!
\sum_{i=0}^{n-k-2} \!\!\! p_1^{(i)} \circ \SSS_1^{k+1}
\Delta_{p_1,p_2,W,i+k+1} p_2^{(n-k-2-i)} \circ \SSS_2^{i+k+2} ,
\nonumber \\
& \null\hspace{0.4cm}
+
\sum_{k=0}^{n-1} \!\pi_1^{(k)}  \partial_\theta p_1 \circ \SSS_1^k \!\!\!
\sum_{i=0}^{n-k-2} \!\!\! p_1^{(i)} \circ \SSS_1^{k+1}
\rho \gota_1 \circ\SSS_1^{i+k+1} \gotf \circ A_0^{i+k+1} p_2^{(n-k-2-i)} \circ \SSS_2^{i+k+2} ,
\nonumber \\
& \null\hspace{-0.6cm}
\partial_\theta q_1\circ \SSS_1^n \!\! - \! \partial_\theta q_2\circ \SSS_2^n  = 
\Delta_{\partial_\theta q_1, \partial_\theta q_2,W,n} + \rho \, \gota_5 \circ \SSS_1^n \, \gotf \circ A_0^n , 
\label{derp1np2n-4}
\end{align}
\end{subequations}
%
%
%
%
%
%
with $\gota_3(\theta) :=\gota_1(\theta)+\partial_\theta c_0(\theta)$,
$\gota_4(\theta) :=\partial_\theta\gota_1(\theta)$ and $\gota_5(\theta)=\partial_\theta \gota_2 (\theta)$.
Finally, in \eqref{derh1-derh2} and \eqref{derp1np2n-a} 
we express $q_1 \circ \SSS_1^ n  - q_2 \circ \SSS_2^n$ and $p_1 \circ \SSS_1^k  - p_2 \circ \SSS_2^k$, respectively,
according to \eqref{p1p2q1q2}, while in \eqref{derp1np2n-b} 
we apply Lemma \ref{lem:new} to the summands containing the functions $\gota_1$, $\gota_3$ and $\gota_4$.

\begin{rmk} \label{rmk:exshortenh2}
\emph{
Bounds analogous to \eqref{shorten-inf} hold also for 
$\Delta_{\pi_1,\pi_2,W,n}$,
$\Delta_{\partial_\theta p_1, \partial_\theta p_2,W,n}$ and
$\Delta_{\partial_\theta q_1, \partial_\theta q_2,W,n}$. Thus, for any
$\Delta,\Delta' \!\in\!\{\Delta_{p_2,W,n},\Delta_{q_2,W,n},
\Delta_{p_1,p_2,W,n},
\Delta_{q_1,q_2,W,n},
\Delta_{\pi_1,\pi_2,W,n},
\Delta_{\partial_\theta p_1, \partial_\theta p_2,W,n},
\Delta_{\partial_\theta q_1, \partial_\theta q_2,W,n} \}$
and for any $k\ge0 $, one has
$\norm \Delta \norm_{\al_0,3}^- \le C \rho$, 
$\aver{ | \Delta \, W \circ A_0^k | } \le C \rho^2$ and 
$\aver{ | \Delta \Delta ' | } \le C\rho^3$.
}
\end{rmk}

By collecting together the expansions arising from \eqref{derh1-derh2} and \eqref{omnia}, 
we write $\partial_\theta h_1 - \partial_\theta h_2$ 
as a sum of 28 contributions, which, despite being long and tedious to estimate,
can all be treated along similar lines. For example, the sum of the contributions
$( p_1^{(n)} - p_2^{(n)} ) \, (\partial_\theta  q_1 \circ \SSS_1^n  - \partial_\theta  q_2 \circ \SSS_2^n )$
in the second line of \eqref{derh1-derh2}, by taking into account \eqref{derp1np2n-a}, \eqref{p1p2q1q2} and \eqref{derp1np2n-4}, leads to
\[
\begin{aligned}
& \sum_{n=0}^{\io} \sum_{k=0}^{n-1} p_1^{(k)}
\, \Delta_{p_1,p_2,W,k}
\, p_2^{(n-k-1)} \circ \SSS_2^{k+1} \,
\Delta_{\partial_\theta q_1, \partial_\theta q_2,W,n} \\
& \quad
+ \sum_{k=0}^{\io} \sum_{n=0}^{\io}
p_1^{(k-1)} \bigl( \gotD_{\gota_1,k,0} \circ \SSS_1^k \, p_2^{(n)} \circ \SSS_2^{k+1} \Delta_{\partial_\theta q_1, \partial_\theta q_2,W,n+k+1} 
\!+\! \de_{k,n} \gota_1 \circ \SSS_1^{n-1} \Delta_{\partial_\theta q_1, \partial_\theta q_2,W,n} \bigr) W \circ A_0^k \\
%
%
& \quad
+ \sum_{k=0}^{\io} \sum_{n=0}^{\io}
p_1^{(k)} \, \Delta_{p_1,p_2,W,k} \, p_2^{(n)} \circ \SSS_2^{k+1} \, \rho \, \gota_5 \circ \SSS_1^{n+k+1} \, \gotf \circ A_0^{n+k+1} \\
& \quad
+ \sum_{k=0}^{\io} \sum_{n=0}^{\io} 
p_1^{(k-1)} \rho \bigl( \gotD_{\gota_1,k,0} \circ \SSS_1^k p_2^{(n)} \circ \SSS_2^{k+1} \gota_5 \circ \SSS_1^{n+k+1}\gotf \circ A_0^{n+k+1} 
\!+\! \de_{k,n} \gota_1 \circ \SSS_1^{n-1} \gota_5 \circ \SSS_1^{n} \gotf \circ A_0^{n} \bigr) W \circ A_0^n
\end{aligned}
\]
where $\de_{k,n}$is  the Kronecker delta. The average of all contributions is bounded by $C\rho$:
in the first two lines, directly because of Remark \ref{rmk:exshortenh2} and \eqref{deriv}; 
in the third line, because of Remark \ref{rmk:exshortenh2}, \eqref{new-bound} and \eqref{deriv},
after writing $p_2^{(n)} \circ\SSS_2^{k+1} =
p_2^{(n)} \circ\SSS_1^{k+1}+(p_2^{(n)} \circ\SSS_2^{k+1}-p_2^{(n)} \circ\SSS_1^{k+1} )$,
expanding $p_2^{(n)} \circ\SSS_2^{k+1}-p_2^{(n)} \circ\SSS_1^{k+1}$ as in \eqref{pnS1pnS2}
to apply Lemma~\ref{S1-S2} and using Lemma~\ref{lem:new1} to obtain
a factor $\gotD_{1,\gota_5,n}\circ\SSS_1^n$ from the terms with $p_2^{(n)} \circ\SSS_1^{k+1}$;
in the fourth line, because of \eqref{Wf=WW} for the terms where only the sum over $n$ appears and
once more because of Remark \ref{rmk:exshortenh2} for all the other terms,
after decomposing $\gota_5 \circ \SSS_1^{n+k+1} = \gota_5 \circ \SSS_2^{n+k+1}+ ( \gota_5 \circ \SSS_1^{n+k+1}-\gota_5 \circ \SSS_2^{n+k+1})$
and applying Lemma~\ref{lem:new1} to obtain a factor $\gotD_{5,\gota_2,n+k+1}$ from the terms containing $\gota_5 \circ \SSS_2^{n+k+1}$
and Lemma~\ref{S1-S2} to extract a further function $W$ from the terms with the difference.
All the other contributions can be dealt with in a similar way and eventually a bound $C\rho$ is found for all of them,
so that the first bound in \eqref{fluhder2} follows.

Next, consider $(\partial_\theta h_1 - \partial_\theta h_2)\,(\partial_\theta h_1 - \partial_\theta h_2)$
and write both differences as in \eqref{derh1-derh2}.
Then, we observe that all the contributions obtained from \eqref{derh1-derh2} either
are $O(1)$ in $\rho$ or contain at least a function $W$.
This yields that
%
$| \average{ \AAA_i \AAA_j}| \le C \rho$ for all $i,j =1,\ldots ,8$,
so that the second bound in \eqref{fluhder2} follows as well.

\subsection{Second derivative of the conjugation:~proof of Lemma~\ref{lem:fluhderder}} \label{proof:fluhderder}

The proof of Lemma~\ref{lem:fluhderder} follows the same scheme as Propositions~\ref{prop:hh2} and \ref{prop:fluhder2}.
In particular, we prove that
%
$|\aver{ (\partial_\theta^2 h_2(\theta,\cdot)  - \partial_\theta^2 \ovl h(\theta) )^2 } | \leq C\rho$
and
$\aver{(\partial_\theta^2 h_1(\theta,\cdot) - \partial_\theta^2 h_2(\theta,\cdot))^2} \leq C\rho$.
%
To this end, we write
\vspace{-.1cm}
\begin{equation*}
\begin{aligned}
\partial_\theta^2 ( p_2^{(n)} \, q_2 \circ\SSS_2^n) - \partial_\theta^2 (\ovl p^{(n)} \, \ovl q \circ \ovl G^n) & =
\partial_\theta^2 p_2^{(n)} \, q_2 \circ \SSS_2^n -  \partial_\theta^2 \ovl p^{(n)} \, \ovl q \circ \ovl G^n  \\
& \hspace{-2cm} +
2 \partial_\theta p_2^{(n)} \, \partial_\theta  q_2 \circ \SSS_2^n  -  2 \partial_\theta \ovl p^{(n)} \, \partial_\theta  \ovl q \circ \ovl G^n 
+ p_2^{(n)} \, \partial_\theta^2  q_2 \circ \SSS_2^n  -  \ovl p^{(n)} \, \partial_\theta^2  \ovl q \circ \ovl G^n , \\
\partial_\theta^2 (p_1^{(n)} \, q_1 \circ \SSS_1^{n}) - \partial_\theta^2 (p_2^{(n)} \, q_2 \circ\SSS_2^n)
& =
\partial_\theta^2 p_1^{(n)} \, q_1 \circ \SSS_1^n -  \partial_\theta^2 p_2^{(n)} \, q_2 \circ \SSS_2^n  \\
& \hspace{-2cm} +
2 \partial_\theta p_1^{(n)} \, \partial_\theta  q_1 \circ \SSS_1^n  -  2 \partial_\theta p_2^{(n)} \, \partial_\theta  q_2 \circ \SSS_2^n 
+ p_1^{(n)} \, \partial_\theta^2  q_1 \circ \SSS_1^n  -  p_2^{(n)} \, \partial_\theta^2  q_2 \circ \SSS_2^n ,
\end{aligned}
\vspace{-.2cm}
\end{equation*}
so that the terms in the second and fourth lines can be studied as in Appendices~\ref{proof:fluhder1} and \ref{proof:fluhder2},
with $q_1$, $q_2$ and $\ovl q$ replaced, respectively, either with
$\partial_\theta q_1$, $\partial_\theta q_2$ and $\partial_\theta \ovl q$ or 
$\partial_\theta^2 q_1$, $\partial_\theta^2 q_2$ and $\partial_\theta^2 \ovl q$.
The terms in the first and third lines can be dealt with by first writing, according to \eqref{deripn2},
\vspace{-.2cm}
\begin{equation} \label{p2nderder}
\begin{aligned}
\partial_\theta^2 & p_2^{(n)} =
\sum_{k=0}^{n-1} ( p_2 \, (\partial_\theta G_2)^2)^{(k)} \, \partial_\theta^2 p_2 \circ \SSS_2^k \, p_2^{(n-k-1)} \circ \SSS_2^{k+1} \\
& + \sum_{k=0}^{n-1} \sum_{i=0}^{k-1} ( p_2 \, (\partial_\theta G_2)^2)^{(i)} \, (\partial_\theta p_2 \partial_\theta G_2)\circ \SSS_2^i \, 
(p_2\partial_\theta G_2)^{(k-i-1)}\circ\SSS_2^{i+1} 
\partial_\theta p_2 \circ \SSS_2^k \, p_2^{(n-k-1)} \circ \SSS_2^{k+1} \\
& + \sum_{k=0}^{n-1} \sum_{i=0}^{k-1} ( p_2 \, (\partial_\theta G_2)^2)^{(i)} \, (p_2 \, \partial_\theta^2 G_2)\circ\SSS_2^i \,
(p_2\partial_\theta G_2)^{(k-i-1)}\circ\SSS_2^{i+1} 
\partial_\theta p_2 \circ \SSS_2^k \, p_2^{(n-k-1)} \circ \SSS_2^{k+1} \\
& + \sum_{k=0}^{n-1} \sum_{i=0}^{n-k-2} ( p_2 \, \partial_\theta G_2)^{(k)} \, \partial_\theta p_2 \circ \SSS_2^k \, 
(p_2 \partial_\theta G_2)^{(i)}\circ\SSS_2^{k+1} 
\partial_\theta p_2 \circ \SSS_2^{k+1+i} \, p_2^{(n-k-i-2)} \circ \SSS_2^{k+2+i} ,
\end{aligned}
\vspace{-.2cm}
\end{equation}
and analogous expressions for $\partial_\theta^2 \ovl p^{(n)}$ and $\partial_\theta^2 p_1^{(n)}$,
with $\ovl \SSS$ and $\SSS_1$ instead of $\SSS_2$, respectively, and then expanding
$\partial_\theta^2 p_2^{(n)} \, q_2 \circ \SSS_2^n -  \partial_\theta^2 \ovl p^{(n)} \, \ovl q \circ \ovl G^n$
and
$\partial_\theta^2 p_1^{(n)} \, q_1 \circ \SSS_1^n -  \partial_\theta^2 p_2^{(n)} \, q_2 \circ \SSS_2^n$
by proceeding as in Appendices~\ref{proof:fluhder1} and \ref{proof:fluhder2}, with respect to which
besides the functions $\pi_1 = p_1 \partial_\theta G_1$, $\pi_2 = p_2 \partial_\theta G_2$ and
$\ovl p \, \partial_\theta \ovl G$, the functions
$p_1 \partial_\theta^2 G_1$, $\partial_\theta p_1 \partial_\theta G_1$, $p_1 (\partial_\theta G_1)^2$,
$p_2 \partial_\theta^2 G_2$, $\partial_\theta p_2 \partial_\theta G_2$, $p_2 (\partial_\theta G_2)^2$
$\ovl p \, \partial_\theta^2 \ovl G$, $\partial_\theta \ovl p \, \partial_\theta \ovl G$ and $\ovl p \, (\partial_\theta \ovl G)^2$ also appear. 
So, when considering $\partial_\theta^2 p_1^{(n)} \, q_1 \circ \SSS_1^n - \partial_\theta^2 p_2^{(n)} \, q_2 \circ \SSS_2^n$,
together with the differences in \eqref{omnia}, we have to expand also the differences
$(p_1 \partial_\theta^2 G_1)^{(n)}-(p_2 \partial_\theta^2 G_2)^{(n)}$, 
$(\partial_\theta p_1 \partial_\theta G_1)^{(n)} \!-\! (\partial_\theta p_2 \partial_\theta G_2)^{(n)}$ and 
$(p_1 (\partial_\theta G_1)^2)^{(n)} \!-\! (p_2 (\partial_\theta G_2)^2)^{(n)}$,
while, inserting \eqref{p2nderder} and the analogous expression for $\partial_\theta^2 \ovl p^{(n)}$ into
$\partial_\theta^2 p_2^{(n)} \, q_2 \circ \SSS_2^n \!-\!  \partial_\theta^2 \ovl p^{(n)} \, \ovl q \circ \ovl G^n$,
we obtain a sum of contributions with the same structure as \eqref{tobebounded}.

Comparing \eqref{deripn2} with \eqref{deripn1} we note that,
because of the presence of an extra derivative with respect to $\theta$, a further sum may appear.
However, when this happens, the two derivatives act on two distinct factors; this produces a factor $\rho$,
which compensates the factor $1/\rho$ arising from the sum.



\end{document}